\documentclass[aps,prmaterials,twocolumn,superscriptaddress,longbibliography]{revtex4-2}
\usepackage{amsmath,amssymb}
\usepackage[pdftex]{hyperref,graphicx}
\hypersetup{colorlinks = true, urlcolor = blue, linkcolor = blue, citecolor = blue}
\usepackage{physics}
\usepackage{xcolor}
\usepackage{bm}
\usepackage[section]{placeins}

\newcommand{\comment}[1]{{}}

\begin{document}

\title{Estimating disorder and its adverse effects in semiconductor Majorana nanowires}

\author{Seongjin Ahn}
\thanks{These authors contributed equally to this work.}
\affiliation{Condensed Matter Theory Center and Joint Quantum Institute, Department of Physics, University of Maryland, College Park, Maryland 20742, USA}

\author{Haining Pan}
\thanks{These authors contributed equally to this work.}
\affiliation{Condensed Matter Theory Center and Joint Quantum Institute, Department of Physics, University of Maryland, College Park, Maryland 20742, USA}

\author{Benjamin Woods}
\thanks{These authors contributed equally to this work.}
\affiliation{Department of Physics and Astronomy, West Virginia University, Morgantown, West Virginia 26506, USA}

\author{Tudor D. Stanescu}
\affiliation{Department of Physics and Astronomy, West Virginia University, Morgantown, West Virginia 26506, USA}

\author{Sankar Das Sarma}
\affiliation{Condensed Matter Theory Center and Joint Quantum Institute, Department of Physics, University of Maryland, College Park, Maryland 20742, USA}

\begin{abstract}
We use the available transport measurements in the literature to develop a dataset for the likely amount of disorder in semiconductor (InAs and InSb) materials which are used in fabricating the superconductor-semiconductor nanowire samples in the experimental search for Majorana zero modes.  Using the estimated disorder in direct Majorana simulations, we conclude that the current level of disorder in semiconductor Majorana nanowires is at least an order of magnitude higher than that necessary for the emergence of topological Majorana zero modes.  In agreement with existing results, we find that our estimated disorder leads to the occasional emergence of trivial zero modes, which can be post-selected and then further fine-tuned by varying system parameters (e.g., tunnel barrier), leading to trivial zero-bias conductance peaks in tunneling spectroscopy with $ \sim 2e^2/h $  magnitude.  Most calculated tunnel spectra in these disordered systems, however, manifest essentially no significant features, which is also consistent with the current experimental status, where zero-bias peaks are found only occasionally in some samples under careful fine-tuning.
\end{abstract}

\maketitle

\section{Introduction}\label{sec:introduction}

Majorana zero modes (MZMs) are neutral zero energy defect-bound localized excitations emerging in one- and two-dimensional condensed matter systems and having the property that they are their own antiparticles ``by definition'', because their associated creation/annihilation operators are the same as a result of a self-adjoint property~\cite{sarma2015majorana,sau2021majorana}. These excitations are ``topological'', with an intrinsic ground state quantum degeneracy, belonging to the $(\text{SU}_2)_2$  algebra and obeying non-Abelian anyonic braiding statistics~\cite{nayak2008nonabelian}.  The topological degeneracy and the associated non-Abelian braiding statistics enable fault-tolerant topological quantum computation by suitably braiding isolated MZMs around each other. Obviously, owing to these remarkable properties, there is a great deal of interest in the subject across many communities in physics and beyond (e.g., mathematics, computer science, engineering, materials science). As a result, MZMs have been studied intensively over the last 20 years~\cite{read2000paired,kitaev2001unpaired,freedman2003topological,nayak2008nonabelian}.   The current theoretical work focuses on the materials aspects of realizing MZMs in the laboratory, in particular, on the critical question regarding the role of disorder in compromising the experimental search for MZMs.

The most extensive experimental MZM search over the past 10 years has focused on ``Majorana nanowires'' --- 1D semiconductor nanowires proximity coupled to superconductors in superconductor-semiconductor (SC-SM) hybrid structures --- fueled by a number of precise theoretical predictions made in 2010~\cite{sau2010generic,lutchyn2010majorana,oreg2010helical,sau2010nonabelian} and by the ``convenience'' of the semiconductor materials platform. The theory 
 not only proposed a specific MZM platform (a 1D semiconductor nanowire made of materials, such as InSb or InAs, with strong Rashba spin-orbit coupling and a large Land\'e $g$-factor in contact with a parent SC, e.g., Al or Nb, providing proximity effect and in the presence of an applied magnetic field, in order to create a Zeeman spin splitting in the nanowire), but also provided a specific protocol (see, e.g., Fig. 14 in Ref.~\onlinecite{sau2010nonabelian}) to find experimental MZM signatures in a normal metal-superconductor (NS) tunnel spectroscopy experiment, where the proximitized nanowire acts as the SC.  
 A large number of experiments has scrupulously followed the protocol proposed in Fig. 14 of Ref.~\onlinecite{sau2010nonabelian}, carrying out NS tunnel spectroscopy measurements in InSb or InAs nanowires with Al or Nb as the parent SC, with numerous reports~\cite{mourik2012signatures,das2012zerobias,deng2012anomalous,churchill2013superconductornanowire,finck2013anomalous, deng2016majorana,nichele2017scaling,zhang2017ballistic,kammhuber2017conductance,gul2018ballistic,vaitiekenas2018effective,moor2018electric,zhang2018quantizeda,bommer2019spinorbit,grivnin2019concomitant,anselmetti2019endtoend,menard2020conductancematrix,puglia2021closing,pan2020situ,zhang2021large,song2021large}  by multiple different groups claiming evidence of MZMs based on the observation of zero-bias conductance peaks (ZBCPs) in the tunneling experiment, as expected based on the theoretical predictions  (see, e.g., Fig. 15 in Ref.~\onlinecite{sau2010nonabelian}).  
  It has been known for a long time that MZMs manifest perfect Andreev reflection, which leads to a tunneling ZBCP with a quantized conductance of $ 2e^2/h $  at zero temperature under ideal conditions~\cite{sengupta2001midgap,law2009majorana,flensberg2010tunneling,wimmer2011quantum}.   However, under realistic tunneling conditions involving finite temperatures and finite tunnel barriers, the precise Majorana quantization may not apply~\cite{setiawan2017electron,lin2012zerobias}, as was already apparent in the predicted tunnel conductance results (see, e.g.,  Fig. 15 in Ref.~\onlinecite{sau2010nonabelian}), and thus the experimental observation of (nonquantized) ZBCPs was extensively touted as ``signature'' or ``evidence'' for MZM rather uncritically.  Most of the early reported ZBCPs were very small in magnitude ($ \ll 2e^2/h $ ), but very recently large ZBCPs with conductance $ \sim 2e^2/h $  have been reported, often with great fanfare~\cite{zhang2018quantizeda,zhang2021retraction,zhang2021large,nichele2017scaling}.

We have asserted in several recent
publications~\cite{pan2020generic,pan2020physical,pan2021threeterminal,pan2021disorder,woods2020electrostatic,lai2021theory,pan2021quantized,dassarma2021disorderinduced,woods2021charge}   that the experimentally observed phenomenology, including the reported ZBCPs claimed as MZM signatures, are in fact generated by nontopological (i.e., trivial) disorder-induced tunneling features occurring close to zero energy in SC-SM systems in the presence of spin-orbit coupling and Zeeman splitting (sometimes alluded to as class D systems in the literature).  We call these disorder-induced trivial ZBCPs ``ugly'', to be contrasted with the predicted topological ``good'' ZBCPs arising from MZMs~\cite{pan2020physical}. Our work shows~\cite{dassarma2021disorderinduced} that even the recently reported large ZBCPs~\cite{zhang2021large} could arise generically as disorder-induced ugly peaks.  Other recent works have come to similar conclusions about the key importance of disorder in controlling the properties of Majorana nanowires~\cite{pan2021quantized,woods2021charge,zeng2021partiallyseparated}.  It is useful to point out in this context that the possible relevance of disorder in interpreting Majorana nanowire experiments was pointed out early in the development of the subject~\cite{brouwer2011probability,bagrets2012class,pikulin2012zerovoltage,sau2013density,sau2012experimental}, but the fact that disorder may actually be the dominant mechanism that controls the low-energy physics and may even produce large ZBCPs was not realized until very recently. 
 It is now mostly accepted that disorder is the most important impediment to the experimental realization of topological MZMs in the laboratory. An early important success in controlling disorder was the development of a hard zero-field superconducting proximity gap~\cite{chang2015hard}, following specific theoretical predictions~\cite{takei2013soft}, but finite magnetic field still tends to produce a soft gap and disorder-induced ``ugly'' ZBCPs.  The most serious consequence of the recent developments is that all observed ZBCPs in hybrid nanowires~\cite{mourik2012signatures,das2012zerobias,deng2012anomalous,churchill2013superconductornanowire,finck2013anomalous, deng2016majorana,nichele2017scaling,zhang2017ballistic,kammhuber2017conductance,gul2018ballistic,vaitiekenas2018effective,moor2018electric,zhang2018quantizeda,bommer2019spinorbit,grivnin2019concomitant,anselmetti2019endtoend,menard2020conductancematrix,puglia2021closing,pan2020situ,zhang2021large}, which have been previously claimed to be signatures and evidence for Majorana zero modes, are now thought to be associated with trivial Andreev bound states (ABS) arising from disorder~\cite{pan2020physical,pan2020generic,dassarma2021disorderinduced,pan2021quantized,woods2021charge}. Hence, eliminating disorder in the nanowire samples is the primary obstacle to further progress in the field. In fact, experimental claims of Majorana observation are no longer meaningful unless it can be decisively shown that the relevant samples are disorder-free.

In spite of the established importance of disorder in Majorana experiments, almost no direct quantitative information is available about the actual amount of \textit{in situ} disorder present in the hybrid superconductor-semiconductor nanowire samples used in the Majorana experiments.  For example, no transport study has reported any mobility measurement in the nanowire devices used in the Majorana tunneling spectroscopic experiments.  In fact, basic parameters, such as the carrier density, or the Fermi level in the nanowires, or how many subbands are occupied, are unknown.  Basic transport measurements reporting carrier mobilities are unavailable not only for the hybrid SC-SM structures, but even for the corresponding isolated InSb or InAs 1D nanowires going into making the hybrid Majorana device.  By contrast, a compelling body of numerical simulations of the tunnel spectroscopic measurements in the SC-SM nanowire devices clearly shows that a considerable amount of disorder is present in the system, preventing the emergence of the topological Majorana zero modes.  Already at a qualitative level the dominant role of disorder in these hybrid devices is obvious from the following experimental facts: (1) most devices do not manifest zero bias peaks; (2) most observed zero bias peaks are weak and unstable; (3) no end-to-end ZBCP correlations are ever observed; (4) there is no evidence for a re-opening of a bulk SC gap; (5) the  induced SC proximity gap becomes soft and very small in the presence of the applied field; (6) there is strong direct evidence for substantial subgap ABS at finite magnetic field; (7) the predicted Majorana oscillations are never observed, even when the length of the nanowire is rather short; (8) the ZBCPs are often irreproducible following any thermal cycling, even in the same sample; (9) nominally identical samples manifest generically different tunnel spectra with no sample-to-sample reproducibility; and (10) many generic irreproducible features of the tunneling data are consistent with the presence of substantial disorder in a SC system, where both spin symmetry and time reversal invariance are broken (the so-called class D behavior). 

Considering the dominant role of disorder in the Majorana nanowires and the lack of direct quantitative information regarding the actual amount of \textit{in situ} disorder, in this work, we have taken an indirect route to estimate the amount of disorder in the samples. Then, using model simulations, we have determined how this estimated amount of disorder would affect the topological Majorana properties, so that we can provide guidance on how much materials development and improvement are necessary for the eventual practical Majorana realization. 
 In view of the absence of direct quantitative information on the disorder amount in the nanowires, we do the next best thing and simulate transport properties of the corresponding 2D InSb and InAs materials from the same materials groups that produce the Majorana nanowire samples under similar conditions and in the same growth chambers \cite{gazibegovic2019bottomup, pauka2020repairing, beznasyuk2021role}.  It is reasonable to assume that the corresponding 2D semiconductor materials provide a stringent lower limit on the likely amount of disorder in the SC-SM hybrid platforms used in Majorana experiments.  This is because the nanowire samples in the SC-SM structures undergo many more processing steps than the 2D materials and are certainly more disordered than the 2D systems. However, the 2D systems do provide us with a valuable estimate of the minimum possible disorder in the nanowire Majorana platforms.  Given that there is no available experimental information on the 1D nanowire mobility, our procedure for estimating the effective disorder by fitting our transport theory to the measured 2D mobility in the same materials grown in the same laboratories under similar circumstances would have to do at this point, until direct information becomes available for the disorder in the 1D nanowires.

We emphasize that being able to properly characterize and, eventually, reduce the effects of disorder in superconductor-semiconductor hybrid structures and, more generally, in solid state-based quantum nanostructures is a requirement of crucial importance for the development of Majorana qubits and, in general, of solid state-based quantum technologies.  Satisfying this requirement will involve systematic and sustained efforts in materials growth, device engineering and experiment, and theory. The characterization component of this effort includes three distinct but interrelated critical tasks: (i) identify and characterize the physical sources of disorder. This implies identifying the type of disorder that is relevant in a given structure (e.g., charge impurities, point defects, atomic vacancies, surface roughness, patterning imperfections, etc.) and determining the relevant disorder parameters (e.g., impurity concentration, spatial distribution, etc.). (ii) Given a specific (physical) source of disorder, determine the corresponding effective disorder potential. This involves taking into account screening effects associated with the electrostatic environment (e.g., screening by the parent superconductor and the metallic gates, as well as the free charge in the semiconductor wire) and determining the effect of the screened potential on the specific modes that control the low-energy physics (e.g., determining the transverse profile of the topmost occupied subband in a Majorana wire and calculating the corresponding matrix elements of the screened potential). (iii) Given a specific effective disorder potential, characterize the low-energy properties of the system within the corresponding effective model. This involves calculating the low-energy BdG spectrum and the corresponding eigenstates, as well as relevant measurable quantities (e.g., the charge tunnel conductance) that might contain information regarding the low-energy modes and the underlying disorder.  

Task (i) involves a massive experimental effort, which is yet to be accomplished. It is, in fact, surprising how little experimental information is available about the quality and disorder content of the nanowires used in the hybrid SC-SM structures for Majorana experiments. In the absence of detailed experimental data, we provide here a contribution to accomplishing this task using an indirect route to estimating the (physical) disorder in Majorana wires based on available 2D transport data in similar semiconductor systems (see Sec. \ref{sec:transport}). Task (ii) was addressed in Ref. \onlinecite{woods2021charge} for charge impurities randomly distributed within the semiconductor nanowire, assuming a low/intermediate impurity density~\cite{woods2021charge}. Here, in Sec. \ref{sec:nw}, we use the methodology of Ref. \onlinecite{woods2021charge} together with the estimates of physical disorder in Sec. \ref{sec:transport} to evaluate the expected strength of the effective potential consistent with experimentally available samples. In addition, we address task (ii) in Sec. \ref{SCD}, in the context of InAs nanowires with surface charge impurities. This evaluation of the effective disorder potential is based on an estimate of physical disorder consistent with, but independent of the results in Sec. \ref{sec:transport}. This calculation also addresses the low/intermediate impurity density situations discussed in Ref. \onlinecite{woods2021charge}. Finally, task (iii) was addressed in numerous works, but starting with ad-hoc, essentially arbitrary model effective disorder potentials. This has clearly established that disorder is highly detrimental for Majorana physics if the (effective) disorder potential is strong-enough, yet determining whether or not the potential characterizing experimentally available structures is ``strong-enough'' (while also being realistic enough) remained an outstanding problem in the absence of quantitative results associated with tasks (i) and (ii). Here, we accomplish task (iii) based on explicit quantitative estimates of the effective disorder potential consistent with the available experimental data. In Sec.  \ref{sec:nw} we perform model simulations of the differential conductance using an effective model potential determined based on the estimates of physical disorder in Sec. \ref{sec:transport} and the realistic results of Ref. \onlinecite{woods2021charge}, while in Sec.~\ref{SCD}, we study the low-energy physics of a hybrid structure in the presence of surface charge impurities using an effective disorder potential calculated self-consistently within the same section. Both calculations provide conclusive evidence that the level of disorder likely to be present in experimentally available superconductor-semiconductor structures is inconsistent with the presence of topological superconductivity and the emergence of MZMs. 

The remainder of this paper is organized as follows.  In Sec.~\ref{sec:transport}, we present our 2D transport calculations, comparing them with the 2D transport measurements on semiconductor nanowire materials (i.e., InSb and InAs) available in the literature. We obtain a rough estimate of the relevant impurity density (i.e., effective disorder) to be used in the Majorana simulations.  To keep the number of parameters to a minimum we fit the experimental 2D mobility data to our transport theory using one effective charge impurity density, which is then used in the Majorana simulations.  In Sec.~\ref{sec:nw}, we first use the estimated disorder extracted in Sec.~\ref{sec:transport} and the self-consistent results of Ref.  \onlinecite{woods2021charge} to determine the effective disorder potential corresponding to a 1D minimal model of the wire. Then, using this effective potential, we perform a  model Majorana simulation to obtain the tunneling spectra corresponding to the SC-SM hybrid structures in the presence of (strong) disorder.  In Sec.~\ref{sec:SC}, we carry out a semi-realistic self-consistent simulation of the nanowire in the presence of realistic surface disorder, consistent with that estimated in Sec.~\ref{sec:transport},  calculating the effective disorder potential and investigating its impact on the low-energy physics. 
We conclude in Sec.~\ref{sec:conclusion} by providing a critical discussion of the prospects for the realization of topological Majorana zero modes in semiconductor-superconductor nanowires, based on our disorder estimates, and emphasizing the necessary materials improvement, which is essential for future progress in the field.  A set of appendices provides the technical details for the transport theory, the minimal model Majorana theory, and the self-consistent hybrid wire theory, while the main text focuses on the results of the calculations and their physical implications for the practical laboratory realization of the topological Majorana zero modes with non-Abelian braiding properties.  Some additional results that complement the main results are also presented in the appendices.

\section{Estimating disorder based on 2D transport properties}\label{sec:transport}

In this section, we develop a minimal transport theory involving scattering by random quenched impurities as the only resistive carrier scattering mechanism for 2D carriers confined in semiconductor (InAs and InSb) layers and compare our results with the available experimental information on 2D systems,
 which are structurally close to the 1D nanowire samples used in the fabrication of  SC-SM Majorana systems.  Since the large disorder scenario is detrimental to topological superconductivity, our goal is to obtain the most optimistic disorder estimates that are also realistic at some level of practicality. Therefore we discard any part of the carrier density-dependent 2D mobility data where the mobility is decreasing with increasing carrier density, which indicates the activation of additional scattering mechanisms (e.g., strong interface roughness scattering, inter-subband scattering as the Fermi level is pushed into the second 2D subband, etc.) causing the effective disorder (mobility) to increase (decrease).  
 More specifically, our goal is to model the peak 2D sample mobility as accurately and faithfully as possible using very few (in fact, just one) disorder parameters, so that the transport theory and the Majorana simulations do not degenerate into hopeless detailed device simulations, where the physics disappears into a bunch of unknown (and often, unknowable) fit parameters.  
 Such a multiple parameter fit approach would be completely unhelpful in the current situation, where little is known about the details of the SC-SM hybrid systems, except that disorder is playing a key role in suppressing the topological MZMs and producing trivial ZBCP.  We are therefore aiming, as much as possible, at finding a single impurity parameter that describes the disorder in the system accurately enough.  We note that the transport model can be easily generalized to include many disorder parameters representing multiple resistive scattering mechanisms, but at this point in the development of the subject such a detailed modeling is an overkill and is completely unnecessary.  As we show below, a single disorder parameter, namely, a bulk charged impurity density, is capable of giving reasonable fits to the available (and highly limited) transport data.  We use the published 2D mobility data in the literature from the materials groups at Eindhoven (InSb), Purdue (InAs), and Copenhagen (InAs), which are also the growers that produce the 1D nanowire samples being used in the Majorana SC-SM hybrid systems. Our conjecture that the estimated disorder in the 2D samples grown in these laboratories is a likely lower bound on the realistic disorder in the 1D nanowires used in the SC-SM hybrid structures for Majorana search is consistent with how the growers themselves see the materials situation at this point in the development of the subject.

We use the Boltzmann theory at $ T=0 $  to obtain the 2D mobility as a function of carrier density following the well-established (and highly successful) procedure for the calculation of the disorder-limited 2D transport, which has been extensively used in the literature for the calculation of 2D transport in Si- and GaAs-based 2D systems~\cite{dassarma2015screening,ando1982electronic}. For completeness, we provide the details of the transport theory in Appendix~\ref{app:A}.

To model the mobility as a function of carrier density, we start by first assuming that there are both short-range and long-range impurity scattering centers randomly distributed in the 2D layer, as well as at interfaces and inside the surrounding layers (when such sample details are available). 
We also include impurities located at the interface with the dielectric, since they often act as a strong scattering source for the shallow 2D materials used for producing 1D nanowires.  However,  such a detailed model necessitates having many free parameters, which is pointless in the absence of additional material information, as discussed above.
Our goal is not to attempt a precise quantitative fit to the sample-dependent experimental 2D mobility, but to obtain a rough estimate of the effective disorder level, which we can then use for our nanowire Majorana simulations. 
 It is, therefore, a huge relief that a reasonable quantitative theoretical transport fit to the measured 2D mobility can be obtained using just a single disorder parameter, namely, the 3D charged impurity density in the 2D layer, (or an equivalent 2D or 1D impurity density). 
 As our results presented below show, there is one caveat to this reasonable fitting with one disorder parameter, which is that the one-parameter fitting works well only in the intermediate carrier density regime where the mobility is close to its peak value. This is, of course, our regime of interest. We focus on the realistic minimal disorder scenario, since very large disorder would completely suppress topological superconductivity and MZMs anyway. Once the sample quality improves substantially in future devices, it may be necessary to do more quantitatively precise simulations for specific devices in specific laboratory setups using a multiparameter transport simulation focused on specific samples.

\begin{figure*}[!htb]
    \centering
    \includegraphics[width=\linewidth]{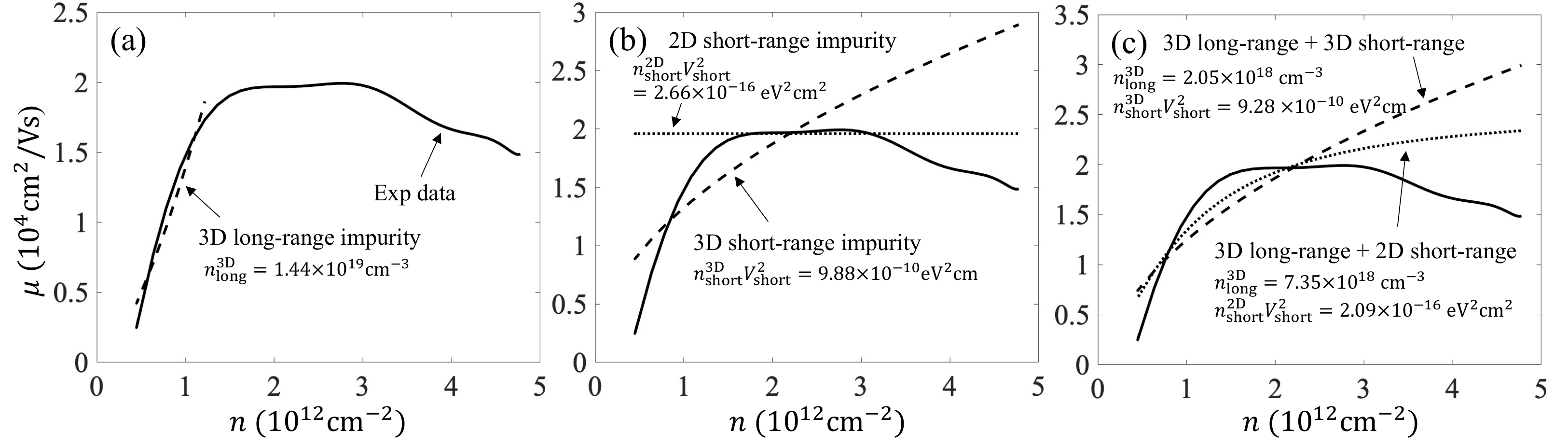}
    \caption{Experimental mobility of an InSb quantum well from the group of Bakkers \cite{gazibegovic2019bottomup} (solid line) plotted as a function of carrier density and theoretical fits using the Boltzmann transport theory. (a) The best fit to the increasing part of the mobility assuming only charge impurities randomly distributed in the InSb quantum well (dashed line, referred to as 3D long-range impurity). (b) The best fit to the flat part assuming short-range scattering by impurities randomly distributed either in the 2D layer itself (dotted line, referred to as 2D short-range impurity) or throughout the whole sample (dashed line, referred to as 3D short-range impurity). Note that the 2D short-range impurity model yields a much better fit than the 3D short-range impurity model. (c) The best fit over the entire range of mobility data using two fitting parameters (one for long-range scattering and the other for short-range scattering).
    Here $V_\mathrm{short}$ is the short-range impurity disorder potential and $n_\mathrm{long(short)}$ is the long-range (short-range) impurity density. We use the background dielectric constant $\kappa_\mathrm{InSb}=18$, the InSb quantum well width $a=80$nm, and the effective mass $m=0.013m_e$ where $m_e$ is the bare electron mass. The experimental data are smoothed for visual clarity. The actual fit is performed to the noisy original data.  }
    \label{fig:mobility_fit_InSb_1}
  \end{figure*}
  
\begin{figure*}[!htb]
    \centering
    \includegraphics[width=\linewidth]{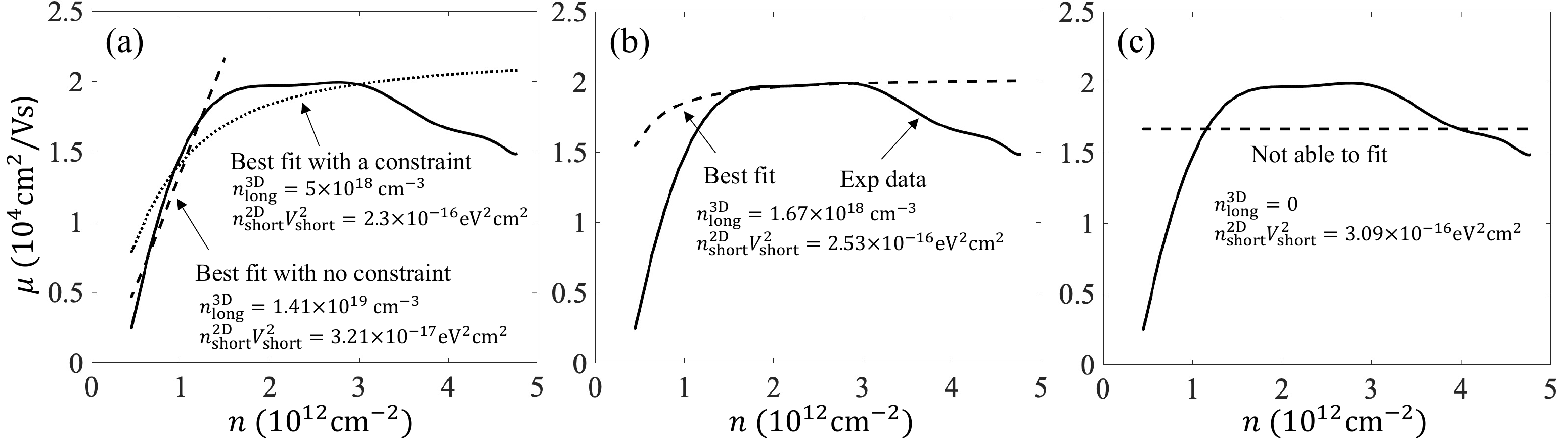}
    \caption{Same as Fig.~\ref{fig:mobility_fit_InSb_1}, but using a two-parameter fitting procedure, with $n^\mathrm{3D}_\mathrm{short}$ and $n^\mathrm{2D}_\mathrm{short}V_\mathrm{short}^2$ being the effective disorder parameters. (a) Best fit to the linearly increasing part of the measured mobility. The dashed line is the best fit without any constraint on the fitting parameters, while the dotted line is obtained by imposing the 
  constraint  $n_\mathrm{long}^\mathrm{3D}< 10^{19}$cm$^{-3}$. The constraint reduces the quality of the fit, which shows that $n_\mathrm{long}^\mathrm{3D}$ needs to be larger than $10^{19}$ cm$^{-3}$ for a reliable fit. (b) The best fit to the flat part of the data. (c) The best fit to the decreasing part of the mobility at high density  ($n>3\times 10^{12}$cm$^{-2}$).}
    \label{fig:mobility_fit_InSb_2}
  \end{figure*}

\begin{figure*}[!htb]
    \centering
    \includegraphics[width=\linewidth]{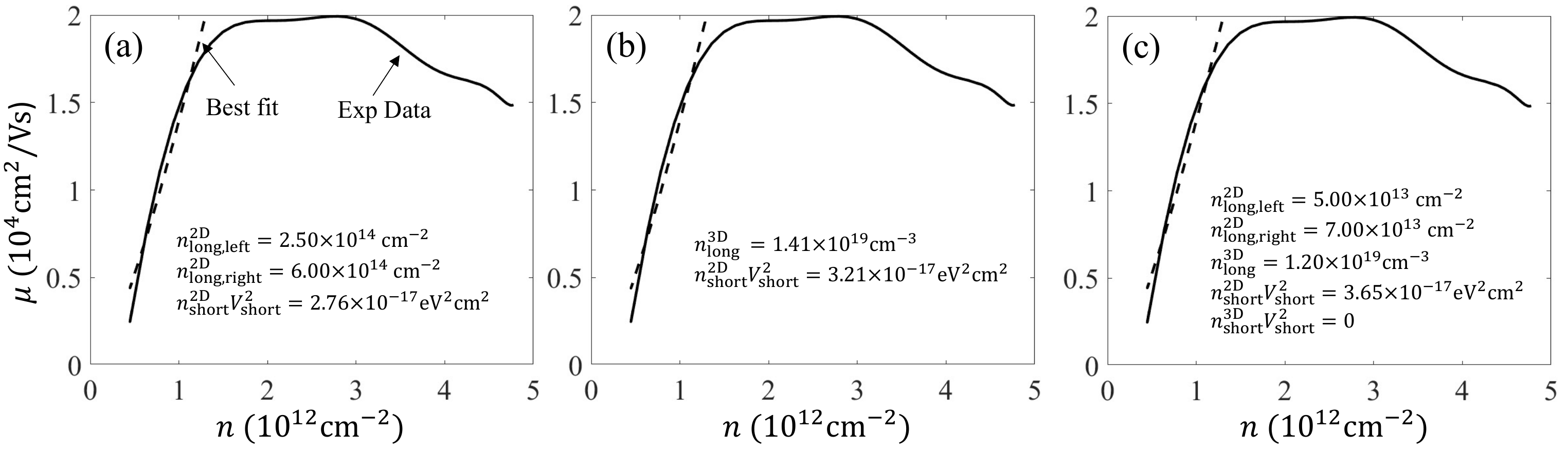}
    \caption{Same as Fig.~\ref{fig:mobility_fit_InSb_1}, but focusing on the increasing part of the measured mobility and assuming scattering by (a) 2D remote charged impurities at the interfaces of the quantum well and short-range impurities, (b) 3D charged and 2D short-range impurities, and (c) all of the mobility-limiting impurities discussed throughout Figs.~\ref{fig:mobility_fit_InSb_1}-\ref{fig:mobility_fit_InSb_3}. }
    \label{fig:mobility_fit_InSb_3}
  \end{figure*}

\begin{figure}[!htb]
    \centering
    \includegraphics[width=\linewidth]{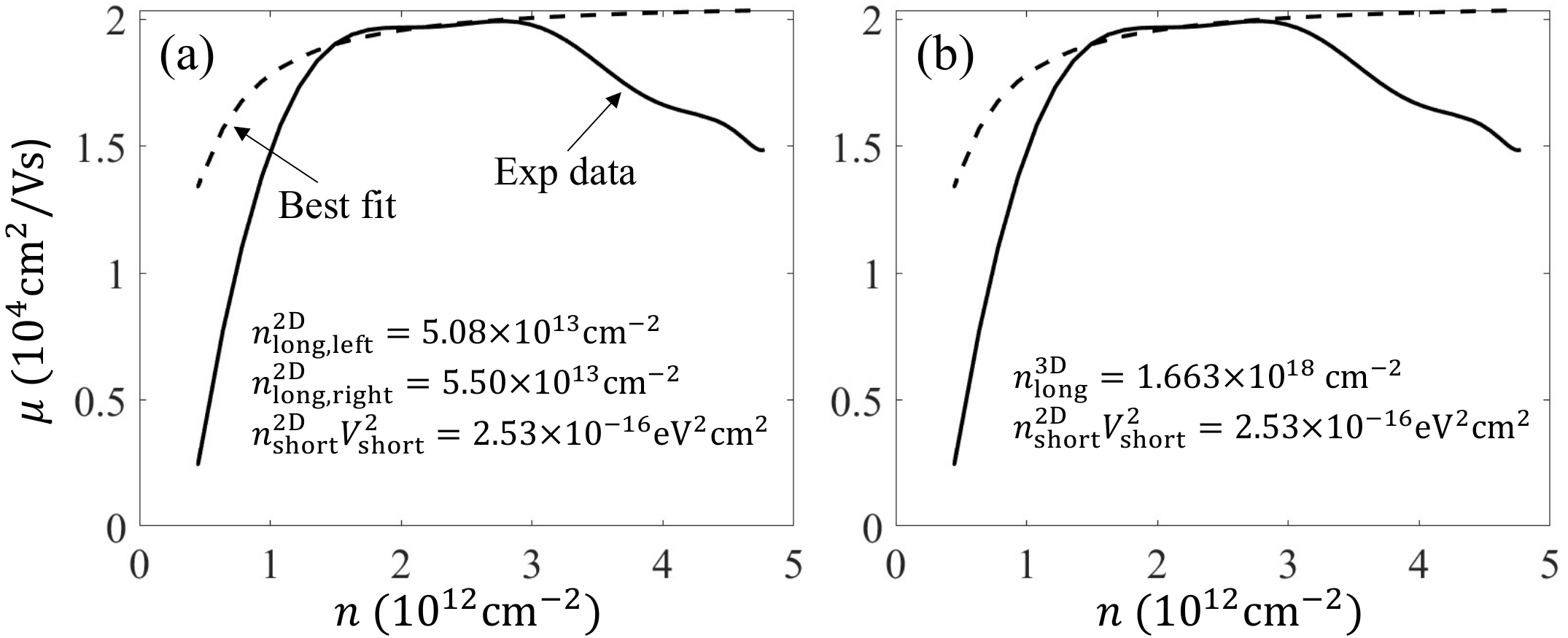}
    \caption{Same as Figs.~\ref{fig:mobility_fit_InSb_3}(a) and~\ref{fig:mobility_fit_InSb_3}(b), except that here we fit the flat part of the measured mobility.}
    \label{fig:mobility_fit_InSb_4}
  \end{figure}

 We start with the measured InSb mobility as presented in Fig. 6 of Ref.~\onlinecite{gazibegovic2019bottomup}  We show in Figs.~\ref{fig:mobility_fit_InSb_1}-\ref{fig:mobility_fit_InSb_4} different theoretical fits to the experimental InSb mobility as a function of carrier density, taking all the sample parameters from Ref.~\onlinecite{gazibegovic2019bottomup}  and other parameters as appropriate for 2D InSb. The values of these parameters are provided in the figures and the corresponding captions. We use transport models involving up to five different scattering mechanisms, which may be operational at various levels. More specifically, we consider both long-range charged impurities and short-range defects,  impurities both in the 2D layer itself and in the materials surrounding it, as well as impurities localized at interfaces.
It turns out that the important intermediate carrier density regime,  where the mobility is increasing toward the peak value, can be well-described by our theory using a single parameter, the 3D density of effective long-range impurities randomly distributed within the (quasi) 2D layer. This property is revealed by the fit in Figs. \ref{fig:mobility_fit_InSb_1}(a) and confirmed by the comparison with other fits shown in  Figs. \ref{fig:mobility_fit_InSb_1}-\ref{fig:mobility_fit_InSb_4} involving different scattering scenarios.  Note that this satisfies our need for characterizing the physical disorder using a single (effective) parameter,  in this case, the 3D density of long-range (charge) impurities.
  
 We note that, generically, the mobility decreases at high carrier density (both for InSb and InAs 2D samples), and this is known to arise from inter-subband scattering processes, which become operational as the 2D Fermi level pushes into higher subbands with increasing density.  This can be simulated within our charged impurity scattering model by generalizing the theory to a multisubband situation. However, this would not provide any new disorder parameter;   basically, new scattering channels are being triggered, as the same charged impurities can now cause scattering between different subbands,  suppressing the mobility.  Therefore we do not include this higher density regime in our transport modeling.

At very low density, when the carrier density is comparable to the effective charged impurity density, the system becomes insulating due to a percolation transition driven by the failure of screening leading to an inhomogeneous density landscape beyond the validity of the Boltzmann transport theory, as is well-established for 2D semiconductor systems~\cite{dassarma2013twodimensional,pudalov2020experimental,li2019evidence,manfra2007transport,lilly2003resistivity,dassarma2005twodimensional,tracy2009observation,leturcq2003resistance,tracy2006surface,wilamowski2001screening,he1998new,ilani2001microscopic,dassarma2014twodimensional,meir1999percolationtype,knap2014transport,shabani2014apparent}.  Our Boltzmann transport theory is obviously not applicable in this low density percolative insulating regime where the conductivity vanishes below a sample-dependent critical or threshold density. In Appendix~\ref{app:B} we provide the details of our low-density analysis of the 2D mobility data, extracting the percolation density and showing that it correlates approximately with the peak mobility value, since the peak mobility and the percolation critical density are both determined by the effective charged impurity density in the system.  This percolation fit provides an additional justification for our single parameter disorder analysis of the 2D mobility data, reinforcing the basic idea that unintentional background doping by random charged impurities is the main disorder mechanism in the Majorana nanowires.  We note that short-range disorder would not lead to such a percolative transition since nonlinear screening and the failure of screening leading to density inhomogeneity are intrinsic to long-range charged Coulomb disorder potential.  We reiterate that the Boltzmann theory applies above the percolation transition, hence our theoretical fit is used to extract the effective impurity density in the intermediate regime where the carrier density is neither too low nor too high, i.e., the density regime leading up to the peak mobility.  In this regime, the mobility should be approximately linear in carrier density because of the dominant role of charged impurity scattering \cite{dassarma2013universal}, as we find theoretically in agreement with the experimental data.  

We emphasize that the basic qualitative features of a mobility peak (at high density) as a function of increasing carrier density and of a percolative insulating transition (at low carrier density) as a function of decreasing density are generic in all the 2D systems analyzed here (i.e., systems from all three laboratories and covering both InSb and InAs).  Our Boltzmann transport theory applies only in the intermediate density and peak mobility region between these high- and low-density regimes. Fortunately, however, this is precisely the regime of interest for estimating the disorder content through theoretical transport calculations.  We use this same intermediate density data fitting strategy to analyze all three data sets (from Eindhoven, Copenhagen, Purdue).    
  
Figures~\ref{fig:mobility_fit_InSb_1} (a) and~\ref{fig:mobility_fit_InSb_1}(b) show that our single-parameter theoretical fits to the increasing and flat parts of the measured mobility are in good agreement with the experimental data. For simplicity, we refer to impurities distributed two-dimensionally (three-dimensionally) as 2D (3D) impurities. The best fit to the entire range of the measured mobility presented in Fig.~\ref{fig:mobility_fit_InSb_1} (c)  is manifestly worse than the fit to the increasing part presented in Fig.~\ref{fig:mobility_fit_InSb_1}(a), since different scattering mechanisms are involved at different density regimes, as discussed above.
In Fig.~\ref{fig:mobility_fit_InSb_2}, we show our results based on two fitting parameters $n_\mathrm{long}^\mathrm{3D}$ and $n^\mathrm{2D}_\mathrm{short}V^2_\mathrm{short}$. By comparing Figs.~\ref{fig:mobility_fit_InSb_2} (a) and~\ref{fig:mobility_fit_InSb_2}(b) with Fig.~\ref{fig:mobility_fit_InSb_1} (a) and~\ref{fig:mobility_fit_InSb_1}(b) 
one notices that  the quality of the two-parameter fit is almost the same as that of the one-parameter fit, giving similar estimated background impurity densities. This indicates that our one-fitting-parameter model is essentially good enough to capture the transport physics of the 2D InSb sample. Also note that our transport model does not fit the decreasing mobility [see Figs.~\ref{fig:mobility_fit_InSb_1} (c) and \ref{fig:mobility_fit_InSb_2}(c)], because we do not include scattering mechanisms responsible for the decreasing mobility behavior in the high-density regime (such as inter-subband scattering).

  In Figs.~\ref{fig:mobility_fit_InSb_3} and \ref{fig:mobility_fit_InSb_4}, we consider an additional realistic scattering scenario where the mobility is limited by the remote 2D charged impurities at the quantum well interface. In the intermediate density regime where the mobility increases linearly, which is our main focus, the mobilities limited by 2D remote charged impurities [Fig.~\ref{fig:mobility_fit_InSb_3} (a)] and 3D long-range impurities [Fig.~\ref{fig:mobility_fit_InSb_3} (b)] have almost the same carrier density dependence, $\mu\sim n^{3/2}$, since the system is in the strongly screened limit due to a small effective mass (i.e., $q_\mathrm{TF}>k_\mathrm{F}$) \cite{dassarma2013universal}. Thus the experimental mobility data are fit almost equally well within both scenarios. 
 Note, however, that the estimated 2D interface charged impurity density obtained in Fig.~\ref{fig:mobility_fit_InSb_3} (a) is unrealistically large ($\sim 5\times 10^{14} \mathrm{cm}^{-2}$), implying that remote 2D charged impurities cannot be the only scattering source in the sample. In Fig.~\ref{fig:mobility_fit_InSb_3} (c), we present the results using our most realistic transport model that includes all the scattering mechanisms discussed above, with a constraint that $n_\mathrm{long}^\mathrm{2D}$ lies within a reasonable range ($<10^{14}\mathrm{cm}^{-2}$). Note that even using this transport model with five-fitting parameters, our results show that $n_\mathrm{long}^\mathrm{3D}\sim 10^{19}\mathrm{cm}^{-3}$, which is consistent with our one fitting parameter result. 

For the InSb sample from the Bakkers group in Eindhoven (Figs. \ref{fig:mobility_fit_InSb_1}-\ref{fig:mobility_fit_InSb_4} in this paper), our extensive theoretical fits provide an effective 3D background charged impurity density of $ 10^{18}-10^{19} $  per cm$ ^3 $, with the larger number for the impurity density definitely being a better fit parameter. This is consistent with the measured peak mobility of $ \sim $  20000 cm$ ^2 $/V$ \cdot $ s, which corresponds to a rather large level broadening, $ \Gamma=\frac{\hbar}{2\tau}\sim 2 $ meV, where $ \tau $  is the scattering time appearing in the mobility, $ \tau= m \mu/e $,  with $ m $  being the carrier effective mass. This is a rather large broadening for MZM studies, given that the topological SC gap is likely to be $ <0.1 $  meV.  In addition, our extracted impurity density of $ > 10^{18} $  per cm$ ^3 $ is more than three orders of magnitude larger than the limit of $ \sim 10^{15} $  per cm$ ^3 $  recently provided in Ref.~\onlinecite{woods2021charge} as necessary for the manifestation of topological SC in SC-SM hybrid structures. We note that our estimated disorder levels are consistent with the rough estimates made by the experimentalists themselves based on their knowledge of the compensation levels in the InSb materials~\cite{bakkers}. 

  \begin{figure}[!htb]
    \centering
    \includegraphics[width=\linewidth]{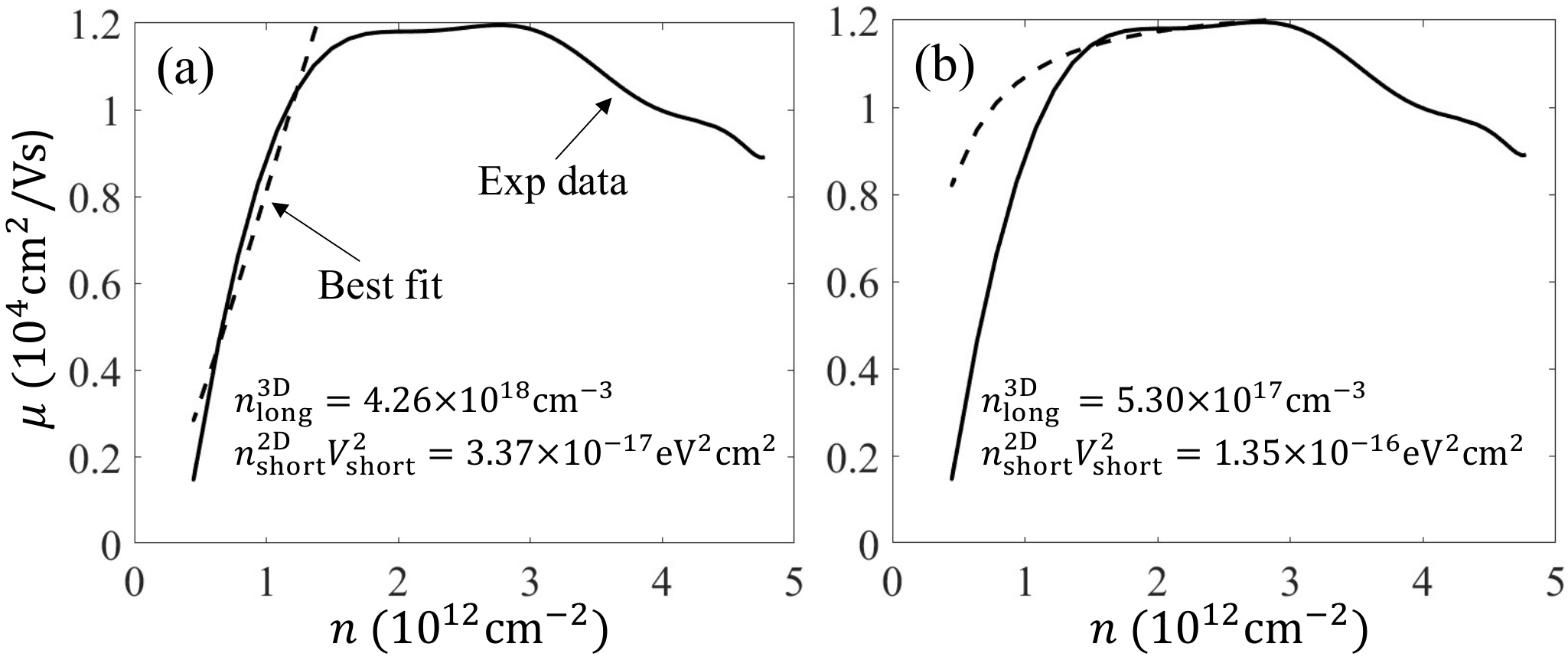}
    \caption{Best fit (dashed line) to the experimental InAs mobility data (solid) reported in the recent paper by the Krogstrup group \cite{beznasyuk2021role}. The carrier density dependence of the mobility is assumed to be the same as that of the Bakkers's InSb mobility, except that the peak mobility is set to be $1.2\times 10^2$cm$^2$/Vs, which is the maximum mobility reported in Ref.~\onlinecite{beznasyuk2021role}. Here we use a background dielectric constant $\kappa_\mathrm{InAs}=15$, an effective mass $m=0.023m_e$, and an InAs quantum well width $a=30$nm.}
    \label{fig:mobility_fit_InAs_Copenhagen}
  \end{figure}  

Next, we analyze the available transport data for 2D InAs samples to obtain rough estimates of the effective disorder in  InAs-based  SC-SM platforms.  In contrast to the InSb mobility data, where only one data set is available from the Eindhoven Bakkers group (i.e., the data analyzed in Figs.~\ref{fig:mobility_fit_InSb_1}-\ref{fig:mobility_fit_InSb_4}), for 2D InAs structures, experimental mobility data are available from two different Majorana materials growers (Copenhagen and Purdue).  We discuss each data set separately below.
  
 \begin{figure*}[!htb]
    \centering
    \includegraphics[width=0.95\linewidth]{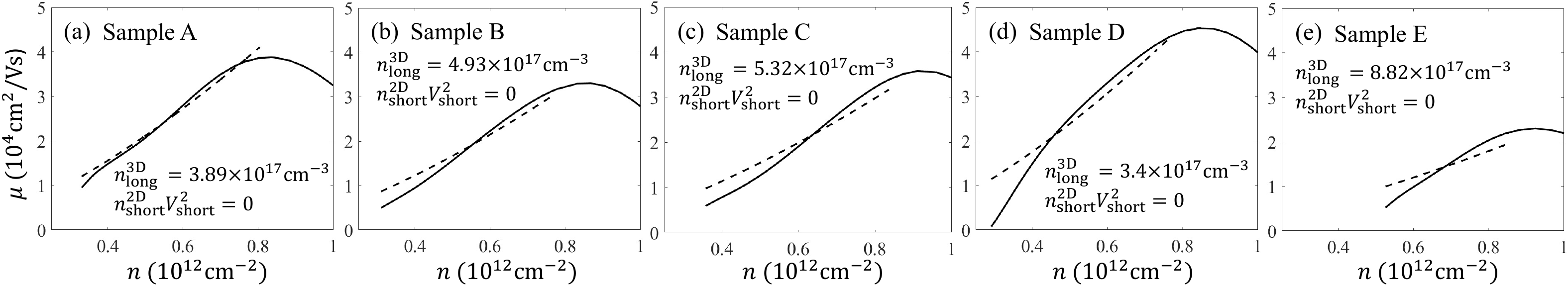}
    \caption{Experimental InAs mobility from the Purdue group \cite{pauka2020repairing} plotted as a function of carrier density for five different samples, which are labeled according to Table I of Ref.~\onlinecite{pauka2020repairing}. The dashed line is the best fit to the linearly increasing part of the measured mobility using the Boltzmann transport theory and assuming two scattering mechanisms associated with 3D long-range impurities in the quantum well of width $a=30~\mathrm{nm}$ and 2D short-range impurities. See Fig.~1(a) of Ref.~\onlinecite{pauka2020repairing} for details of the sample structure. The experimental data are smoothed for visual clarity. The actual fit is performed to the noisy original data.}
    \label{fig:mobility_fit_InAs_Purdue_1}
  \end{figure*}
  
  \begin{figure*}[!htb]
    \centering
    \includegraphics[width=0.95\linewidth]{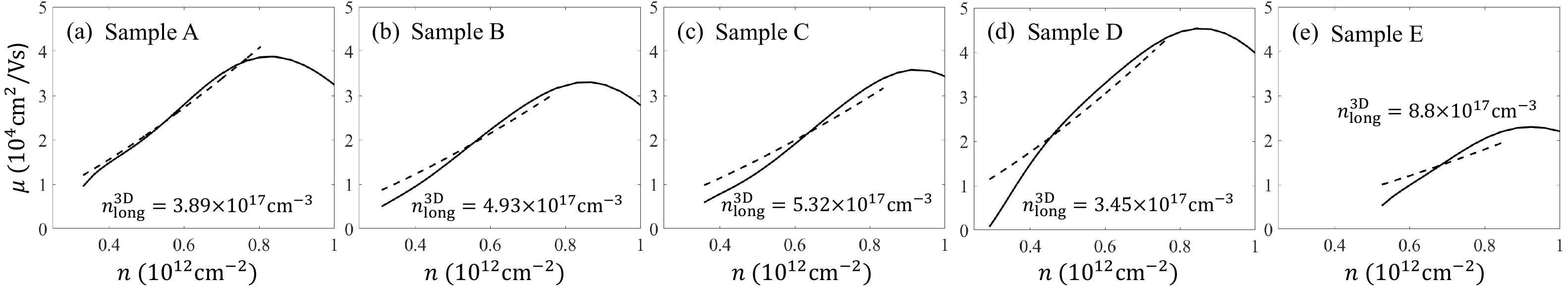}
    \caption{Same as Fig.~\ref{fig:mobility_fit_InAs_Purdue_1}, but  considering only background 3D charged impurities randomly distributed within the InAs quantum well. }
    \label{fig:mobility_fit_InAs_Purdue_2}
  \end{figure*}
    
  \begin{figure*}[!htb]
    \centering
    \includegraphics[width=0.95\linewidth]{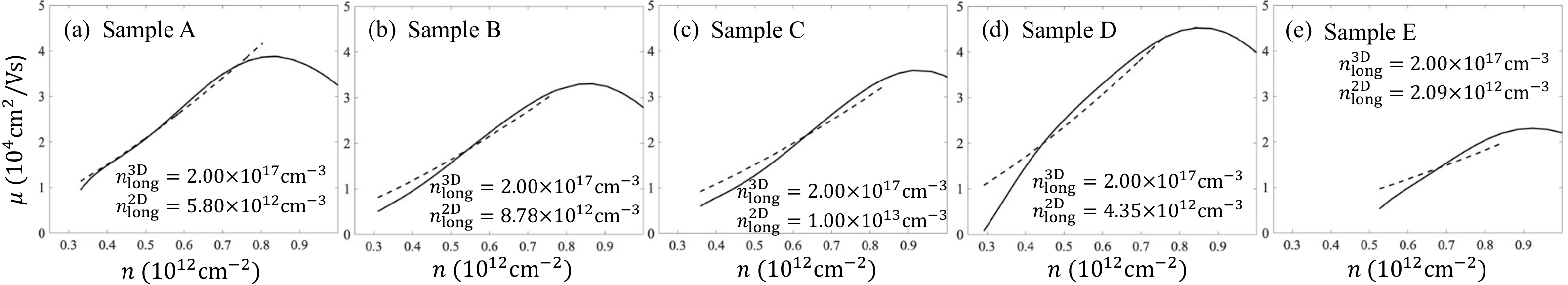}
    \caption{Same as Fig.~\ref{fig:mobility_fit_InAs_Purdue_1}, but considering background 3D impurities randomly distributed within the InAs quantum well and 2D long-range remote charged impurities at the dielectric interface separated by $d=10~$nm from the surface of the quantum well. }
    \label{fig:mobility_fit_InAs_Purdue_3}
  \end{figure*}
  
  \begin{figure*}[!htb]
    \centering
    \includegraphics[width=6.8in]{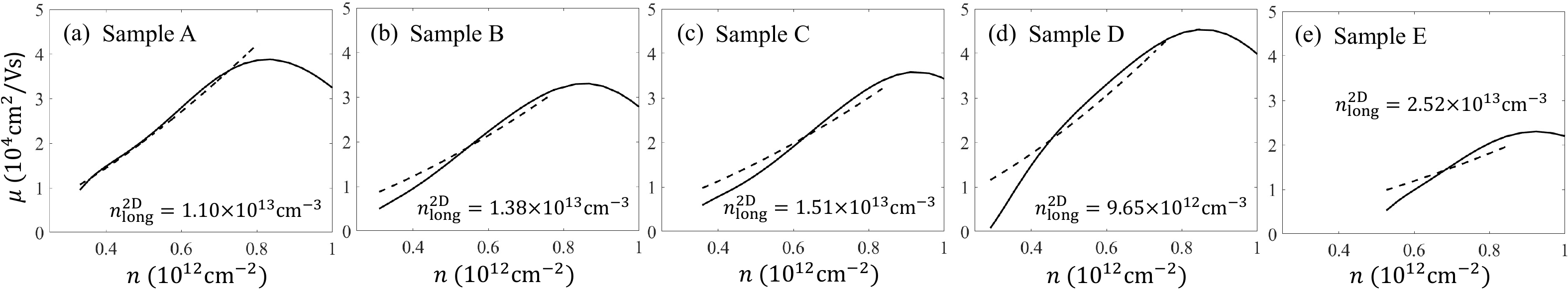}
    \caption{Same as Fig.~\ref{fig:mobility_fit_InAs_Purdue_1}, but considering only 2D long-range remote charged impurities at the dielectric interface separated by $d=10$nm from the surface of the quantum well.}
    \label{fig:mobility_fit_InAs_Purdue_4}
  \end{figure*}
  
  \begin{figure*}[!htb]
    \centering
    \includegraphics[width=0.95\linewidth]{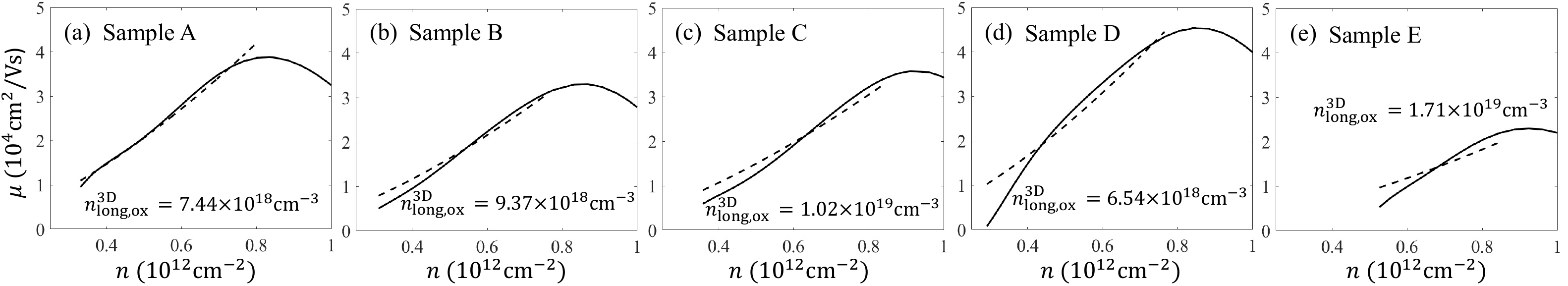}
    \caption{Same as Fig.~\ref{fig:mobility_fit_InAs_Purdue_1}, but considering only remote impurities distributed three dimensionally (i.e., 3D long-range impurities) in the dielectric layer of thickness $a_\mathrm{oxide}=8$nm located $10~$nm away from the interface of the quantum well.}
    \label{fig:mobility_fit_InAs_Purdue_5}
  \end{figure*}

We start with the recent InAs transport data from the Copenhagen group of Krogstrup as presented in Ref,~\onlinecite{beznasyuk2021role}.  In Fig.~\ref{fig:mobility_fit_InAs_Copenhagen}, we show our theoretical fit to this experiment following the same procedure as that described above for InSb, except for using system parameters corresponding to the InAs sample measured in Ref.~\onlinecite{beznasyuk2021role}. 
 As shown in Fig.~\ref{fig:mobility_fit_InAs_Copenhagen}, the best fit  produces a background 3D charged impurity density of $ \sim 5\times10^{17} $  to $ 4\times10^{18} $ per cm$ ^3 $,  depending on whether the fit emphasizes the peak mobility itself or the intermediate density regime leading to the peak mobility.  We note that the fit in Fig.~\ref{fig:mobility_fit_InAs_Copenhagen}  involves some short-range disorder.  If we assume an effective disorder arising entirely from random Coulomb disorder, the likely background charged impurity density would be slightly higher, making it comparable to that in the InSb system analyzed in Figs.~\ref{fig:mobility_fit_InSb_1}-\ref{fig:mobility_fit_InSb_4}.  The peak mobility of 12000 cm$ ^2 $/V$ \cdot $ s in InAs, with its higher effective mass (as compared to InSb) corresponds to essentially the same level broadening of $ \sim 2-3 $  meV as for the InSb Eindhoven sample discussed above.  We mention that very recent unpublished work from Copenhagen finds a direct experimental level broadening of $ \sim 2~$meV in the 1D subbands of InAs nanowires, providing strong support for our approach toward estimating disorder by analyzing 2D sample mobility~\cite{private_Cui}.  Similar to our analysis of the Eindhoven InSb data, the Copenhagen InAs data can be reasonably well explained by assuming a background charged impurity density, which is about three orders of magnitude larger than the level of quality necessary for the practical realization of topological MZMs in nanowires in this system (i.e., typical InAs-Al or InSb-Al SC-SM hybrid structures). 

Finally, we consider the InAs 2D samples in Ref.~\onlinecite{pauka2020repairing}, which are the most extensive transport data available in the context of Majorana nanowire materials growth.  We present our extensive theoretical analysis of the Purdue data in Figs.~\ref{fig:mobility_fit_InAs_Purdue_1}-\ref{fig:mobility_fit_InAs_Purdue_5}, focusing on the best fitting in the intermediate carrier density regime, where our transport theory applies well. The Purdue experiment involves extensive processing of samples with various techniques modifying the effective mobility, providing an additional variable (i.e., processing) that directly affects the peak mobility. The goal of the growers here is to identify ideal processing to suppress disorder and enhance mobility, but from our theoretical perspective, the processing provides a test for our characterization of the effective sample disorder through modeling. 
As described below in detail and as shown in Figs.~\ref{fig:mobility_fit_InAs_Purdue_1}-\ref{fig:mobility_fit_InAs_Purdue_5}, we find that random long-range charged impurity scattering dominates the transport properties, our fitting showing a clear correlation between the extracted charged impurity density and the measured peak mobility. More specifically, we find that, depending on the processing details, an effective background unintentional charged impurity density of 2-9 $\times 10^{17} $  per cm$ ^3 $  provides a very good fit to the measured density-dependent mobility in the intermediate density regime.  This corresponds to a peak mobility of $ \sim $  25000 – 50000 cm$ ^2 $/V$ \cdot $s--- the lower impurity density (and the higher peak mobility) of the Purdue samples (as compared with the ones from Eindhoven and Copenhagen) implying a higher sample quality associated with the MBE technique used in growing the Purdue 2D samples.  Nonetheless, the effective disorder level, as reflected in the extracted charge impurity density, is still much higher (by a factor $ >100 $ ) than the $ 10^{15} $  per cm$ ^3 $ level necessary for the realization of the topological MZMs. 

In Fig.~\ref{fig:mobility_fit_InAs_Purdue_1}, we present the best-fit results to the intermediate regime of the mobility using the Boltzmann transport theory including 3D charged impurities in the quantum well and 2D short-range disorders. During the fitting procedure, we find that the short-range fitting parameter $n^\mathrm{2D}_\mathrm{short}$ is driven to a very small value close to zero, and thus the best fit results are identical to those obtained with only 3D charged impurity, which are shown in Fig.~\ref{fig:mobility_fit_InAs_Purdue_2}. This implies that the long-range charged impurity scattering is the dominant scattering mechanism in the intermediate density regime leading to the peak mobility. 
Since each sample is processed using a different technique,  which may affect the remote charged impurities at the interface or inside the surrounding dielectric material, we present our best fits involving scattering by those types of impurities in Figs.~\ref{fig:mobility_fit_InAs_Purdue_3}-\ref{fig:mobility_fit_InAs_Purdue_5}. We find that the overall best fit for all samples (even though differences in fitting quality are not significant) corresponds to the background charged impurity density ($n^\mathrm{3D}_\mathrm{long}$) being the same for all samples,  with only the remote charged surface impurity density ($n^\mathrm{2D}_\mathrm{long}$) varying from sample to sample. The estimated background charged impurity density  is $n^\mathrm{3D}_\mathrm{long}\sim2\times 10^{17} \mathrm{cm^{-3}}$ (see Fig.~\ref{fig:mobility_fit_InAs_Purdue_3}).

\begin{table*}[t]
    
	\caption{Summary of the estimated impurity densities presented through Figs.~\ref{fig:mobility_fit_InSb_1}-\ref{fig:mobility_fit_InAs_Purdue_5} for the InSb (from Eindhoven \cite{gazibegovic2019bottomup}) and InAs (from Copenhagen \cite{beznasyuk2021role} and Purdue \cite{pauka2020repairing}) samples . Here $a$ is the width of the corresponding quantum well.
	$*$ InAs sample from Copenhagen.
	$**$ InAs sample from Purdue. }
	\label{table:summary_disorder}
	\includegraphics[width=6.8in]{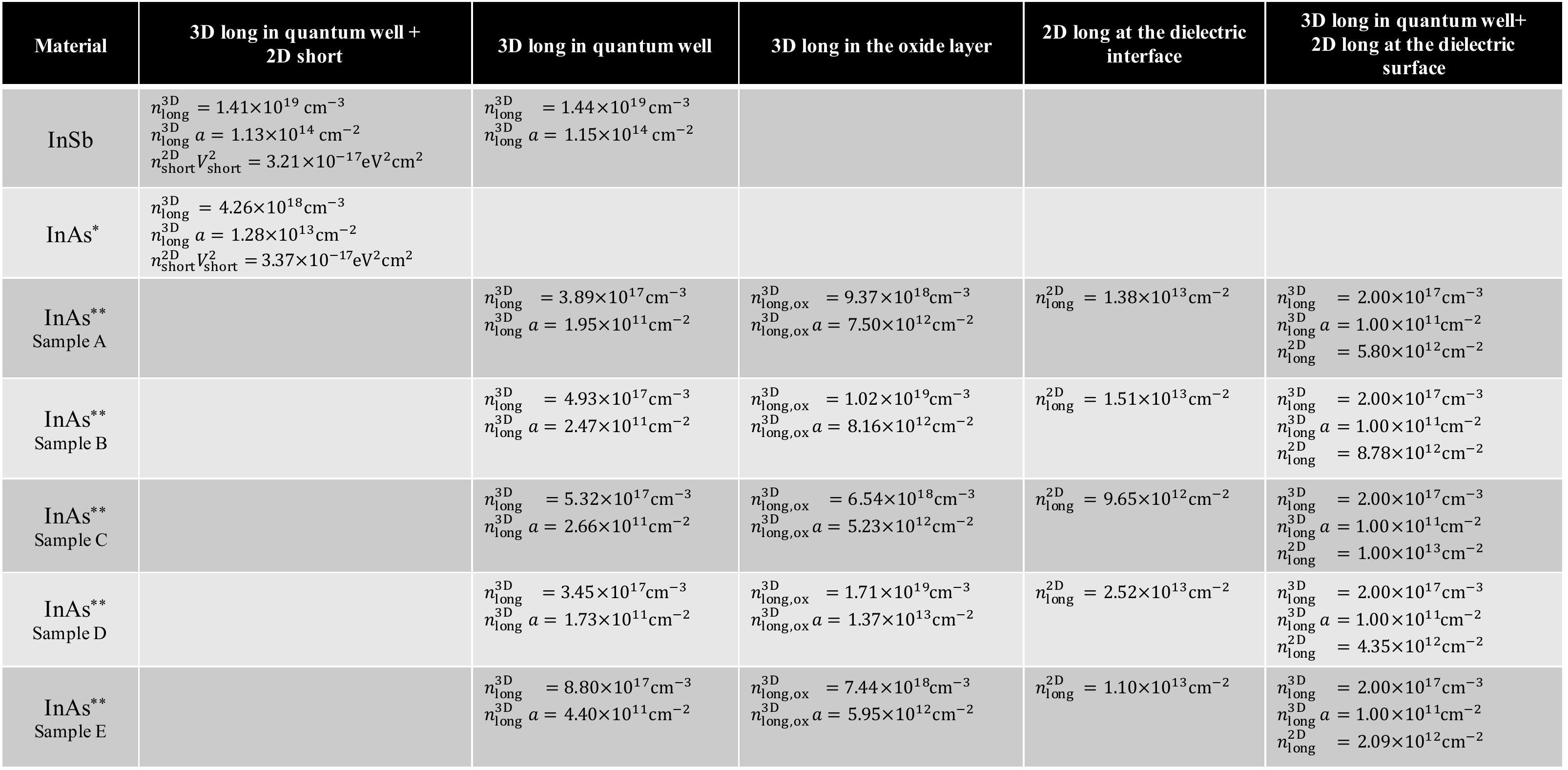}
\end{table*}

We have also carried out an extensive analysis of the Purdue data by estimating the percolation density and the density scaling exponent of mobility, which are presented in-depth in Appendix~\ref{app:B}.  Basically, the percolation density decreases with increasing peak mobility, since higher peak mobility implies lower background impurity density. 

We note that a related measure of disorder, the single-particle broadening or the quantum relaxation time, is sometimes discussed in the literature~\cite{dassarma1985singleparticle, hwang2008singleparticle, dassarma2014mobility}.  This is the single-particle level broadening associated with the imaginary part of the self-energy and corresponds to a relaxation process including forward scattering (i.e., without the vertex correction, $ 1-\cos\theta $, factor in the scattering rate, which suppresses forward scattering relaxation contributions to the resistivity).  For pure short-range $ s $-wave disorder, the scattering is isotropic, and the transport relaxation is the same as the single-particle relaxation, and hence in regular 3D metals, the single-particle level broadening is simply given by the transport scattering rate since impurity scattering in metals is primarily of short-range nature.  However, in semiconductor systems of our interest, the main disorder scattering arises from highly anisotropic long-ranged charged impurity scattering which is poorly screened by the carriers in InSb and InAs by virtue of the very small electron effective mass making the screening wave vector very small.  Therefore the single-particle quantum scattering rate should be much larger than the transport relaxation rate determining the mobility, thus enhancing the quantum level broadening substantially above the transport level broadening ($\sim$2-4 meV) estimated above.  This is indeed true as we find that the calculated single-particle quantum level broadening in the 2D InSb and InAs samples of interest in the current work is substantially (by more than a factor of 10) larger than the transport broadening entering the mobility calculation. Such large estimated single-particle broadening values, which should manifest in large experimental Dingle temperatures, are another stark reminder of the poor quality of the currently utilized Majorana materials. These results are presented in Appendix~\ref{app:C}. 

In Table~\ref{table:summary_disorder}, we summarize our results for the InSb (from Eindhoven) and InAs (from Copenhagen and Purdue) 2D samples, providing disorder estimates obtained from the comparison between our theory and the experimental mobility data, as described above.  The effective disorder in these samples can be modeled by a background random 3D charged impurity density of $ 1.4\times10^{17} $  to $ 3.4\times10^{19} $  per cm$ ^3 $ , which strictly on dimensional grounds is equivalent to a charge impurity density of $ 10\times10^5 $  cm$ ^{-1} $  to $ 70\times10^5 $  cm$ ^{-1} $  for a 1D system.  In a nanowire, this disorder range corresponds, roughly, to 100 charged impurities to 600 charged impurities per micron, far too high for the realization of topological Majorana zero modes, as we explicitly show in our simulations below. This reinforces the view that signatures of topological MZMs have not yet been seen in hybrid SC-SM systems because the necessary condition for system purity has not yet been achieved. The disorder needs to come down below 10 charged impurities per micron for topological MZMs to emerge in nanowires~\cite{woods2021charge}.

The estimates of the (physical) disorder level summarized in Table~\ref{table:summary_disorder} represent the main result of this section. The next critical task is to evaluate the corresponding effective disorder potential to be used in our simulations of SC-SM nanowires.  
We note in this context that the quantum level broadening, rather than the 2D transport broadening,  may appear as the appropriate quantitative measure of the strength of the effective disorder potential in nanowires, since essentially all transport scattering in 1D systems is forward scattering. This possibility would be rather disturbing, as our calculated level broadening (see Appendix~\ref{app:C}) is 20-100 times larger than the transport broadening, which is itself $2-5~$meV in the 2D samples, as quantified by the peak mobility. We do, however, believe that the mobility broadening rather than the single-particle broadening is the appropriate measure for the nanowire quality since in 1D systems, the impurity scattering is always in the forward direction with no vertex correction.  In any case, the physical quantity characterizing the disorder is the impurity density which is uniquely determined by the measured mobility. Also, we have to take into account the fact that in hybrid superconductor-semiconductor structures additional screening of charge impurities distributed throughout the semiconductor nanowire (or on its surface)  is provided by the parent superconductor, as well as the nearby metallic gates. In addition, the effective potential used in model calculations is not the screened potential itself,  but corresponds to matrix elements of the screened potential with (transverse) wave functions associated with the low-energy subbands, as explained in detail below. Consequently, neither the 2D transport broadening nor the corresponding quantum level broadening can provide good estimates of the effective potential strength. What really matters is the estimated impurity density, which serves as the key disorder parameter. The actual task of evaluating the effective disorder potential associated with a given level of physical disorder (i.e., impurity concentration) has been carried out in Ref. \onlinecite{woods2021charge} for charge impurities randomly distributed throughout the semiconductor nanowire and below, in Sec. \ref{SCD}, for charge impurities on the surface of the semiconductor nanowire. 
In the next two sections we provide simulations of SC-SM nanowires based on a minimal 1D model in the presence of realistic disorder. In  Sec.~\ref{sec:nw} the effective disorder potential is evaluated based on the estimates of the physical disorder level obtained above and the self-consistent results of  Ref. \onlinecite{woods2021charge}. In Sec.~\ref{sec:SC} the effective disorder potential is calculated self-consistently starting from a microscopic model of the hybrid device. For our Majorana simulations using the estimated disorder in the rest of this paper, we use InAs nanowire parameters simply because the current disorder is lower in InAs than in InSb as discussed above (and also because InAs is the experimental focus right now for Majorana search).  The results for InSb would look qualitatively identical, and producing results just by changing the parameters to those of InSb is a useless overkill at this point.

\section{Majorana simulations based on the minimal 1D nanowire model in the presence of strong disorder}\label{sec:nw}

In this section, we evaluate the effect of disorder on the low-energy physics of hybrid semiconductor-superconductor devices based on a minimal 1D nanowire model with random onsite disorder. (The disorder range is approximately incorporated in the theory through a judicious choice of the lattice spacing in the 1D model, so the onsite disorder model roughly corresponds to the appropriately screened Coulomb disorder.) In Sec. \ref{SSec3A} we estimate the strength of the effective onsite disorder potential based on (i) the estimates of the physical disorder levels (i.e., impurity concentrations) in Sec. \ref{sec:transport}, (ii) the results of Ref. \onlinecite{woods2021charge}, where the effective disorder potential associated with charge impurities has been determined explicitly using a self-consistent approach, and (iii) the concept of ``equivalent disorder potentials'', which is introduced below. In Sec. \ref{SSec3B}, we calculate conductance spectra as functions of the applied Zeeman field for different disorder realizations and disorder strengths comparable to or lower than our estimates in Sec. \ref{SSec3A}. The results suggest that the estimated disorder strength corresponding to experimentally available hybrid wires is inconsistent with the presence of topological superconductivity and with topological MZMs localized at the ends of the system. The most likely low field features emerging in these systems are (relatively rare and essentially random) disorder-induced, topologically trivial zero-bias conductance peaks (ZBCPs) generated by (trivial ) Andreev bound states. These findings are strengthened by the results in Sec.   \ref{SSec3C}, where we calculate the ``phase diagrams''  corresponding to the zero-bias conductance as a function of Zeeman field and chemical potential. We believe that it is likely that all existing experimental Majorana observations are reporting these strong disorder-induced trivial subgap Andreev features.

\subsection{The effective disorder potential} \label{SSec3A} 

The low-energy physics of a (clean) semiconductor-superconductor hybrid structure can be accurately described using a multiorbital 1D effective model with ``orbitals'' given by the transverse wave functions $\varphi_\alpha$ associated with the confinement-induced subbands \cite{woods2018effective}. These ``orbitals'' incorporate electrostatic effects generated by the environment (e.g.,  potential gates,  superconducting layer, etc.) and by the free charge. Within this framework,  disorder can be incorporated as a subband-dependent effective potential, $V_{eff}^{\alpha\beta}(z) = \langle \varphi_\alpha|\phi_{dis}|\varphi_\beta\rangle$, where $z$ represents the position along the wire and $\phi_{dis}({\bm r})$ is the (screened) potential generated by the ``physical'' sources of disorder, e.g., by charge impurities. A major simplification occurs when the inter-subband spacing is large compared to the effective disorder potential, since the off-diagonal terms $V_{eff}^{\alpha\beta}$, with $\alpha\neq\beta$, can be neglected and the subbands become independent. Furthermore, since Majorana physics is controlled by the subband $\alpha_o$ closest to the chemical potential, one can focus on the relevant subband and reduce the model to a single orbital effective model, i.e., the well-known minimal 1D nanowire model. Within this approximation, the effective disorder potential becomes $V_{dis} \equiv   V_{eff}^{\alpha_o\alpha_o}$.  
We add here that, if many subbands participate in transport, with inter-subband scattering being important, the effective disorder is substantially enhanced, since the presence of inter-subband coupling acts as an additional source of randomness. Thus neglecting inter-subband scattering and focusing on a single subband is in the appropriate spirit of our focus on the most optimistic realistic disorder model. {We also note that including orbital effects \cite{nijholt2016orbital,manolescu2017majorana,nowak2018renormalization,serra2020evidence,lei2021majorana} (in addition to Zeeman splitting) typically results in a reduction of the topological gap\cite{nijholt2016orbital}, which makes the system more susceptible to disorder. For consistency with our general approach of considering the effects of disorder within otherwise optimal conditions, we do not include orbital effects.}
\begin{figure}[t]
\begin{center}
\includegraphics[width=0.49\textwidth]{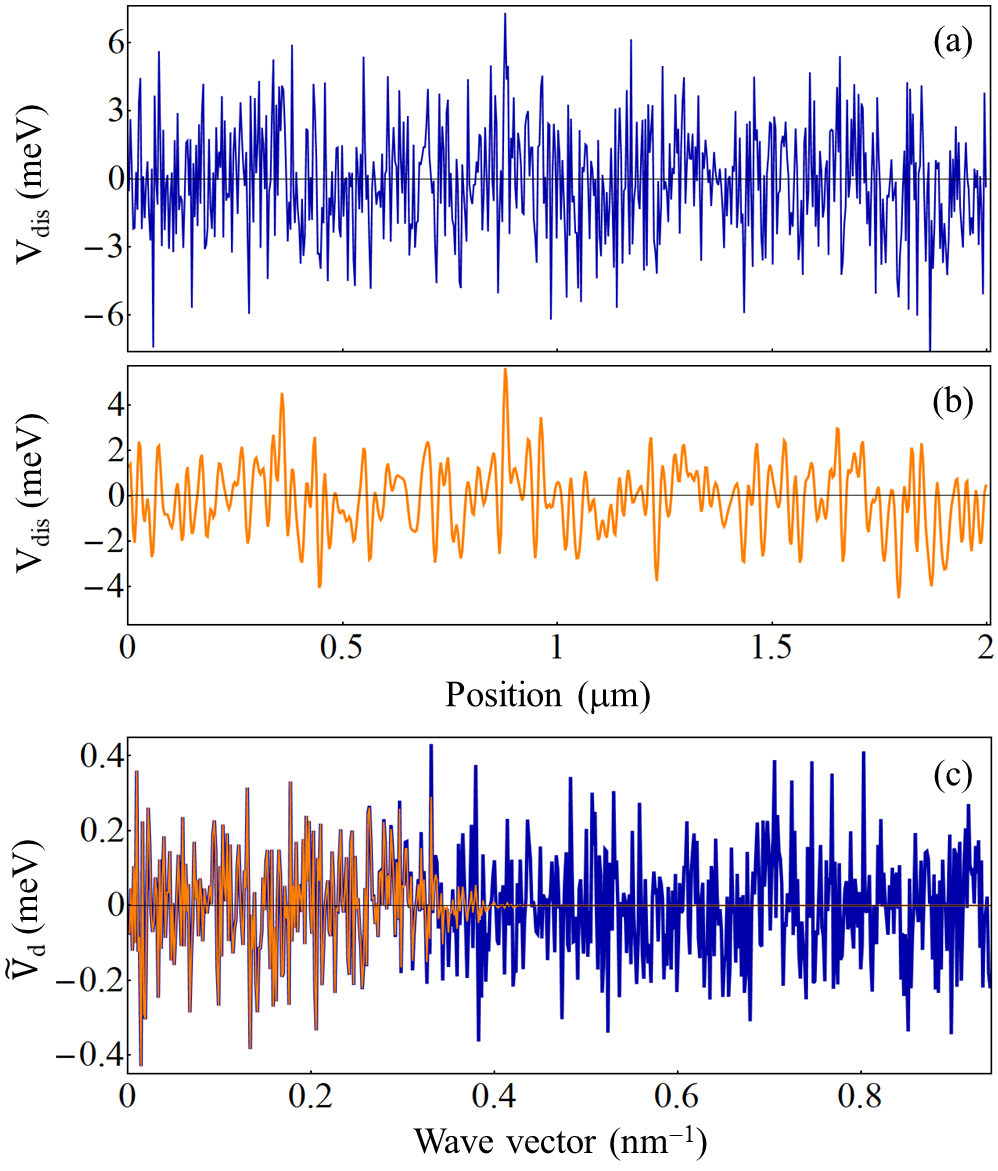}
\end{center}
\vspace{-3mm}
\caption{``Equivalent'' disorder potentials. (a) Onsite random potential drawn from an uncorrelated Gaussian distribution with zero mean and standard deviation $\sigma_\mu = 2.5~$meV. (b) Smoother potential obtained by removing the highly oscillating components of the potential in (a). (c) Fourier components of the two potentials as given by Eq. (\ref{tildeV}). Note that the low wave vector (i.e., long wavelength) components of the two potentials coincide, i.e., the orange and dark blue lines are on top of each other for wave vectors smaller than about $0.35~$nm$^{-1}$. The low-energy BdG spectra generated in the presence of the two ``equivalent'' disorder potentials are shown in Fig. \ref{FigTS2}.}
\label{FigTS1}
\vspace{-1mm}
\end{figure}

The first problem that we address concerns the features of $V_{dis}$ that are most relevant in relation to Majorana physics. In essence, since Majorana physics is controlled by the topmost occupied subband (i.e., the subband closest to the chemical potential), the low-energy states have relatively small characteristic momenta, i.e.,  long-wavelength oscillatory features, typically on the order of tens to hundreds of nanometers.  More specifically, 
we can define the characteristic Majorana ``oscillation length'' $\lambda_M \sim \pi \hbar/\sqrt{2 m^* \epsilon_o}$, where $\epsilon_o$, the characteristic energy associated with Majorana physics, is on the order of $1~$meV. We note that $\lambda_M$ is different from (and should not be confused with) the Majorana ``localization length'' $\xi$, which characterizes the (exponential) decay of the wave function describing the Majorana bound state. 
For an effective mass $m^*=0.026m_o$, the characteristic Majorana length is $\lambda_M \approx 25-60~$nm. In general, $V_{dis}$ has features characterized by multiple length scales.
The key property of the disorder potential is that only the components characterized by length scales comparable to or larger than $\lambda_M$ are relevant for Majorana physics. Indeed, the rapidly varying components of $V_{dis}$  having characteristic length scales smaller than $\lambda_M$ get ``averaged out'' and have a minimal impact on the low-energy physics \cite{zeng2021partiallyseparated}. By contrast, components of the disorder potential with characteristic length scale comparable to or larger than $\lambda_M$ have a major impact on the low-energy physics once their amplitude becomes comparable to or exceeds the Majorana energy scale $\epsilon_o$. In particular, long-wavelength components having length scales much larger than $\lambda_M$ can act as effective ``smooth potentials'' \cite{kells2012nearzeroenergy,stanescu2019robust}, which leads to local Majorana physics and the emergence of partially separated Majorana modes, instead of a topological superconducting phase supporting well separated MZMs \cite{zeng2021partiallyseparated}. Furthermore, if the dominant component of the disorder potential has a length scale comparable to $\lambda_M$, even the emergence of local Majorana physics can be suppressed. 

\begin{figure}[t]
\begin{center}
\includegraphics[width=0.49\textwidth]{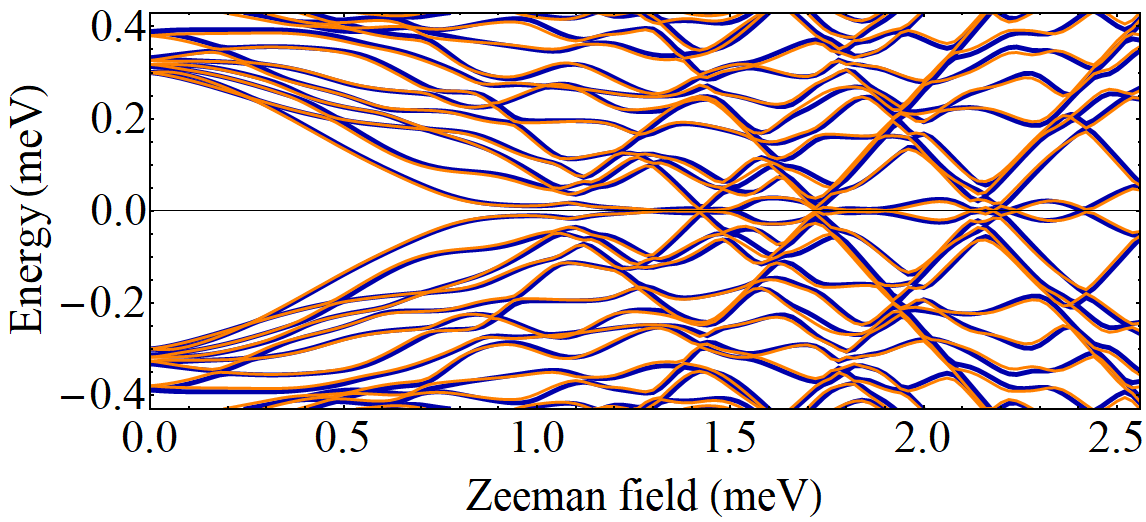}
\end{center}
\vspace{-3mm}
\caption{Low-energy spectra as  functions of the Zeeman field for a system with chemical potential $\mu=1.5~$meV and effective disorder potentials given by Fig. \ref{FigTS1}(a) (dark blue lines) and  Fig. \ref{FigTS1}(b) (orange lines). Note that the two spectra are, practically, on top of each other, which establishes the ``equivalence'' of the two disorder potentials. Also note that, while in a clean system the bulk gap closes (and then reopens) at a critical Zeeman field slightly above $1.5~$meV, simultaneously with the emergence of a MZM, in the disordered system a near-zero energy (topologically trivial) mode emerges at a significantly lower field.}
\label{FigTS2}
\vspace{-1mm}
\end{figure}

To illustrate this key  property of the disorder potential, we consider a system of length $L=2~\mu$m described by the minimal 1D model [see, for example, Eq. (\ref{HamBdG})] with $V_{dis}$ corresponding to an onsite random potential drawn from an uncorrelated Gaussian distribution with zero mean and standard deviation $\sigma_\mu = 2.5~$meV. The (fixed) model parameter values are: effective mass $m^* = 0.03 m_e$ (where $m_e$ is the bare electron mass), Rashba spin-orbit coupling coefficient $\alpha_R = 250~$meV$\cdot$\AA, and  superconducting pairing amplitude $\Delta = 0.3~$meV. The system is discretized on a lattice with lattice constant $a = 3.33~$nm. The position dependence of $V_{dis}$ corresponding to a specific disorder realization is shown in Fig. \ref{FigTS1}(a). The Fourier components of the effective disorder potential defined as
\begin{equation}
\widetilde{V}_d(k_n) = \frac{2a}{L}\sum_{i}V_{dis}(z_i) \sin(k_n z_i), \label{tildeV}
\end{equation}
where $k_n = n\pi/L$ is the wave vector and $z_i$ is the position corresponding to lattice site $i$, are given by the blue line in Fig. \ref{FigTS1}(c). Note that the edge of the Brillouin zone is at $\pi/a \approx 0.94~$nm$^{-1}$ and that the typical values of  $\widetilde{V}_d(k_n)$ are independent of $k_n$, reflecting the local nature of the disorder potential. 

Next, we remove the components of the Fourier spectrum with $k_n$ larger than about $0.35~$nm$^{-1}$, while retaining the low-wave vector components, as shown by the orange line in  Fig. \ref{FigTS1}(c). The position dependence of the corresponding (real space) disorder potential is shown in 
  Fig. \ref{FigTS1}(b). Using this procedure, we obtain two rather different looking disorder potentials [see Fig.  \ref{FigTS1}, panels (a) and (b)] characterized by identical low-$k$ Fourier spectra. We then solve the BdG equation corresponding to the two disorder potentials and calculate the low-energy spectrum as a function of the Zeeman field for a system with chemical potential $\mu=1.5~$meV. The results are shown in Fig. \ref{FigTS2}, with the blue and orange lines corresponding to the two disorder potentials from Figs. \ref{FigTS1}(a) and~\ref{FigTS1}(b), respectively. Note that the two spectra are, practically, on top of each other, revealing the fact that the two (rather different) disorder potentials have the same effect on low-energy physics. Hence, we can view the disorder potentials as being ``equivalent'' from the point of view of their impact on low-energy physics.
  Furthermore, if we focus on the qualitative features relevant for Majorana physics (e.g., the emergence of well-separated MZMs, versus the presence of partially separated Majorana modes, or of ``standard'' Andreev bound states), without considering the quantitative details (e.g., the exact value of the Zeeman field associated with the emergence of a low-energy mode), we can relax this definition of ``equivalent disorder potentials'' by requiring the identity of the Fourier spectra for $k_n$ less than about $\pi/\lambda_M \lesssim 0.1~$nm$^{-1}$. Hence, from the perspective of their impact on the Majorana physics, the effective disorder potentials can be divided into equivalence classes defined by the property that two equivalent potentials have the same Fourier spectrum for wave numbers lower than the inverse Majorana ``oscillation length'', i.e., lower than the characteristic Fermi momentum.   

The second problem that we want to address concerns the relationship between the spectral properties of the effective disorder potential and the characteristics of the physical source of disorder, e.g., the impurity concentration and characteristic length scale associated with the single-impurity potential. We note that this type of analysis has to be carried out for each type of disorder. Here, we focus on a hybrid system with charge impurities randomly distributed throughout the semiconductor nanowire,  for which the critical task of determining the effective disorder potential associated with a given type of physical source of disorder has been accomplished in Ref. \onlinecite{woods2021charge}.  
For simplicity, we consider a phenomenological modeling of charge impurity-induced disorder corresponding to the effective disorder potential \cite{woods2021charge}
\begin{equation}
V_{dis}(z) = \sum_{j} A_j V_{imp}(z, z_j),  \label{Vchargeimp}
\end{equation}
where $A_{imp}$ is an amplitude having random sign, average absolute value $1.7~$meV, and variance of the absolute value $0.7~$meV, while $V_{imp}(z, z_j) = \exp(-|z-z_j|/\lambda)$, with $\lambda = 15~$nm and $z_j$  representing the position along the wire of the j$^{\rm th}$ impurity. The values of these parameters are based on the numerical results of Ref. \onlinecite{woods2021charge}. A specific disorder realization corresponds to a set of impurity locations, $\{z_j\}$, and single-impurity amplitudes, $\{A_j\}$, where  $A_j$ carries information about the type of impurity (i.e., positive or negative charge) and its transverse location (which determines the magnitude of  $|A_j|$ \cite{woods2021charge}). The number of impurities is determined by the linear impurity concentration $n_{imp}$, i.e., the number of impurities per unit length. Here, we assume charge neutrality, i.e.,  an equal number of positive and negative charge impurities. 

\begin{figure}[t]
\begin{center}
\includegraphics[width=0.49\textwidth]{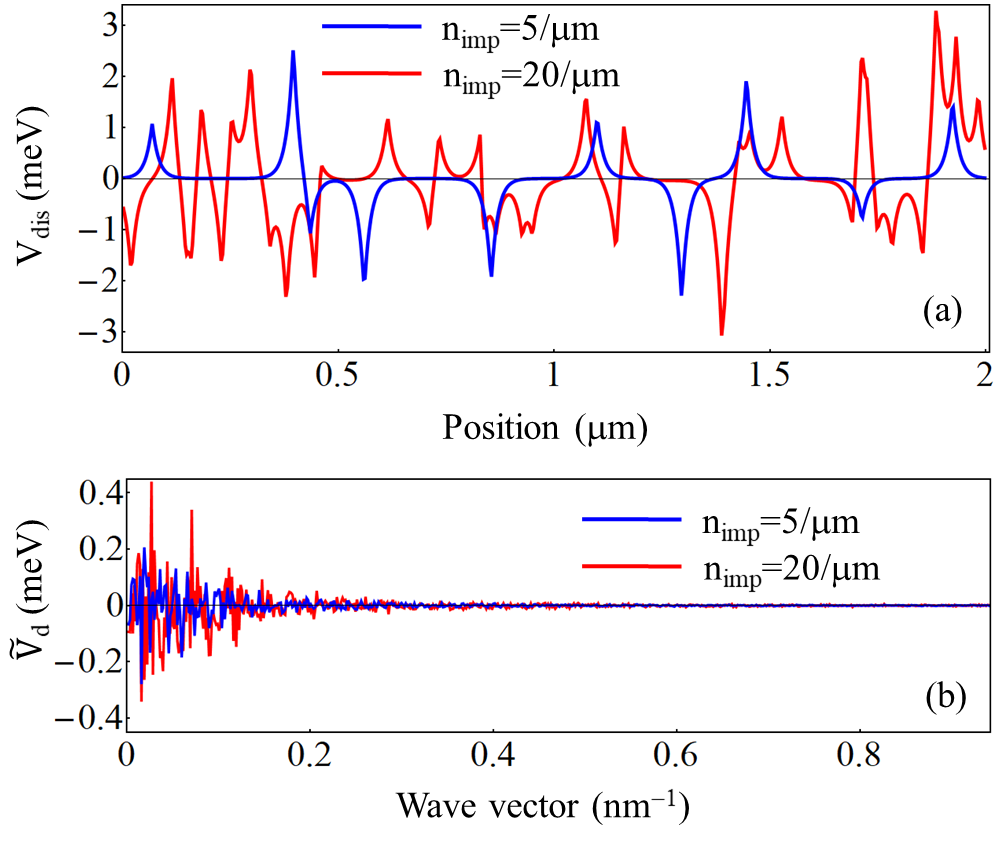}
\end{center}
\vspace{-3mm}
\caption{(a) Position dependence of the effective disorder potential generated by charge impurities  for  two specific disorder realizations corresponding to impurity densities $n_{imp}=5/\mu$m (blue line) and  $n_{imp}=20/\mu$m (red line). The general form of this type of effective disorder potential is given by Eq. (\ref{Vchargeimp}).}
\label{FigTS3}
\vspace{-1mm}
\end{figure}

Two examples of effective disorder potentials generated by charge impurities that  correspond to impurity concentrations $n_{imp}=5/\mu$m (blue line) and  $n_{imp}=20/\mu$m (red line) are shown in Fig. \ref{FigTS3}(a). The corresponding Fourier transforms are shown in Fig. \ref{FigTS3}(b). Note that, in striking contrast with the Fourier spectrum of the onsite random potential [dark blue line in Fig. \ref{FigTS1}(c)], the components of the charge impurity potential with wave numbers larger than about $0.2~$nm$^{-1}$ are negligible. This reflects the presence of a finite length scale ($\lambda=15~$nm) associated with this type of potential. Also notice that the typical amplitude of $\widetilde{V}_d$ corresponding to $n_{imp}=5/\mu$m [blue line in Fig. \ref{FigTS3}(b)] is manifestly smaller than the typical amplitude of the potential corresponding to $n_{imp}=20/\mu$m  [red line in Fig. \ref{FigTS3}(b)]. To make this observation more quantitative, it is useful to define the disorder-averaged absolute value of the Fourier transformed disorder potential, $\langle|\widetilde{V}_d(k_n)|\rangle$. Note that this quantity characterizes the {\em type} of disorder under consideration, rather than a specific disorder realization, and can be viewed as a {\em spectral signature} of that type of disorder. 

\begin{figure}[t]
\begin{center}
\includegraphics[width=0.49\textwidth]{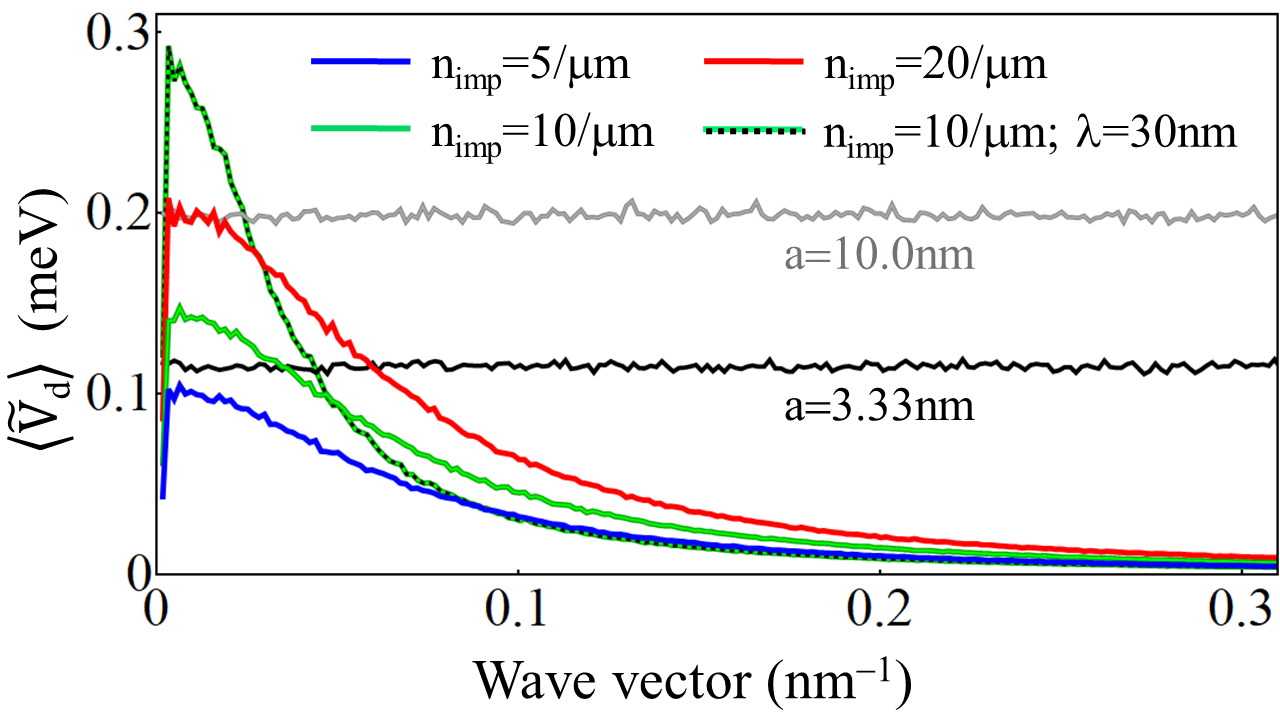}
\end{center}
\vspace{-3mm}
\caption{Spectral signatures, $\langle|\widetilde{V}_d(k_n)|\rangle$, corresponding to different types of disorder. The disorder averaging was done using 3000 disorder realizations for a system of length $L=2~\mu$m. The blue, green, and red lines correspond to the charge impurity-induced disorder given by Eq. (\ref{Vchargeimp}) with $\lambda=15~$nm and impurity concentrations $n_{imp}=5/\mu$m, $10/\mu$m, and $20/\mu$m, respectively, while the dashed green/black line corresponds to $\lambda=30~$nm and $n_{imp}=10/\mu$m. Note that all these curves collapse onto a single curve when using the scaling given by Eq. (\ref{scal}).  The (horizontal) black and gray lines correspond to an onsite random potential drawn from an uncorrelated Gaussian distribution with zero mean and standard deviation $\sigma_\mu = 2.5~$meV. Note the dependence on the lattice constant $a$.}
\label{FigTS4}
\vspace{-1mm}
\end{figure}

Examples of spectral signatures corresponding to different types of disorder are provided in Fig. \ref{FigTS4}. First, let us focus on the blue, green, and red lines, which represent the spectral signatures of effective disorder potentials generated by charge impurities with $\lambda=15$nm and concentrations $n_{imp}= 5/\mu$m, $10/\mu$m, and $20/\mu$m, respectively. Let us notice that $\langle|\widetilde{V}_d|\rangle$ has a peak at low values of the wavenumber and becomes negligible for $k_n$ larger than about $0.2~$nm$^{-1}$, which is consistent with the features characterizing the specific disorder realizations shown in Fig. \ref{FigTS3}(b). As discussed above, the values of $\langle|\widetilde{V}_d|\rangle$ at low wave vectors (below approximately $0.1~$nm$^{-1}$) determine the impact of disorder on the Majorana physics.
An important property revealed by the spectral signatures is that the strength of the disorder potential and, implicitly, its effect on the low-energy physics, increase with the impurity concentration. Specifically, $\langle|\widetilde{V}_d|\rangle$ scales with 
$\sqrt{n_{imp}}$. Of course, deviations from this simple scaling law are expected in the limit of large impurity concentrations, where interference effects become important. A direct consequence of the practical importance of this scaling property is that reducing the amplitude of the charge impurity-induced disorder potential by a factor $f$ requires lowering the impurity concentration by a factor $f^2$. 

Next, we consider the dashed green/black line in  Fig. \ref{FigTS4}, which corresponds to an impurity concentration $n_{imp}= 10/\mu$m (same as the green line), but involves a single impurity potential $V_{imp}$ with a larger characteristic length, $\lambda = 30~$nm. The values of $\langle|\widetilde{V}_d|\rangle$ at low wave vectors increase by roughly a factor of two as compared with the green line ($\lambda = 15~$nm). This implies that, given an impurity concentration,  the effect of the disorder potential on the Majorana physics becomes stronger as one increases the characteristic length scale $\lambda$. We note that the dependence of the spectral signature on the characteristic length $\lambda$ also follows a scaling law. More specifically, and taking into account the dependence on $n_{imp}$ discussed above, as well as the dependence on the wire length $L$, one can collapse all spectral signatures associated with charge impurity disorder (including the blue, green, red, and dashed green/black lines in Fig. \ref{FigTS4}) into a single curve using the scaling
\begin{equation}
\frac{\lambda_0}{\lambda}\sqrt{\frac{n_{imp}^0 L}{n_{imp} L_0}}\bigg\langle\bigg\vert\widetilde{V}_d\left(\frac{\lambda}{\lambda_0}k_n\right)\bigg\vert\bigg\rangle,    \label{scal}
\end{equation}
where $\lambda_0$, $L_0$, and $n_0$ are (fixed) reference values for the characteristic length scale, wire length, and impurity concentration, respectively. Finally, we note that the Fourier components of the charge impurity-induced disorder potential defined by Eq. (\ref{tildeV}), as well as the corresponding spectral signature  $\langle|\widetilde{V}_d|\rangle$ are independent of the lattice constant $a$, provided it is small enough to correctly capture the relevant short length scale physics, which, of course, is a general requirement for Majorana simulations.  

The third problem addressed in this section concerns the legitimacy of using a random onsite potential as a model for the effective disorder potential. We note that, while this type of disorder potential is widely used in the literature, its connection with a specific disorder mechanism (e.g., charge impurities,  point defects, atomic vacancies, etc.) and, ultimately, its physical relevance remain unclear. We start by noticing that the random onsite potential has a $k$ -independent spectral signature, consistent with its purely local nature. Two examples are shown in Fig. \ref{FigTS4} (black and gray lines). Note the dependence on the lattice constant $a$, in sharp contrast with the spectral signatures associated with charge impurity-induced disorder. The ``universal'' spectral signature of the onsite random Gaussian potential is a horizontal line of height $\sigma_\mu^0/\sigma_\mu\sqrt{(L a_0)/(L_0 a)}~\!\langle|\widetilde{V}_d|\rangle$, where $\sigma_\mu^0$, $L_0$, and $a_0$ are (fixed) reference values for the variance , wire length, and lattice constant, respectively.  
Next, we exploit the ``equivalence'' relation between different disorder potentials and the properties of the charge impurity-induced disorder discussed above. Using the examples shown in Fig. \ref{FigTS4}, we note that for small wave vector values the gray line provides a good approximation for the red line. More generally,  the range over which an onsite random potential can reasonably approximate the long-wavelength (short $k_n$) features of an impurity-induced potential increases with decreasing $\lambda$. If this range is comparable to $\pi/\lambda_M$, the two types of disorder can generate ``equivalent'' disorder potentials. In other words, the onsite random potential represents a good model for short-range disorder having characteristic length scale(s) smaller than the Majorana oscillation length $\lambda_M$. For longer range disorder, we expect the onsite random potential to still capture some important qualitative features, since one can always match the low-$k_n$ components of the two types of disorder, which are of critical importance for the low-energy physics, but it loses its quantitative relevance. Finally, we emphasize that the actual strength of the random onsite potential depends not only on the variance $\sigma_\mu$, but also on the size of unit cell $a$ used in the discretization procedure. Using the example in Fig. \ref{FigTS4}, an onsite potential with $\sigma_\mu=2.5~$meV and $a=3.33~$nm is approximately equivalent to a charge impurity-induced potential with characteristic length scale $\lambda=15~$nm and impurity concentration $n_{imp}\approx 6-7/\mu$m. By contrast, an onsite potential having the same variance, but on a lattice with $a=10~$nm is approximately equivalent with a charge impurity-induced potential corresponding to an impurity concentration $n_{imp}\approx 18-20/\mu$m , which has a significantly stronger effect on the low-energy physics. 

\begin{figure*}[t]
    \centering
    \includegraphics[width=6.8in]{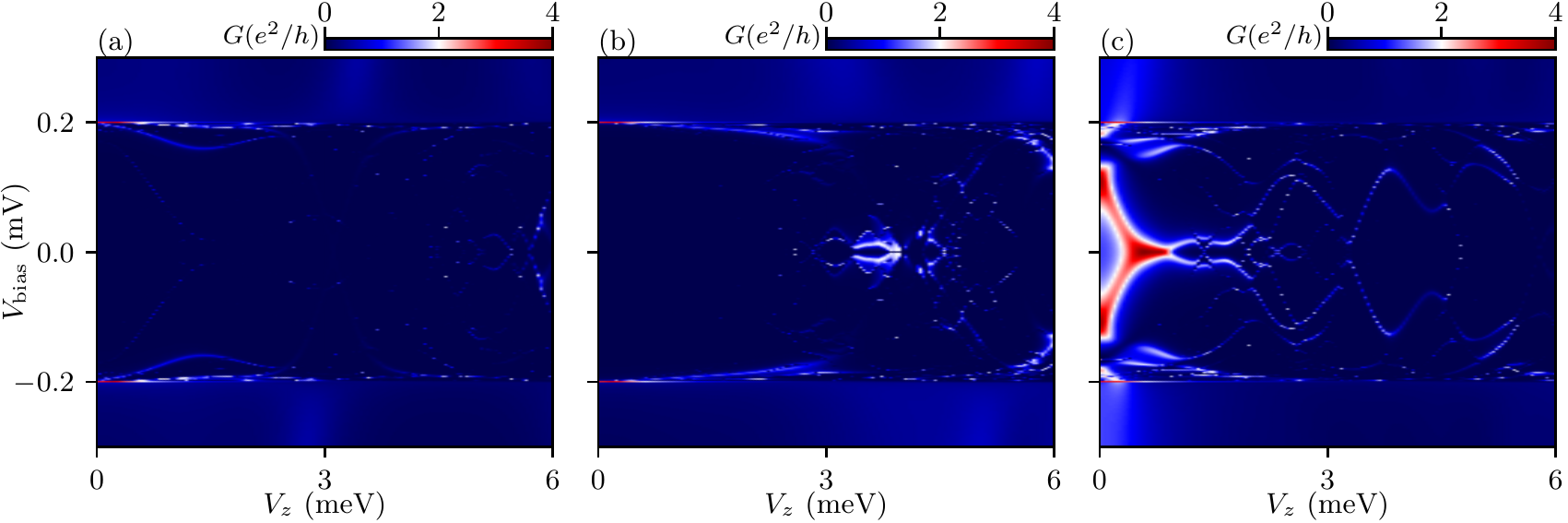}
    \caption{Conductance as a function of the applied  Zeeman field and bias voltage in the presence of three different disorder realizations with $\sigma_\mu=5$ meV. These examples are selected from a set of 120 different disorder realizations. Most ``samples'' (i.e., disorder realizations) show no low-energy features [similar to panel (a)], while occasionally one notices some disorder-induced, essentially random subgap features [see panel (b)].  For three disorder realizations times (out of 120) we have obtained low-field  ZBCPs, one example is shown in panel (c). 
    The  parameters used in the calculation are: chemical potential $\mu=5$ meV, parent superconductor gap $\Delta_0=0.2 $meV, superconductor-semiconductor coupling strength $\gamma=0.2$ meV, spin-orbit coupling strength $0.5$ eV\AA, wire length $L=20~\mu$m, barrier heights 20 meV. The details of calculation are provided in Appendix~\ref{app:D}.}
    \label{fig:11_muVar5}
\end{figure*}

We conclude this section with an estimate of the variance $\sigma_\mu$ characterizing an onsite random potential that would be consistent (i.e., approximately ``equivalent'') with a disorder potential generated by charge impurities with densities given by the estimates obtained in Sec. \ref{sec:transport}. We assume that the impurities are well screened, so that the typical amplitude of the single impurity potential is about $1~$meV and its characteristic length scale $\lambda = 7.5~$nm, both values being near the lower end of the ranges calculated in Ref. \onlinecite{woods2021charge}. Such a short-range potential would also justify the use of the onsite random potential model, as explained above. If we consider now the estimates of the impurity density obtained in  Sec. \ref{sec:transport} (see Table \ref{table:summary_disorder}), focusing on the 3D densities for InAs (in the quantum well), and assuming a wire geometry similar to that in Ref. \onlinecite{woods2021charge}, we obtain $n_{imp}$ values that are larger by a factor of 50-700 than the impurity density associated with the red line in Fig. \ref{FigTS4}. Finally, under the assumption that the scaling relation (\ref{scal}) still holds at large impurity densities, we estimate the variance characterizing the approximately ``equivalent'' onsite random potential on a lattice with lattice constant $a=10~$nm as being $\sigma_\mu \approx 5-20~$meV. As shown below, this places the experimentally available SC-SM hybrid structures in the strong (or even extreme) disorder regime.

\subsection{Charge tunneling spectra in the presence of strong disorder} \label{SSec3B}    

Our estimates of the impurity density based on the 2D samples analyzed in Sec.~\ref{sec:transport} suggest that the levels of physical disorder (e.g., charge impurities) present in experimentally available  Majorana nanostructures could be up to three orders of magnitude higher than the ``intermediate'' disorder regime discussed in Ref. \onlinecite{woods2021charge}. Moreover, the SC-SM hybrid structures might have even higher disorder than the 2D systems because of additional processing and the existence of SC-SM interfaces, which generate additional sources of disorder. This situation makes it imperative to understand in detail the impact of strong disorder on the low-energy physics of the hybrid structures. What low-energy phenomenology should one expect in the presence of strong disorder? Here, we address this question by performing model calculations based on the well-established 1D Majorana nanowire model in the presence of strong disorder. Since details regarding the exact nature of the sources of disorder, as well as the properties of the corresponding effective disorder potential are not available, we work under the (rather optimistic) assumption that the relevant type of disorder is short-range disorder (e.g., point defects, well-screened charge impurities, etc.). Under this assumption, it is appropriate to model the effective disorder potential as random onsite disorder. This model describes accurately the short-range disorder regime, as discussed in Sec. \ref{SSec3A}. 

\begin{figure*}[t]
    \centering
    \includegraphics[width=6.8in]{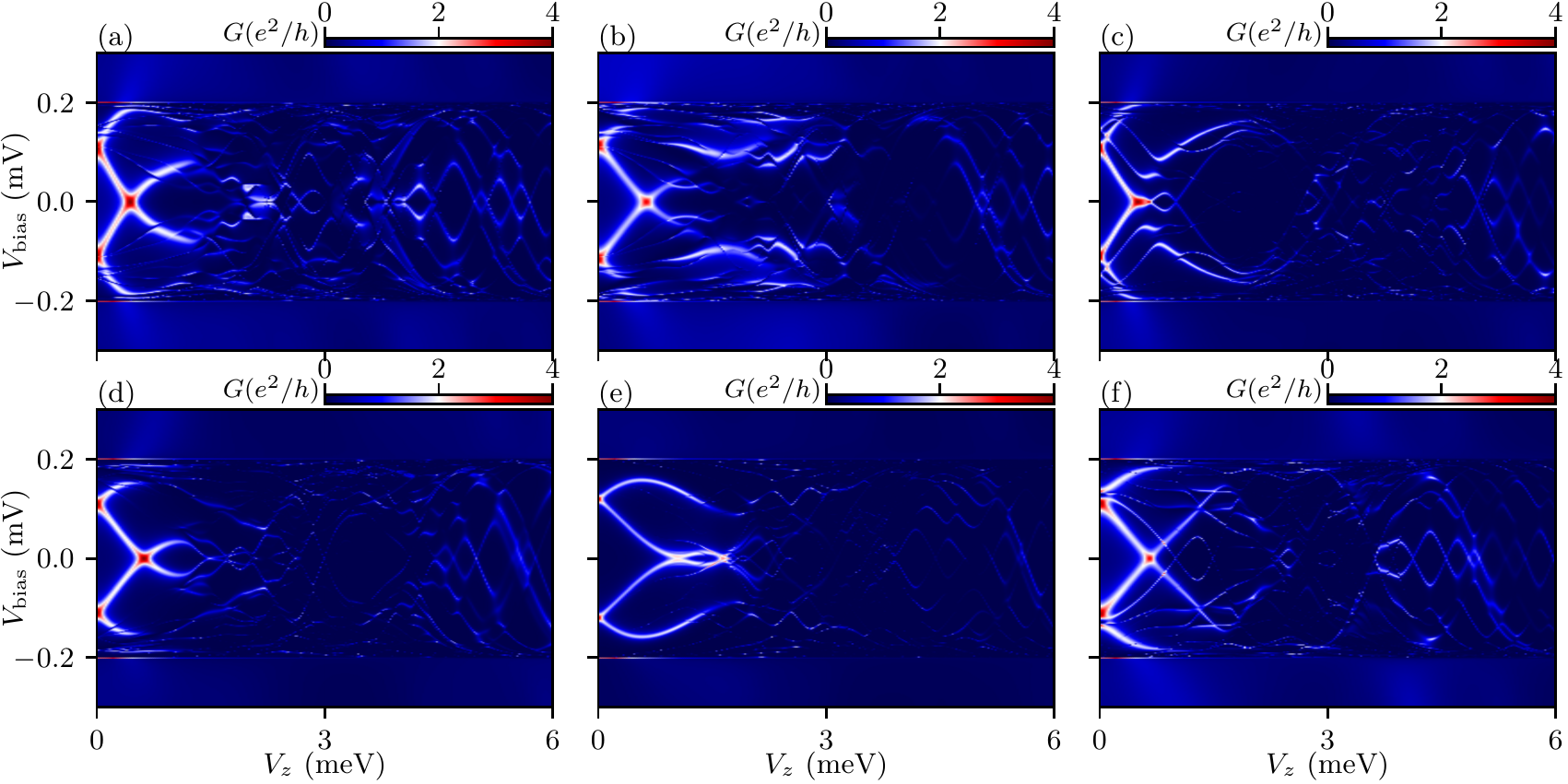}
    \caption{Conductance as a function of the applied  Zeeman field and bias voltage in the presence of six different disorder realizations with $\sigma_\mu=3$ meV. Only these six disorder configurations (out of 120) support a low-field ZBCP (at either the left or the right end of the system). 
   The chemical potential is $\mu=3~$meV, while the other parameters are the same as in Fig.~\ref{fig:11_muVar5}.}
    \label{fig:12_muVar3}
\end{figure*}

The theory (for details see Appendix~\ref{app:D}) includes the ``standard'' ingredients, i.e., proximity-induced superconductivity,  spin-orbit coupling, and Zeeman splitting, as well as onsite disorder modeled by a random Gaussian potential with zero mean and variance $\sigma_\mu$. We focus on the lower end ($5-8~$meV) of the range estimated in Sec. \ref{SSec3A} based on the results of Sec. \ref{sec:transport}. In addition, we also consider lower values of $\sigma_\mu$ ($1-4~$meV) to make a connection with the known results corresponding to intermediate disorder. 
We note that the weaker disorder situation (with disorder broadening $ < 1~$meV) has already been studied extensively~\cite{brouwer2011topological,lutchyn2011search,akhmerov2011quantized,sau2012experimental,liu2012zerobias,hui2015bulk,sau2013density,liu2017andreev,haim2019benefits,pan2020physical,pan2021threeterminal,pan2021disorder,dassarma2021disorderinduced}.
We emphasize that the calculation is exact within the free fermion BdG theory (see Appendix~\ref{app:D}) and provides the eigenstates and eigenenergies of the system.  Without any disorder, the pristine results (not shown, since they are well-known) manifest topological MZMs at the wire ends for Zeeman fields larger than a critical value associated with the topological quantum phase transition (TQPT), where the bulk gap closes.  
Finite systems manifest (end-to-end-correlated) MZM oscillations~\cite{dassarma2012splitting}, but these features were never seen experimentally, most likely because the systems are not clean enough. Note that the parent superconducting gap gets quenched by external magnetic fields exceeding a certain value $B^*$. This field-induced bulk SC gap collapse, most likely arising from the orbital effect of the applied field penetrating the parent superconductor, is a persistent problem in all nanowire experiments, preventing the high-field regime from being experimentally accessible. We do not include this bulk gap collapse in the theoretical simulations, since it is a nonessential effect that has little to do with disorder in this context. However, we note that the Zeeman field $V_z^*$ corresponding to $B^*$ sets the upper bound for the disorder strength consistent with the emergence of MZMs since $B>B^*$  is experimentally inaccessible. More specifically, any effective disorder potential having the amplitude of the relevant long-wavelength components (see Sec.  \ref{SSec3A}) larger than $V_z^*$ is inconsistent with the presence of genuine MZMs localized at the ends of the system. An optimistic estimate of $B^*=2~$T for a system with effective g-factor $g=45$ gives $V_z^* \approx 5~$meV. For most experimentally available hybrid structures $V_z^*$ is probably on the order of $1~$meV. 

Consistent with the existing Majorana nanowire experimental studies, we focus on tunnel spectroscopy, where the appearance of stable quantized $ 2e^2/h $ zero-bias conductance peaks (ZBCPs) is expected to represent a signature of topological MZMs. Note, however, that all observed ZBCPs may very well be generated by disorder-induced trivial Andreev bound states, as none of them has passed a quantifiable stability requirement, end-to-end correlation requirement, or any other of the more MZM-specific criteria. 
Starting with a disorder potential with $\sigma_\mu = 5~$meV, we calculate the tunneling spectrum as a function of the Zeeman field for fixed chemical potential, $\mu=5~$meV, and different disorder realizations. Three cases are shown in Fig.~\ref{fig:11_muVar5}. We note that the corresponding clean system is characterized by a finite gap for Zeeman fields below the critical value, $V_z < 5~$meV, and by the emergence of an MZM-induced ZBCP at higher fields, although the high-field regime is probably irrelevant because $V_z > V_z^*$, i.e., the parent superconducting gap collapses.  
In the presence of disorder, the tunneling spectra typically show no low-energy features, as illustrated in  Figs.~\ref{fig:11_muVar5}(a) and~\ref{fig:11_muVar5}(b). This is consistent with experiments, where most samples manifest no sub-gap features.  Once in a while, there may be some disorder-induced, essentially random subgap features, like the feature in Fig.~\ref{fig:11_muVar5}(b) near $V_z\approx 4~$meV. However, these rare features are neither stable nor generic, i.e., they do not occur inside well-defined regions of the parameter space.
Rarely, some low-field zero bias features may manifest, as in Fig.~\ref{fig:11_muVar5}(c), along with ``gap closing'' features associated with Andreev bound states coming together. Such disorder-induced ZBCPs are rare, unstable, and typically have magnitudes different from $ 2e^2/h $,  although they may be fine-tuned to $2e^2/h $  by varying the tunnel barrier.  These ZBCPs are neither topological (since they occur outside the nominally topological region) nor nonlocal, and never emerge simultaneously when tunneling from both ends.  In the simulations leading to Fig.~\ref{fig:11_muVar5}, only 3 out of 120 disordered ``samples'' (i.e., disordered configurations with the same variance) have manifested any kind of observable ZBCPs (from either the left or the right end). This situation is strikingly similar to the experimental situation, where most samples show nothing in their tunnel spectra, while finding features that mimic Majorana physics requires sample selection and considerable fine-tuning. 

\begin{figure}[h]
    \centering
    \includegraphics[width=0.4\textwidth]{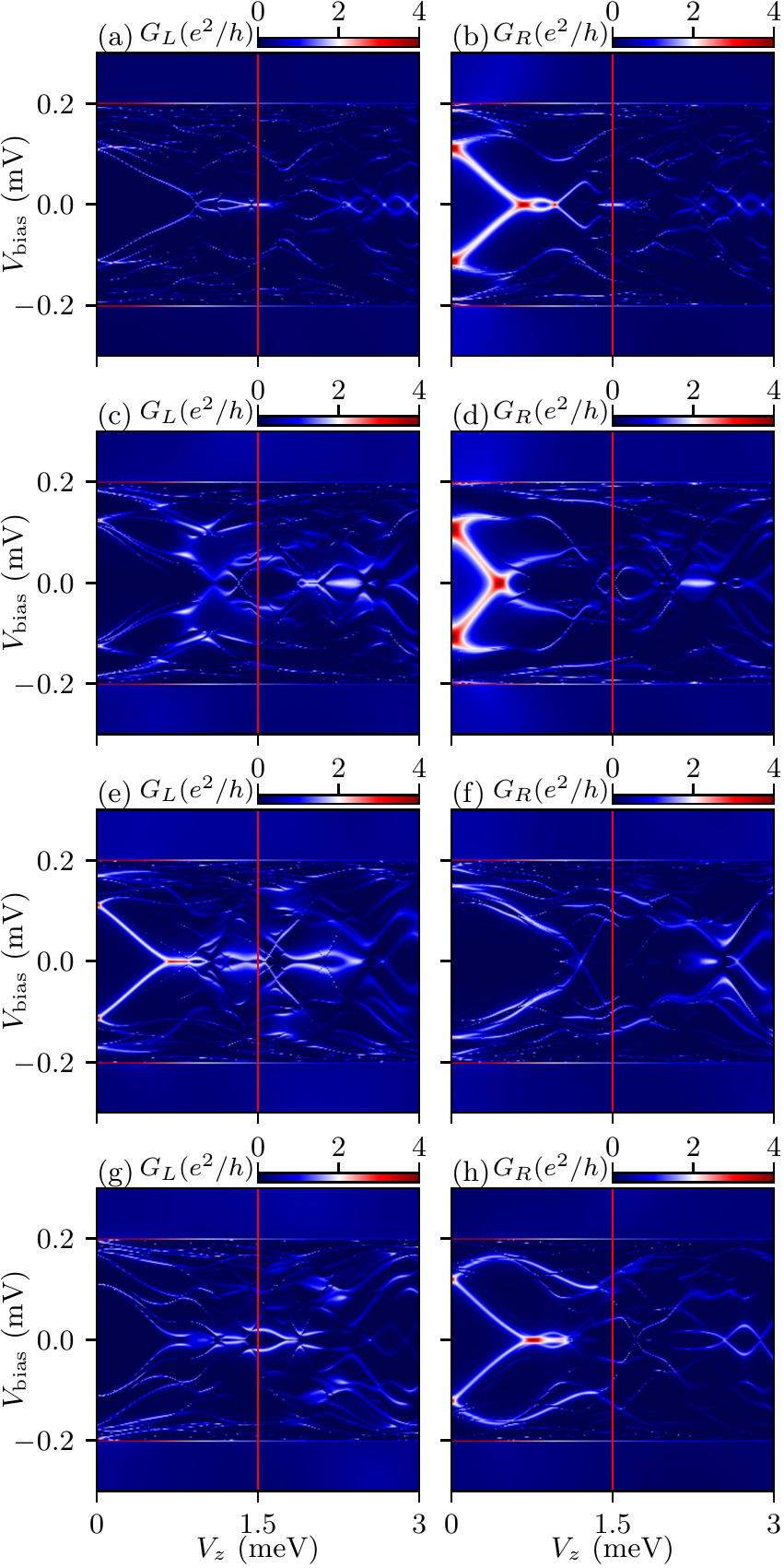}
    \caption{Conductance traces measured from the left end (left column) and the right end (right column) of the hybrid system as functions of the Zeeman field and bias voltage in the presence of disorder with $\sigma_\mu=2$ meV. The pairs of spectra (a)-(b), (c)-(d), (e)-(f), (g)-(h) correspond to different disorder realizations. The parameters used in the calculation are: chemical potential $\mu=1.5~$meV, parent superconductor gap $\Delta_0=0.2~$meV, superconductor-semiconductor coupling strength $\gamma=0.2~$meV, spin-orbit coupling strength $0.5~$eV\AA, wire length $L=3~\mu$m, and tunnel barrier heights $10~$meV.}
    \label{fig:13-16_muVar2}
\end{figure}

These results suggest that $\sigma_\mu = 5~$meV already represents strong disorder, which is inconsistent with the emergence of MZMs localized at the ends of the wire. This conclusion is further supported by the results presented in  Sec. \ref{SSec3C}, where we explore the dependence of the low-energy features on the chemical potential and Zeeman field. Since our estimate of the disorder strength based on the  2D transport calculations of Sec. \ref{sec:transport} is $\sigma_\mu \approx 5-20~$meV, we conclude that existing SC-SM hybrid structures are in all likelihood deep inside the strong disorder regime. 

Next, we lower the disorder strength below the range estimated in Sec. \ref{SSec3A}, to make a connection with the intermediate/low disorder regime and determine the maximum level of disorder consistent with the realization of topological superconductivity and Majorana zero modes. 
However, for $\sigma_\mu \sim 3~$ meV the situation hardly changes, with most samples still showing almost no zero bias features and only  6 out of 120 configurations manifesting some (nongeneric) low-field ZBCPs,  as shown in Fig.~\ref{fig:12_muVar3}. Again, these are nontopological disorder-induced ZBCP features that are nongeneric, unstable, and characterized by a random ZBCP strength. Note that we never find ZBCPs when tunneling from both ends in a given system, as must happen for topological MZMs.  All these ZBCPs are below the TQPT (associated with the pristine wire), and most ($>90$ \%) of the tunnel spectra manifest no ZBCPs at all. 

Further reducing the disorder strength to $\sigma_\mu = 2~$meV leads, occasionally, to the possibility of ZBCPs appearing above the TQPT, as shown in Fig. \ref{fig:13-16_muVar2}. We note that the ZBCPs appearing above the TQPT, although rather weak and not persistent as a function of the Zeeman field, are correlated from both ends of the wire (see Fig. \ref{fig:13-16_muVar2}), which is a clear signature of Majorana physics. 
 We emphasize, however, that almost all the low field ZBCPs are still topologically trivial ZBCPs occurring below the TQPT, with no correlations from the two ends. Also, most tunnel spectra are still random, with little zero bias features; note that in Fig. \ref{fig:13-16_muVar2} we have selected a few spectra that do manifest some ZBCP features in the simulations. Nonetheless, since the possibility exists for topological ZBCPs to manifest once in a while, we establish $\sigma_\mu = 2~$meV as the (approximate) upper bound of the disorder strength consistent with Majorana physics in this system (i.e., typical InAs-Al or InSb-Al SC-SM hybrid structures). 

\begin{figure}[t]
    \centering
    \includegraphics[width=0.4\textwidth]{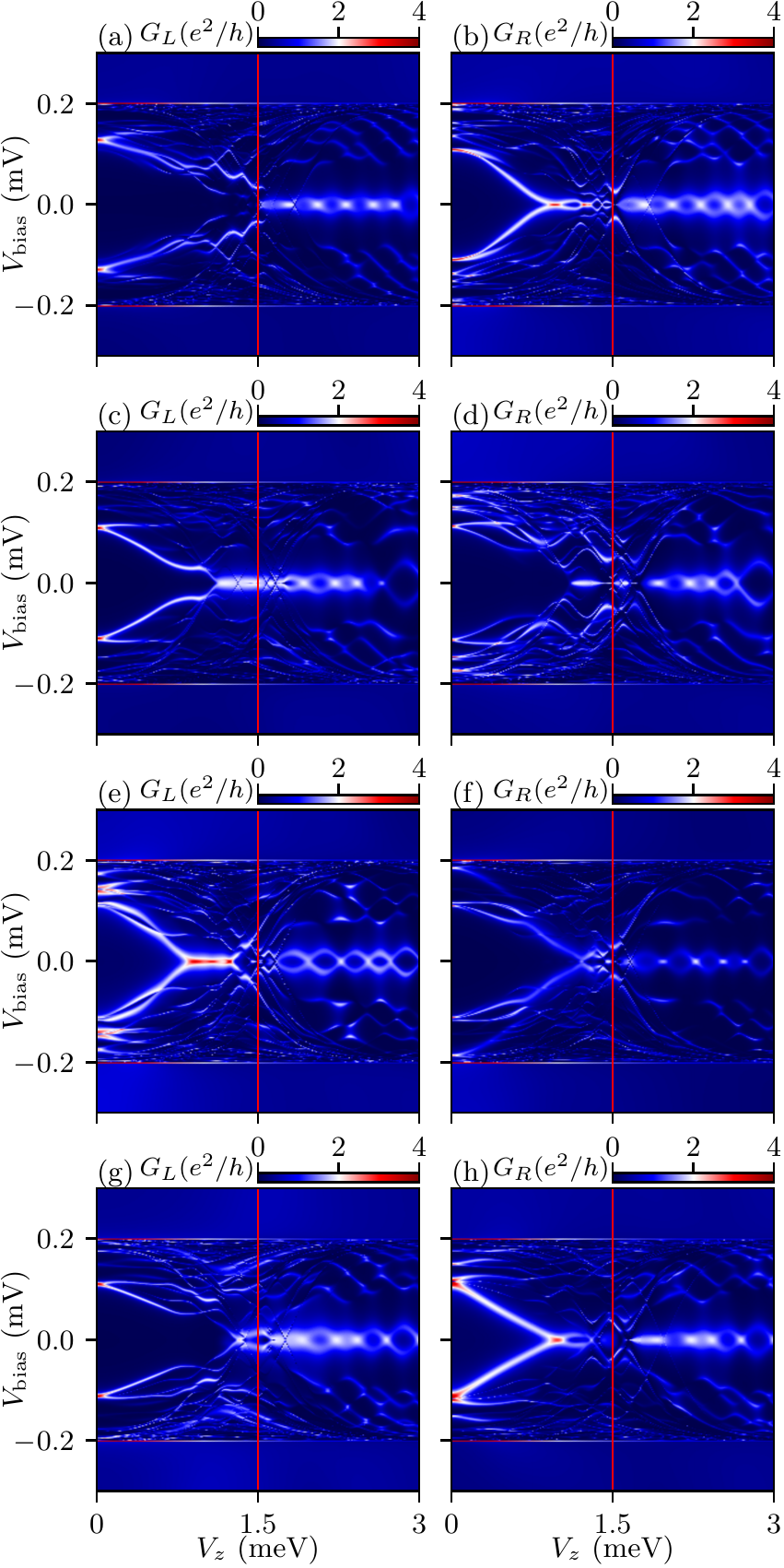}
    \caption{Conductances measured from the left end (left column) and the right end (right column) as a function of the Zeeman field and the bias voltage in the presence of disorder with $\sigma_\mu=1$ meV. The parameters are: chemical potential $\mu=1.5$ meV, parent superconductor gap $\Delta_0=0.2 $meV, superconductor-semiconductor coupling strength $\gamma=0.2$ meV, spin-orbit coupling strength $0.5$ eV\AA, wire length $L=3~\mu$m, barrier heights at the interface of the lead and nanowire is 10 meV.}
    \label{fig:17-20_muVar1}
\end{figure}

Finally, we consider the case $\sigma_\mu = 1~$meV, with a few representative spectra being shown in  Fig. \ref{fig:17-20_muVar1}. Note that topologically trivial ZBCPs are still present below the TQPT, but correlated features associated with the presence of MZMs emerge consistently in the topological regime. The low-field conductance features emerging in the trivial regime (i.e., below the TQPT) have a striking resemblance to similar features characterizing higher disorder samples (see Figs. \ref{fig:11_muVar5}-\ref{fig:13-16_muVar2}) and to the best available experimental Majorana nanowire tunneling data~\cite{zhang2021large,nichele2017scaling}. 
Indeed, large ZBCPs with conductance $>2e^2/h$ have been reported.  Typically, these observations involve considerable post-selection and fine-tuning of the control parameters, consistent with our estimated low probability of having such  ZBCPs in high-disorder samples. Furthermore, the observed features are uncorrelated from the two ends, nongeneric, and unstable, existing only over narrow regimes of magnetic field sweeps, which suggests that these features are topologically trivial, like the corresponding features in our calculation. On the other hand,  features similar to the ZBCPs emerging above the TQPT in our theoretical results shown in Fig. \ref{fig:17-20_muVar1} have never been reported in the experimental literature.  In particular, no hint of Majorana oscillations (apparent in some of our results above the TQPT) has ever been reported experimentally. Also, there has been no report of the experimental observation of end-to-end correlated low-energy features, or of stable ZBCPs with conductance $\sim 2e^2/h$.  
Our analysis suggests that the fundamental reason for not being able to observe these basic Majorana features is that the experimentally available nanowire samples are in 
the strong disorder regime corresponding to $\sigma_\mu \gtrsim 3~$meV, consistent with our estimate in Sec. \ref{SSec3A}. 
We note that nanowire samples are expected to have higher disorder than the corresponding 2D materials. Also, we emphasize that, although in principle one could access the topological regime even in the presence of relatively strong disorder by sufficiently increasing the Zeeman field (and going to sufficiently low temperatures), this possibility is limited by the persistent experimental problem associated with the high-field collapse of the bulk Al superconductivity.

The simulations presented in this section, which are based on a minimal model of the hybrid device, show that in currently available samples disorder is strong enough so that it prevents the system from achieving topological superconductivity. The typical low-energy features that emerge in currently available samples are likely to be (occasional) ZBCPs associated with disorder-induced trivial Andreev bound state. Our strong disorder simulations are completely consistent with the experimental claims of occasional large fine-tuned trivial ZBCPs, which are neither stable nor correlated from the two wire ends and never manifest Majorana oscillations.

\subsection{Zero-bias conductance ``phase diagrams'' } \label{SSec3C} 

In the previous section, we have analyzed tunneling spectra as functions of the Zeeman field for fixed values of the chemical potential. The natural question is whether or not our conclusions regarding the presence/absence of Majorana-induced features in samples characterized by a certain disorder strength hold for arbitrary chemical potential values within the relevant range consistent with the emergence of topological superconductivity in clean samples. To address this question, we focus on the zero-bias conductance and investigate its dependence on Zeeman field and chemical potential, which generates ``phase diagrams'' that, in the clean limit,  converge toward the well-known topological phase diagram of the hybrid system.

\begin{figure}[h]
    \centering
    \includegraphics[width=0.38\textwidth]{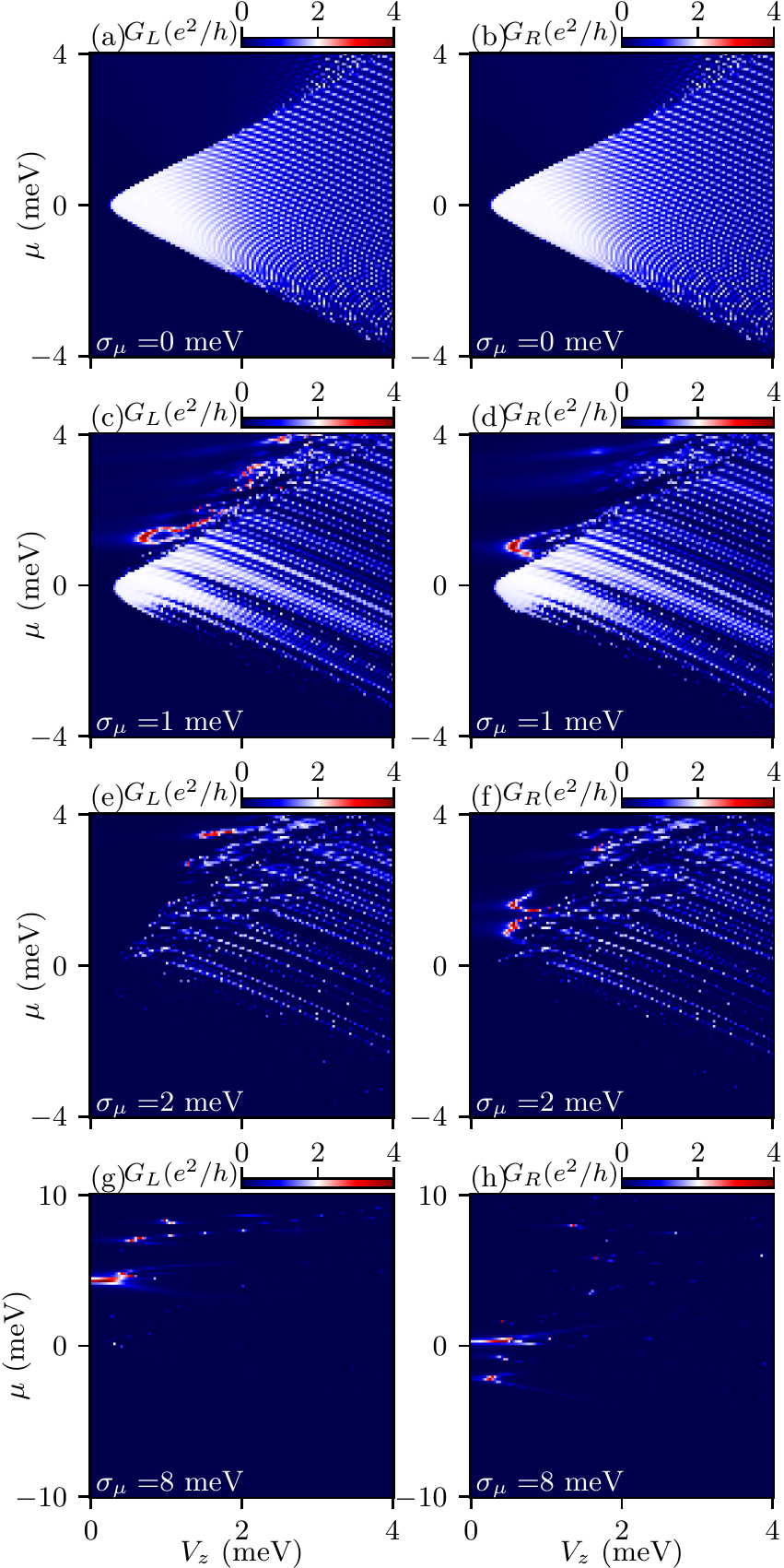}
    \caption{``Phase diagrams'' representing  zero-bias conductance maps as functions of the Zeeman field ($V_z$) and chemical potential ($\mu$) for systems with increasing disorder strength. The left and right columns correspond to tunneling into the left and right ends of the system, respectively.   
   (a)-(b) correspond to the pristine case,  $\sigma_\mu=0$; (c)-(d) correspond to a low-disorder sample, $\sigma_\mu=1$ meV; (e)-(f) illustrate the intermediate disorder case, $\sigma_\mu=2$ meV; (g)-(h) represent an example of strong disorder, $\sigma_\mu=8$ meV. The other parameters used in the calculation are: parent superconductor gap $\Delta_0=0.2~$meV, superconductor-semiconductor coupling strength $\gamma=0.2~$meV, spin-orbit coupling strength $0.5~$eV\AA, wire length $L=3~\mu$m, tunnel barrier height $10~$meV above the chemical potential in the nanowire.}
    \label{fig:21}
\end{figure}

Indeed, as shown in  Fig.~\ref{fig:21} (a) and (b), the zero-bias conductance of a pristine wire is exponentially small in the trivial regime  (i.e., for $V_z \lesssim\sqrt{\mu^2+\gamma^2}$; see Appendix~\ref{app:D} for details) and reaches the quantized value, $2 e^2/h$ (white in Fig. \ref{fig:21}),  in the topological regime ($V_z\gtrsim\sqrt{\mu^2+\gamma^2}$). Note that for $V_z > 2~$meV the quantized region breaks into stripy features that disperse down in $\mu$ with increasing $V_z$. This is the effect of finite size-induced Majorana oscillations; as a result of these oscillations, the ZBCP splits and the zero-bias conductance is quantized only in the vicinity of the ``nodes'', where the energy of the in-gap Majorana mode vanishes. The stripy (white) features simply trace the position of these nodes in the $V_z-\mu$ plane. Also note that there is a perfect correlation between the features characterizing the left and right conductance, which clearly indicates that they are generated by MZMs localized at the two ends of the system.   We emphasize that the ``phase diagrams'' in Fig.~\ref{fig:21} (a) and (b) reveal three basic features associated with Majorana physics, none of which was observed experimentally: robust ZBCP quantization (associated with the presence of finite, relatively large white areas in the phase diagram), Majorana oscillations (associated with the stripy features), and perfect end-to-end correlation (associated with the perfect correlation between the features characterizing the left and right conductance). 

Introducing some weak disorder corresponding to $\sigma_\mu=1~$meV modifies the phase diagram, but does not destroy the basic features associated with Majorana physics, as shown in Figs.~\ref{fig:21}(c) and~\ref{fig:21}(d).
In particular, one can clearly identify the signatures associated with robust ZBCP quantization, Majorana oscillations, and end-to-end correlation. However, we notice that these features are reduced or even absent in certain areas of the nominally topological region. In addition, new features emerge in the topologically trivial regime. Some of these features correspond to small islands with conductance exceeding the quantized value (and occasionally approaching $4e^2/h$); these features are generated by disorder-induced trivial Andreev bound states consisting of strongly overlapping Majorana components. We also notice the presence of a few small quantized (white) islands, which indicate the presence of disorder-induced partially separated Majorana modes that mimic the local properties of MZMs  \cite{zeng2021partiallyseparated}. We emphasize that these trivial features are not end-to-end correlated, which reveals their essentially local nature. 

Further increasing the disorder strength to $\sigma_\mu=2~$meV has a major impact on the phase diagram, as shown in  Figs.~\ref{fig:21}(e) and~\ref{fig:21}(f). On the one hand, the low-field region with $0\lesssim \mu\lesssim 4~$meV is dominated by disorder-induced, uncorrelated, topologically trivial features. On the other hand, for high-enough $V_z$ one can still observe correlated stripy features indicative of Majorana oscillations and topological superconductivity. Note, however, that for a (realistic) value of the maximum field associated with the collapse of the parent superconducting gap  $V_z^*=2~$meV, only a small region in the vicinity of $V_z\approx 1.5~$meV, $\mu\approx 0$  would contain such Majorana features. Furthermore, this property is not generic, in the sense that for certain disorder realizations the Majorana features occur only above $V_z^*$ and, consequently, are not observable. These properties are consistent with our results in Sec. \ref{SSec3B} (see, in particular, Fig. \ref{fig:13-16_muVar2} and the accompanying text) and justify our identification of $\sigma_\mu=2~$meV as the (approximate) upper bound of the disorder strength consistent with Majorana physics, or, in other words, as representing the  ``intermediate'' disorder regime, where signatures of topological superconductivity and Majorana physics may or may not be present, depending on the specific disorder realization (i.e.,  nanowire sample). 

\begin{figure}[t]
    \centering
    \includegraphics[width=0.4\textwidth]{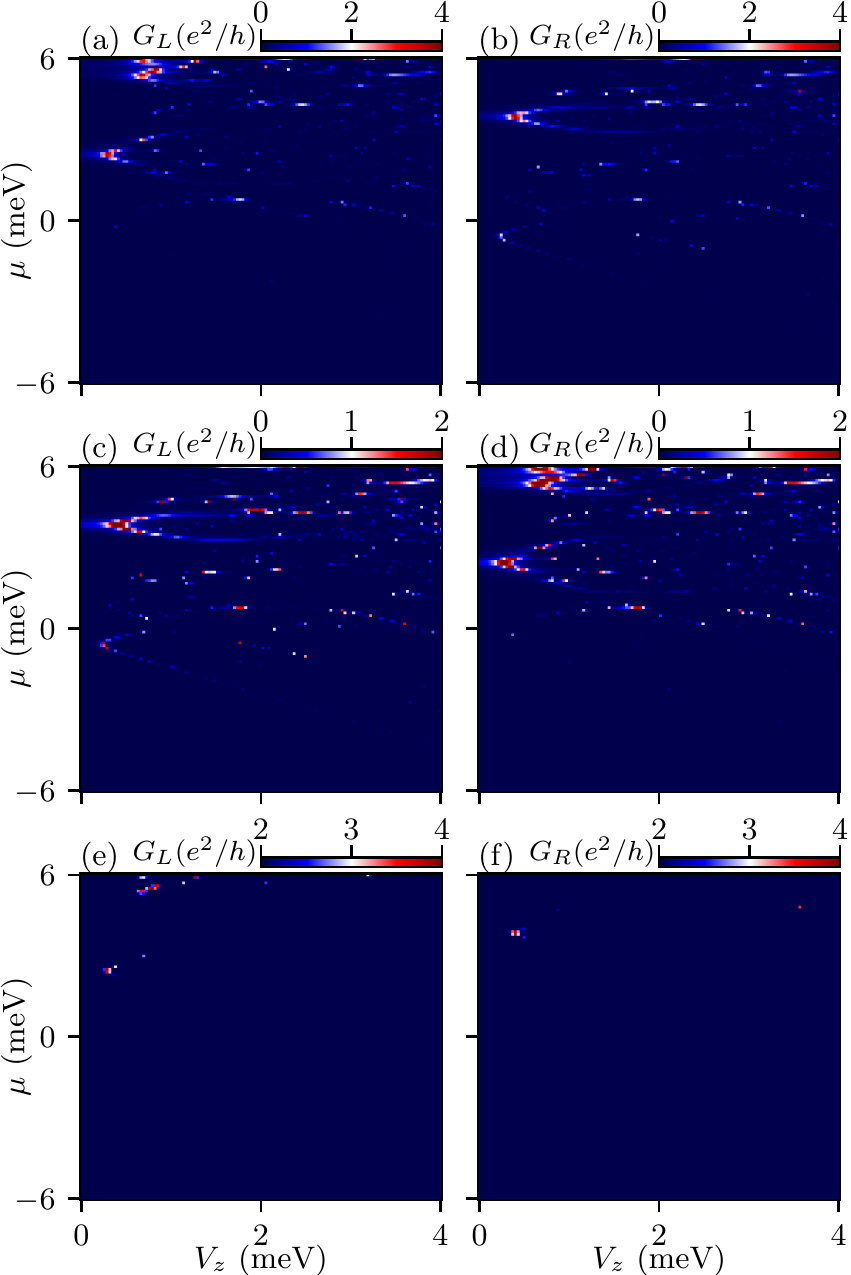}
    \caption{``Phase diagrams'' representing  zero-bias conductance maps as functions of the Zeeman field ($V_z$) and chemical potential ($\mu$) for a disordered system with $\sigma_\mu=4$ meV. The left and right columns correspond to tunneling into the left and right ends of the system, respectively.   
    (a)-(b) maps generated using a conductance range from 0 to $4e^2/h$; (c)-(d) same data represented using a conductance range from 0 to $2e^2/h$ (with conductance values larger than $2e^2/h$ saturated to red); (e)-(f)  same data represented using a  conductance range from $2e^2/h$ to $4e^2/h$ (with conductance smaller than $2e^2/h$ saturated to blue).
    The system parameters are the same as in Fig.~\ref{fig:21}.}
    \label{fig:22}
\end{figure}

Next, we delve deep into the strong disorder regime and consider an example corresponding to $\sigma_\mu=8~$meV. Note that this is still within the lower half of our estimated disorder strength based on the 2D transport calculations in Sec. \ref{sec:transport} (see Sec. \ref{SSec3A} for details). The ``phase diagrams'' for the left and right conductances are shown in Figs. \ref{fig:21}(g) and~\ref{fig:21}(h), respectively. There is absolutely no feature associated with topological superconductivity and Majorana physics. Furthermore, the only high-conductance (and low-field) features are a few small, isolated islands characterized by conductance values typically exceeding $2e^2/h$, which are clearly associated with disorder-generated, topologically trivial Andreev bound states. We emphasize that fine-tuning the control parameters (i.e., $V_z$ and $\mu$) near the boundary of such an island can always generate a quantized ZBCP with a value $\sim 2e^2/h$, particularly since the tunnel barrier can be fine-tuned to change the conductance at low temperatures. Of course, this has nothing to do with Majorana physics, not even with local Majorana physics, which is associated with the presence of partially separated Majorana modes and leads to the emergence of quantized islands in the ``phase diagram'' \cite{zeng2021partiallyseparated}. On the other hand, the structure of the ``phase diagram'' is consistent with our findings in Sec. \ref{SSec3B}, in particular with the small probability of finding a (low-field) ZBCP in the presence of strong disorder, and with the experimental situation, which involves sample selection and fine-tuning. Unfortunately, these features are also consistent with all the experimentally reported ZBCPs.

Finally, we consider a case of disorder strength $\sigma_\mu=4~$meV, which corresponds to the lower end of the strong disorder regime. The results are shown in Fig.~\ref{fig:22} using three different scales for the magnitude of the zero-bias conductance: a ``regular'' scale, $ 0 - 4e^2/h$, in panels (a) and (b), a low-conductance scale, $ 0 - 2e^2/h$ (with conductance values larger than $2e^2/h$ saturated to red),  in panels (c) and (d), and a high-conductance scale,  $2 - 4 e^2/h$ (with conductance smaller than $2e^2/h$ saturated to blue), in panels (e) and (f). The results have the same general characteristics of the strong coupling regime discussed above: no clear signature of Majorana physics and a few small, uncorrelated (essentially random) islands of large conductivity. Most of the features have conductivity values lower than $2e^2/h$ [see Figs.~\ref{fig:22}(c) and~\ref{fig:22}(d)], with very few maxima exceeding the quantized value [see Figs.~\ref{fig:22}(e) and~\ref{fig:22}(f)]. A few tiny quantized islands may signal the presence of local (quasi-Majorana) physics. This is similar to the experimental situation, where only a small fraction of the parameter space consisting of (essentially random) islands is consistent with large conductance values. Observing such features requires a great amount of effort to fine-tune various gate voltages to find them. On a quantitative note, we point out that the density of zero-bias conductivity features in Fig.~\ref{fig:22} is higher than in Figs.  \ref{fig:21}(g) and~\ref{fig:21}(h), which is consistent with an increased probability of having low-field ZBCPs as disorder decreases. Qualitatively though,  Fig.~\ref{fig:22} is characteristic of the strong disorder regime, with all low-field features being associated with disorder-induced, topologically trivial Andreev bound states. 

We conclude this section with a summary of the main results and a comment regarding their immediate relevance to the experimental effort. We have introduced the concept of ``equivalent'' disorder potentials, based on their long-wavelength spectral properties. Using this concept, we have argued that the random onsite disorder model is ``equivalent'' to short-range disorder, e.g., disorder generated by well-screened charge impurities. Based on the estimates of physical disorder (i.e., impurity densities) from Sec. \ref{sec:transport} and the results of Ref.  \onlinecite{woods2021charge}, we have evaluated the strength of the ``equivalent'' random onsite potential corresponding to experimentally available hybrid nanostructures as being $\sigma_\mu \approx 5-20~$meV. This places the experimental samples inside the strong disorder regime, which is characterized by disorder-induced, topologically trivial conductance features that emerge as small, isolated and essentially random island in the parameter space. Observing such features would require sample and data post-selection and fine-tuning, which is consistent with the actual experimental situation, further confirming our disorder strength estimate. 

Considering this situation, one should make the experimental investigation of disorder, together with a systematic effort to reduce it, the top priorities in this field. In this context, we note that reaching the intermediate disorder regime (which would enable the observation of some Majorana features) would require a reduction of the effective disorder potential amplitude by a factor of $2.5-10$, which implies reducing the impurity density by $1-2$ orders of magnitude. Finally, we note that generating experimental ``phase diagrams'' similar to those in Figs. \ref{fig:21} and \ref{fig:22} for multiple nominally identical devices, which are completely within the existing technical capabilities, would provide valuable information regarding the actual disorder strength in the available samples. In this context, even finding that $98\%$ of the samples show nothing (i.e., produce basically featureless phase diagrams), would represent a significant result. By contrast, showing (only) some ``interesting'' ZBCPs that occur in $2\%$ of the samples (after significant fine-tuning) is not only potentially misleading, but provides no information regarding the underlying disorder. One definitive conclusion of our results is that experimentalists should be strongly discouraged from just publishing claims of MZM observation through large ZBCPs, which are never generic and always first post-selected and further fine-tuned, but should be strongly encouraged to publish (or at least make available) all their data including all tunnel conductance spectra showing no subgap features.

\section{Self-consistent theory of nanostructures with disorder induced by surface charge impurities} \label{SCD}

In this section we perform two critical tasks: (i) we evaluate the effective disorder potential generated by the presence of charge impurities on the surface of the semiconductor nanowire and (ii) we investigate the fate of Majorana physics in a hybrid system with surface charge disorder. The first task is accomplished using a microscopic model of the nanostructure that incorporates the electrostatic environment (e.g., gate potentials, free charge, etc.) by solving the corresponding Schr\"odinger-Poisson problem self-consistently. Next, based on the effective disorder potential calculated self-consistently, we determine the {\em Majorana separation length}~\cite{woods2021charge} defined as the minimum distance between the leftmost Majorana mode and the other Majorana modes associated with low-energy BdG states within a certain energy window.  We find that, for typical values of the surface charge density characterizing InAs nanowires, the Majorana separation length is comparable to the characteristic length scale of the Majorana modes over the relevant range of control parameters (i.e., Zeeman field and chemical potential). This implies that the Majorana modes are strongly overlapping and shows that Majorana physics is inconsistent with this level of surface charge-induced disorder. We note that the calculation of the charge impurity-induced effective disorder potential presented below goes beyond the independent impurity approximation used in Ref. \onlinecite{woods2021charge}. More specifically, the potential generated by multiple impurities is not calculated as a sum of single-impurity potentials (calculated separately), but is determined directly for a finite wire containing many surface impurities. While the independent impurity approximation is expected to be accurate in the low/intermediate disorder regime, the nonperturbative multi-impurity calculation presented here is the appropriate approach to strong disorder. The results confirm our conclusion in Sec. \ref{sec:nw}, namely that experimentally available samples are most likely in the strong disorder regime.

\subsection{Modeling} \label{sec:SC} 

\begin{figure}[t]
    \begin{center}
    \includegraphics[width=0.48\textwidth]{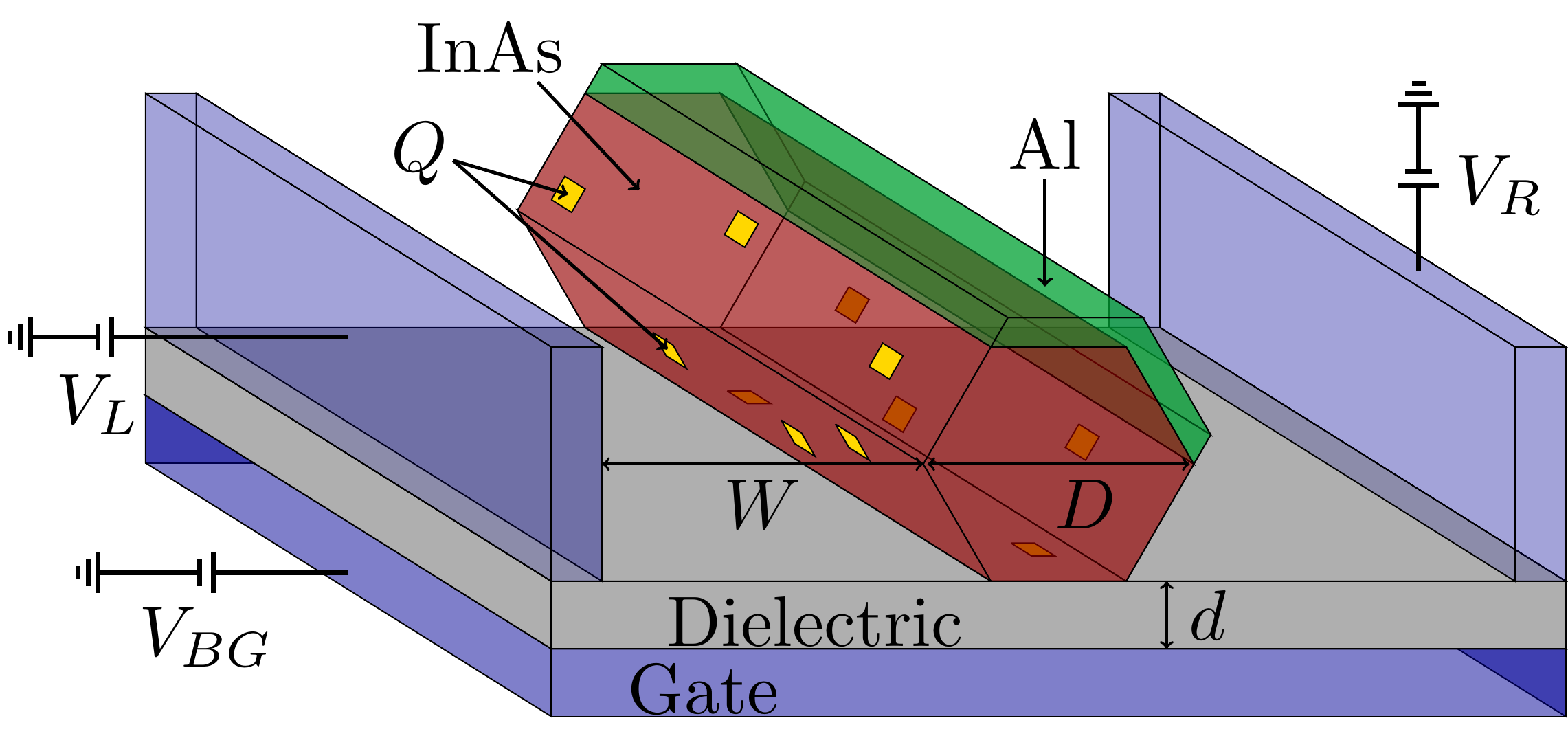}
    \end{center}
    \vspace{-0.5cm}
    \caption{Schematic representation of a hybrid device consisting of a semiconductor (InAs) nanowire (red) of diameter $D$ proximity coupled to a thin superconductor (Al) layer (green). Metallic gates (blue) control the electrostatic environment. Charge impurities (yellow squares), each carrying charge $Q$, are randomly placed on the facets of the nanowire that are not covered by Al and produce a nonuniform potential along the length of the device.}
    \label{FIG_SCD1}
    \vspace{-1mm}
\end{figure}

We consider the semiconductor-superconductor hybrid device shown schematically in Fig. \ref{FIG_SCD1}, which consists of a semiconductor (InAs) nanowire (red) of diameter $D$ proximity coupled to a thin superconductor layer (Al; green) deposited on two facets of the nanowire. Three metallic gates (blue) are used to control the electrostatic environment, specifically to tune the band edges of the low-energy subbands of the InAs nanowire close to the Fermi level, so that Majorana physics may occur. This includes a bottom gate, which is separated from the InAs nanowire by a dielectric of thickness $d$, and two side gates placed at a distance $W$ from the nearest vertices of the nanowire. With the exception of small details, this setup corresponds to the hybrid semiconductor-superconductor devices most frequently used in the experimental study of Majorana physics \cite{mourik2012signatures,deng2012anomalous,das2012zerobias,chang2015hard,albrecht2016exponential,chen2017experimental,moor2018electric,lee2019selectivearea,bommer2019spinorbit,shen2021full,yu2021nonmajorana}. The final and most important ingredient included in the model are charge impurities (yellow squares) placed randomly on the facets of the InAs nanowire that are not covered by Al. These impurities form a charge accumulation layer that is known to occur on the surfaces of InAs nanowires \cite{olsson1996charge} and are believed to be generated by either ionized hydrogen impurities attaching to the surface or native point defects \cite{weber2010intrinsic,castleton2013hydrogen}. We note that a surface charge has been included in a few previous theoretical investigations of Majorana nanostructures \cite{winkler2019unified,escribano2019effects,woods2020subband,liu2021electronic}. However, these studies assumed a uniform surface charge density on the uncovered facets of the device, which yields a translation-invariant potential along the length of the wire. In stark contrast, our use of randomly placed impurities to model the surface charge disorder produces an electrostatic potential with large fluctuations along the length of the system. These potential fluctuations act effectively as disorder within the device, which can be detrimental to Majorana physics \cite{stanescu2011majorana,bagrets2012class,liu2012zerobias,lutchyn2012momentum,rainis2013realistic,sau2013density,degottardi2013majorana,adagideli2014effects,cole2016proximity,woods2019zeroenergy,pan2020physical,woods2021charge,pan2021crossover,dassarma2021disorderinduced,zeng2021partiallyseparated}. 

The first natural question that we address concerns the magnitude and characteristic length scale of the effective disorder potential that one can expect in a typical hybrid device containing an InAs nanowire due to the presence of surface charge impurities. This is done by self-consistently solving the Schr{\"o}dinger-Poisson equations of the 3D hybrid system shown in Fig. \ref{FIG_SCD1}. 
The InAs nanowire is modeled using an effective mass Hamiltonian,
\begin{equation}
	H = 
	-\frac{\hbar^2}{2 m^*} 
	\nabla^2 
	- e \phi\left(\mathbf{r}\right), \label{SM_Ham}
\end{equation}
where $m^*$ is the effective mass, $\nabla^2$ is the Laplacian operator in 3-dimensional space, and $\phi$ is the electrostatic potential inside the wire. The electrostatic potential satisfies the Poisson equation,
\begin{equation}
   	 \nabla \cdot \left[\epsilon(\mathbf{r}) \nabla\phi(\mathbf{r})\right] =
    	  -\rho(\mathbf{r}), \label{Pois}
\end{equation}
where $\rho$ is the charge density and $\epsilon$ is a material dependent dielectric constant taking different values inside the dielectric, InAs nanowire, and surrounding vacuum. Additionally, the electrostatic potential $\phi$ satisfies Dirichlet boundary conditions on the metallic gates: $\phi = V_L, V_R, V_{BG}$ on the surface of the left, right, and bottom gates, respectively. In addition, the Dirichlet boundary condition $\phi = V_{SC}$ at the InAs-Al interface is used to account for the band bending induced by the work function difference between the two materials \cite{vuik2016effects,antipov2018effects,mikkelsen2018hybridization,woods2018effective}. The charge density $\rho$ can be written as the sum of two terms,
\begin{equation}
    \rho(\mathbf{r}) = \rho_f(\mathbf{r}) + \rho_{imp}(\mathbf{r}), \label{Poisson}
\end{equation}
where $\rho_f$ is the free charge density within the wire and $\rho_{imp}$ is the charge density generated by the surface  impurities. In turn, the free charge depends upon the occupied states and is explicitly given by
\begin{equation}
    \rho_f(\mathbf{r}) = -2 e \sum_{n} \left|
    \psi_n(\mathbf{r})
    \right|^2 f\left(E_n\right), \label{freeCharge}
\end{equation}
where $\psi_n$ is the n\textsuperscript{th} eigenstate of the Hamiltonian in Eq. (\ref{SM_Ham}) with energy $E_n$, $f$ is the Fermi function, and the factor of $2$ accounts for spin-degeneracy. Eqs. (\ref{SM_Ham}), (\ref{Pois}), and (\ref{freeCharge}) are collectively referred to as the Schr{\"o}dinger-Poisson equations and require a self-consistent solution \cite{vuik2016effects,woods2018effective}. Additional details regarding the model, along with the method for self-consistently solving the Schr{\"o}dinger-Poisson equations, can be found in Appendix \ref{app:E}. 

A key element of our modeling is the assumption that the charge impurities are randomly distributed over the (uncovered) surface of the semiconductor nanowire, which implies that the impurity charge density $\rho_{imp}$ explicitly breaks translation invariance along the wire (which we will take as the $z$ direction). It is convenient to separate the surface charge density into the (translation invariant) average density, $\bar{\rho}_{imp}(x,y)$, and a fluctuation component, $\rho_{imp}^\prime(\mathbf{r})$, i.e., $ \rho_{imp}(\mathbf{r}) = \bar{\rho}_{imp}(x,y) + \rho_{imp}^\prime(\mathbf{r})$. Thus the total charge density can be rewritten as 
\begin{equation}
    \rho(\mathbf{r}) = \rho_o(x,y) + \rho_{imp}^\prime(\mathbf{r}) + \rho_{red}(\mathbf{r}),
\end{equation}
where $\rho_o$ is the (translation invariant) total charge density corresponding to $\rho_{imp}^\prime = 0$, i.e., the charge density for a system with uniform surface charge and specified electrostatic environment (e.g., gate voltage values, work function difference, etc.), while $\rho_{red}$ accounts for the redistribution of free charge due to the presence of surface charge fluctuations, $\rho_{imp}^\prime$, (i.e., it represents the {\em screening charge}).
Note that $\rho_o$ includes both $\bar{\rho}_{imp}(x,y)$ and a translation-invariant component of the free charge. The solution of the Schr{\"o}dinger-Poisson equations yields all these components of the total charge density, as well as the corresponding components of the electrostatic potential. To investigate the effects of charge impurity-induced disorder, the relevant quantities are the matrix elements of the potential fluctuation with the transverse wave functions corresponding to the uniform system  (see Appendix \ref{app:E} for details), 
\begin{equation}
    V_{eff}^{\alpha \beta}(z) = 
    \int \varphi_{\alpha}^*(x,y)
    \left[
    \phi_{imp}^\prime(\mathbf{r}) 
    + \phi_{red}(\mathbf{r}) 
    \right]
    \varphi_{\beta}(x,y) \, dxdy, \label{EffPot0}
\end{equation}
where $\phi_{imp}^\prime$ and $\phi_{red}$ are the components of the potential corresponding to $ \rho_{imp}^\prime$ and $\rho_{red}$, respectively, and $\varphi_{\alpha}$ is the normalized transverse orbital of the $\alpha$ subband for a uniform system (i.e., a system with with  $\rho = \rho_o$). In particular, neglecting the inter-band coupling, we can define the {\em effective disorder potential} as  $ V_{dis}(z) = V_{eff}^{\alpha,\alpha}(z)$, with $\alpha$ corresponding to the subband nearest to the Fermi level.  Note that the average disorder potential is approximately zero, i.e., $\langle V_{dis} \rangle \approx 0$.
Using the disorder potential provided by the self-consistent solution of the Schr{\"o}dinger-Poisson equations for a 3D structure with a specific (random) distribution of surface impurities, 
we can map the problem into an effective 1D model  \cite{lutchyn2010majorana,oreg2010helical} defined by the BdG Hamiltonian,
\begin{equation}
    \begin{split}
    H_{\text{BdG}} =& \left(-\frac{\hbar^2}{2m^*} \partial_z^2 - \mu - i\alpha_R\sigma_y\partial_z + V_z\sigma_z 
    \right) \tau_z \\
    &+ \Delta \sigma_y \tau_y 
    + V_{dis}(z) \tau_z.
    \end{split} \label{HamBdG}
\end{equation}
Here, $\mu$ is the chemical potential, $\alpha_R$ is the Rashba spin-orbit coefficient, $V_z$ is the Zeeman energy, $\Delta$ is the superconducting pairing amplitude, $V_{dis}$ is the (effective) disorder potential, and $\sigma_i$ and $\tau_i$, with $i = \{x,y,z\}$, are Pauli matrices acting on the spin and particle-hole spaces, respectively. To better understand the  effects of disorder, we will also consider the ``reduced disorder'' problem defined by the effective  potential  $V_{dis}(z; \beta) = \beta V_{eff}^{\alpha \alpha}(z)$, with $\beta\leq 1$ being an adjustable parameter. Note that $\beta=0$ corresponds to a clean system, while $\beta=1$ corresponds to the ``actual'' disorder potential of a system with a given distribution of surface charge impurities, which is determined by the self-consistent solution of the Schr{\"o}dinger-Poisson problem, as described above.

\subsection{Results} \label{SCD_Results} 

In this section, we solve numerically the  Schr{\"o}dinger-Poisson problem corresponding to the system represented  in Fig. \ref{FIG_SCD1}, calculate the effective disorder potential generated by the surface impurities, and investigate the low-energy physics described by the effective 1D Hamiltonian given by Eq. (\ref{HamBdG}) in the presence of this surface disorder. We focus on a hybrid system with the following parameters: effective mass  $m^* = 0.026 m_e$, with $m_e$ being the bare electron mass, wire diameter $D = 100~\text{nm}$, lateral gate spacing $W = 30~\text{nm}$, dielectric thickness $d = 50~\text{nm}$, charge $Q = e$ for each surface impurity,  supercell length $L = 1~\mu\text{m}$. Additionally, the relative dielectric constants for the InAs nanowire, SiO$_2$  dielectric, and surrounding vacuum are $\epsilon_{InAs}=15.15$, $\epsilon_{diel}=3.9$, and $\epsilon_{vac} = 1$, respectively. The band-bending at the InAs-Al interface corresponds to $V_{SC} = 0.25~\text{V}$, which is similar to the value  reported in Ref.~\onlinecite{schuwalow2019band}. The average surface charge density is $\sigma = 0.5 \times 10^{12}~\text{e cm}^{-2}$, which represents the lower end of the range found in the literature \cite{olsson1996charge} and used in other theoretical studies of Majorana nanowires \cite{winkler2019unified,escribano2019effects,woods2020subband,liu2021electronic}. This translates into a linear charge density of $\lambda = 10^{3}~ e ~ \mu\text{m}^{-1}$ and a volume charge density of $\rho = 1.54 \times 10^{17}~e~\text{cm}^{-3}$. We emphasize that this represents an optimistic estimate of the impurity density, if we take into account  the previously used values, as well as our results in Sec.~\ref{sec:transport}. This relatively ``low'' impurity density puts this situation at the most optimistic end of our estimated disorder based on the 2D mobility analysis of Sec.~\ref{sec:transport} .

\begin{figure}[t]
    \begin{center}
    \includegraphics[width=0.48\textwidth]{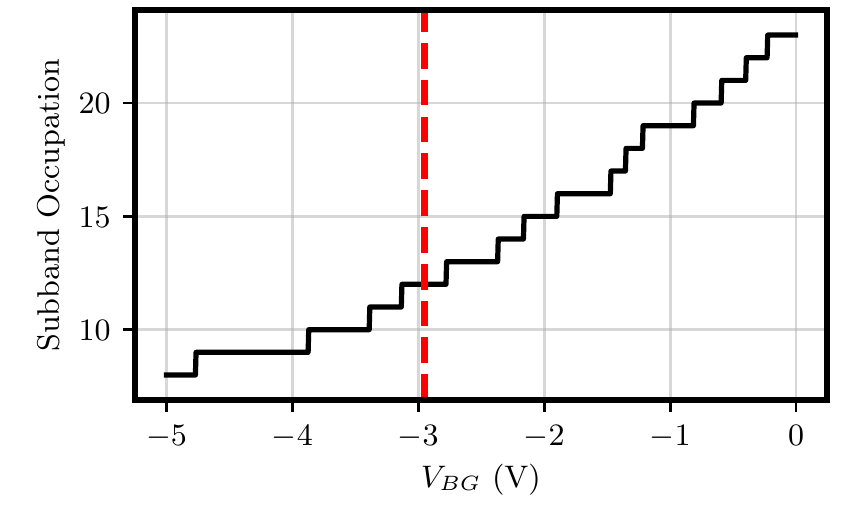}
    \end{center}
    \vspace{-.5cm}
    \caption{Subband occupation as a function of back gate voltage $V_{BG}$ for a uniform system with  $V_L = V_R = 0$, $V_{SC} = 0.25~\text{V}$, and surface charge density $\sigma = 0.5 \times 10^{12}~\text{e cm}^{-2}$, which translates to a volume charge density $\rho = 1.54 \times 10^{17}~\text{e cm}^{-3}$. Note that in this calculation the surface impurity density fluctuations are assumed to be zero, $\rho^\prime_{imp}=0$. The dashed red line indicates the expected voltage associated with the onset of holes near the bottom of the wire.}
    \label{FIG_SCD2}
    \vspace{-1mm}
\end{figure}

We begin with a uniform system (i.e., a system with no charge density fluctuations, $\rho^{\prime}_{imp} = 0$) and calculate the subband occupation as a function of the applied back gate voltage $V_{BG}$. The potential of the side gates is set to zero, $V_L = V_R = 0$. Since $\rho_{imp}=\bar{\rho}_{imp}(x,y)$ is independent of $z$, the system is translation invariant and the subbands are well defined. The results are shown in Fig. \ref{FIG_SCD2} for a voltage range $-5~\text{V} \leq V_{BG} \leq 0$. Note that for $V_{BG} = 0$ (no applied back gate voltage) there are $23$  occupied subbands (each being double spin degenerate), clearly placing the system in the high-occupancy regime \cite{woods2020subband}. This is due to the presence of positive surface charges, 
as well as the band bending at the semiconductor-superconductor interface.
The number of occupied subbands can be reduced by applying a negative gate voltage. However, there is a limit to how much the occupancy can be reduced (i.e., a maximum $|V_{BG}|$ value) before the emergence of holes localized near the bottom of the nanowire (i.e., close to the back gate).
Roughly, this occurs when the electrostatic potential becomes sufficiently negative to overcome the band gap between the conduction and valence bands, i.e. for $E_{gap} \approx -e\phi(\mathbf{r})$, where $E_{gap} = 0.418~\text{eV}$ \cite{winkler2003spinorbit} and $\mathbf{r}$ is a position near the bottom of the InAs nanowire.
The dashed red line in Fig. \ref{FIG_SCD2} indicates this limit; for more negative voltage values holes are expected to emerge, creating a  dissipative normal channel parallel to the conduction subbands designed to harbor Majorana physics. Note that, unlike electrons, the holes are localized away from the semiconductor-superconductor interface and are not proximitized by the superconductor. Hence, the dissipative nature of the hole channel. 

\begin{figure}[ht]
    \begin{center}
    \includegraphics[width=0.48\textwidth]{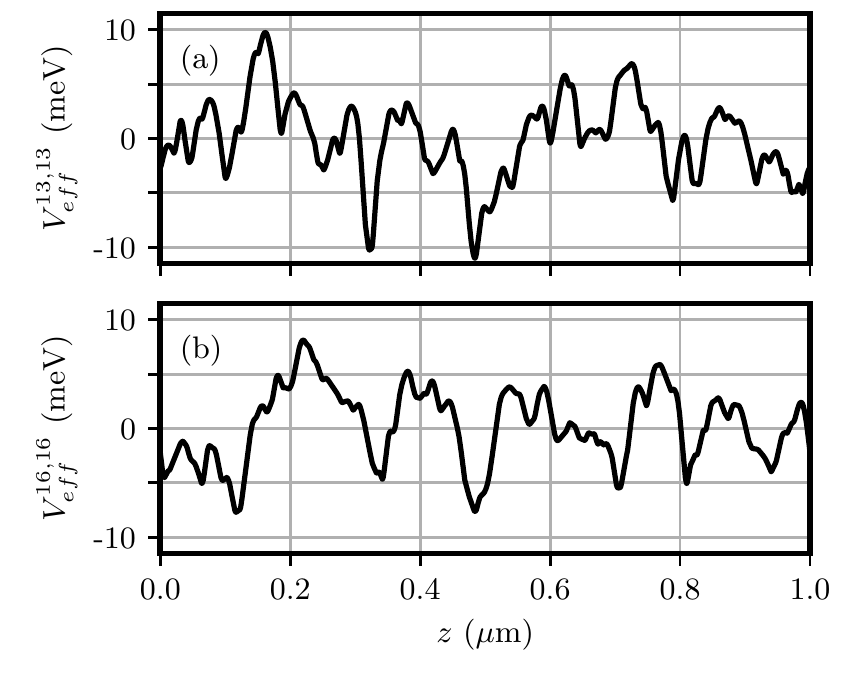}
    \end{center}
    \vspace{-.5cm}
    \caption{Effective disorder potential $V_{dis}(z) = V_{eff}^{\alpha,\alpha}(z)$ for a system with the bottom of the $\alpha$ subband tuned near the Fermi level, where (a) $\alpha = 13$, which corresponds to a back gate voltage  $V_{BG} = -2.78~\text{V}$, and (b) $\alpha = 16$, which corresponds to $V_{BG} = -1.90~\text{V}$. The side gates are at zero voltage, $V_L = V_R = 0$.} 
    \label{FIG_SCD3}
    \vspace{-1mm}
\end{figure}

Next, we tune $V_{BG}$ so that the bottom of one of the subbands is at the Fermi level and solve the Schr{\"o}dinger-Poisson equations again, this time explicitly including the surface charge density fluctuations, $\rho_{imp}^\prime$, associated with the (random) nonuniform distribution of surface charge impurities. 
The effective disorder potential is then calculated from the self-consistent solution using Eq. (\ref{EffPot0}) with $\alpha=\beta$ corresponding to the top occupied subband. Two examples of effective disorder potentials for a system with the $13^\text{th}$ and the $16^\text{th}$ subbands tuned near the Fermi level are shown in Fig. \ref{FIG_SCD3}, panels (a) and (b), respectively. The first relevant feature is the relatively large magnitude of the effective potential, as compared to the typical energy scale associated with Majorana physics. Indeed, the potentials shown in Fig. \ref{FIG_SCD3}, panels (a) and (b), have maximum amplitudes of approximately $10$ and $8~\text{meV}$, respectively. This is larger than the pairing potential $\Delta$ by a factor of $33$ and $27$, respectively. 
In addition, the root mean square values $\langle ( V_{eff}^{\alpha,\alpha})^2\rangle^{1/2}$ of the effective potentials in Figs. \ref{FIG_SCD3}(a) and (b) are $3.72 ~\text{meV}$ and $3.47 ~\text{meV}$, respectively, larger than $\Delta$ by a factor of about $12$. Clearly, such a strong disorder potential should have a major impact on the Majorana physics.
The second key feature of the disorder potential generated by surface charge impurities is its characteristic length scale. To quantify this property, we calculate the correlation length $l_c$ defined as the full width at half maximum for the correlation function,
\begin{equation}
    C_\alpha(\delta) = \sqrt{\bigg\langle V^{\alpha\alpha}_{eff}(z) V^{\alpha\alpha}_{eff}(z + \delta) \bigg\rangle_z},
\end{equation}
where $\alpha$ is the index of the subband tuned to the Fermi level and $\langle \dots\rangle_z$ represents averaging over the position $z$ along the wire for a given disorder realization. 
For the examples in Fig. \ref{FIG_SCD3} (a) and (b), the  correlation length is $l_c = 64 ~\text{nm}$ and $l_c = 70~\text{nm}$, respectively. Note that this is roughly double the correlation length found in Ref.~\onlinecite{woods2021charge} for a similar system with an InAs nanowire of diameter $D = 70~\text{nm}$ having charge impurities placed within the wire. 

At this point, we would like to reemphasize the key importance of the characteristic length scale of the disorder potential in determining the impact of disorder on the low-energy physics of the hybrid device. As discussed in Sec. \ref{SSec3A}, 
the components of the disorder potential with characteristic length scale comparable to or larger than the characteristic Majorana ``oscillation length'', 
$\lambda_M\approx 25-60~$nm,  have a major impact on the low-energy physics once their amplitude becomes comparable to or exceeds the Majorana energy scale, $\epsilon^*\sim 1~$meV. From this perspective, the disorder potentials shown in Fig. \ref{FIG_SCD3}, which are characterized by length scales comparable to $\lambda_M$ and large amplitudes (well above $\epsilon^*$) are expected to have catastrophic effects on the Majorana physics. The model calculations discussed below confirm this expectation.     

Before we investigate the impact of surface-induced disorder on low-energy physics, we address an important technical aspect regarding the calculated effective disorder potential, namely the role of screening by the free charge in suppressing the potential fluctuations. To quantify this effect, we define the screening factor,
\begin{equation}
    \mathcal{Z}_\alpha = \sqrt{
  \frac{\bigg\langle \left(V^{\alpha\alpha}_{eff}\right)^2 \bigg\rangle_z
    }
    {\bigg\langle \left(V^{\alpha\alpha}_{imp}\right)^2 \bigg\rangle_z}
    },
\end{equation}
where $V^{\alpha\alpha}_{imp}$ is the effective potential given by Eq. (\ref{EffPot0}) with $\phi_{red}=0$, which does not take into account the contribution from the redistribution of free charge. Note that $V^{\alpha\alpha}_{imp}$ includes the screening by the superconductor and the metallic gates. For the effective potentials in Figs. \ref{FIG_SCD3} (a) and (b), the screening factors are $\mathcal{Z}_{13} = 0.37$ and $\mathcal{Z}_{16} = 0.32$, respectively. Basically, the screening by the free charge reduces the potential fluctuations by about $60$-$70$\%, but even the screened charge fluctuations are strong. Finally, we note that similar values for the effective potential amplitude, correlation length, and screening factor were found for other disorder realizations that we considered. 

\begin{figure}[t]
    \begin{center}
    \includegraphics[width=0.48\textwidth]{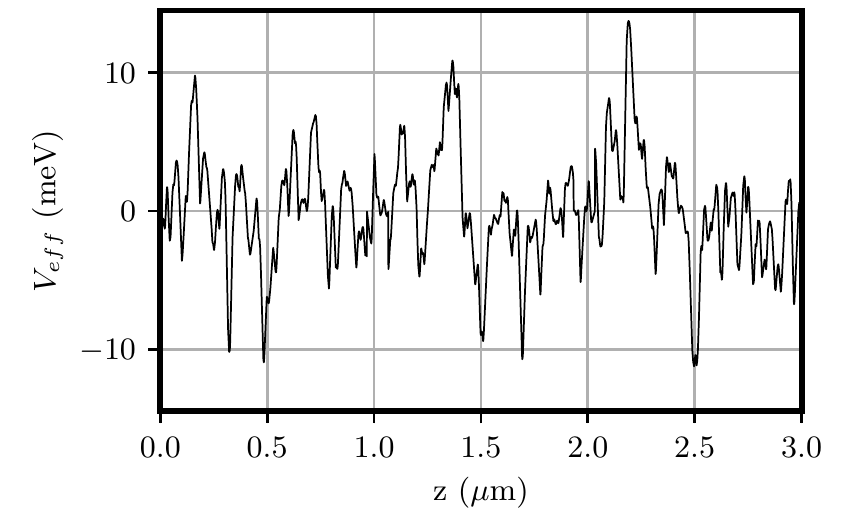}
    \end{center}
    \vspace{-0.5cm}
    \caption{Effective potential profile $V_{eff}(z)$ for a wire of length $L_z=3~\mu$m having a surface charge density $\sigma = 0.5 \times 10^{12}~\text{e cm}^{-2}$ in the presence of a back gate potential $V_{BG}=-2.78~$V, which corresponds to the chemical potential being at the bottom of the $13^\text{th}$ subband. The results in Fig. \ref{FIG_SCD5} are calculated using this profile and different values of the ``reduction coefficient'' $\beta$, i.e., the disorder potential $V_{dis}(z;\beta)=\beta~\!V_{eff}(z)$.}
    \label{FIG_SCD4}
    \vspace{-1mm}
\end{figure}

Turning now our attention toward the low-energy physics of the hybrid device in the presence of surface disorder, we incorporate the effective potential calculated self-consistently into the 1D minimal model described by the BdG Hamiltonian in Eq. (\ref{HamBdG}) and determine the properties of the low-energy modes. 
The additional system parameters characterizing the disordered finite wire are the Rashba spin-orbit coupling, $\alpha_R = 200~\text{meV} \cdot \text{\AA}$, the induced superconducting pairing, $\Delta = 0.3~\text{meV}$, and the length of the wire, $L_z = 3~\mu\text{m}$. Note that the values for $\alpha_R$ and $\Delta$ are rather optimistic for InAs/Al hybrid structures. The effective disorder potential $V_{eff}$ used in the finite wire calculations is shown in Fig. \ref{FIG_SCD4}.
We note that this potential profile is obtained by ``stitching'' together the effective potentials given by three separate self-consistent calculations with a supercell of length $L = 1~\mu\text{m}$ and corresponding to different impurity distributions. The system has the $13^\text{th}$ subband tuned near the Fermi level. 
To avoid sudden jumps associated with stitching together the effective potentials, they are overlapped within ``boundary'' regions of length $100~\text{nm}$; the resulting potential transitions linearly from one potential realization to the other within each boundary region. Note that this construction of the effective disorder potential is based on the implicit assumption that the potential is determined by the ``local'' distribution of charge impurities within a neighborhood no larger than $1~\mu$m.   

To characterize quantitatively the Majorana physics in the presence of surface charge disorder, we define the \textit{Majorana separation length}, $\ell_{sep}$, as a measure of the spatial separation of the Majorana modes associated with the low-energy BdG states. More specifically, let $\psi_\varepsilon$ be an eigenstate of the BdG Hamiltonian in Eq. (\ref{HamBdG}) with energy $\varepsilon > 0$. As a result of particle-hole symmetry, $\psi_{-\varepsilon} = \tau_x \psi_{\varepsilon}^*$ is guaranteed to be an eigenstate of the BdG Hamiltonian having energy $-\varepsilon$. We define the Majorana wave functions, 
\begin{align}
    \chi_{\epsilon1} &= \frac{1}{\sqrt{2}} \left(
    \psi_\varepsilon + \psi_{-\varepsilon}\right), \label{Maj1} \\
    \chi_{\epsilon2} &= \frac{i}{\sqrt{2}} \left(
    \psi_\varepsilon - \psi_{-\varepsilon}\right), \label{Maj2}
\end{align}
which, by construction, satisfy the Majorana condition, $\tau_x \chi_{\varepsilon \nu}^* = \chi_{\varepsilon \nu}$. Note that, in the language of the second quantization, this condition translates into $\gamma_{\varepsilon \nu}^\dagger = \gamma_{\varepsilon \nu}$, where $\gamma_{\varepsilon \nu}^\dagger$ is a Majorana creation operator for the mode $\chi_{\varepsilon \nu}$. Next, we define the Majorana {\em center of mass} (MCM), 
\begin{equation}
    \langle z_{\varepsilon \nu} \rangle = 
    \sum_{i \sigma \tau} z_i \left|
    \left(\chi_{\varepsilon \nu}\right)_{i \sigma \tau}
    \right|^2,
\end{equation}
where $i$, $\sigma$ and $\tau$ are position (i.e., lattice site), spin, and particle-hole indices, respectively. We calculate the MCMs $\langle z_{\varepsilon \nu} \rangle$ of all Majorana modes corresponding to $\varepsilon < \varepsilon_{\text{cut}}$, where $\varepsilon_{\text{cut}}<\Delta$ is a cutoff energy, and order them so that $\langle z\rangle_1 \leq \langle z\rangle_2 \leq \dots$, where $\langle z \rangle_i$ is the i\textsuperscript{th} smallest $\langle z_{\varepsilon \nu} \rangle$ value. Finally, we define the \textit{Majorana separation length} as $\ell_{sep} = \langle z \rangle_2 - \langle z \rangle_1$. In other words, the Majorana separation length is defined as the minimum distance between the leftmost Majorana mode with $\varepsilon < \varepsilon_\text{cut}$ and all other Majorana modes within the cutoff energy window. Note that if a BdG state with $|\varepsilon| < \varepsilon_\text{cut}$ does not exist, the Majorana separation length is simply undefined. 
Also, note that there is an ambiguity in the definition of the Majorana wave functions in Eqs. (\ref{Maj1}) and (\ref{Maj2}), since multiplying $\psi_{\varepsilon}$ by an arbitrary phase creates two new Majorana wave functions that are generically a superposition of the two original Majorana wave functions. We choose the phase so that $|\langle z_{\varepsilon 1} \rangle - \langle z_{\varepsilon 2} \rangle|$ is maximized, i.e., the distance between the two MCMs is largest. 

The physical reason for defining $\ell_{sep}$ in this manner is the following. First, by including $\langle z \rangle_1$ in the definition we make sure that any low-energy state capable of generating a zero-bias conductance peak (ZBCP) for charge tunneling into the left end of the system will be characterized by a well-defined value of $\ell_{sep}$. Note, on the other hand, that $\ell_{sep}$ being well-defined does not guarantee the emergence of a ZBCP (or even a split ZBCP) in charge tunneling, e.g., if the leftmost Majorana mode is localized away from the end of the wire and does not couple to the normal lead. Also note that one can define a similar Majorana separation length involving the rightmost Majorana mode and its ``nearest neighbor'', which would be relevant for charge tunneling into the right end of the wire. Second, this definition of $\ell_{sep}$ provides a convenient tool for distinguishing topologically protected Majorana zero modes, which are characterized by $\ell_{sep}\sim L_z$ (i.e., the MZMs are localized at the opposite ends of the system), partially separated Majorana modes associated with local (remnant) Majorana physics, which correspond to $\xi \lesssim \ell_{sep} < L_z$ (i.e., the Majorana separation is smaller than the length of the system, but larger than the Majorana localization length, $\xi$), and trivial Andreev bound states, which are characterized by $\ell_{sep}<\xi$ (i.e., the two leftmost Majorana modes are, practically, on top of each other).  

\begin{figure}[ht]
    \begin{center}
    \includegraphics[width=0.48\textwidth]{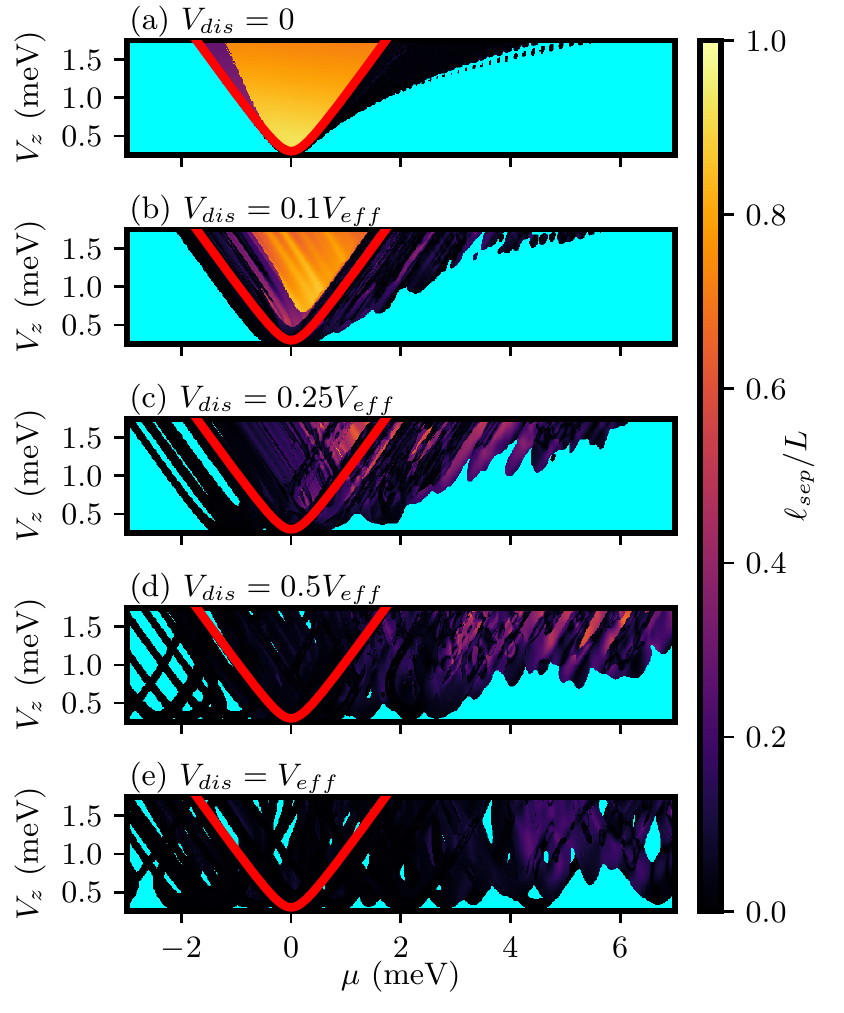}
    \end{center}
    \vspace{-0.5cm}
    \caption{Majorana separation length as a function of chemical potential $\mu$ and Zeeman energy $V_z$ for a system of length $L_z=3~\mu\text{m}$ in the presence of a disorder potential $V_{dis}(z; \beta) = \beta V_{eff}(z)$, where $V_{eff}$ is the effective potential shown in Fig. \ref{FIG_SCD4}, which was calculated self-consistently for a wire with surface charge density $\sigma = 0.5 \times 10^{12}~\text{e cm}^{-2}$. The disorder reduction coefficient takes the values: (a) $\beta=0$ (clean system), (b) $\beta=0.1$, (c) $\beta=0.25$, (d) $\beta=0.5$, and (b) $\beta=1$ (full disorder strength).   
     The red line in each panel corresponds to the topological phase boundary of the clean system, which is given by the condition $V_z = \sqrt{\mu^2 + |\Delta|^2}$. Light blue regions indicate the absence of any BdG state within the  low-energy window $|\varepsilon| < \varepsilon_{\text{cut}} = 0.08~\text{meV}$, which makes the Majorana separation length undefined. Yellow ($\ell_{sep}\sim L_z$) indicates the presence of well-separated (topological) Majorana zero modes, black ($\ell_{sep}<0.2L_z$) corresponds to trivial low-energy Andreev bound states, while the intermediate regime corresponds to the presence of partially separated Majorana modes associated with remnant (local) Majorana physics.}
    \label{FIG_SCD5}
    \vspace{-1mm}
\end{figure}

Equipped with this definition of the Majorana separation  length, we investigate the presence of Majorana physics (or remnant/local Majorana physics) in a hybrid system with disorder induced by surface charge impurities by calculating $\ell_{sep}$ as a function of Zeeman field and chemical potential using the 1D effective model given by Eq. (\ref{HamBdG}) with a disorder potential $V_{dis}(z;\beta) =\beta V_{eff}$, where $V_{eff}$ is the self-consistent effective potential shown in Fig. \ref{FIG_SCD4}. The results corresponding to $\beta=0, 0.1, 0.25, 0.5$, and $1$ are shown in Fig. \ref{FIG_SCD5} (a)-(e), respectively. In each panel the topological phase boundary of the clean system, which is given by the condition $V_z = \sqrt{\mu^2 + |\Delta|^2}$, is marked by a red line. The cutoff energy used in the definition of $\ell_{sep}$ is $\varepsilon_{\text{cut}} = 0.08~\text{meV}$. 

First, we point out that for the clean system [see Fig. \ref{FIG_SCD5} (a)], the Majorana separation length clearly distinguishes between the topological superconducting phase, which is characterized by values of $\ell_{sep}/L_z$ of order $1$ (yellow region), and the trivial phase, which is characterized by the absence of low-energy states ($\ell_{sep}$ undefined, cyan region) or by the presence of topologically trivial in-gap Andreev bound states (also called intrinsic Andreev bond states \cite{huang2018metamorphosis}) that generate low values of $\ell_{sep}$ (black region). 
Next, we turn on the disorder to one-tenth of its actual strength, $\beta = 0.1$. As shown in Fig. \ref{FIG_SCD5} (b), the region characterized by large Majorana separation (yellow) shrinks, while small ``islands'' of remnant (local) Majorana physics characterized by $\ell_{sep}/L_z \sim 0.2-0.5$ emerge both inside the nominally topological region as well as outside it. Further increasing the disorder strength to $\beta = 0.25$ [see Fig. \ref{FIG_SCD5} (c)] and  $\beta = 0.5$ [see Fig. \ref{FIG_SCD5} (d)] results in the disappearance of well-separated Majorana modes characterized by $\ell_{sep}\sim L_z$, while the ``islands'' of remnant (local) Majorana physics tend to migrate outside the nominally topological region (marked by the red line) into the trivial region, toward larger values of the chemical potential. Finally, in the presence of the full strength disorder potential [$\beta=1$; see Fig. \ref{FIG_SCD5} (e)]  even the ``islands'' of remnant (local) Majorana physics associated with the presence of partially separated Majorana modes shrink and practically disappear, signaling the absence of Majorana physics. Also, note that almost the entire parameter space shown in Fig. \ref{FIG_SCD5} (e) corresponds to small values of $\ell_{sep}$ (dark region), indicating the ubiquitous presence of disorder-induced low-energy trivial Andreev bound states consisting of nearly overlapping Majorana modes.   The maps in Fig. \ref{FIG_SCD5} should be compared with the ``phase diagrams'' in Fig. \ref{fig:21}. Note that in a clean system [Figs. \ref{FIG_SCD5}(a) and \ref{fig:21}(a-b)],  or in the presence of weak disorder [Figs. \ref{FIG_SCD5}(b) and \ref{fig:21}(c-d)], the area characterized by large Majorana separation length corresponds to conductance features associated with Majorana physics (i.e., stable conductance quantization, Majorana oscillations, and end-to-end correlations). By contrast, the experimentally relevant strong disorder regime [Figs. \ref{FIG_SCD5}(e) and \ref{fig:21}(g-h)] is characterized by the ubiquitous presence of disorder-induced (trivial) Andreev bound states with $\ell_{sep}<\xi$; when the leftmost Andreev bound state happens to be close enough to the end of the system to couple to the tunneling probe (which is rather rare and purely random), a large conductance feature emerges. 

The analysis presented in this section has two significant outcomes. First, we have calculated the effective disorder potential generated by the random distribution of surface charge impurities in a semiconductor-superconductor device based on the self-consistent solution of a microscopic model of the hybrid system. We found that the effective disorder potential has amplitudes of the order of $10~$meV, well above the typical Majorana energy scale, and characteristic length scales of the order of $50-80~$nm, comparable to the typical Majorana oscillation length. These results were based on a rather optimistic estimate of the surface charge density ($\sigma = 0.5 \times 10^{12}~\text{e cm}^{-2}$). Second, by determining the Majorana separation length over a large window of control parameters, we have shown that the calculated effective disorder potential is inconsistent with the presence of Majorana physics. We note that this result was obtained based on rather optimistic estimates of the spin-orbit coupling and induced pairing potential. Moreover, our results suggest that the presence of disorder induced by surface impurities generates (topologically trivial) low-energy Andreev bound states that are ubiquitous within the relevant parameter space. By contrast, identifying an ``accidental'' Majorana island supporting local Majorana physics and partially separated Majorana modes [similar to the ``intermediate'' separation length features in Figs. \ref{FIG_SCD5}(c) and (d)] may require very significant sample selection and parameter fine-tuning, if possible at all. Based on our results, a hybrid system characterized by the systematic presence of remnant Majorana islands requires the reduction of the disorder potential amplitude by a factor of about $2-3$ (i.e., amplitudes of the order of $4-5~$meV), assuming that the characteristic length scale remains the same. Finally, obtaining genuine Majorana zero modes would require the reduction of the disorder potential amplitude by a factor of 5, or more, i.e., maximum amplitudes below $2~$meV. We mention that even the full disorder case here with $\beta=1$  is an optimistic underestimate of the currently prevailing InAs and InSb disorder as estimated in Sec.~\ref{sec:transport} of this paper, and thus the emergence of topological Majorana likely necessitates at least two orders of magnitude reduction in the sample disorder.

\section{Conclusion}\label{sec:conclusion}

We have carried out an evaluation of the disorder that characterizes experimentally available superconductor-semiconductor hybrid structures and a characterization of its effects on the low-energy physics based on a multipronged approach that addresses the three critical tasks associated with this type of effort: estimating the (physical) disorder based on available experimental data, calculating the corresponding effective disorder potential, and simulating the low-energy physics in the presence of disorder. The physical disorder estimates were obtained rather indirectly, from an analysis of the transport data for corresponding 2D semiconductor materials grown by the same growers who grow the nanowires used in Majorana experiments (and using the same growth chambers and processes).  Since there is no direct experimental information available on the amount of disorder in the SC-SM Majorana structures, using our estimated disorder based on the mobility data of the corresponding 2D materials is the best one can do at this point in trying to develop some quantitative feel for the realistic disorder level in hybrid superconductor-semiconductor structures. In turn,  this enables a more realistic evaluation of the effects of disorder on the possible emergence of topological superconductivity and Majorana modes in semiconductor nanowire systems.  To this end,  we have carried out two complementary Majorana simulations: the first is based on the highly successful and widely used minimal 1D BdG model, with an effective disorder potential evaluated based on our estimates of physical disorder (i.e., charge impurity concentrations) provided by the 2D transport analysis and on the results of Ref. \onlinecite{woods2021charge} (regarding the effective potential associated with charge impurities randomly distributed within the semiconductor nanowire), while the second represents a semi-realistic self-consistent calculation of the effective disorder potential associated with the presence of surface charge impurities in a hybrid Majorana structure, as well as the implications on the low-energy physics. 

Our findings are sobering. Even assuming that the level of disorder in the nanowires is the same as that estimated from the measured 2D mobility (which is likely to be an optimistic assumption, since the SC-SM structures should have much more disorder arising from additional processing and interfaces necessary in the multilayer SC-SM structures), we conclude that topological superconductivity and Majorana zero modes cannot emerge in systems with the currently prevailing materials quality.  We find that disorder-induced trivial zero-bias conductance peaks associated with subgap fermionic Andreev bound states do emerge in the nanowires and occasionally mimic the local properties of Majorana zero modes, but such trivial zero-bias peaks are neither generic nor stable as a function of the control parameters, in contrast to the Majorana zero modes.  Observing these trivial zero-bias peaks require post-selection and fine-tuning and they never appear in tunneling from both ends, as nonlocal Majorana zero modes do.  Also, these disorder-induced trivial ZBCPs do not manifest Majorana oscillations and do not manifest stable $2e^2/h$  quantization (as a function of the control parameters). This phenomenology is characteristic of what we dubbed as the strong disorder regime and is strikingly similar to the experimental situation, which strongly suggests that our analysis does not overestimate the actual disorder strength, as explicitly intended in this study.  

In connecting our work with the existing theoretical work in the literature, we mention that all the low-field zero-bias peaks that we find in the strong disorder regime are so-called ``ugly'' peaks, as dubbed in Ref.~\onlinecite{pan2020physical}, which correspond to disorder-induced trivial Andreev bound states.  These peaks can have large-conductance values, as observed in recent experiments~\cite{zhang2021large,nichele2017scaling} and discussed in recent theoretical papers~\cite{dassarma2021disorderinduced,pan2021quantized,woods2021charge}. Even in the most disordered samples, where most simulations (i.e., simulations corresponding to most disorder realizations) find nothing of interest, there are some trivial ZBCPs showing up occasionally, which could be postselected (and perhaps fine-tuned by adjusting the tunnel barrier potential and other parameters) to produce $\sim 2e^2/h$  zero-bias peaks. Note, however, that such peaks are never generic or stable and they are local, without manifesting in tunneling from both ends. Also, these trivial peaks do not manifest any Majorana oscillation patterns, which are a hallmark of MZMs at high Zeeman fields and/or short nanowires, as used mostly in current experiments. 
We also point out that partially separated Majorana modes \cite{stanescu2019robust,zeng2021partiallyseparated}, which are associated with local Majorana physics and interpolate continuously between trivial ABSs and MZMs, are expected to emerge in the ``intermediate'' disorder regime, i.e., for impurity concentrations significantly lower than our current estimates, but higher than the levels required for the presence of topological superconductivity and MZMs. The features associated with these modes \cite{zeng2021partiallyseparated} could provide practical landmarks that may be useful in assisting the effort of reducing the disorder. In this context, we point out that disorder potentials dominated by long-wavelength components (like, e.g., the effective potential calculated in Sec. \ref{SCD}) tend to favor the emergence of partially separated Majorana modes, while short-range potentials (e.g., the onsite random potential model) tend to destabilize these modes. Consequently, investigating theoretically the features associated with partially separated Majorana modes, which are expected to be observed once the intermediate disorder regime becomes accessible, should be based on properly determined effective potentials. We emphasize, however, that our current work shows definitively that the disorder in the currently available samples is far above this intermediate disorder regime, and no existing experiment has ever accessed the regime of partially separated Majorana modes, let alone approaching the weak disorder regime where the topological MZMs exist.

For InAs- or InSb-based SC-SM hybrid structures with Al as the parent superconductor, it has recently been shown~\cite{woods2021charge} that topological MZM realization requires a disorder no stronger than a few ($\sim$ 10) charged impurities per micron length of the nanowire, which translates to an equivalent 3D impurity density of about $ 10^{15} $  cm$ ^{-3} $.  Of course, the real systems are complex, and impurities/disorder arise from various sources and at various interfaces, with no simple one-parameter disorder characterization being an exact description. But, in view of a complete lack of any available disorder characterization information available for the nanowire samples in the SC-SM platforms, our current work shows, through a detailed mobility analysis of the corresponding 2D semiconductor systems grown under similar conditions, that a single parameter disorder characterization using an effective 3D unintentional random background charged impurity density of the semiconductor is an effective disorder diagnostic capable of an accurate description of the 2D carrier density dependence of the measured mobility.  The effective disorder we extract from our detailed mobility analysis corresponds to a random background charged impurity density ranging between $10^{17}$ and $10^{19} $  cm$ ^{-3}$, which is 2 to 4 orders of magnitude larger than the limit consistent with the realization of topological MZMs in InAs/Al and InSb/Al SC-SM platforms. Dimensional conversion of our 3D impurity density to the 1D nanowires yields an estimated random charged impurity concentration of $50-2000$ impurities per micron, which is a factor of $10-400$ larger than the expected limit for MZM realization. In terms of the 2D mobility numbers, as measured in the experiments analyzed in detail in the current work, we need InAs and InSb 2D mobility to increase to $100000-300000~$cm$ ^2 $/V$ \cdot$s, compared with the current 2D mobility of $10000-30000~$cm$ ^2 $/V$ \cdot$s which we analyzed in Sec.~\ref{sec:transport}. This of course involves the key assumption that in going from 2D samples to 1D nanowires the effective disorder remains the same, which is highly unlikely as the additional processing steps and the additional interfaces in the SC-SM platforms (as compared with the 2D systems) are likely to degrade the sample quality in 1D nanowires.  Our numbers for the necessary 2D mobility values should therefore be taken as an optimistic lower limit.  An earlier analytical work~\cite{sau2012experimental} came to a minimum 1D mobility estimate of $ \sim $ 100000 cm$ ^2 $/V$ \cdot $s using the rather na\"ive criterion that the transport level broadening should be less than the induced topological gap estimated to be 0.1 meV, and our detailed theory is approximately consistent with this simple estimate as well.  Perhaps the best single disorder number to keep in mind is that the charged impurity density should be no more than 10 per micron of the nanowire, and the current samples appear to have, based on our estimates in the current work, $ \sim 50-2000$ such impurities per nanowire, necessitating at least an order of magnitude improvement in the nanowire quality before topological MZMs are realized in the SC-SM platforms.

Very recent developments in Eindhoven have led to InSb nanowire growth with mobilities $\sim 4\times10^4 $cm$ ^2 $/V$ \cdot$s  for wires of approximate 1-2 $\mu$m length~\cite{badawy2019high}.  Unfortunately, the carrier densities in these nanowires are unknown, and hence it is not possible for us to figure out the sample quality or the actual impurity content of these wires.  Assuming the best possible scenario, these new InSb wires are likely to have a 3D impurity concentration of $10^{17}$  cm$^{-3}$  bringing them in more or less the same level of cleanliness as the InAs samples from Purdue~\cite{pauka2020repairing}, both still roughly a factor of 100 too dirty for the Majorana manifestation.  It is, however, encouraging that there have been improvements in growth conditions bringing both InSb and InAs materials within a factor $\sim $ 100 of the requisite impurity level.  Given that ultra-high mobility GaAs samples exist with 3D impurity concentration$ < 10^{13}$ cm$^{-3}$~\cite{chung2021ultrahighquality}, we can be optimistic that further improvement in materials growth will lead to SC-SM hybrid Majorana platforms with sufficient cleanliness leading to the manifestation of the non-Abelian topological MZM modes.

{Before concluding, we mention that our work leaves out many effects, which may be relevant for the Majorana realization in nanowires, as we focus on what is generally believed to be the most significant obstacle hindering the experimental observation of topological Majorana zero modes, namely random unintentional background disorder in the system.  Thus orbital effects of the magnetic field are ignored in our theory, although it is possible, perhaps even likely, that such orbital effects lead to the eventual quenching of the bulk superconductivity in the parent superconductor itself.  If so, then this orbital effect would lead to a stronger manifestation of disorder than found in our current work because the induced superconducting gap due to proximity effect would be much lower, consequently making the disorder effects more detrimental.  Our results should thus be construed as the most optimistic estimate of the minimal disorder necessary for the experimental realization of the topological Majorana modes in the currently used InAs or InSb based Majorana nanowires.  The fact that this minimal estimated disorder is already much higher than that existing in current samples indicates that a substantial materials effort is necessary leading to much cleaner nanowires for further progress.  In this context, it is encouraging that the disorder content in the best MBE-grown GaAs layers is at the level of $\sim 10^{13}$  cm$^{-3}$~\cite{dassarma2015transport}, which is roughly two orders of magnitude lower than our estimated minimal disorder necessary for the practical realization of the Majorana modes, providing considerable optimism that the laboratory realization of topological Majorana zero modes is highly probable in the near future.  We also mention that most of the random disorder is likely to reside on the nanowire surface, and therefore, it is possible that the superconductor itself may screen some of this disorder, effectively reducing the disorder effect.  Calculating the importance of such screening by the superconductor is not possible at this point since even the impurity density in the nanowire is unknown, let alone their locations on the surface (and obviously, the details of such screening would depend on the impurity distribution on the nanowire surface), but the possibility of such a screening certainly exists.  On the other hand, it is also possible that the superconductor will introduce effective additional disorder from the metal through a ``disorder proximity effect''~\cite{cole2016proximity}, somewhat nullifying the screening effect.  These details of the effects of the metal as well as the orbital effect of the magnetic field can be included in future theories once more details are available about the actual impurity content and its spatial distribution in the nanowire.  We believe that right now our detailed quantitative considerations provided in the current work should serve as a useful guide to Majorana experiments with a goal of reducing the disorder content down to the equivalent of $10^{15}$ cm$^{-3}$ charged impurity concentration in the system.}

Our work establishes that the current InAs/Al and InSb/Al SC-SM nanowire samples are far too disordered for the observation of Majorana signatures, and all experimentally observed zero-bias conductance peaks so far arise from disorder-induced Andreev bound states through careful postselection and fine-tuning.  In some sense, this conclusion, although disappointing,  is not disastrous because we now clearly know the reason underlying the absence of topological Majorana zero modes in experiments. It is because the samples are far too disordered, and fortunately, this is a soluble problem with improved materials, growth, and fabrication. An order of magnitude or so decrease in the number of effective impurities (to 10 or below per micron) in the nanowire should be achievable and all experimental activity should focus on studying improved samples. Our work shows that one can always find impressive-looking zero-bias conductance peaks in nanowires, no matter how disordered they are, through postselection of the data taken on many samples. This is because even in the presence of extreme disorder, where most samples produce no subgap features, some samples will manifest zero-bias peaks which may occasionally mimic Majorana zero modes, although they will always lack the key requirements of stability, nonlocality, Majorana oscillations, and generic appearance without fine-tuning.  Further studies of such trivial zero-bias peaks by themselves, without sample improvement, will not take us closer to the goal of realizing true topological Majorana zero modes.  In this context, we point out that generating experimental ``phase diagrams'' over extended regions of the parameter space and for multiple nominally identical devices, which is completely within the existing technical capabilities, would provide valuable information regarding the actual disorder strength in the available samples.
In addition to improving disorder, experimental efforts must focus on enabling the applied magnetic field to increase further without the collapse of the parent superconductivity, as happens now universally in all Majorana nanowire experiments.  The serious issue with this bulk gap collapse is that we have no way of guaranteeing that such a vanishing of all superconductivity is not happening already below the topological quantum phase transition critical field, and if so, no Majorana will ever arise in the system, even in a pristine clean sample with no disorder,  since the topological regime is then simply inaccessible.  There are several ways of assuring that the topological regime with a critical field above the TQPT point is accessible: (1) Better experimental design avoiding the bulk orbital penetration of the applied magnetic field in the parent superconductor; (2) using parent superconductors with very large $ H_{c2} $ values; (3) increasing the induced SC gap by better interface engineering or by using a parent SC with a larger gap; (4) increase the $ g $-factor of the nanowire material so that the TQPT critical field decreases.

Finally, we mention that although we focus on the SC-SM nanowire Majorana platforms because of the obvious reason that it is by far the most-studied and the most-successful of all MZM searches, our general results on the role of disorder apply to all Majorana platforms.  Typically, semiconductors have much better electronic quality and much higher mobility than metallic systems, and therefore, we believe that other Majorana platforms such as ferromagnetic chains or Fe-based superconductors, being metal-based, have hopelessly huge amount of disorder with no hope of sample improvement as there is for semiconductor-based Majorana devices.  We believe that the disorder problem is far worse in any metal-based Majorana platform than the SC-SM structures we focus on in the current work.

This work is supported by the Laboratory for Physical Sciences and the Microsoft Corporation.

\bibliography{Paper_disorder_NW}

\begin{thebibliography}{134}%
\makeatletter
\providecommand \@ifxundefined [1]{%
 \@ifx{#1\undefined}
}%
\providecommand \@ifnum [1]{%
 \ifnum #1\expandafter \@firstoftwo
 \else \expandafter \@secondoftwo
 \fi
}%
\providecommand \@ifx [1]{%
 \ifx #1\expandafter \@firstoftwo
 \else \expandafter \@secondoftwo
 \fi
}%
\providecommand \natexlab [1]{#1}%
\providecommand \enquote  [1]{``#1''}%
\providecommand \bibnamefont  [1]{#1}%
\providecommand \bibfnamefont [1]{#1}%
\providecommand \citenamefont [1]{#1}%
\providecommand \href@noop [0]{\@secondoftwo}%
\providecommand \href [0]{\begingroup \@sanitize@url \@href}%
\providecommand \@href[1]{\@@startlink{#1}\@@href}%
\providecommand \@@href[1]{\endgroup#1\@@endlink}%
\providecommand \@sanitize@url [0]{\catcode `\\12\catcode `\$12\catcode
  `\&12\catcode `\#12\catcode `\^12\catcode `\_12\catcode `\%12\relax}%
\providecommand \@@startlink[1]{}%
\providecommand \@@endlink[0]{}%
\providecommand \url  [0]{\begingroup\@sanitize@url \@url }%
\providecommand \@url [1]{\endgroup\@href {#1}{\urlprefix }}%
\providecommand \urlprefix  [0]{URL }%
\providecommand \Eprint [0]{\href }%
\providecommand \doibase [0]{https://doi.org/}%
\providecommand \selectlanguage [0]{\@gobble}%
\providecommand \bibinfo  [0]{\@secondoftwo}%
\providecommand \bibfield  [0]{\@secondoftwo}%
\providecommand \translation [1]{[#1]}%
\providecommand \BibitemOpen [0]{}%
\providecommand \bibitemStop [0]{}%
\providecommand \bibitemNoStop [0]{.\EOS\space}%
\providecommand \EOS [0]{\spacefactor3000\relax}%
\providecommand \BibitemShut  [1]{\csname bibitem#1\endcsname}%
\let\auto@bib@innerbib\@empty
\bibitem [{\citenamefont {Sarma}\ \emph {et~al.}(2015)\citenamefont {Sarma},
  \citenamefont {Freedman},\ and\ \citenamefont {Nayak}}]{sarma2015majorana}%
  \BibitemOpen
  \bibfield  {author} {\bibinfo {author} {\bibfnamefont {S.~D.}\ \bibnamefont
  {Sarma}}, \bibinfo {author} {\bibfnamefont {M.}~\bibnamefont {Freedman}},\
  and\ \bibinfo {author} {\bibfnamefont {C.}~\bibnamefont {Nayak}},\ }\bibfield
   {title} {\bibinfo {title} {Majorana zero modes and topological quantum
  computation},\ }\href {https://doi.org/10.1038/npjqi.2015.1} {\bibfield
  {journal} {\bibinfo  {journal} {npj Quantum Information}\ }\textbf {\bibinfo
  {volume} {1}},\ \bibinfo {pages} {15001} (\bibinfo {year}
  {2015})}\BibitemShut {NoStop}%
\bibitem [{\citenamefont {Sau}\ and\ \citenamefont
  {Tewari}(2021)}]{sau2021majorana}%
  \BibitemOpen
  \bibfield  {author} {\bibinfo {author} {\bibfnamefont {J.}~\bibnamefont
  {Sau}}\ and\ \bibinfo {author} {\bibfnamefont {S.}~\bibnamefont {Tewari}},\
  }\bibfield  {title} {\bibinfo {title} {From {{Majorana}} fermions to
  topological quantum computation in semiconductor/superconductor
  heterostructures},\ }\href {http://arxiv.org/abs/2105.03769} {\bibfield
  {journal} {\bibinfo  {journal} {arXiv:2105.03769}\ } (\bibinfo {year}
  {2021})}\BibitemShut {NoStop}%
\bibitem [{\citenamefont {Nayak}\ \emph {et~al.}(2008)\citenamefont {Nayak},
  \citenamefont {Simon}, \citenamefont {Stern}, \citenamefont {Freedman},\ and\
  \citenamefont {Das~Sarma}}]{nayak2008nonabelian}%
  \BibitemOpen
  \bibfield  {author} {\bibinfo {author} {\bibfnamefont {C.}~\bibnamefont
  {Nayak}}, \bibinfo {author} {\bibfnamefont {S.~H.}\ \bibnamefont {Simon}},
  \bibinfo {author} {\bibfnamefont {A.}~\bibnamefont {Stern}}, \bibinfo
  {author} {\bibfnamefont {M.}~\bibnamefont {Freedman}},\ and\ \bibinfo
  {author} {\bibfnamefont {S.}~\bibnamefont {Das~Sarma}},\ }\bibfield  {title}
  {\bibinfo {title} {Non-{{Abelian}} anyons and topological quantum
  computation},\ }\href {https://doi.org/10.1103/RevModPhys.80.1083} {\bibfield
   {journal} {\bibinfo  {journal} {Reviews of Modern Physics}\ }\textbf
  {\bibinfo {volume} {80}},\ \bibinfo {pages} {1083} (\bibinfo {year}
  {2008})}\BibitemShut {NoStop}%
\bibitem [{\citenamefont {Read}\ and\ \citenamefont
  {Green}(2000)}]{read2000paired}%
  \BibitemOpen
  \bibfield  {author} {\bibinfo {author} {\bibfnamefont {N.}~\bibnamefont
  {Read}}\ and\ \bibinfo {author} {\bibfnamefont {D.}~\bibnamefont {Green}},\
  }\bibfield  {title} {\bibinfo {title} {Paired states of fermions in two
  dimensions with breaking of parity and time-reversal symmetries and the
  fractional quantum {{Hall}} effect},\ }\href
  {https://doi.org/10.1103/PhysRevB.61.10267} {\bibfield  {journal} {\bibinfo
  {journal} {Phys. Rev. B}\ }\textbf {\bibinfo {volume} {61}},\ \bibinfo
  {pages} {10267} (\bibinfo {year} {2000})}\BibitemShut {NoStop}%
\bibitem [{\citenamefont {Kitaev}(2001)}]{kitaev2001unpaired}%
  \BibitemOpen
  \bibfield  {author} {\bibinfo {author} {\bibfnamefont {A.~Y.}\ \bibnamefont
  {Kitaev}},\ }\bibfield  {title} {\bibinfo {title} {Unpaired {{Majorana}}
  fermions in quantum wires},\ }\href
  {https://doi.org/10.1070/1063-7869/44/10S/S29} {\bibfield  {journal}
  {\bibinfo  {journal} {Phys.-Usp.}\ }\textbf {\bibinfo {volume} {44}},\
  \bibinfo {pages} {131} (\bibinfo {year} {2001})}\BibitemShut {NoStop}%
\bibitem [{\citenamefont {Freedman}\ \emph {et~al.}(2003)\citenamefont
  {Freedman}, \citenamefont {Kitaev}, \citenamefont {Larsen},\ and\
  \citenamefont {Wang}}]{freedman2003topological}%
  \BibitemOpen
  \bibfield  {author} {\bibinfo {author} {\bibfnamefont {M.}~\bibnamefont
  {Freedman}}, \bibinfo {author} {\bibfnamefont {A.}~\bibnamefont {Kitaev}},
  \bibinfo {author} {\bibfnamefont {M.}~\bibnamefont {Larsen}},\ and\ \bibinfo
  {author} {\bibfnamefont {Z.}~\bibnamefont {Wang}},\ }\bibfield  {title}
  {\bibinfo {title} {Topological quantum computation},\ }\href
  {https://doi.org/10.1090/S0273-0979-02-00964-3} {\bibfield  {journal}
  {\bibinfo  {journal} {Bull. Amer. Math. Soc.}\ }\textbf {\bibinfo {volume}
  {40}},\ \bibinfo {pages} {31} (\bibinfo {year} {2003})}\BibitemShut {NoStop}%
\bibitem [{\citenamefont {Sau}\ \emph {et~al.}(2010{\natexlab{a}})\citenamefont
  {Sau}, \citenamefont {Lutchyn}, \citenamefont {Tewari},\ and\ \citenamefont
  {Das~Sarma}}]{sau2010generic}%
  \BibitemOpen
  \bibfield  {author} {\bibinfo {author} {\bibfnamefont {J.~D.}\ \bibnamefont
  {Sau}}, \bibinfo {author} {\bibfnamefont {R.~M.}\ \bibnamefont {Lutchyn}},
  \bibinfo {author} {\bibfnamefont {S.}~\bibnamefont {Tewari}},\ and\ \bibinfo
  {author} {\bibfnamefont {S.}~\bibnamefont {Das~Sarma}},\ }\bibfield  {title}
  {\bibinfo {title} {Generic {{New Platform}} for {{Topological Quantum
  Computation Using Semiconductor Heterostructures}}},\ }\href
  {https://doi.org/10.1103/PhysRevLett.104.040502} {\bibfield  {journal}
  {\bibinfo  {journal} {Phys. Rev. Lett.}\ }\textbf {\bibinfo {volume} {104}},\
  \bibinfo {pages} {040502} (\bibinfo {year} {2010}{\natexlab{a}})}\BibitemShut
  {NoStop}%
\bibitem [{\citenamefont {Lutchyn}\ \emph {et~al.}(2010)\citenamefont
  {Lutchyn}, \citenamefont {Sau},\ and\ \citenamefont
  {Das~Sarma}}]{lutchyn2010majorana}%
  \BibitemOpen
  \bibfield  {author} {\bibinfo {author} {\bibfnamefont {R.~M.}\ \bibnamefont
  {Lutchyn}}, \bibinfo {author} {\bibfnamefont {J.~D.}\ \bibnamefont {Sau}},\
  and\ \bibinfo {author} {\bibfnamefont {S.}~\bibnamefont {Das~Sarma}},\
  }\bibfield  {title} {\bibinfo {title} {Majorana {{Fermions}} and a
  {{Topological Phase Transition}} in {{Semiconductor-Superconductor
  Heterostructures}}},\ }\href {https://doi.org/10.1103/PhysRevLett.105.077001}
  {\bibfield  {journal} {\bibinfo  {journal} {Phys. Rev. Lett.}\ }\textbf
  {\bibinfo {volume} {105}},\ \bibinfo {pages} {077001} (\bibinfo {year}
  {2010})}\BibitemShut {NoStop}%
\bibitem [{\citenamefont {Oreg}\ \emph {et~al.}(2010)\citenamefont {Oreg},
  \citenamefont {Refael},\ and\ \citenamefont {{von Oppen}}}]{oreg2010helical}%
  \BibitemOpen
  \bibfield  {author} {\bibinfo {author} {\bibfnamefont {Y.}~\bibnamefont
  {Oreg}}, \bibinfo {author} {\bibfnamefont {G.}~\bibnamefont {Refael}},\ and\
  \bibinfo {author} {\bibfnamefont {F.}~\bibnamefont {{von Oppen}}},\
  }\bibfield  {title} {\bibinfo {title} {Helical {{Liquids}} and {{Majorana
  Bound States}} in {{Quantum Wires}}},\ }\href
  {https://doi.org/10.1103/PhysRevLett.105.177002} {\bibfield  {journal}
  {\bibinfo  {journal} {Phys. Rev. Lett.}\ }\textbf {\bibinfo {volume} {105}},\
  \bibinfo {pages} {177002} (\bibinfo {year} {2010})}\BibitemShut {NoStop}%
\bibitem [{\citenamefont {Sau}\ \emph {et~al.}(2010{\natexlab{b}})\citenamefont
  {Sau}, \citenamefont {Tewari}, \citenamefont {Lutchyn}, \citenamefont
  {Stanescu},\ and\ \citenamefont {Das~Sarma}}]{sau2010nonabelian}%
  \BibitemOpen
  \bibfield  {author} {\bibinfo {author} {\bibfnamefont {J.~D.}\ \bibnamefont
  {Sau}}, \bibinfo {author} {\bibfnamefont {S.}~\bibnamefont {Tewari}},
  \bibinfo {author} {\bibfnamefont {R.~M.}\ \bibnamefont {Lutchyn}}, \bibinfo
  {author} {\bibfnamefont {T.~D.}\ \bibnamefont {Stanescu}},\ and\ \bibinfo
  {author} {\bibfnamefont {S.}~\bibnamefont {Das~Sarma}},\ }\bibfield  {title}
  {\bibinfo {title} {Non-{{Abelian}} quantum order in spin-orbit-coupled
  semiconductors: {{Search}} for topological {{Majorana}} particles in
  solid-state systems},\ }\href {https://doi.org/10.1103/PhysRevB.82.214509}
  {\bibfield  {journal} {\bibinfo  {journal} {Phys. Rev. B}\ }\textbf {\bibinfo
  {volume} {82}},\ \bibinfo {pages} {214509} (\bibinfo {year}
  {2010}{\natexlab{b}})}\BibitemShut {NoStop}%
\bibitem [{\citenamefont {Mourik}\ \emph {et~al.}(2012)\citenamefont {Mourik},
  \citenamefont {Zuo}, \citenamefont {Frolov}, \citenamefont {Plissard},
  \citenamefont {Bakkers},\ and\ \citenamefont
  {Kouwenhoven}}]{mourik2012signatures}%
  \BibitemOpen
  \bibfield  {author} {\bibinfo {author} {\bibfnamefont {V.}~\bibnamefont
  {Mourik}}, \bibinfo {author} {\bibfnamefont {K.}~\bibnamefont {Zuo}},
  \bibinfo {author} {\bibfnamefont {S.~M.}\ \bibnamefont {Frolov}}, \bibinfo
  {author} {\bibfnamefont {S.~R.}\ \bibnamefont {Plissard}}, \bibinfo {author}
  {\bibfnamefont {E.~P. A.~M.}\ \bibnamefont {Bakkers}},\ and\ \bibinfo
  {author} {\bibfnamefont {L.~P.}\ \bibnamefont {Kouwenhoven}},\ }\bibfield
  {title} {\bibinfo {title} {Signatures of {{Majorana Fermions}} in {{Hybrid
  Superconductor-Semiconductor Nanowire Devices}}},\ }\href
  {https://doi.org/10.1126/science.1222360} {\bibfield  {journal} {\bibinfo
  {journal} {Science}\ }\textbf {\bibinfo {volume} {336}},\ \bibinfo {pages}
  {1003} (\bibinfo {year} {2012})}\BibitemShut {NoStop}%
\bibitem [{\citenamefont {Das}\ \emph {et~al.}(2012)\citenamefont {Das},
  \citenamefont {Ronen}, \citenamefont {Most}, \citenamefont {Oreg},
  \citenamefont {Heiblum},\ and\ \citenamefont {Shtrikman}}]{das2012zerobias}%
  \BibitemOpen
  \bibfield  {author} {\bibinfo {author} {\bibfnamefont {A.}~\bibnamefont
  {Das}}, \bibinfo {author} {\bibfnamefont {Y.}~\bibnamefont {Ronen}}, \bibinfo
  {author} {\bibfnamefont {Y.}~\bibnamefont {Most}}, \bibinfo {author}
  {\bibfnamefont {Y.}~\bibnamefont {Oreg}}, \bibinfo {author} {\bibfnamefont
  {M.}~\bibnamefont {Heiblum}},\ and\ \bibinfo {author} {\bibfnamefont
  {H.}~\bibnamefont {Shtrikman}},\ }\bibfield  {title} {\bibinfo {title}
  {Zero-bias peaks and splitting in an {{Al}}\textendash{{InAs}} nanowire
  topological superconductor as a signature of {{Majorana}} fermions},\ }\href
  {https://www.nature.com/articles/nphys2479} {\bibfield  {journal} {\bibinfo
  {journal} {Nature Physics}\ }\textbf {\bibinfo {volume} {8}},\ \bibinfo
  {pages} {887} (\bibinfo {year} {2012})}\BibitemShut {NoStop}%
\bibitem [{\citenamefont {Deng}\ \emph {et~al.}(2012)\citenamefont {Deng},
  \citenamefont {Yu}, \citenamefont {Huang}, \citenamefont {Larsson},
  \citenamefont {Caroff},\ and\ \citenamefont {Xu}}]{deng2012anomalous}%
  \BibitemOpen
  \bibfield  {author} {\bibinfo {author} {\bibfnamefont {M.~T.}\ \bibnamefont
  {Deng}}, \bibinfo {author} {\bibfnamefont {C.~L.}\ \bibnamefont {Yu}},
  \bibinfo {author} {\bibfnamefont {G.~Y.}\ \bibnamefont {Huang}}, \bibinfo
  {author} {\bibfnamefont {M.}~\bibnamefont {Larsson}}, \bibinfo {author}
  {\bibfnamefont {P.}~\bibnamefont {Caroff}},\ and\ \bibinfo {author}
  {\bibfnamefont {H.~Q.}\ \bibnamefont {Xu}},\ }\bibfield  {title} {\bibinfo
  {title} {Anomalous {{Zero-Bias Conductance Peak}} in a
  {{Nb}}\textendash{{InSb Nanowire}}\textendash{{Nb Hybrid Device}}},\ }\href
  {https://doi.org/10.1021/nl303758w} {\bibfield  {journal} {\bibinfo
  {journal} {Nano Letters}\ }\textbf {\bibinfo {volume} {12}},\ \bibinfo
  {pages} {6414} (\bibinfo {year} {2012})}\BibitemShut {NoStop}%
\bibitem [{\citenamefont {Churchill}\ \emph {et~al.}(2013)\citenamefont
  {Churchill}, \citenamefont {Fatemi}, \citenamefont {{Grove-Rasmussen}},
  \citenamefont {Deng}, \citenamefont {Caroff}, \citenamefont {Xu},\ and\
  \citenamefont {Marcus}}]{churchill2013superconductornanowire}%
  \BibitemOpen
  \bibfield  {author} {\bibinfo {author} {\bibfnamefont {H.~O.~H.}\
  \bibnamefont {Churchill}}, \bibinfo {author} {\bibfnamefont {V.}~\bibnamefont
  {Fatemi}}, \bibinfo {author} {\bibfnamefont {K.}~\bibnamefont
  {{Grove-Rasmussen}}}, \bibinfo {author} {\bibfnamefont {M.~T.}\ \bibnamefont
  {Deng}}, \bibinfo {author} {\bibfnamefont {P.}~\bibnamefont {Caroff}},
  \bibinfo {author} {\bibfnamefont {H.~Q.}\ \bibnamefont {Xu}},\ and\ \bibinfo
  {author} {\bibfnamefont {C.~M.}\ \bibnamefont {Marcus}},\ }\bibfield  {title}
  {\bibinfo {title} {Superconductor-nanowire devices from tunneling to the
  multichannel regime: {{Zero-bias}} oscillations and magnetoconductance
  crossover},\ }\href {https://doi.org/10.1103/PhysRevB.87.241401} {\bibfield
  {journal} {\bibinfo  {journal} {Phys. Rev. B}\ }\textbf {\bibinfo {volume}
  {87}},\ \bibinfo {pages} {241401} (\bibinfo {year} {2013})}\BibitemShut
  {NoStop}%
\bibitem [{\citenamefont {Finck}\ \emph {et~al.}(2013)\citenamefont {Finck},
  \citenamefont {Van~Harlingen}, \citenamefont {Mohseni}, \citenamefont
  {Jung},\ and\ \citenamefont {Li}}]{finck2013anomalous}%
  \BibitemOpen
  \bibfield  {author} {\bibinfo {author} {\bibfnamefont {A.~D.~K.}\
  \bibnamefont {Finck}}, \bibinfo {author} {\bibfnamefont {D.~J.}\ \bibnamefont
  {Van~Harlingen}}, \bibinfo {author} {\bibfnamefont {P.~K.}\ \bibnamefont
  {Mohseni}}, \bibinfo {author} {\bibfnamefont {K.}~\bibnamefont {Jung}},\ and\
  \bibinfo {author} {\bibfnamefont {X.}~\bibnamefont {Li}},\ }\bibfield
  {title} {\bibinfo {title} {Anomalous {{Modulation}} of a {{Zero-Bias Peak}}
  in a {{Hybrid Nanowire-Superconductor Device}}},\ }\href
  {https://doi.org/10.1103/PhysRevLett.110.126406} {\bibfield  {journal}
  {\bibinfo  {journal} {Phys. Rev. Lett.}\ }\textbf {\bibinfo {volume} {110}},\
  \bibinfo {pages} {126406} (\bibinfo {year} {2013})}\BibitemShut {NoStop}%
\bibitem [{\citenamefont {Deng}\ \emph {et~al.}(2016)\citenamefont {Deng},
  \citenamefont {Vaitiek{\.e}nas}, \citenamefont {Hansen}, \citenamefont
  {Danon}, \citenamefont {Leijnse}, \citenamefont {Flensberg}, \citenamefont
  {Nyg{\aa}rd}, \citenamefont {Krogstrup},\ and\ \citenamefont
  {Marcus}}]{deng2016majorana}%
  \BibitemOpen
  \bibfield  {author} {\bibinfo {author} {\bibfnamefont {M.}~\bibnamefont
  {Deng}}, \bibinfo {author} {\bibfnamefont {S.}~\bibnamefont
  {Vaitiek{\.e}nas}}, \bibinfo {author} {\bibfnamefont {E.~B.}\ \bibnamefont
  {Hansen}}, \bibinfo {author} {\bibfnamefont {J.}~\bibnamefont {Danon}},
  \bibinfo {author} {\bibfnamefont {M.}~\bibnamefont {Leijnse}}, \bibinfo
  {author} {\bibfnamefont {K.}~\bibnamefont {Flensberg}}, \bibinfo {author}
  {\bibfnamefont {J.}~\bibnamefont {Nyg{\aa}rd}}, \bibinfo {author}
  {\bibfnamefont {P.}~\bibnamefont {Krogstrup}},\ and\ \bibinfo {author}
  {\bibfnamefont {C.~M.}\ \bibnamefont {Marcus}},\ }\bibfield  {title}
  {\bibinfo {title} {Majorana bound state in a coupled quantum-dot
  hybrid-nanowire system},\ }\href
  {http://science.sciencemag.org/content/354/6319/1557.abstract?casa_token=iM4dNJjIvEwAAAAA:EZj32k4K_Wj6ZicTGaO0AGQQlMuRZZ8wypAaqXRZyZgY66JCXd0w_rc_dC1Y2SO26oa2JO66xzFy}
  {\bibfield  {journal} {\bibinfo  {journal} {Science}\ }\textbf {\bibinfo
  {volume} {354}},\ \bibinfo {pages} {1557} (\bibinfo {year}
  {2016})}\BibitemShut {NoStop}%
\bibitem [{\citenamefont {Nichele}\ \emph {et~al.}(2017)\citenamefont
  {Nichele}, \citenamefont {Drachmann}, \citenamefont {Whiticar}, \citenamefont
  {O'Farrell}, \citenamefont {Suominen}, \citenamefont {Fornieri},
  \citenamefont {Wang}, \citenamefont {Gardner}, \citenamefont {Thomas},
  \citenamefont {Hatke}, \citenamefont {Krogstrup}, \citenamefont {Manfra},
  \citenamefont {Flensberg},\ and\ \citenamefont
  {Marcus}}]{nichele2017scaling}%
  \BibitemOpen
  \bibfield  {author} {\bibinfo {author} {\bibfnamefont {F.}~\bibnamefont
  {Nichele}}, \bibinfo {author} {\bibfnamefont {A.~C.~C.}\ \bibnamefont
  {Drachmann}}, \bibinfo {author} {\bibfnamefont {A.~M.}\ \bibnamefont
  {Whiticar}}, \bibinfo {author} {\bibfnamefont {E.~C.~T.}\ \bibnamefont
  {O'Farrell}}, \bibinfo {author} {\bibfnamefont {H.~J.}\ \bibnamefont
  {Suominen}}, \bibinfo {author} {\bibfnamefont {A.}~\bibnamefont {Fornieri}},
  \bibinfo {author} {\bibfnamefont {T.}~\bibnamefont {Wang}}, \bibinfo {author}
  {\bibfnamefont {G.~C.}\ \bibnamefont {Gardner}}, \bibinfo {author}
  {\bibfnamefont {C.}~\bibnamefont {Thomas}}, \bibinfo {author} {\bibfnamefont
  {A.~T.}\ \bibnamefont {Hatke}}, \bibinfo {author} {\bibfnamefont
  {P.}~\bibnamefont {Krogstrup}}, \bibinfo {author} {\bibfnamefont {M.~J.}\
  \bibnamefont {Manfra}}, \bibinfo {author} {\bibfnamefont {K.}~\bibnamefont
  {Flensberg}},\ and\ \bibinfo {author} {\bibfnamefont {C.~M.}\ \bibnamefont
  {Marcus}},\ }\bibfield  {title} {\bibinfo {title} {Scaling of {{Majorana
  Zero-Bias Conductance Peaks}}},\ }\href
  {https://doi.org/10.1103/PhysRevLett.119.136803} {\bibfield  {journal}
  {\bibinfo  {journal} {Phys. Rev. Lett.}\ }\textbf {\bibinfo {volume} {119}},\
  \bibinfo {pages} {136803} (\bibinfo {year} {2017})}\BibitemShut {NoStop}%
\bibitem [{\citenamefont {Zhang}\ \emph {et~al.}(2017)\citenamefont {Zhang},
  \citenamefont {G{\"u}l}, \citenamefont {{Conesa-Boj}}, \citenamefont {Nowak},
  \citenamefont {Wimmer}, \citenamefont {Zuo}, \citenamefont {Mourik},
  \citenamefont {{de Vries}}, \citenamefont {{van Veen}}, \citenamefont {{de
  Moor}}, \citenamefont {Bommer}, \citenamefont {{van Woerkom}}, \citenamefont
  {Car}, \citenamefont {Plissard}, \citenamefont {Bakkers}, \citenamefont
  {{Quintero-P{\'e}rez}}, \citenamefont {Cassidy}, \citenamefont {Koelling},
  \citenamefont {Goswami}, \citenamefont {Watanabe}, \citenamefont
  {Taniguchi},\ and\ \citenamefont {Kouwenhoven}}]{zhang2017ballistic}%
  \BibitemOpen
  \bibfield  {author} {\bibinfo {author} {\bibfnamefont {H.}~\bibnamefont
  {Zhang}}, \bibinfo {author} {\bibfnamefont {{\"O}.}~\bibnamefont {G{\"u}l}},
  \bibinfo {author} {\bibfnamefont {S.}~\bibnamefont {{Conesa-Boj}}}, \bibinfo
  {author} {\bibfnamefont {M.~P.}\ \bibnamefont {Nowak}}, \bibinfo {author}
  {\bibfnamefont {M.}~\bibnamefont {Wimmer}}, \bibinfo {author} {\bibfnamefont
  {K.}~\bibnamefont {Zuo}}, \bibinfo {author} {\bibfnamefont {V.}~\bibnamefont
  {Mourik}}, \bibinfo {author} {\bibfnamefont {F.~K.}\ \bibnamefont {{de
  Vries}}}, \bibinfo {author} {\bibfnamefont {J.}~\bibnamefont {{van Veen}}},
  \bibinfo {author} {\bibfnamefont {M.~W.~A.}\ \bibnamefont {{de Moor}}},
  \bibinfo {author} {\bibfnamefont {J.~D.~S.}\ \bibnamefont {Bommer}}, \bibinfo
  {author} {\bibfnamefont {D.~J.}\ \bibnamefont {{van Woerkom}}}, \bibinfo
  {author} {\bibfnamefont {D.}~\bibnamefont {Car}}, \bibinfo {author}
  {\bibfnamefont {S.~R.}\ \bibnamefont {Plissard}}, \bibinfo {author}
  {\bibfnamefont {E.~P. A.~M.}\ \bibnamefont {Bakkers}}, \bibinfo {author}
  {\bibfnamefont {M.}~\bibnamefont {{Quintero-P{\'e}rez}}}, \bibinfo {author}
  {\bibfnamefont {M.~C.}\ \bibnamefont {Cassidy}}, \bibinfo {author}
  {\bibfnamefont {S.}~\bibnamefont {Koelling}}, \bibinfo {author}
  {\bibfnamefont {S.}~\bibnamefont {Goswami}}, \bibinfo {author} {\bibfnamefont
  {K.}~\bibnamefont {Watanabe}}, \bibinfo {author} {\bibfnamefont
  {T.}~\bibnamefont {Taniguchi}},\ and\ \bibinfo {author} {\bibfnamefont
  {L.~P.}\ \bibnamefont {Kouwenhoven}},\ }\bibfield  {title} {\bibinfo {title}
  {Ballistic superconductivity in semiconductor nanowires},\ }\href
  {https://doi.org/10.1038/ncomms16025} {\bibfield  {journal} {\bibinfo
  {journal} {Nature Communications}\ }\textbf {\bibinfo {volume} {8}},\
  \bibinfo {pages} {16025} (\bibinfo {year} {2017})}\BibitemShut {NoStop}%
\bibitem [{\citenamefont {Kammhuber}\ \emph {et~al.}(2017)\citenamefont
  {Kammhuber}, \citenamefont {Cassidy}, \citenamefont {Pei}, \citenamefont
  {Nowak}, \citenamefont {Vuik}, \citenamefont {G{\"u}l}, \citenamefont {Car},
  \citenamefont {Plissard}, \citenamefont {Bakkers}, \citenamefont {Wimmer},\
  and\ \citenamefont {Kouwenhoven}}]{kammhuber2017conductance}%
  \BibitemOpen
  \bibfield  {author} {\bibinfo {author} {\bibfnamefont {J.}~\bibnamefont
  {Kammhuber}}, \bibinfo {author} {\bibfnamefont {M.~C.}\ \bibnamefont
  {Cassidy}}, \bibinfo {author} {\bibfnamefont {F.}~\bibnamefont {Pei}},
  \bibinfo {author} {\bibfnamefont {M.~P.}\ \bibnamefont {Nowak}}, \bibinfo
  {author} {\bibfnamefont {A.}~\bibnamefont {Vuik}}, \bibinfo {author}
  {\bibfnamefont {{\"O}.}~\bibnamefont {G{\"u}l}}, \bibinfo {author}
  {\bibfnamefont {D.}~\bibnamefont {Car}}, \bibinfo {author} {\bibfnamefont
  {S.~R.}\ \bibnamefont {Plissard}}, \bibinfo {author} {\bibfnamefont {E.~P.
  a.~M.}\ \bibnamefont {Bakkers}}, \bibinfo {author} {\bibfnamefont
  {M.}~\bibnamefont {Wimmer}},\ and\ \bibinfo {author} {\bibfnamefont {L.~P.}\
  \bibnamefont {Kouwenhoven}},\ }\bibfield  {title} {\bibinfo {title}
  {Conductance through a helical state in an {{Indium}} antimonide nanowire},\
  }\href {https://doi.org/10.1038/s41467-017-00315-y} {\bibfield  {journal}
  {\bibinfo  {journal} {Nature Communications}\ }\textbf {\bibinfo {volume}
  {8}},\ \bibinfo {pages} {478} (\bibinfo {year} {2017})}\BibitemShut {NoStop}%
\bibitem [{\citenamefont {G{\"u}l}\ \emph {et~al.}(2018)\citenamefont
  {G{\"u}l}, \citenamefont {Zhang}, \citenamefont {Bommer}, \citenamefont
  {de~Moor}, \citenamefont {Car}, \citenamefont {Plissard}, \citenamefont
  {Bakkers}, \citenamefont {Geresdi}, \citenamefont {Watanabe}, \citenamefont
  {Taniguchi},\ and\ \citenamefont {Kouwenhoven}}]{gul2018ballistic}%
  \BibitemOpen
  \bibfield  {author} {\bibinfo {author} {\bibfnamefont {{\"O}.}~\bibnamefont
  {G{\"u}l}}, \bibinfo {author} {\bibfnamefont {H.}~\bibnamefont {Zhang}},
  \bibinfo {author} {\bibfnamefont {J.~D.~S.}\ \bibnamefont {Bommer}}, \bibinfo
  {author} {\bibfnamefont {M.~W.~A.}\ \bibnamefont {de~Moor}}, \bibinfo
  {author} {\bibfnamefont {D.}~\bibnamefont {Car}}, \bibinfo {author}
  {\bibfnamefont {S.~R.}\ \bibnamefont {Plissard}}, \bibinfo {author}
  {\bibfnamefont {E.~P. A.~M.}\ \bibnamefont {Bakkers}}, \bibinfo {author}
  {\bibfnamefont {A.}~\bibnamefont {Geresdi}}, \bibinfo {author} {\bibfnamefont
  {K.}~\bibnamefont {Watanabe}}, \bibinfo {author} {\bibfnamefont
  {T.}~\bibnamefont {Taniguchi}},\ and\ \bibinfo {author} {\bibfnamefont
  {L.~P.}\ \bibnamefont {Kouwenhoven}},\ }\bibfield  {title} {\bibinfo {title}
  {Ballistic {{Majorana}} nanowire devices},\ }\href
  {https://doi.org/10.1038/s41565-017-0032-8} {\bibfield  {journal} {\bibinfo
  {journal} {Nature Nanotechnology}\ }\textbf {\bibinfo {volume} {13}},\
  \bibinfo {pages} {192} (\bibinfo {year} {2018})}\BibitemShut {NoStop}%
\bibitem [{\citenamefont {Vaitiek{\.e}nas}\ \emph {et~al.}(2018)\citenamefont
  {Vaitiek{\.e}nas}, \citenamefont {Deng}, \citenamefont {Nyg{\aa}rd},
  \citenamefont {Krogstrup},\ and\ \citenamefont
  {Marcus}}]{vaitiekenas2018effective}%
  \BibitemOpen
  \bibfield  {author} {\bibinfo {author} {\bibfnamefont {S.}~\bibnamefont
  {Vaitiek{\.e}nas}}, \bibinfo {author} {\bibfnamefont {M.-T.}\ \bibnamefont
  {Deng}}, \bibinfo {author} {\bibfnamefont {J.}~\bibnamefont {Nyg{\aa}rd}},
  \bibinfo {author} {\bibfnamefont {P.}~\bibnamefont {Krogstrup}},\ and\
  \bibinfo {author} {\bibfnamefont {C.~M.}\ \bibnamefont {Marcus}},\ }\bibfield
   {title} {\bibinfo {title} {Effective g {{Factor}} of {{Subgap States}} in
  {{Hybrid Nanowires}}},\ }\href
  {https://doi.org/10.1103/PhysRevLett.121.037703} {\bibfield  {journal}
  {\bibinfo  {journal} {Phys. Rev. Lett.}\ }\textbf {\bibinfo {volume} {121}},\
  \bibinfo {pages} {037703} (\bibinfo {year} {2018})}\BibitemShut {NoStop}%
\bibitem [{\citenamefont {de~Moor}\ \emph {et~al.}(2018)\citenamefont
  {de~Moor}, \citenamefont {Bommer}, \citenamefont {Xu}, \citenamefont
  {Winkler}, \citenamefont {Antipov}, \citenamefont {Bargerbos}, \citenamefont
  {Wang}, \citenamefont {van Loo}, \citenamefont {het Veld}, \citenamefont
  {Gazibegovic}, \citenamefont {Car}, \citenamefont {Logan}, \citenamefont
  {Pendharkar}, \citenamefont {Lee}, \citenamefont {Bakkers}, \citenamefont
  {Palmstr{\o}m}, \citenamefont {Lutchyn}, \citenamefont {Kouwenhoven},\ and\
  \citenamefont {Zhang}}]{moor2018electric}%
  \BibitemOpen
  \bibfield  {author} {\bibinfo {author} {\bibfnamefont {M.~W.~A.}\
  \bibnamefont {de~Moor}}, \bibinfo {author} {\bibfnamefont {J.~D.~S.}\
  \bibnamefont {Bommer}}, \bibinfo {author} {\bibfnamefont {D.}~\bibnamefont
  {Xu}}, \bibinfo {author} {\bibfnamefont {G.~W.}\ \bibnamefont {Winkler}},
  \bibinfo {author} {\bibfnamefont {A.~E.}\ \bibnamefont {Antipov}}, \bibinfo
  {author} {\bibfnamefont {A.}~\bibnamefont {Bargerbos}}, \bibinfo {author}
  {\bibfnamefont {G.}~\bibnamefont {Wang}}, \bibinfo {author} {\bibfnamefont
  {N.}~\bibnamefont {van Loo}}, \bibinfo {author} {\bibfnamefont {R.~L. M.~O.}\
  \bibnamefont {het Veld}}, \bibinfo {author} {\bibfnamefont {S.}~\bibnamefont
  {Gazibegovic}}, \bibinfo {author} {\bibfnamefont {D.}~\bibnamefont {Car}},
  \bibinfo {author} {\bibfnamefont {J.~A.}\ \bibnamefont {Logan}}, \bibinfo
  {author} {\bibfnamefont {M.}~\bibnamefont {Pendharkar}}, \bibinfo {author}
  {\bibfnamefont {J.~S.}\ \bibnamefont {Lee}}, \bibinfo {author} {\bibfnamefont
  {E.~P. A.~M.}\ \bibnamefont {Bakkers}}, \bibinfo {author} {\bibfnamefont
  {C.~J.}\ \bibnamefont {Palmstr{\o}m}}, \bibinfo {author} {\bibfnamefont
  {R.~M.}\ \bibnamefont {Lutchyn}}, \bibinfo {author} {\bibfnamefont {L.~P.}\
  \bibnamefont {Kouwenhoven}},\ and\ \bibinfo {author} {\bibfnamefont
  {H.}~\bibnamefont {Zhang}},\ }\bibfield  {title} {\bibinfo {title} {Electric
  field tunable superconductor-semiconductor coupling in {{Majorana}}
  nanowires},\ }\href {https://doi.org/10.1088/1367-2630/aae61d} {\bibfield
  {journal} {\bibinfo  {journal} {New J. Phys.}\ }\textbf {\bibinfo {volume}
  {20}},\ \bibinfo {pages} {103049} (\bibinfo {year} {2018})}\BibitemShut
  {NoStop}%
\bibitem [{\citenamefont {Zhang}\ \emph {et~al.}(2018)\citenamefont {Zhang},
  \citenamefont {Liu}, \citenamefont {Gazibegovic}, \citenamefont {Xu},
  \citenamefont {Logan}, \citenamefont {Wang}, \citenamefont {{van Loo}},
  \citenamefont {Bommer}, \citenamefont {{de Moor}}, \citenamefont {Car},
  \citenamefont {{Op het Veld}}, \citenamefont {{van Veldhoven}}, \citenamefont
  {Koelling}, \citenamefont {Verheijen}, \citenamefont {Pendharkar},
  \citenamefont {Pennachio}, \citenamefont {Shojaei}, \citenamefont {Lee},
  \citenamefont {Palmstr{\o}m}, \citenamefont {Bakkers}, \citenamefont
  {Sarma},\ and\ \citenamefont {Kouwenhoven}}]{zhang2018quantizeda}%
  \BibitemOpen
  \bibfield  {author} {\bibinfo {author} {\bibfnamefont {H.}~\bibnamefont
  {Zhang}}, \bibinfo {author} {\bibfnamefont {C.-X.}\ \bibnamefont {Liu}},
  \bibinfo {author} {\bibfnamefont {S.}~\bibnamefont {Gazibegovic}}, \bibinfo
  {author} {\bibfnamefont {D.}~\bibnamefont {Xu}}, \bibinfo {author}
  {\bibfnamefont {J.~A.}\ \bibnamefont {Logan}}, \bibinfo {author}
  {\bibfnamefont {G.}~\bibnamefont {Wang}}, \bibinfo {author} {\bibfnamefont
  {N.}~\bibnamefont {{van Loo}}}, \bibinfo {author} {\bibfnamefont {J.~D.~S.}\
  \bibnamefont {Bommer}}, \bibinfo {author} {\bibfnamefont {M.~W.~A.}\
  \bibnamefont {{de Moor}}}, \bibinfo {author} {\bibfnamefont {D.}~\bibnamefont
  {Car}}, \bibinfo {author} {\bibfnamefont {R.~L.~M.}\ \bibnamefont {{Op het
  Veld}}}, \bibinfo {author} {\bibfnamefont {P.~J.}\ \bibnamefont {{van
  Veldhoven}}}, \bibinfo {author} {\bibfnamefont {S.}~\bibnamefont {Koelling}},
  \bibinfo {author} {\bibfnamefont {M.~A.}\ \bibnamefont {Verheijen}}, \bibinfo
  {author} {\bibfnamefont {M.}~\bibnamefont {Pendharkar}}, \bibinfo {author}
  {\bibfnamefont {D.~J.}\ \bibnamefont {Pennachio}}, \bibinfo {author}
  {\bibfnamefont {B.}~\bibnamefont {Shojaei}}, \bibinfo {author} {\bibfnamefont
  {J.~S.}\ \bibnamefont {Lee}}, \bibinfo {author} {\bibfnamefont {C.~J.}\
  \bibnamefont {Palmstr{\o}m}}, \bibinfo {author} {\bibfnamefont {E.~P. A.~M.}\
  \bibnamefont {Bakkers}}, \bibinfo {author} {\bibfnamefont {S.~D.}\
  \bibnamefont {Sarma}},\ and\ \bibinfo {author} {\bibfnamefont {L.~P.}\
  \bibnamefont {Kouwenhoven}},\ }\bibfield  {title} {\bibinfo {title}
  {Quantized majorana conductance},\ }\href
  {https://doi.org/10.1038/nature26142} {\bibfield  {journal} {\bibinfo
  {journal} {[Retracted] Nature}\ }\textbf {\bibinfo {volume} {556}},\ \bibinfo
  {pages} {74} (\bibinfo {year} {2018})},\ \Eprint
  {https://arxiv.org/abs/1710.10701} {arXiv:1710.10701} \BibitemShut {NoStop}%
\bibitem [{\citenamefont {Bommer}\ \emph {et~al.}(2019)\citenamefont {Bommer},
  \citenamefont {Zhang}, \citenamefont {G{\"u}l}, \citenamefont {Nijholt},
  \citenamefont {Wimmer}, \citenamefont {Rybakov}, \citenamefont {Garaud},
  \citenamefont {Rodic}, \citenamefont {Babaev}, \citenamefont {Troyer},
  \citenamefont {Car}, \citenamefont {Plissard}, \citenamefont {Bakkers},
  \citenamefont {Watanabe}, \citenamefont {Taniguchi},\ and\ \citenamefont
  {Kouwenhoven}}]{bommer2019spinorbit}%
  \BibitemOpen
  \bibfield  {author} {\bibinfo {author} {\bibfnamefont {J.~D.~S.}\
  \bibnamefont {Bommer}}, \bibinfo {author} {\bibfnamefont {H.}~\bibnamefont
  {Zhang}}, \bibinfo {author} {\bibfnamefont {{\"O}.}~\bibnamefont {G{\"u}l}},
  \bibinfo {author} {\bibfnamefont {B.}~\bibnamefont {Nijholt}}, \bibinfo
  {author} {\bibfnamefont {M.}~\bibnamefont {Wimmer}}, \bibinfo {author}
  {\bibfnamefont {F.~N.}\ \bibnamefont {Rybakov}}, \bibinfo {author}
  {\bibfnamefont {J.}~\bibnamefont {Garaud}}, \bibinfo {author} {\bibfnamefont
  {D.}~\bibnamefont {Rodic}}, \bibinfo {author} {\bibfnamefont
  {E.}~\bibnamefont {Babaev}}, \bibinfo {author} {\bibfnamefont
  {M.}~\bibnamefont {Troyer}}, \bibinfo {author} {\bibfnamefont
  {D.}~\bibnamefont {Car}}, \bibinfo {author} {\bibfnamefont {S.~R.}\
  \bibnamefont {Plissard}}, \bibinfo {author} {\bibfnamefont {E.~P. A.~M.}\
  \bibnamefont {Bakkers}}, \bibinfo {author} {\bibfnamefont {K.}~\bibnamefont
  {Watanabe}}, \bibinfo {author} {\bibfnamefont {T.}~\bibnamefont
  {Taniguchi}},\ and\ \bibinfo {author} {\bibfnamefont {L.~P.}\ \bibnamefont
  {Kouwenhoven}},\ }\bibfield  {title} {\bibinfo {title} {Spin-{{Orbit
  Protection}} of {{Induced Superconductivity}} in {{Majorana Nanowires}}},\
  }\href {https://doi.org/10.1103/PhysRevLett.122.187702} {\bibfield  {journal}
  {\bibinfo  {journal} {Phys. Rev. Lett.}\ }\textbf {\bibinfo {volume} {122}},\
  \bibinfo {pages} {187702} (\bibinfo {year} {2019})}\BibitemShut {NoStop}%
\bibitem [{\citenamefont {Grivnin}\ \emph {et~al.}(2019)\citenamefont
  {Grivnin}, \citenamefont {Bor}, \citenamefont {Heiblum}, \citenamefont
  {Oreg},\ and\ \citenamefont {Shtrikman}}]{grivnin2019concomitant}%
  \BibitemOpen
  \bibfield  {author} {\bibinfo {author} {\bibfnamefont {A.}~\bibnamefont
  {Grivnin}}, \bibinfo {author} {\bibfnamefont {E.}~\bibnamefont {Bor}},
  \bibinfo {author} {\bibfnamefont {M.}~\bibnamefont {Heiblum}}, \bibinfo
  {author} {\bibfnamefont {Y.}~\bibnamefont {Oreg}},\ and\ \bibinfo {author}
  {\bibfnamefont {H.}~\bibnamefont {Shtrikman}},\ }\bibfield  {title} {\bibinfo
  {title} {Concomitant opening of a bulk-gap with an emerging possible
  {{Majorana}} zero mode},\ }\href {https://doi.org/10.1038/s41467-019-09771-0}
  {\bibfield  {journal} {\bibinfo  {journal} {Nat Commun}\ }\textbf {\bibinfo
  {volume} {10}},\ \bibinfo {pages} {1940} (\bibinfo {year}
  {2019})}\BibitemShut {NoStop}%
\bibitem [{\citenamefont {Anselmetti}\ \emph {et~al.}(2019)\citenamefont
  {Anselmetti}, \citenamefont {Martinez}, \citenamefont {M{\'e}nard},
  \citenamefont {Puglia}, \citenamefont {Malinowski}, \citenamefont {Lee},
  \citenamefont {Choi}, \citenamefont {Pendharkar}, \citenamefont
  {Palmstr{\o}m}, \citenamefont {Marcus}, \citenamefont {Casparis},\ and\
  \citenamefont {Higginbotham}}]{anselmetti2019endtoend}%
  \BibitemOpen
  \bibfield  {author} {\bibinfo {author} {\bibfnamefont {G.~L.~R.}\
  \bibnamefont {Anselmetti}}, \bibinfo {author} {\bibfnamefont {E.~A.}\
  \bibnamefont {Martinez}}, \bibinfo {author} {\bibfnamefont {G.~C.}\
  \bibnamefont {M{\'e}nard}}, \bibinfo {author} {\bibfnamefont
  {D.}~\bibnamefont {Puglia}}, \bibinfo {author} {\bibfnamefont {F.~K.}\
  \bibnamefont {Malinowski}}, \bibinfo {author} {\bibfnamefont {J.~S.}\
  \bibnamefont {Lee}}, \bibinfo {author} {\bibfnamefont {S.}~\bibnamefont
  {Choi}}, \bibinfo {author} {\bibfnamefont {M.}~\bibnamefont {Pendharkar}},
  \bibinfo {author} {\bibfnamefont {C.~J.}\ \bibnamefont {Palmstr{\o}m}},
  \bibinfo {author} {\bibfnamefont {C.~M.}\ \bibnamefont {Marcus}}, \bibinfo
  {author} {\bibfnamefont {L.}~\bibnamefont {Casparis}},\ and\ \bibinfo
  {author} {\bibfnamefont {A.~P.}\ \bibnamefont {Higginbotham}},\ }\bibfield
  {title} {\bibinfo {title} {End-to-end correlated subgap states in hybrid
  nanowires},\ }\href {https://doi.org/10.1103/PhysRevB.100.205412} {\bibfield
  {journal} {\bibinfo  {journal} {Phys. Rev. B}\ }\textbf {\bibinfo {volume}
  {100}},\ \bibinfo {pages} {205412} (\bibinfo {year} {2019})}\BibitemShut
  {NoStop}%
\bibitem [{\citenamefont {M{\'e}nard}\ \emph {et~al.}(2020)\citenamefont
  {M{\'e}nard}, \citenamefont {Anselmetti}, \citenamefont {Martinez},
  \citenamefont {Puglia}, \citenamefont {Malinowski}, \citenamefont {Lee},
  \citenamefont {Choi}, \citenamefont {Pendharkar}, \citenamefont
  {Palmstr{\o}m}, \citenamefont {Flensberg}, \citenamefont {Marcus},
  \citenamefont {Casparis},\ and\ \citenamefont
  {Higginbotham}}]{menard2020conductancematrix}%
  \BibitemOpen
  \bibfield  {author} {\bibinfo {author} {\bibfnamefont {G.~C.}\ \bibnamefont
  {M{\'e}nard}}, \bibinfo {author} {\bibfnamefont {G.~L.~R.}\ \bibnamefont
  {Anselmetti}}, \bibinfo {author} {\bibfnamefont {E.~A.}\ \bibnamefont
  {Martinez}}, \bibinfo {author} {\bibfnamefont {D.}~\bibnamefont {Puglia}},
  \bibinfo {author} {\bibfnamefont {F.~K.}\ \bibnamefont {Malinowski}},
  \bibinfo {author} {\bibfnamefont {J.~S.}\ \bibnamefont {Lee}}, \bibinfo
  {author} {\bibfnamefont {S.}~\bibnamefont {Choi}}, \bibinfo {author}
  {\bibfnamefont {M.}~\bibnamefont {Pendharkar}}, \bibinfo {author}
  {\bibfnamefont {C.~J.}\ \bibnamefont {Palmstr{\o}m}}, \bibinfo {author}
  {\bibfnamefont {K.}~\bibnamefont {Flensberg}}, \bibinfo {author}
  {\bibfnamefont {C.~M.}\ \bibnamefont {Marcus}}, \bibinfo {author}
  {\bibfnamefont {L.}~\bibnamefont {Casparis}},\ and\ \bibinfo {author}
  {\bibfnamefont {A.~P.}\ \bibnamefont {Higginbotham}},\ }\bibfield  {title}
  {\bibinfo {title} {Conductance-{{Matrix Symmetries}} of a {{Three-Terminal
  Hybrid Device}}},\ }\href {https://doi.org/10.1103/PhysRevLett.124.036802}
  {\bibfield  {journal} {\bibinfo  {journal} {Phys. Rev. Lett.}\ }\textbf
  {\bibinfo {volume} {124}},\ \bibinfo {pages} {036802} (\bibinfo {year}
  {2020})}\BibitemShut {NoStop}%
\bibitem [{\citenamefont {Puglia}\ \emph {et~al.}(2021)\citenamefont {Puglia},
  \citenamefont {Martinez}, \citenamefont {M{\'e}nard}, \citenamefont
  {P{\"o}schl}, \citenamefont {Gronin}, \citenamefont {Gardner}, \citenamefont
  {Kallaher}, \citenamefont {Manfra}, \citenamefont {Marcus}, \citenamefont
  {Higginbotham},\ and\ \citenamefont {Casparis}}]{puglia2021closing}%
  \BibitemOpen
  \bibfield  {author} {\bibinfo {author} {\bibfnamefont {D.}~\bibnamefont
  {Puglia}}, \bibinfo {author} {\bibfnamefont {E.~A.}\ \bibnamefont
  {Martinez}}, \bibinfo {author} {\bibfnamefont {G.~C.}\ \bibnamefont
  {M{\'e}nard}}, \bibinfo {author} {\bibfnamefont {A.}~\bibnamefont
  {P{\"o}schl}}, \bibinfo {author} {\bibfnamefont {S.}~\bibnamefont {Gronin}},
  \bibinfo {author} {\bibfnamefont {G.~C.}\ \bibnamefont {Gardner}}, \bibinfo
  {author} {\bibfnamefont {R.}~\bibnamefont {Kallaher}}, \bibinfo {author}
  {\bibfnamefont {M.~J.}\ \bibnamefont {Manfra}}, \bibinfo {author}
  {\bibfnamefont {C.~M.}\ \bibnamefont {Marcus}}, \bibinfo {author}
  {\bibfnamefont {A.~P.}\ \bibnamefont {Higginbotham}},\ and\ \bibinfo {author}
  {\bibfnamefont {L.}~\bibnamefont {Casparis}},\ }\bibfield  {title} {\bibinfo
  {title} {Closing of the induced gap in a hybrid superconductor-semiconductor
  nanowire},\ }\href {https://doi.org/10.1103/PhysRevB.103.235201} {\bibfield
  {journal} {\bibinfo  {journal} {Phys. Rev. B}\ }\textbf {\bibinfo {volume}
  {103}},\ \bibinfo {pages} {235201} (\bibinfo {year} {2021})}\BibitemShut
  {NoStop}%
\bibitem [{\citenamefont {Pan}\ \emph {et~al.}(2020{\natexlab{a}})\citenamefont
  {Pan}, \citenamefont {Song}, \citenamefont {Zhang}, \citenamefont {Liu},
  \citenamefont {Wen}, \citenamefont {Liao}, \citenamefont {Zhuo},
  \citenamefont {Wang}, \citenamefont {Zhang}, \citenamefont {Yang},
  \citenamefont {Ying}, \citenamefont {Miao}, \citenamefont {Li}, \citenamefont
  {Shang}, \citenamefont {Zhang},\ and\ \citenamefont {Zhao}}]{pan2020situ}%
  \BibitemOpen
  \bibfield  {author} {\bibinfo {author} {\bibfnamefont {D.}~\bibnamefont
  {Pan}}, \bibinfo {author} {\bibfnamefont {H.}~\bibnamefont {Song}}, \bibinfo
  {author} {\bibfnamefont {S.}~\bibnamefont {Zhang}}, \bibinfo {author}
  {\bibfnamefont {L.}~\bibnamefont {Liu}}, \bibinfo {author} {\bibfnamefont
  {L.}~\bibnamefont {Wen}}, \bibinfo {author} {\bibfnamefont {D.}~\bibnamefont
  {Liao}}, \bibinfo {author} {\bibfnamefont {R.}~\bibnamefont {Zhuo}}, \bibinfo
  {author} {\bibfnamefont {Z.}~\bibnamefont {Wang}}, \bibinfo {author}
  {\bibfnamefont {Z.}~\bibnamefont {Zhang}}, \bibinfo {author} {\bibfnamefont
  {S.}~\bibnamefont {Yang}}, \bibinfo {author} {\bibfnamefont {J.}~\bibnamefont
  {Ying}}, \bibinfo {author} {\bibfnamefont {W.}~\bibnamefont {Miao}}, \bibinfo
  {author} {\bibfnamefont {Y.}~\bibnamefont {Li}}, \bibinfo {author}
  {\bibfnamefont {R.}~\bibnamefont {Shang}}, \bibinfo {author} {\bibfnamefont
  {H.}~\bibnamefont {Zhang}},\ and\ \bibinfo {author} {\bibfnamefont
  {J.}~\bibnamefont {Zhao}},\ }\bibfield  {title} {\bibinfo {title} {In {{Situ
  Epitaxy}} of {{Pure Phase Ultra-Thin InAs-Al Nanowires}} for {{Quantum
  Devices}}},\ }\href {http://arxiv.org/abs/2011.13620} {\bibfield  {journal}
  {\bibinfo  {journal} {arXiv:2011.13620}\ } (\bibinfo {year}
  {2020}{\natexlab{a}})}\BibitemShut {NoStop}%
\bibitem [{\citenamefont {Zhang}\ \emph
  {et~al.}(2021{\natexlab{a}})\citenamefont {Zhang}, \citenamefont {{de Moor}},
  \citenamefont {Bommer}, \citenamefont {Xu}, \citenamefont {Wang},
  \citenamefont {{van Loo}}, \citenamefont {Liu}, \citenamefont {Gazibegovic},
  \citenamefont {Logan}, \citenamefont {Car}, \citenamefont {het Veld},
  \citenamefont {{van Veldhoven}}, \citenamefont {Koelling}, \citenamefont
  {Verheijen}, \citenamefont {Pendharkar}, \citenamefont {Pennachio},
  \citenamefont {Shojaei}, \citenamefont {Lee}, \citenamefont {Palmstr{\o}m},
  \citenamefont {Bakkers}, \citenamefont {Sarma},\ and\ \citenamefont
  {Kouwenhoven}}]{zhang2021large}%
  \BibitemOpen
  \bibfield  {author} {\bibinfo {author} {\bibfnamefont {H.}~\bibnamefont
  {Zhang}}, \bibinfo {author} {\bibfnamefont {M.~W.~A.}\ \bibnamefont {{de
  Moor}}}, \bibinfo {author} {\bibfnamefont {J.~D.~S.}\ \bibnamefont {Bommer}},
  \bibinfo {author} {\bibfnamefont {D.}~\bibnamefont {Xu}}, \bibinfo {author}
  {\bibfnamefont {G.}~\bibnamefont {Wang}}, \bibinfo {author} {\bibfnamefont
  {N.}~\bibnamefont {{van Loo}}}, \bibinfo {author} {\bibfnamefont {C.-X.}\
  \bibnamefont {Liu}}, \bibinfo {author} {\bibfnamefont {S.}~\bibnamefont
  {Gazibegovic}}, \bibinfo {author} {\bibfnamefont {J.~A.}\ \bibnamefont
  {Logan}}, \bibinfo {author} {\bibfnamefont {D.}~\bibnamefont {Car}}, \bibinfo
  {author} {\bibfnamefont {R.~L. M.~O.}\ \bibnamefont {het Veld}}, \bibinfo
  {author} {\bibfnamefont {P.~J.}\ \bibnamefont {{van Veldhoven}}}, \bibinfo
  {author} {\bibfnamefont {S.}~\bibnamefont {Koelling}}, \bibinfo {author}
  {\bibfnamefont {M.~A.}\ \bibnamefont {Verheijen}}, \bibinfo {author}
  {\bibfnamefont {M.}~\bibnamefont {Pendharkar}}, \bibinfo {author}
  {\bibfnamefont {D.~J.}\ \bibnamefont {Pennachio}}, \bibinfo {author}
  {\bibfnamefont {B.}~\bibnamefont {Shojaei}}, \bibinfo {author} {\bibfnamefont
  {J.~S.}\ \bibnamefont {Lee}}, \bibinfo {author} {\bibfnamefont {C.~J.}\
  \bibnamefont {Palmstr{\o}m}}, \bibinfo {author} {\bibfnamefont {E.~P. A.~M.}\
  \bibnamefont {Bakkers}}, \bibinfo {author} {\bibfnamefont {S.~D.}\
  \bibnamefont {Sarma}},\ and\ \bibinfo {author} {\bibfnamefont {L.~P.}\
  \bibnamefont {Kouwenhoven}},\ }\bibfield  {title} {\bibinfo {title} {Large
  zero-bias peaks in {{InSb-Al}} hybrid semiconductor-superconductor nanowire
  devices},\ }\href {http://arxiv.org/abs/2101.11456} {\bibfield  {journal}
  {\bibinfo  {journal} {arXiv:2101.11456}\ } (\bibinfo {year}
  {2021}{\natexlab{a}})}\BibitemShut {NoStop}%
\bibitem [{\citenamefont {Song}\ \emph {et~al.}(2021)\citenamefont {Song},
  \citenamefont {Zhang}, \citenamefont {Pan}, \citenamefont {Liu},
  \citenamefont {Wang}, \citenamefont {Cao}, \citenamefont {Liu}, \citenamefont
  {Wen}, \citenamefont {Liao}, \citenamefont {Zhuo}, \citenamefont {Liu},
  \citenamefont {Shang}, \citenamefont {Zhao},\ and\ \citenamefont
  {Zhang}}]{song2021large}%
  \BibitemOpen
  \bibfield  {author} {\bibinfo {author} {\bibfnamefont {H.}~\bibnamefont
  {Song}}, \bibinfo {author} {\bibfnamefont {Z.}~\bibnamefont {Zhang}},
  \bibinfo {author} {\bibfnamefont {D.}~\bibnamefont {Pan}}, \bibinfo {author}
  {\bibfnamefont {D.}~\bibnamefont {Liu}}, \bibinfo {author} {\bibfnamefont
  {Z.}~\bibnamefont {Wang}}, \bibinfo {author} {\bibfnamefont {Z.}~\bibnamefont
  {Cao}}, \bibinfo {author} {\bibfnamefont {L.}~\bibnamefont {Liu}}, \bibinfo
  {author} {\bibfnamefont {L.}~\bibnamefont {Wen}}, \bibinfo {author}
  {\bibfnamefont {D.}~\bibnamefont {Liao}}, \bibinfo {author} {\bibfnamefont
  {R.}~\bibnamefont {Zhuo}}, \bibinfo {author} {\bibfnamefont {D.~E.}\
  \bibnamefont {Liu}}, \bibinfo {author} {\bibfnamefont {R.}~\bibnamefont
  {Shang}}, \bibinfo {author} {\bibfnamefont {J.}~\bibnamefont {Zhao}},\ and\
  \bibinfo {author} {\bibfnamefont {H.}~\bibnamefont {Zhang}},\ }\bibfield
  {title} {\bibinfo {title} {Large zero bias peaks and dips in a four-terminal
  thin {{InAs-Al}} nanowire device},\ }\href {http://arxiv.org/abs/2107.08282}
  {\bibfield  {journal} {\bibinfo  {journal} {arXiv:2107.08282}\ } (\bibinfo
  {year} {2021})}\BibitemShut {NoStop}%
\bibitem [{\citenamefont {Sengupta}\ \emph {et~al.}(2001)\citenamefont
  {Sengupta}, \citenamefont {{\v Z}uti{\'c}}, \citenamefont {Kwon},
  \citenamefont {Yakovenko},\ and\ \citenamefont
  {Das~Sarma}}]{sengupta2001midgap}%
  \BibitemOpen
  \bibfield  {author} {\bibinfo {author} {\bibfnamefont {K.}~\bibnamefont
  {Sengupta}}, \bibinfo {author} {\bibfnamefont {I.}~\bibnamefont {{\v
  Z}uti{\'c}}}, \bibinfo {author} {\bibfnamefont {H.-J.}\ \bibnamefont {Kwon}},
  \bibinfo {author} {\bibfnamefont {V.~M.}\ \bibnamefont {Yakovenko}},\ and\
  \bibinfo {author} {\bibfnamefont {S.}~\bibnamefont {Das~Sarma}},\ }\bibfield
  {title} {\bibinfo {title} {Midgap edge states and pairing symmetry of
  quasi-one-dimensional organic superconductors},\ }\href
  {https://doi.org/10.1103/PhysRevB.63.144531} {\bibfield  {journal} {\bibinfo
  {journal} {Phys. Rev. B}\ }\textbf {\bibinfo {volume} {63}},\ \bibinfo
  {pages} {144531} (\bibinfo {year} {2001})}\BibitemShut {NoStop}%
\bibitem [{\citenamefont {Law}\ \emph {et~al.}(2009)\citenamefont {Law},
  \citenamefont {Lee},\ and\ \citenamefont {Ng}}]{law2009majorana}%
  \BibitemOpen
  \bibfield  {author} {\bibinfo {author} {\bibfnamefont {K.~T.}\ \bibnamefont
  {Law}}, \bibinfo {author} {\bibfnamefont {P.~A.}\ \bibnamefont {Lee}},\ and\
  \bibinfo {author} {\bibfnamefont {T.~K.}\ \bibnamefont {Ng}},\ }\bibfield
  {title} {\bibinfo {title} {Majorana {{Fermion Induced Resonant Andreev
  Reflection}}},\ }\href {https://doi.org/10.1103/PhysRevLett.103.237001}
  {\bibfield  {journal} {\bibinfo  {journal} {Phys. Rev. Lett.}\ }\textbf
  {\bibinfo {volume} {103}},\ \bibinfo {pages} {237001} (\bibinfo {year}
  {2009})}\BibitemShut {NoStop}%
\bibitem [{\citenamefont {Flensberg}(2010)}]{flensberg2010tunneling}%
  \BibitemOpen
  \bibfield  {author} {\bibinfo {author} {\bibfnamefont {K.}~\bibnamefont
  {Flensberg}},\ }\bibfield  {title} {\bibinfo {title} {Tunneling
  characteristics of a chain of {{Majorana}} bound states},\ }\href
  {https://doi.org/10.1103/PhysRevB.82.180516} {\bibfield  {journal} {\bibinfo
  {journal} {Phys. Rev. B}\ }\textbf {\bibinfo {volume} {82}},\ \bibinfo
  {pages} {180516} (\bibinfo {year} {2010})}\BibitemShut {NoStop}%
\bibitem [{\citenamefont {Wimmer}\ \emph {et~al.}(2011)\citenamefont {Wimmer},
  \citenamefont {Akhmerov}, \citenamefont {Dahlhaus},\ and\ \citenamefont
  {Beenakker}}]{wimmer2011quantum}%
  \BibitemOpen
  \bibfield  {author} {\bibinfo {author} {\bibfnamefont {M.}~\bibnamefont
  {Wimmer}}, \bibinfo {author} {\bibfnamefont {A.~R.}\ \bibnamefont
  {Akhmerov}}, \bibinfo {author} {\bibfnamefont {J.~P.}\ \bibnamefont
  {Dahlhaus}},\ and\ \bibinfo {author} {\bibfnamefont {C.~W.~J.}\ \bibnamefont
  {Beenakker}},\ }\bibfield  {title} {\bibinfo {title} {Quantum point contact
  as a probe of a topological superconductor},\ }\href
  {https://doi.org/10.1088/1367-2630/13/5/053016} {\bibfield  {journal}
  {\bibinfo  {journal} {New J. Phys.}\ }\textbf {\bibinfo {volume} {13}},\
  \bibinfo {pages} {053016} (\bibinfo {year} {2011})}\BibitemShut {NoStop}%
\bibitem [{\citenamefont {Setiawan}\ \emph {et~al.}(2017)\citenamefont
  {Setiawan}, \citenamefont {Liu}, \citenamefont {Sau},\ and\ \citenamefont
  {Das~Sarma}}]{setiawan2017electron}%
  \BibitemOpen
  \bibfield  {author} {\bibinfo {author} {\bibfnamefont {F.}~\bibnamefont
  {Setiawan}}, \bibinfo {author} {\bibfnamefont {C.-X.}\ \bibnamefont {Liu}},
  \bibinfo {author} {\bibfnamefont {J.~D.}\ \bibnamefont {Sau}},\ and\ \bibinfo
  {author} {\bibfnamefont {S.}~\bibnamefont {Das~Sarma}},\ }\bibfield  {title}
  {\bibinfo {title} {Electron temperature and tunnel coupling dependence of
  zero-bias and almost-zero-bias conductance peaks in {{Majorana}} nanowires},\
  }\href {https://doi.org/10.1103/PhysRevB.96.184520} {\bibfield  {journal}
  {\bibinfo  {journal} {Phys. Rev. B}\ }\textbf {\bibinfo {volume} {96}},\
  \bibinfo {pages} {184520} (\bibinfo {year} {2017})}\BibitemShut {NoStop}%
\bibitem [{\citenamefont {Lin}\ \emph {et~al.}(2012)\citenamefont {Lin},
  \citenamefont {Sau},\ and\ \citenamefont {Das~Sarma}}]{lin2012zerobias}%
  \BibitemOpen
  \bibfield  {author} {\bibinfo {author} {\bibfnamefont {C.-H.}\ \bibnamefont
  {Lin}}, \bibinfo {author} {\bibfnamefont {J.~D.}\ \bibnamefont {Sau}},\ and\
  \bibinfo {author} {\bibfnamefont {S.}~\bibnamefont {Das~Sarma}},\ }\bibfield
  {title} {\bibinfo {title} {Zero-bias conductance peak in {{Majorana}} wires
  made of semiconductor/superconductor hybrid structures},\ }\href
  {https://doi.org/10.1103/PhysRevB.86.224511} {\bibfield  {journal} {\bibinfo
  {journal} {Phys. Rev. B}\ }\textbf {\bibinfo {volume} {86}},\ \bibinfo
  {pages} {224511} (\bibinfo {year} {2012})}\BibitemShut {NoStop}%
\bibitem [{\citenamefont {Zhang}\ \emph
  {et~al.}(2021{\natexlab{b}})\citenamefont {Zhang}, \citenamefont {Liu},
  \citenamefont {Gazibegovic}, \citenamefont {Xu}, \citenamefont {Logan},
  \citenamefont {Wang}, \citenamefont {{van Loo}}, \citenamefont {Bommer},
  \citenamefont {{de Moor}}, \citenamefont {Car}, \citenamefont {{Op het
  Veld}}, \citenamefont {{van Veldhoven}}, \citenamefont {Koelling},
  \citenamefont {Verheijen}, \citenamefont {Pendharkar}, \citenamefont
  {Pennachio}, \citenamefont {Shojaei}, \citenamefont {Lee}, \citenamefont
  {Palmstr{\o}m}, \citenamefont {Bakkers}, \citenamefont {Das~Sarma},\ and\
  \citenamefont {Kouwenhoven}}]{zhang2021retraction}%
  \BibitemOpen
  \bibfield  {author} {\bibinfo {author} {\bibfnamefont {H.}~\bibnamefont
  {Zhang}}, \bibinfo {author} {\bibfnamefont {C.-X.}\ \bibnamefont {Liu}},
  \bibinfo {author} {\bibfnamefont {S.}~\bibnamefont {Gazibegovic}}, \bibinfo
  {author} {\bibfnamefont {D.}~\bibnamefont {Xu}}, \bibinfo {author}
  {\bibfnamefont {J.~A.}\ \bibnamefont {Logan}}, \bibinfo {author}
  {\bibfnamefont {G.}~\bibnamefont {Wang}}, \bibinfo {author} {\bibfnamefont
  {N.}~\bibnamefont {{van Loo}}}, \bibinfo {author} {\bibfnamefont {J.~D.~S.}\
  \bibnamefont {Bommer}}, \bibinfo {author} {\bibfnamefont {M.~W.~A.}\
  \bibnamefont {{de Moor}}}, \bibinfo {author} {\bibfnamefont {D.}~\bibnamefont
  {Car}}, \bibinfo {author} {\bibfnamefont {R.~L.~M.}\ \bibnamefont {{Op het
  Veld}}}, \bibinfo {author} {\bibfnamefont {P.~J.}\ \bibnamefont {{van
  Veldhoven}}}, \bibinfo {author} {\bibfnamefont {S.}~\bibnamefont {Koelling}},
  \bibinfo {author} {\bibfnamefont {M.~A.}\ \bibnamefont {Verheijen}}, \bibinfo
  {author} {\bibfnamefont {M.}~\bibnamefont {Pendharkar}}, \bibinfo {author}
  {\bibfnamefont {D.~J.}\ \bibnamefont {Pennachio}}, \bibinfo {author}
  {\bibfnamefont {B.}~\bibnamefont {Shojaei}}, \bibinfo {author} {\bibfnamefont
  {J.~S.}\ \bibnamefont {Lee}}, \bibinfo {author} {\bibfnamefont {C.~J.}\
  \bibnamefont {Palmstr{\o}m}}, \bibinfo {author} {\bibfnamefont {E.~P. A.~M.}\
  \bibnamefont {Bakkers}}, \bibinfo {author} {\bibfnamefont {S.}~\bibnamefont
  {Das~Sarma}},\ and\ \bibinfo {author} {\bibfnamefont {L.~P.}\ \bibnamefont
  {Kouwenhoven}},\ }\bibfield  {title} {\bibinfo {title} {Retraction {{Note}}:
  {{Quantized Majorana}} conductance},\ }\bibfield  {journal} {\bibinfo
  {journal} {Nature}\ }\href {https://doi.org/10.1038/s41586-021-03373-x}
  {10.1038/s41586-021-03373-x} (\bibinfo {year}
  {2021}{\natexlab{b}})\BibitemShut {NoStop}%
\bibitem [{\citenamefont {Pan}\ \emph {et~al.}(2020{\natexlab{b}})\citenamefont
  {Pan}, \citenamefont {Cole}, \citenamefont {Sau},\ and\ \citenamefont
  {Das~Sarma}}]{pan2020generic}%
  \BibitemOpen
  \bibfield  {author} {\bibinfo {author} {\bibfnamefont {H.}~\bibnamefont
  {Pan}}, \bibinfo {author} {\bibfnamefont {W.~S.}\ \bibnamefont {Cole}},
  \bibinfo {author} {\bibfnamefont {J.~D.}\ \bibnamefont {Sau}},\ and\ \bibinfo
  {author} {\bibfnamefont {S.}~\bibnamefont {Das~Sarma}},\ }\bibfield  {title}
  {\bibinfo {title} {Generic quantized zero-bias conductance peaks in
  superconductor-semiconductor hybrid structures},\ }\href
  {https://doi.org/10.1103/PhysRevB.101.024506} {\bibfield  {journal} {\bibinfo
   {journal} {Phys. Rev. B}\ }\textbf {\bibinfo {volume} {101}},\ \bibinfo
  {pages} {024506} (\bibinfo {year} {2020}{\natexlab{b}})}\BibitemShut
  {NoStop}%
\bibitem [{\citenamefont {Pan}\ and\ \citenamefont
  {Das~Sarma}(2020)}]{pan2020physical}%
  \BibitemOpen
  \bibfield  {author} {\bibinfo {author} {\bibfnamefont {H.}~\bibnamefont
  {Pan}}\ and\ \bibinfo {author} {\bibfnamefont {S.}~\bibnamefont
  {Das~Sarma}},\ }\bibfield  {title} {\bibinfo {title} {Physical mechanisms for
  zero-bias conductance peaks in {{Majorana}} nanowires},\ }\href
  {https://doi.org/10.1103/PhysRevResearch.2.013377} {\bibfield  {journal}
  {\bibinfo  {journal} {Phys. Rev. Research}\ }\textbf {\bibinfo {volume}
  {2}},\ \bibinfo {pages} {013377} (\bibinfo {year} {2020})}\BibitemShut
  {NoStop}%
\bibitem [{\citenamefont {Pan}\ \emph {et~al.}(2021{\natexlab{a}})\citenamefont
  {Pan}, \citenamefont {Sau},\ and\ \citenamefont
  {Das~Sarma}}]{pan2021threeterminal}%
  \BibitemOpen
  \bibfield  {author} {\bibinfo {author} {\bibfnamefont {H.}~\bibnamefont
  {Pan}}, \bibinfo {author} {\bibfnamefont {J.~D.}\ \bibnamefont {Sau}},\ and\
  \bibinfo {author} {\bibfnamefont {S.}~\bibnamefont {Das~Sarma}},\ }\bibfield
  {title} {\bibinfo {title} {Three-terminal nonlocal conductance in
  {{Majorana}} nanowires: {{Distinguishing}} topological and trivial in
  realistic systems with disorder and inhomogeneous potential},\ }\href
  {https://doi.org/10.1103/PhysRevB.103.014513} {\bibfield  {journal} {\bibinfo
   {journal} {Phys. Rev. B}\ }\textbf {\bibinfo {volume} {103}},\ \bibinfo
  {pages} {014513} (\bibinfo {year} {2021}{\natexlab{a}})}\BibitemShut
  {NoStop}%
\bibitem [{\citenamefont {Pan}\ and\ \citenamefont
  {Das~Sarma}(2021{\natexlab{a}})}]{pan2021disorder}%
  \BibitemOpen
  \bibfield  {author} {\bibinfo {author} {\bibfnamefont {H.}~\bibnamefont
  {Pan}}\ and\ \bibinfo {author} {\bibfnamefont {S.}~\bibnamefont
  {Das~Sarma}},\ }\bibfield  {title} {\bibinfo {title} {Disorder effects on
  {{Majorana}} zero modes: {{Kitaev}} chain versus semiconductor nanowire},\
  }\href {https://doi.org/10.1103/PhysRevB.103.224505} {\bibfield  {journal}
  {\bibinfo  {journal} {Phys. Rev. B}\ }\textbf {\bibinfo {volume} {103}},\
  \bibinfo {pages} {224505} (\bibinfo {year} {2021}{\natexlab{a}})}\BibitemShut
  {NoStop}%
\bibitem [{\citenamefont {Woods}\ and\ \citenamefont
  {Stanescu}(2020)}]{woods2020electrostatic}%
  \BibitemOpen
  \bibfield  {author} {\bibinfo {author} {\bibfnamefont {B.~D.}\ \bibnamefont
  {Woods}}\ and\ \bibinfo {author} {\bibfnamefont {T.~D.}\ \bibnamefont
  {Stanescu}},\ }\bibfield  {title} {\bibinfo {title} {Electrostatic effects
  and topological superconductivity in semiconductor-superconductor-magnetic
  insulator hybrid wires},\ }\href {http://arxiv.org/abs/2011.01933} {\bibfield
   {journal} {\bibinfo  {journal} {arXiv:2011.01933}\ } (\bibinfo {year}
  {2020})}\BibitemShut {NoStop}%
\bibitem [{\citenamefont {Lai}\ \emph {et~al.}(2021)\citenamefont {Lai},
  \citenamefont {Das~Sarma},\ and\ \citenamefont {Sau}}]{lai2021theory}%
  \BibitemOpen
  \bibfield  {author} {\bibinfo {author} {\bibfnamefont {Y.-H.}\ \bibnamefont
  {Lai}}, \bibinfo {author} {\bibfnamefont {S.}~\bibnamefont {Das~Sarma}},\
  and\ \bibinfo {author} {\bibfnamefont {J.~D.}\ \bibnamefont {Sau}},\
  }\bibfield  {title} {\bibinfo {title} {Theory of {{Coulomb}} blockaded
  transport in realistic {{Majorana}} nanowires},\ }\href
  {https://doi.org/10.1103/PhysRevB.104.085403} {\bibfield  {journal} {\bibinfo
   {journal} {Phys. Rev. B}\ }\textbf {\bibinfo {volume} {104}},\ \bibinfo
  {pages} {085403} (\bibinfo {year} {2021})}\BibitemShut {NoStop}%
\bibitem [{\citenamefont {Pan}\ \emph {et~al.}(2021{\natexlab{b}})\citenamefont
  {Pan}, \citenamefont {Liu}, \citenamefont {Wimmer},\ and\ \citenamefont
  {Das~Sarma}}]{pan2021quantized}%
  \BibitemOpen
  \bibfield  {author} {\bibinfo {author} {\bibfnamefont {H.}~\bibnamefont
  {Pan}}, \bibinfo {author} {\bibfnamefont {C.-X.}\ \bibnamefont {Liu}},
  \bibinfo {author} {\bibfnamefont {M.}~\bibnamefont {Wimmer}},\ and\ \bibinfo
  {author} {\bibfnamefont {S.}~\bibnamefont {Das~Sarma}},\ }\bibfield  {title}
  {\bibinfo {title} {Quantized and unquantized zero-bias tunneling conductance
  peaks in {{Majorana}} nanowires: {{Conductance}} below and above
  {$2{e}^{2}/h$}},\ }\href {https://doi.org/10.1103/PhysRevB.103.214502}
  {\bibfield  {journal} {\bibinfo  {journal} {Phys. Rev. B}\ }\textbf {\bibinfo
  {volume} {103}},\ \bibinfo {pages} {214502} (\bibinfo {year}
  {2021}{\natexlab{b}})}\BibitemShut {NoStop}%
\bibitem [{\citenamefont {Das~Sarma}\ and\ \citenamefont
  {Pan}(2021)}]{dassarma2021disorderinduced}%
  \BibitemOpen
  \bibfield  {author} {\bibinfo {author} {\bibfnamefont {S.}~\bibnamefont
  {Das~Sarma}}\ and\ \bibinfo {author} {\bibfnamefont {H.}~\bibnamefont
  {Pan}},\ }\bibfield  {title} {\bibinfo {title} {Disorder-induced zero-bias
  peaks in {{Majorana}} nanowires},\ }\href
  {https://doi.org/10.1103/PhysRevB.103.195158} {\bibfield  {journal} {\bibinfo
   {journal} {Phys. Rev. B}\ }\textbf {\bibinfo {volume} {103}},\ \bibinfo
  {pages} {195158} (\bibinfo {year} {2021})}\BibitemShut {NoStop}%
\bibitem [{\citenamefont {Woods}\ \emph {et~al.}(2021)\citenamefont {Woods},
  \citenamefont {Sarma},\ and\ \citenamefont {Stanescu}}]{woods2021charge}%
  \BibitemOpen
  \bibfield  {author} {\bibinfo {author} {\bibfnamefont {B.~D.}\ \bibnamefont
  {Woods}}, \bibinfo {author} {\bibfnamefont {S.~D.}\ \bibnamefont {Sarma}},\
  and\ \bibinfo {author} {\bibfnamefont {T.~D.}\ \bibnamefont {Stanescu}},\
  }\bibfield  {title} {\bibinfo {title} {Charge impurity effects in hybrid
  {{Majorana}} nanowires},\ }\href {http://arxiv.org/abs/2103.06880} {\bibfield
   {journal} {\bibinfo  {journal} {arXiv:2103.06880}\ } (\bibinfo {year}
  {2021})}\BibitemShut {NoStop}%
\bibitem [{\citenamefont {Zeng}\ \emph {et~al.}(2021)\citenamefont {Zeng},
  \citenamefont {Sharma}, \citenamefont {Tewari},\ and\ \citenamefont
  {Stanescu}}]{zeng2021partiallyseparated}%
  \BibitemOpen
  \bibfield  {author} {\bibinfo {author} {\bibfnamefont {C.}~\bibnamefont
  {Zeng}}, \bibinfo {author} {\bibfnamefont {G.}~\bibnamefont {Sharma}},
  \bibinfo {author} {\bibfnamefont {S.}~\bibnamefont {Tewari}},\ and\ \bibinfo
  {author} {\bibfnamefont {T.}~\bibnamefont {Stanescu}},\ }\bibfield  {title}
  {\bibinfo {title} {Partially-separated {{Majorana}} modes in a disordered
  medium},\ }\href {http://arxiv.org/abs/2105.06469} {\bibfield  {journal}
  {\bibinfo  {journal} {arXiv:2105.06469}\ } (\bibinfo {year}
  {2021})}\BibitemShut {NoStop}%
\bibitem [{\citenamefont {Brouwer}\ \emph
  {et~al.}(2011{\natexlab{a}})\citenamefont {Brouwer}, \citenamefont
  {Duckheim}, \citenamefont {Romito},\ and\ \citenamefont {{von
  Oppen}}}]{brouwer2011probability}%
  \BibitemOpen
  \bibfield  {author} {\bibinfo {author} {\bibfnamefont {P.~W.}\ \bibnamefont
  {Brouwer}}, \bibinfo {author} {\bibfnamefont {M.}~\bibnamefont {Duckheim}},
  \bibinfo {author} {\bibfnamefont {A.}~\bibnamefont {Romito}},\ and\ \bibinfo
  {author} {\bibfnamefont {F.}~\bibnamefont {{von Oppen}}},\ }\bibfield
  {title} {\bibinfo {title} {Probability {{Distribution}} of {{Majorana
  End-State Energies}} in {{Disordered Wires}}},\ }\href
  {https://doi.org/10.1103/PhysRevLett.107.196804} {\bibfield  {journal}
  {\bibinfo  {journal} {Phys. Rev. Lett.}\ }\textbf {\bibinfo {volume} {107}},\
  \bibinfo {pages} {196804} (\bibinfo {year} {2011}{\natexlab{a}})}\BibitemShut
  {NoStop}%
\bibitem [{\citenamefont {Bagrets}\ and\ \citenamefont
  {Altland}(2012)}]{bagrets2012class}%
  \BibitemOpen
  \bibfield  {author} {\bibinfo {author} {\bibfnamefont {D.}~\bibnamefont
  {Bagrets}}\ and\ \bibinfo {author} {\bibfnamefont {A.}~\bibnamefont
  {Altland}},\ }\bibfield  {title} {\bibinfo {title} {Class {$D$} {{Spectral
  Peak}} in {{Majorana Quantum Wires}}},\ }\href
  {https://doi.org/10.1103/PhysRevLett.109.227005} {\bibfield  {journal}
  {\bibinfo  {journal} {Phys. Rev. Lett.}\ }\textbf {\bibinfo {volume} {109}},\
  \bibinfo {pages} {227005} (\bibinfo {year} {2012})}\BibitemShut {NoStop}%
\bibitem [{\citenamefont {Pikulin}\ \emph {et~al.}(2012)\citenamefont
  {Pikulin}, \citenamefont {Dahlhaus}, \citenamefont {Wimmer}, \citenamefont
  {Schomerus},\ and\ \citenamefont {Beenakker}}]{pikulin2012zerovoltage}%
  \BibitemOpen
  \bibfield  {author} {\bibinfo {author} {\bibfnamefont {D.~I.}\ \bibnamefont
  {Pikulin}}, \bibinfo {author} {\bibfnamefont {J.~P.}\ \bibnamefont
  {Dahlhaus}}, \bibinfo {author} {\bibfnamefont {M.}~\bibnamefont {Wimmer}},
  \bibinfo {author} {\bibfnamefont {H.}~\bibnamefont {Schomerus}},\ and\
  \bibinfo {author} {\bibfnamefont {C.~W.~J.}\ \bibnamefont {Beenakker}},\
  }\bibfield  {title} {\bibinfo {title} {A zero-voltage conductance peak from
  weak antilocalization in a {{Majorana}} nanowire},\ }\href
  {https://doi.org/10.1088/1367-2630/14/12/125011} {\bibfield  {journal}
  {\bibinfo  {journal} {New J. Phys.}\ }\textbf {\bibinfo {volume} {14}},\
  \bibinfo {pages} {125011} (\bibinfo {year} {2012})}\BibitemShut {NoStop}%
\bibitem [{\citenamefont {Sau}\ and\ \citenamefont
  {Das~Sarma}(2013)}]{sau2013density}%
  \BibitemOpen
  \bibfield  {author} {\bibinfo {author} {\bibfnamefont {J.~D.}\ \bibnamefont
  {Sau}}\ and\ \bibinfo {author} {\bibfnamefont {S.}~\bibnamefont
  {Das~Sarma}},\ }\bibfield  {title} {\bibinfo {title} {Density of states of
  disordered topological superconductor-semiconductor hybrid nanowires},\
  }\href {https://doi.org/10.1103/PhysRevB.88.064506} {\bibfield  {journal}
  {\bibinfo  {journal} {Phys. Rev. B}\ }\textbf {\bibinfo {volume} {88}},\
  \bibinfo {pages} {064506} (\bibinfo {year} {2013})}\BibitemShut {NoStop}%
\bibitem [{\citenamefont {Sau}\ \emph {et~al.}(2012)\citenamefont {Sau},
  \citenamefont {Tewari},\ and\ \citenamefont
  {Das~Sarma}}]{sau2012experimental}%
  \BibitemOpen
  \bibfield  {author} {\bibinfo {author} {\bibfnamefont {J.~D.}\ \bibnamefont
  {Sau}}, \bibinfo {author} {\bibfnamefont {S.}~\bibnamefont {Tewari}},\ and\
  \bibinfo {author} {\bibfnamefont {S.}~\bibnamefont {Das~Sarma}},\ }\bibfield
  {title} {\bibinfo {title} {Experimental and materials considerations for the
  topological superconducting state in electron- and hole-doped semiconductors:
  {{Searching}} for non-{{Abelian Majorana}} modes in {{1D}} nanowires and
  {{2D}} heterostructures},\ }\href
  {https://doi.org/10.1103/PhysRevB.85.064512} {\bibfield  {journal} {\bibinfo
  {journal} {Phys. Rev. B}\ }\textbf {\bibinfo {volume} {85}},\ \bibinfo
  {pages} {064512} (\bibinfo {year} {2012})}\BibitemShut {NoStop}%
\bibitem [{\citenamefont {Chang}\ \emph {et~al.}(2015)\citenamefont {Chang},
  \citenamefont {Albrecht}, \citenamefont {Jespersen}, \citenamefont
  {Kuemmeth}, \citenamefont {Krogstrup}, \citenamefont {Nyg{\aa}rd},\ and\
  \citenamefont {Marcus}}]{chang2015hard}%
  \BibitemOpen
  \bibfield  {author} {\bibinfo {author} {\bibfnamefont {W.}~\bibnamefont
  {Chang}}, \bibinfo {author} {\bibfnamefont {S.~M.}\ \bibnamefont {Albrecht}},
  \bibinfo {author} {\bibfnamefont {T.~S.}\ \bibnamefont {Jespersen}}, \bibinfo
  {author} {\bibfnamefont {F.}~\bibnamefont {Kuemmeth}}, \bibinfo {author}
  {\bibfnamefont {P.}~\bibnamefont {Krogstrup}}, \bibinfo {author}
  {\bibfnamefont {J.}~\bibnamefont {Nyg{\aa}rd}},\ and\ \bibinfo {author}
  {\bibfnamefont {C.~M.}\ \bibnamefont {Marcus}},\ }\bibfield  {title}
  {\bibinfo {title} {Hard gap in epitaxial semiconductor\textendash
  superconductor nanowires},\ }\href {https://doi.org/10.1038/nnano.2014.306}
  {\bibfield  {journal} {\bibinfo  {journal} {Nature Nanotechnology}\ }\textbf
  {\bibinfo {volume} {10}},\ \bibinfo {pages} {232} (\bibinfo {year}
  {2015})}\BibitemShut {NoStop}%
\bibitem [{\citenamefont {Takei}\ \emph {et~al.}(2013)\citenamefont {Takei},
  \citenamefont {Fregoso}, \citenamefont {Hui}, \citenamefont {Lobos},\ and\
  \citenamefont {Das~Sarma}}]{takei2013soft}%
  \BibitemOpen
  \bibfield  {author} {\bibinfo {author} {\bibfnamefont {S.}~\bibnamefont
  {Takei}}, \bibinfo {author} {\bibfnamefont {B.~M.}\ \bibnamefont {Fregoso}},
  \bibinfo {author} {\bibfnamefont {H.-Y.}\ \bibnamefont {Hui}}, \bibinfo
  {author} {\bibfnamefont {A.~M.}\ \bibnamefont {Lobos}},\ and\ \bibinfo
  {author} {\bibfnamefont {S.}~\bibnamefont {Das~Sarma}},\ }\bibfield  {title}
  {\bibinfo {title} {Soft {{Superconducting Gap}} in {{Semiconductor Majorana
  Nanowires}}},\ }\href {https://doi.org/10.1103/PhysRevLett.110.186803}
  {\bibfield  {journal} {\bibinfo  {journal} {Phys. Rev. Lett.}\ }\textbf
  {\bibinfo {volume} {110}},\ \bibinfo {pages} {186803} (\bibinfo {year}
  {2013})}\BibitemShut {NoStop}%
\bibitem [{\citenamefont {Gazibegovic}\ \emph {et~al.}(2019)\citenamefont
  {Gazibegovic}, \citenamefont {Badawy}, \citenamefont {Buckers}, \citenamefont
  {Leubner}, \citenamefont {Shen}, \citenamefont {de~Vries}, \citenamefont
  {Koelling}, \citenamefont {Kouwenhoven}, \citenamefont {Verheijen},\ and\
  \citenamefont {Bakkers}}]{gazibegovic2019bottomup}%
  \BibitemOpen
  \bibfield  {author} {\bibinfo {author} {\bibfnamefont {S.}~\bibnamefont
  {Gazibegovic}}, \bibinfo {author} {\bibfnamefont {G.}~\bibnamefont {Badawy}},
  \bibinfo {author} {\bibfnamefont {T.~L.~J.}\ \bibnamefont {Buckers}},
  \bibinfo {author} {\bibfnamefont {P.}~\bibnamefont {Leubner}}, \bibinfo
  {author} {\bibfnamefont {J.}~\bibnamefont {Shen}}, \bibinfo {author}
  {\bibfnamefont {F.~K.}\ \bibnamefont {de~Vries}}, \bibinfo {author}
  {\bibfnamefont {S.}~\bibnamefont {Koelling}}, \bibinfo {author}
  {\bibfnamefont {L.~P.}\ \bibnamefont {Kouwenhoven}}, \bibinfo {author}
  {\bibfnamefont {M.~A.}\ \bibnamefont {Verheijen}},\ and\ \bibinfo {author}
  {\bibfnamefont {E.~P. A.~M.}\ \bibnamefont {Bakkers}},\ }\bibfield  {title}
  {\bibinfo {title} {Bottom-{{Up Grown 2D InSb Nanostructures}}},\ }\href
  {https://doi.org/10.1002/adma.201808181} {\bibfield  {journal} {\bibinfo
  {journal} {Advanced Materials}\ }\textbf {\bibinfo {volume} {31}},\ \bibinfo
  {pages} {1808181} (\bibinfo {year} {2019})}\BibitemShut {NoStop}%
\bibitem [{\citenamefont {Pauka}\ \emph {et~al.}(2020)\citenamefont {Pauka},
  \citenamefont {Witt}, \citenamefont {Allen}, \citenamefont {{Harlech-Jones}},
  \citenamefont {Jouan}, \citenamefont {Gardner}, \citenamefont {Gronin},
  \citenamefont {Wang}, \citenamefont {Thomas}, \citenamefont {Manfra},
  \citenamefont {Gukelberger}, \citenamefont {Gamble}, \citenamefont {Reilly},\
  and\ \citenamefont {Cassidy}}]{pauka2020repairing}%
  \BibitemOpen
  \bibfield  {author} {\bibinfo {author} {\bibfnamefont {S.~J.}\ \bibnamefont
  {Pauka}}, \bibinfo {author} {\bibfnamefont {J.~D.~S.}\ \bibnamefont {Witt}},
  \bibinfo {author} {\bibfnamefont {C.~N.}\ \bibnamefont {Allen}}, \bibinfo
  {author} {\bibfnamefont {B.}~\bibnamefont {{Harlech-Jones}}}, \bibinfo
  {author} {\bibfnamefont {A.}~\bibnamefont {Jouan}}, \bibinfo {author}
  {\bibfnamefont {G.~C.}\ \bibnamefont {Gardner}}, \bibinfo {author}
  {\bibfnamefont {S.}~\bibnamefont {Gronin}}, \bibinfo {author} {\bibfnamefont
  {T.}~\bibnamefont {Wang}}, \bibinfo {author} {\bibfnamefont {C.}~\bibnamefont
  {Thomas}}, \bibinfo {author} {\bibfnamefont {M.~J.}\ \bibnamefont {Manfra}},
  \bibinfo {author} {\bibfnamefont {J.}~\bibnamefont {Gukelberger}}, \bibinfo
  {author} {\bibfnamefont {J.}~\bibnamefont {Gamble}}, \bibinfo {author}
  {\bibfnamefont {D.~J.}\ \bibnamefont {Reilly}},\ and\ \bibinfo {author}
  {\bibfnamefont {M.~C.}\ \bibnamefont {Cassidy}},\ }\bibfield  {title}
  {\bibinfo {title} {Repairing the surface of {{InAs-based}} topological
  heterostructures},\ }\href {https://doi.org/10.1063/5.0014361} {\bibfield
  {journal} {\bibinfo  {journal} {Journal of Applied Physics}\ }\textbf
  {\bibinfo {volume} {128}},\ \bibinfo {pages} {114301} (\bibinfo {year}
  {2020})}\BibitemShut {NoStop}%
\bibitem [{\citenamefont {Beznasyuk}\ \emph {et~al.}(2021)\citenamefont
  {Beznasyuk}, \citenamefont {{Mart{\'i}-S{\'a}nchez}}, \citenamefont {Kang},
  \citenamefont {Tanta}, \citenamefont {Rajpalke}, \citenamefont {Stankevi{\v
  c}}, \citenamefont {Christensen}, \citenamefont {Spadaro}, \citenamefont
  {Bergamaschini}, \citenamefont {Arbiol},\ and\ \citenamefont
  {Krogstrup}}]{beznasyuk2021role}%
  \BibitemOpen
  \bibfield  {author} {\bibinfo {author} {\bibfnamefont {D.~V.}\ \bibnamefont
  {Beznasyuk}}, \bibinfo {author} {\bibfnamefont {S.}~\bibnamefont
  {{Mart{\'i}-S{\'a}nchez}}}, \bibinfo {author} {\bibfnamefont {J.-H.}\
  \bibnamefont {Kang}}, \bibinfo {author} {\bibfnamefont {R.}~\bibnamefont
  {Tanta}}, \bibinfo {author} {\bibfnamefont {M.}~\bibnamefont {Rajpalke}},
  \bibinfo {author} {\bibfnamefont {T.}~\bibnamefont {Stankevi{\v c}}},
  \bibinfo {author} {\bibfnamefont {A.~W.}\ \bibnamefont {Christensen}},
  \bibinfo {author} {\bibfnamefont {M.~C.}\ \bibnamefont {Spadaro}}, \bibinfo
  {author} {\bibfnamefont {R.}~\bibnamefont {Bergamaschini}}, \bibinfo {author}
  {\bibfnamefont {J.}~\bibnamefont {Arbiol}},\ and\ \bibinfo {author}
  {\bibfnamefont {P.}~\bibnamefont {Krogstrup}},\ }\bibfield  {title} {\bibinfo
  {title} {The role of growth temperature on the electron mobility of
  {{InAs}}/{{In}}{$_x$}{{Ga}}{$_{1-x}$}{{As}} selective area grown nanowires},\
  }\href {http://arxiv.org/abs/2103.15971} {\bibfield  {journal} {\bibinfo
  {journal} {arXiv:2103.15971}\ } (\bibinfo {year} {2021})}\BibitemShut
  {NoStop}%
\bibitem [{\citenamefont {Das~Sarma}\ and\ \citenamefont
  {Hwang}(2015)}]{dassarma2015screening}%
  \BibitemOpen
  \bibfield  {author} {\bibinfo {author} {\bibfnamefont {S.}~\bibnamefont
  {Das~Sarma}}\ and\ \bibinfo {author} {\bibfnamefont {E.~H.}\ \bibnamefont
  {Hwang}},\ }\bibfield  {title} {\bibinfo {title} {Screening and transport in
  {{2D}} semiconductor systems at low temperatures},\ }\href
  {https://doi.org/10.1038/srep16655} {\bibfield  {journal} {\bibinfo
  {journal} {Sci Rep}\ }\textbf {\bibinfo {volume} {5}},\ \bibinfo {pages}
  {16655} (\bibinfo {year} {2015})}\BibitemShut {NoStop}%
\bibitem [{\citenamefont {Ando}\ \emph {et~al.}(1982)\citenamefont {Ando},
  \citenamefont {Fowler},\ and\ \citenamefont {Stern}}]{ando1982electronic}%
  \BibitemOpen
  \bibfield  {author} {\bibinfo {author} {\bibfnamefont {T.}~\bibnamefont
  {Ando}}, \bibinfo {author} {\bibfnamefont {A.~B.}\ \bibnamefont {Fowler}},\
  and\ \bibinfo {author} {\bibfnamefont {F.}~\bibnamefont {Stern}},\ }\bibfield
   {title} {\bibinfo {title} {Electronic properties of two-dimensional
  systems},\ }\href {https://doi.org/10.1103/RevModPhys.54.437} {\bibfield
  {journal} {\bibinfo  {journal} {Rev. Mod. Phys.}\ }\textbf {\bibinfo {volume}
  {54}},\ \bibinfo {pages} {437} (\bibinfo {year} {1982})}\BibitemShut
  {NoStop}%
\bibitem [{\citenamefont {Das~Sarma}\ \emph {et~al.}(2013)\citenamefont
  {Das~Sarma}, \citenamefont {Hwang},\ and\ \citenamefont
  {Li}}]{dassarma2013twodimensional}%
  \BibitemOpen
  \bibfield  {author} {\bibinfo {author} {\bibfnamefont {S.}~\bibnamefont
  {Das~Sarma}}, \bibinfo {author} {\bibfnamefont {E.~H.}\ \bibnamefont
  {Hwang}},\ and\ \bibinfo {author} {\bibfnamefont {Q.}~\bibnamefont {Li}},\
  }\bibfield  {title} {\bibinfo {title} {Two-dimensional metal-insulator
  transition as a potential fluctuation driven semiclassical transport
  phenomenon},\ }\href {https://doi.org/10.1103/PhysRevB.88.155310} {\bibfield
  {journal} {\bibinfo  {journal} {Phys. Rev. B}\ }\textbf {\bibinfo {volume}
  {88}},\ \bibinfo {pages} {155310} (\bibinfo {year} {2013})}\BibitemShut
  {NoStop}%
\bibitem [{\citenamefont {Pudalov}\ and\ \citenamefont
  {Gershenson}(2020)}]{pudalov2020experimental}%
  \BibitemOpen
  \bibfield  {author} {\bibinfo {author} {\bibfnamefont {V.~M.}\ \bibnamefont
  {Pudalov}}\ and\ \bibinfo {author} {\bibfnamefont {M.~E.}\ \bibnamefont
  {Gershenson}},\ }\bibfield  {title} {\bibinfo {title} {Experimental
  {{Evidence}} for an {{Inhomogeneous State}} of the {{Correlated
  Two-Dimensional Electron System}} in the {{Vicinity}} of a
  {{Metal}}\textendash{{Insulator Transition}}},\ }\href
  {https://doi.org/10.1134/S0021364020040116} {\bibfield  {journal} {\bibinfo
  {journal} {Jetp Lett.}\ }\textbf {\bibinfo {volume} {111}},\ \bibinfo {pages}
  {225} (\bibinfo {year} {2020})}\BibitemShut {NoStop}%
\bibitem [{\citenamefont {Li}\ \emph {et~al.}(2019)\citenamefont {Li},
  \citenamefont {Zhang}, \citenamefont {Ghaemi},\ and\ \citenamefont
  {Sarachik}}]{li2019evidence}%
  \BibitemOpen
  \bibfield  {author} {\bibinfo {author} {\bibfnamefont {S.}~\bibnamefont
  {Li}}, \bibinfo {author} {\bibfnamefont {Q.}~\bibnamefont {Zhang}}, \bibinfo
  {author} {\bibfnamefont {P.}~\bibnamefont {Ghaemi}},\ and\ \bibinfo {author}
  {\bibfnamefont {M.~P.}\ \bibnamefont {Sarachik}},\ }\bibfield  {title}
  {\bibinfo {title} {Evidence for mixed phases and percolation at the
  metal-insulator transition in two dimensions},\ }\href
  {https://doi.org/10.1103/PhysRevB.99.155302} {\bibfield  {journal} {\bibinfo
  {journal} {Phys. Rev. B}\ }\textbf {\bibinfo {volume} {99}},\ \bibinfo
  {pages} {155302} (\bibinfo {year} {2019})}\BibitemShut {NoStop}%
\bibitem [{\citenamefont {Manfra}\ \emph {et~al.}(2007)\citenamefont {Manfra},
  \citenamefont {Hwang}, \citenamefont {Das~Sarma}, \citenamefont {Pfeiffer},
  \citenamefont {West},\ and\ \citenamefont {Sergent}}]{manfra2007transport}%
  \BibitemOpen
  \bibfield  {author} {\bibinfo {author} {\bibfnamefont {M.~J.}\ \bibnamefont
  {Manfra}}, \bibinfo {author} {\bibfnamefont {E.~H.}\ \bibnamefont {Hwang}},
  \bibinfo {author} {\bibfnamefont {S.}~\bibnamefont {Das~Sarma}}, \bibinfo
  {author} {\bibfnamefont {L.~N.}\ \bibnamefont {Pfeiffer}}, \bibinfo {author}
  {\bibfnamefont {K.~W.}\ \bibnamefont {West}},\ and\ \bibinfo {author}
  {\bibfnamefont {A.~M.}\ \bibnamefont {Sergent}},\ }\bibfield  {title}
  {\bibinfo {title} {Transport and {{Percolation}} in a {{Low-Density
  High-Mobility Two-Dimensional Hole System}}},\ }\href
  {https://doi.org/10.1103/PhysRevLett.99.236402} {\bibfield  {journal}
  {\bibinfo  {journal} {Phys. Rev. Lett.}\ }\textbf {\bibinfo {volume} {99}},\
  \bibinfo {pages} {236402} (\bibinfo {year} {2007})}\BibitemShut {NoStop}%
\bibitem [{\citenamefont {Lilly}\ \emph {et~al.}(2003)\citenamefont {Lilly},
  \citenamefont {Reno}, \citenamefont {Simmons}, \citenamefont {Spielman},
  \citenamefont {Eisenstein}, \citenamefont {Pfeiffer}, \citenamefont {West},
  \citenamefont {Hwang},\ and\ \citenamefont
  {Das~Sarma}}]{lilly2003resistivity}%
  \BibitemOpen
  \bibfield  {author} {\bibinfo {author} {\bibfnamefont {M.~P.}\ \bibnamefont
  {Lilly}}, \bibinfo {author} {\bibfnamefont {J.~L.}\ \bibnamefont {Reno}},
  \bibinfo {author} {\bibfnamefont {J.~A.}\ \bibnamefont {Simmons}}, \bibinfo
  {author} {\bibfnamefont {I.~B.}\ \bibnamefont {Spielman}}, \bibinfo {author}
  {\bibfnamefont {J.~P.}\ \bibnamefont {Eisenstein}}, \bibinfo {author}
  {\bibfnamefont {L.~N.}\ \bibnamefont {Pfeiffer}}, \bibinfo {author}
  {\bibfnamefont {K.~W.}\ \bibnamefont {West}}, \bibinfo {author}
  {\bibfnamefont {E.~H.}\ \bibnamefont {Hwang}},\ and\ \bibinfo {author}
  {\bibfnamefont {S.}~\bibnamefont {Das~Sarma}},\ }\bibfield  {title} {\bibinfo
  {title} {Resistivity of {{Dilute 2D Electrons}} in an {{Undoped GaAs
  Heterostructure}}},\ }\href {https://doi.org/10.1103/PhysRevLett.90.056806}
  {\bibfield  {journal} {\bibinfo  {journal} {Phys. Rev. Lett.}\ }\textbf
  {\bibinfo {volume} {90}},\ \bibinfo {pages} {056806} (\bibinfo {year}
  {2003})}\BibitemShut {NoStop}%
\bibitem [{\citenamefont {Das~Sarma}\ \emph {et~al.}(2005)\citenamefont
  {Das~Sarma}, \citenamefont {Lilly}, \citenamefont {Hwang}, \citenamefont
  {Pfeiffer}, \citenamefont {West},\ and\ \citenamefont
  {Reno}}]{dassarma2005twodimensional}%
  \BibitemOpen
  \bibfield  {author} {\bibinfo {author} {\bibfnamefont {S.}~\bibnamefont
  {Das~Sarma}}, \bibinfo {author} {\bibfnamefont {M.~P.}\ \bibnamefont
  {Lilly}}, \bibinfo {author} {\bibfnamefont {E.~H.}\ \bibnamefont {Hwang}},
  \bibinfo {author} {\bibfnamefont {L.~N.}\ \bibnamefont {Pfeiffer}}, \bibinfo
  {author} {\bibfnamefont {K.~W.}\ \bibnamefont {West}},\ and\ \bibinfo
  {author} {\bibfnamefont {J.~L.}\ \bibnamefont {Reno}},\ }\bibfield  {title}
  {\bibinfo {title} {Two-{{Dimensional Metal-Insulator Transition}} as a
  {{Percolation Transition}} in a {{High-Mobility Electron System}}},\ }\href
  {https://doi.org/10.1103/PhysRevLett.94.136401} {\bibfield  {journal}
  {\bibinfo  {journal} {Phys. Rev. Lett.}\ }\textbf {\bibinfo {volume} {94}},\
  \bibinfo {pages} {136401} (\bibinfo {year} {2005})}\BibitemShut {NoStop}%
\bibitem [{\citenamefont {Tracy}\ \emph {et~al.}(2009)\citenamefont {Tracy},
  \citenamefont {Hwang}, \citenamefont {Eng}, \citenamefont {Ten~Eyck},
  \citenamefont {Nordberg}, \citenamefont {Childs}, \citenamefont {Carroll},
  \citenamefont {Lilly},\ and\ \citenamefont
  {Das~Sarma}}]{tracy2009observation}%
  \BibitemOpen
  \bibfield  {author} {\bibinfo {author} {\bibfnamefont {L.~A.}\ \bibnamefont
  {Tracy}}, \bibinfo {author} {\bibfnamefont {E.~H.}\ \bibnamefont {Hwang}},
  \bibinfo {author} {\bibfnamefont {K.}~\bibnamefont {Eng}}, \bibinfo {author}
  {\bibfnamefont {G.~A.}\ \bibnamefont {Ten~Eyck}}, \bibinfo {author}
  {\bibfnamefont {E.~P.}\ \bibnamefont {Nordberg}}, \bibinfo {author}
  {\bibfnamefont {K.}~\bibnamefont {Childs}}, \bibinfo {author} {\bibfnamefont
  {M.~S.}\ \bibnamefont {Carroll}}, \bibinfo {author} {\bibfnamefont {M.~P.}\
  \bibnamefont {Lilly}},\ and\ \bibinfo {author} {\bibfnamefont
  {S.}~\bibnamefont {Das~Sarma}},\ }\bibfield  {title} {\bibinfo {title}
  {Observation of percolation-induced two-dimensional metal-insulator
  transition in a {{Si MOSFET}}},\ }\href
  {https://doi.org/10.1103/PhysRevB.79.235307} {\bibfield  {journal} {\bibinfo
  {journal} {Phys. Rev. B}\ }\textbf {\bibinfo {volume} {79}},\ \bibinfo
  {pages} {235307} (\bibinfo {year} {2009})}\BibitemShut {NoStop}%
\bibitem [{\citenamefont {Leturcq}\ \emph {et~al.}(2003)\citenamefont
  {Leturcq}, \citenamefont {L'H{\^o}te}, \citenamefont {Tourbot}, \citenamefont
  {Mellor},\ and\ \citenamefont {Henini}}]{leturcq2003resistance}%
  \BibitemOpen
  \bibfield  {author} {\bibinfo {author} {\bibfnamefont {R.}~\bibnamefont
  {Leturcq}}, \bibinfo {author} {\bibfnamefont {D.}~\bibnamefont {L'H{\^o}te}},
  \bibinfo {author} {\bibfnamefont {R.}~\bibnamefont {Tourbot}}, \bibinfo
  {author} {\bibfnamefont {C.~J.}\ \bibnamefont {Mellor}},\ and\ \bibinfo
  {author} {\bibfnamefont {M.}~\bibnamefont {Henini}},\ }\bibfield  {title}
  {\bibinfo {title} {Resistance {{Noise Scaling}} in a {{Dilute Two-Dimensional
  Hole System}} in {{GaAs}}},\ }\href
  {https://doi.org/10.1103/PhysRevLett.90.076402} {\bibfield  {journal}
  {\bibinfo  {journal} {Phys. Rev. Lett.}\ }\textbf {\bibinfo {volume} {90}},\
  \bibinfo {pages} {076402} (\bibinfo {year} {2003})}\BibitemShut {NoStop}%
\bibitem [{\citenamefont {Tracy}\ \emph {et~al.}(2006)\citenamefont {Tracy},
  \citenamefont {Eisenstein}, \citenamefont {Lilly}, \citenamefont {Pfeiffer},\
  and\ \citenamefont {West}}]{tracy2006surface}%
  \BibitemOpen
  \bibfield  {author} {\bibinfo {author} {\bibfnamefont {L.~A.}\ \bibnamefont
  {Tracy}}, \bibinfo {author} {\bibfnamefont {J.~P.}\ \bibnamefont
  {Eisenstein}}, \bibinfo {author} {\bibfnamefont {M.~P.}\ \bibnamefont
  {Lilly}}, \bibinfo {author} {\bibfnamefont {L.~N.}\ \bibnamefont
  {Pfeiffer}},\ and\ \bibinfo {author} {\bibfnamefont {K.~W.}\ \bibnamefont
  {West}},\ }\bibfield  {title} {\bibinfo {title} {Surface acoustic wave
  propagation and inhomogeneities in low-density two-dimensional electron
  systems near the metal\textendash insulator transition},\ }\href
  {https://doi.org/10.1016/j.ssc.2005.10.028} {\bibfield  {journal} {\bibinfo
  {journal} {Solid State Communications}\ }\textbf {\bibinfo {volume} {137}},\
  \bibinfo {pages} {150} (\bibinfo {year} {2006})}\BibitemShut {NoStop}%
\bibitem [{\citenamefont {Wilamowski}\ \emph {et~al.}(2001)\citenamefont
  {Wilamowski}, \citenamefont {Sandersfeld}, \citenamefont {Jantsch},
  \citenamefont {T{\"o}bben},\ and\ \citenamefont
  {Sch{\"a}ffler}}]{wilamowski2001screening}%
  \BibitemOpen
  \bibfield  {author} {\bibinfo {author} {\bibfnamefont {Z.}~\bibnamefont
  {Wilamowski}}, \bibinfo {author} {\bibfnamefont {N.}~\bibnamefont
  {Sandersfeld}}, \bibinfo {author} {\bibfnamefont {W.}~\bibnamefont
  {Jantsch}}, \bibinfo {author} {\bibfnamefont {D.}~\bibnamefont
  {T{\"o}bben}},\ and\ \bibinfo {author} {\bibfnamefont {F.}~\bibnamefont
  {Sch{\"a}ffler}},\ }\bibfield  {title} {\bibinfo {title} {Screening
  {{Breakdown}} on the {{Route}} toward the {{Metal-Insulator Transition}} in
  {{Modulation Doped Si}} {$/$}{{SiGe Quantum Wells}}},\ }\href
  {https://doi.org/10.1103/PhysRevLett.87.026401} {\bibfield  {journal}
  {\bibinfo  {journal} {Phys. Rev. Lett.}\ }\textbf {\bibinfo {volume} {87}},\
  \bibinfo {pages} {026401} (\bibinfo {year} {2001})}\BibitemShut {NoStop}%
\bibitem [{\citenamefont {He}\ and\ \citenamefont {Xie}(1998)}]{he1998new}%
  \BibitemOpen
  \bibfield  {author} {\bibinfo {author} {\bibfnamefont {S.}~\bibnamefont
  {He}}\ and\ \bibinfo {author} {\bibfnamefont {X.~C.}\ \bibnamefont {Xie}},\
  }\bibfield  {title} {\bibinfo {title} {New {{Liquid Phase}} and
  {{Metal-Insulator Transition}} in {{Si MOSFETs}}},\ }\href
  {https://doi.org/10.1103/PhysRevLett.80.3324} {\bibfield  {journal} {\bibinfo
   {journal} {Phys. Rev. Lett.}\ }\textbf {\bibinfo {volume} {80}},\ \bibinfo
  {pages} {3324} (\bibinfo {year} {1998})}\BibitemShut {NoStop}%
\bibitem [{\citenamefont {Ilani}\ \emph {et~al.}(2001)\citenamefont {Ilani},
  \citenamefont {Yacoby}, \citenamefont {Mahalu},\ and\ \citenamefont
  {Shtrikman}}]{ilani2001microscopic}%
  \BibitemOpen
  \bibfield  {author} {\bibinfo {author} {\bibfnamefont {S.}~\bibnamefont
  {Ilani}}, \bibinfo {author} {\bibfnamefont {A.}~\bibnamefont {Yacoby}},
  \bibinfo {author} {\bibfnamefont {D.}~\bibnamefont {Mahalu}},\ and\ \bibinfo
  {author} {\bibfnamefont {H.}~\bibnamefont {Shtrikman}},\ }\bibfield  {title}
  {\bibinfo {title} {Microscopic {{Structure}} of the {{Metal-Insulator
  Transition}} in {{Two Dimensions}}},\ }\href
  {https://doi.org/10.1126/science.1058645} {\bibfield  {journal} {\bibinfo
  {journal} {Science}\ }\textbf {\bibinfo {volume} {292}},\ \bibinfo {pages}
  {1354} (\bibinfo {year} {2001})}\BibitemShut {NoStop}%
\bibitem [{\citenamefont {Das~Sarma}\ and\ \citenamefont
  {Hwang}(2014{\natexlab{a}})}]{dassarma2014twodimensional}%
  \BibitemOpen
  \bibfield  {author} {\bibinfo {author} {\bibfnamefont {S.}~\bibnamefont
  {Das~Sarma}}\ and\ \bibinfo {author} {\bibfnamefont {E.~H.}\ \bibnamefont
  {Hwang}},\ }\bibfield  {title} {\bibinfo {title} {Two-dimensional
  metal-insulator transition as a strong localization induced crossover
  phenomenon},\ }\href {https://doi.org/10.1103/PhysRevB.89.235423} {\bibfield
  {journal} {\bibinfo  {journal} {Phys. Rev. B}\ }\textbf {\bibinfo {volume}
  {89}},\ \bibinfo {pages} {235423} (\bibinfo {year}
  {2014}{\natexlab{a}})}\BibitemShut {NoStop}%
\bibitem [{\citenamefont {Meir}(1999)}]{meir1999percolationtype}%
  \BibitemOpen
  \bibfield  {author} {\bibinfo {author} {\bibfnamefont {Y.}~\bibnamefont
  {Meir}},\ }\bibfield  {title} {\bibinfo {title} {Percolation-{{Type
  Description}} of the {{Metal-Insulator Transition}} in {{Two Dimensions}}},\
  }\href {https://doi.org/10.1103/PhysRevLett.83.3506} {\bibfield  {journal}
  {\bibinfo  {journal} {Phys. Rev. Lett.}\ }\textbf {\bibinfo {volume} {83}},\
  \bibinfo {pages} {3506} (\bibinfo {year} {1999})}\BibitemShut {NoStop}%
\bibitem [{\citenamefont {Knap}\ \emph {et~al.}(2014)\citenamefont {Knap},
  \citenamefont {Sau}, \citenamefont {Halperin},\ and\ \citenamefont
  {Demler}}]{knap2014transport}%
  \BibitemOpen
  \bibfield  {author} {\bibinfo {author} {\bibfnamefont {M.}~\bibnamefont
  {Knap}}, \bibinfo {author} {\bibfnamefont {J.~D.}\ \bibnamefont {Sau}},
  \bibinfo {author} {\bibfnamefont {B.~I.}\ \bibnamefont {Halperin}},\ and\
  \bibinfo {author} {\bibfnamefont {E.}~\bibnamefont {Demler}},\ }\bibfield
  {title} {\bibinfo {title} {Transport in {{Two-Dimensional Disordered
  Semimetals}}},\ }\href {https://doi.org/10.1103/PhysRevLett.113.186801}
  {\bibfield  {journal} {\bibinfo  {journal} {Phys. Rev. Lett.}\ }\textbf
  {\bibinfo {volume} {113}},\ \bibinfo {pages} {186801} (\bibinfo {year}
  {2014})}\BibitemShut {NoStop}%
\bibitem [{\citenamefont {Shabani}\ \emph {et~al.}(2014)\citenamefont
  {Shabani}, \citenamefont {Das~Sarma},\ and\ \citenamefont
  {Palmstr{\o}m}}]{shabani2014apparent}%
  \BibitemOpen
  \bibfield  {author} {\bibinfo {author} {\bibfnamefont {J.}~\bibnamefont
  {Shabani}}, \bibinfo {author} {\bibfnamefont {S.}~\bibnamefont {Das~Sarma}},\
  and\ \bibinfo {author} {\bibfnamefont {C.~J.}\ \bibnamefont {Palmstr{\o}m}},\
  }\bibfield  {title} {\bibinfo {title} {An apparent metal-insulator transition
  in high-mobility two-dimensional {{InAs}} heterostructures},\ }\href
  {https://doi.org/10.1103/PhysRevB.90.161303} {\bibfield  {journal} {\bibinfo
  {journal} {Phys. Rev. B}\ }\textbf {\bibinfo {volume} {90}},\ \bibinfo
  {pages} {161303} (\bibinfo {year} {2014})}\BibitemShut {NoStop}%
\bibitem [{\citenamefont {Das~Sarma}\ and\ \citenamefont
  {Hwang}(2013)}]{dassarma2013universal}%
  \BibitemOpen
  \bibfield  {author} {\bibinfo {author} {\bibfnamefont {S.}~\bibnamefont
  {Das~Sarma}}\ and\ \bibinfo {author} {\bibfnamefont {E.~H.}\ \bibnamefont
  {Hwang}},\ }\bibfield  {title} {\bibinfo {title} {Universal density scaling
  of disorder-limited low-temperature conductivity in high-mobility
  two-dimensional systems},\ }\href
  {https://doi.org/10.1103/PhysRevB.88.035439} {\bibfield  {journal} {\bibinfo
  {journal} {Phys. Rev. B}\ }\textbf {\bibinfo {volume} {88}},\ \bibinfo
  {pages} {035439} (\bibinfo {year} {2013})}\BibitemShut {NoStop}%
\bibitem [{bak()}]{bakkers}%
  \BibitemOpen
  \href@noop {} {\bibinfo {title} {E {{Bakkers}} (private
  communication)}}\BibitemShut {NoStop}%
\bibitem [{pri()}]{private_Cui}%
  \BibitemOpen
  \href@noop {} {\bibinfo  {journal} {A. Cui and P. Krogstrup (private
  communication)}\ }\BibitemShut {NoStop}%
\bibitem [{\citenamefont {Das~Sarma}\ and\ \citenamefont
  {Stern}(1985)}]{dassarma1985singleparticle}%
  \BibitemOpen
\bibfield  {journal} {  }\bibfield  {author} {\bibinfo {author} {\bibfnamefont
  {S.}~\bibnamefont {Das~Sarma}}\ and\ \bibinfo {author} {\bibfnamefont
  {F.}~\bibnamefont {Stern}},\ }\bibfield  {title} {\bibinfo {title}
  {Single-particle relaxation time versus scattering time in an impure electron
  gas},\ }\href {https://doi.org/10.1103/PhysRevB.32.8442} {\bibfield
  {journal} {\bibinfo  {journal} {Phys. Rev. B}\ }\textbf {\bibinfo {volume}
  {32}},\ \bibinfo {pages} {8442} (\bibinfo {year} {1985})}\BibitemShut
  {NoStop}%
\bibitem [{\citenamefont {Hwang}\ and\ \citenamefont
  {Das~Sarma}(2008)}]{hwang2008singleparticle}%
  \BibitemOpen
  \bibfield  {author} {\bibinfo {author} {\bibfnamefont {E.~H.}\ \bibnamefont
  {Hwang}}\ and\ \bibinfo {author} {\bibfnamefont {S.}~\bibnamefont
  {Das~Sarma}},\ }\bibfield  {title} {\bibinfo {title} {Single-particle
  relaxation time versus transport scattering time in a two-dimensional
  graphene layer},\ }\href {https://doi.org/10.1103/PhysRevB.77.195412}
  {\bibfield  {journal} {\bibinfo  {journal} {Phys. Rev. B}\ }\textbf {\bibinfo
  {volume} {77}},\ \bibinfo {pages} {195412} (\bibinfo {year}
  {2008})}\BibitemShut {NoStop}%
\bibitem [{\citenamefont {Das~Sarma}\ and\ \citenamefont
  {Hwang}(2014{\natexlab{b}})}]{dassarma2014mobility}%
  \BibitemOpen
  \bibfield  {author} {\bibinfo {author} {\bibfnamefont {S.}~\bibnamefont
  {Das~Sarma}}\ and\ \bibinfo {author} {\bibfnamefont {E.~H.}\ \bibnamefont
  {Hwang}},\ }\bibfield  {title} {\bibinfo {title} {Mobility versus quality in
  two-dimensional semiconductor structures},\ }\href
  {https://doi.org/10.1103/PhysRevB.90.035425} {\bibfield  {journal} {\bibinfo
  {journal} {Phys. Rev. B}\ }\textbf {\bibinfo {volume} {90}},\ \bibinfo
  {pages} {035425} (\bibinfo {year} {2014}{\natexlab{b}})}\BibitemShut
  {NoStop}%
\bibitem [{\citenamefont {Woods}\ \emph {et~al.}(2018)\citenamefont {Woods},
  \citenamefont {Stanescu},\ and\ \citenamefont
  {Das~Sarma}}]{woods2018effective}%
  \BibitemOpen
  \bibfield  {author} {\bibinfo {author} {\bibfnamefont {B.~D.}\ \bibnamefont
  {Woods}}, \bibinfo {author} {\bibfnamefont {T.~D.}\ \bibnamefont
  {Stanescu}},\ and\ \bibinfo {author} {\bibfnamefont {S.}~\bibnamefont
  {Das~Sarma}},\ }\bibfield  {title} {\bibinfo {title} {Effective theory
  approach to the {{Schr\"odinger-Poisson}} problem in semiconductor
  {{Majorana}} devices},\ }\href {https://doi.org/10.1103/PhysRevB.98.035428}
  {\bibfield  {journal} {\bibinfo  {journal} {Phys. Rev. B}\ }\textbf {\bibinfo
  {volume} {98}},\ \bibinfo {pages} {035428} (\bibinfo {year}
  {2018})}\BibitemShut {NoStop}%
\bibitem [{\citenamefont {Nijholt}\ and\ \citenamefont
  {Akhmerov}(2016)}]{nijholt2016orbital}%
  \BibitemOpen
  \bibfield  {author} {\bibinfo {author} {\bibfnamefont {B.}~\bibnamefont
  {Nijholt}}\ and\ \bibinfo {author} {\bibfnamefont {A.~R.}\ \bibnamefont
  {Akhmerov}},\ }\bibfield  {title} {\bibinfo {title} {Orbital effect of
  magnetic field on the {{Majorana}} phase diagram},\ }\href
  {https://doi.org/10.1103/PhysRevB.93.235434} {\bibfield  {journal} {\bibinfo
  {journal} {Phys. Rev. B}\ }\textbf {\bibinfo {volume} {93}},\ \bibinfo
  {pages} {235434} (\bibinfo {year} {2016})}\BibitemShut {NoStop}%
\bibitem [{\citenamefont {Manolescu}\ \emph {et~al.}(2017)\citenamefont
  {Manolescu}, \citenamefont {Sitek}, \citenamefont {Osca}, \citenamefont
  {Serra}, \citenamefont {Gudmundsson},\ and\ \citenamefont
  {Stanescu}}]{manolescu2017majorana}%
  \BibitemOpen
  \bibfield  {author} {\bibinfo {author} {\bibfnamefont {A.}~\bibnamefont
  {Manolescu}}, \bibinfo {author} {\bibfnamefont {A.}~\bibnamefont {Sitek}},
  \bibinfo {author} {\bibfnamefont {J.}~\bibnamefont {Osca}}, \bibinfo {author}
  {\bibfnamefont {L.}~\bibnamefont {Serra}}, \bibinfo {author} {\bibfnamefont
  {V.}~\bibnamefont {Gudmundsson}},\ and\ \bibinfo {author} {\bibfnamefont
  {T.~D.}\ \bibnamefont {Stanescu}},\ }\bibfield  {title} {\bibinfo {title}
  {Majorana states in prismatic core-shell nanowires},\ }\href
  {https://doi.org/10.1103/PhysRevB.96.125435} {\bibfield  {journal} {\bibinfo
  {journal} {Phys. Rev. B}\ }\textbf {\bibinfo {volume} {96}},\ \bibinfo
  {pages} {125435} (\bibinfo {year} {2017})}\BibitemShut {NoStop}%
\bibitem [{\citenamefont {Nowak}\ and\ \citenamefont
  {W{\'o}jcik}(2018)}]{nowak2018renormalization}%
  \BibitemOpen
  \bibfield  {author} {\bibinfo {author} {\bibfnamefont {M.~P.}\ \bibnamefont
  {Nowak}}\ and\ \bibinfo {author} {\bibfnamefont {P.}~\bibnamefont
  {W{\'o}jcik}},\ }\bibfield  {title} {\bibinfo {title} {Renormalization of the
  {{Majorana}} bound state decay length in a perpendicular magnetic field},\
  }\href {https://doi.org/10.1103/PhysRevB.97.045419} {\bibfield  {journal}
  {\bibinfo  {journal} {Phys. Rev. B}\ }\textbf {\bibinfo {volume} {97}},\
  \bibinfo {pages} {045419} (\bibinfo {year} {2018})}\BibitemShut {NoStop}%
\bibitem [{\citenamefont {Serra}\ and\ \citenamefont
  {Delfanazari}(2020)}]{serra2020evidence}%
  \BibitemOpen
  \bibfield  {author} {\bibinfo {author} {\bibfnamefont {L.}~\bibnamefont
  {Serra}}\ and\ \bibinfo {author} {\bibfnamefont {K.}~\bibnamefont
  {Delfanazari}},\ }\bibfield  {title} {\bibinfo {title} {Evidence for
  {{Majorana}} phases in the magnetoconductance of topological junctions based
  on two-dimensional electron gases},\ }\href
  {https://doi.org/10.1103/PhysRevB.101.115409} {\bibfield  {journal} {\bibinfo
   {journal} {Phys. Rev. B}\ }\textbf {\bibinfo {volume} {101}},\ \bibinfo
  {pages} {115409} (\bibinfo {year} {2020})}\BibitemShut {NoStop}%
\bibitem [{\citenamefont {Lei}\ \emph {et~al.}(2021)\citenamefont {Lei},
  \citenamefont {Khalsa}, \citenamefont {Du},\ and\ \citenamefont
  {MacDonald}}]{lei2021majorana}%
  \BibitemOpen
  \bibfield  {author} {\bibinfo {author} {\bibfnamefont {C.}~\bibnamefont
  {Lei}}, \bibinfo {author} {\bibfnamefont {G.}~\bibnamefont {Khalsa}},
  \bibinfo {author} {\bibfnamefont {J.}~\bibnamefont {Du}},\ and\ \bibinfo
  {author} {\bibfnamefont {A.~H.}\ \bibnamefont {MacDonald}},\ }\bibfield
  {title} {\bibinfo {title} {Majorana zero modes in a cylindrical semiconductor
  quantum wire},\ }\href {https://doi.org/10.1103/PhysRevB.104.035426}
  {\bibfield  {journal} {\bibinfo  {journal} {Phys. Rev. B}\ }\textbf {\bibinfo
  {volume} {104}},\ \bibinfo {pages} {035426} (\bibinfo {year}
  {2021})}\BibitemShut {NoStop}%
\bibitem [{\citenamefont {Kells}\ \emph {et~al.}(2012)\citenamefont {Kells},
  \citenamefont {Meidan},\ and\ \citenamefont
  {Brouwer}}]{kells2012nearzeroenergy}%
  \BibitemOpen
  \bibfield  {author} {\bibinfo {author} {\bibfnamefont {G.}~\bibnamefont
  {Kells}}, \bibinfo {author} {\bibfnamefont {D.}~\bibnamefont {Meidan}},\ and\
  \bibinfo {author} {\bibfnamefont {P.~W.}\ \bibnamefont {Brouwer}},\
  }\bibfield  {title} {\bibinfo {title} {Near-zero-energy end states in
  topologically trivial spin-orbit coupled superconducting nanowires with a
  smooth confinement},\ }\href {https://doi.org/10.1103/PhysRevB.86.100503}
  {\bibfield  {journal} {\bibinfo  {journal} {Phys. Rev. B}\ }\textbf {\bibinfo
  {volume} {86}},\ \bibinfo {pages} {100503} (\bibinfo {year}
  {2012})}\BibitemShut {NoStop}%
\bibitem [{\citenamefont {Stanescu}\ and\ \citenamefont
  {Tewari}(2019)}]{stanescu2019robust}%
  \BibitemOpen
  \bibfield  {author} {\bibinfo {author} {\bibfnamefont {T.~D.}\ \bibnamefont
  {Stanescu}}\ and\ \bibinfo {author} {\bibfnamefont {S.}~\bibnamefont
  {Tewari}},\ }\bibfield  {title} {\bibinfo {title} {Robust low-energy
  {{Andreev}} bound states in semiconductor-superconductor structures:
  {{Importance}} of partial separation of component {{Majorana}} bound
  states},\ }\href {https://doi.org/10.1103/PhysRevB.100.155429} {\bibfield
  {journal} {\bibinfo  {journal} {Phys. Rev. B}\ }\textbf {\bibinfo {volume}
  {100}},\ \bibinfo {pages} {155429} (\bibinfo {year} {2019})}\BibitemShut
  {NoStop}%
\bibitem [{\citenamefont {Brouwer}\ \emph
  {et~al.}(2011{\natexlab{b}})\citenamefont {Brouwer}, \citenamefont
  {Duckheim}, \citenamefont {Romito},\ and\ \citenamefont {{von
  Oppen}}}]{brouwer2011topological}%
  \BibitemOpen
  \bibfield  {author} {\bibinfo {author} {\bibfnamefont {P.~W.}\ \bibnamefont
  {Brouwer}}, \bibinfo {author} {\bibfnamefont {M.}~\bibnamefont {Duckheim}},
  \bibinfo {author} {\bibfnamefont {A.}~\bibnamefont {Romito}},\ and\ \bibinfo
  {author} {\bibfnamefont {F.}~\bibnamefont {{von Oppen}}},\ }\bibfield
  {title} {\bibinfo {title} {Topological superconducting phases in disordered
  quantum wires with strong spin-orbit coupling},\ }\href
  {https://doi.org/10.1103/PhysRevB.84.144526} {\bibfield  {journal} {\bibinfo
  {journal} {Phys. Rev. B}\ }\textbf {\bibinfo {volume} {84}},\ \bibinfo
  {pages} {144526} (\bibinfo {year} {2011}{\natexlab{b}})}\BibitemShut
  {NoStop}%
\bibitem [{\citenamefont {Lutchyn}\ \emph {et~al.}(2011)\citenamefont
  {Lutchyn}, \citenamefont {Stanescu},\ and\ \citenamefont
  {Das~Sarma}}]{lutchyn2011search}%
  \BibitemOpen
  \bibfield  {author} {\bibinfo {author} {\bibfnamefont {R.~M.}\ \bibnamefont
  {Lutchyn}}, \bibinfo {author} {\bibfnamefont {T.~D.}\ \bibnamefont
  {Stanescu}},\ and\ \bibinfo {author} {\bibfnamefont {S.}~\bibnamefont
  {Das~Sarma}},\ }\bibfield  {title} {\bibinfo {title} {Search for {{Majorana
  Fermions}} in {{Multiband Semiconducting Nanowires}}},\ }\href
  {https://doi.org/10.1103/PhysRevLett.106.127001} {\bibfield  {journal}
  {\bibinfo  {journal} {Phys. Rev. Lett.}\ }\textbf {\bibinfo {volume} {106}},\
  \bibinfo {pages} {127001} (\bibinfo {year} {2011})}\BibitemShut {NoStop}%
\bibitem [{\citenamefont {Akhmerov}\ \emph {et~al.}(2011)\citenamefont
  {Akhmerov}, \citenamefont {Dahlhaus}, \citenamefont {Hassler}, \citenamefont
  {Wimmer},\ and\ \citenamefont {Beenakker}}]{akhmerov2011quantized}%
  \BibitemOpen
  \bibfield  {author} {\bibinfo {author} {\bibfnamefont {A.~R.}\ \bibnamefont
  {Akhmerov}}, \bibinfo {author} {\bibfnamefont {J.~P.}\ \bibnamefont
  {Dahlhaus}}, \bibinfo {author} {\bibfnamefont {F.}~\bibnamefont {Hassler}},
  \bibinfo {author} {\bibfnamefont {M.}~\bibnamefont {Wimmer}},\ and\ \bibinfo
  {author} {\bibfnamefont {C.~W.~J.}\ \bibnamefont {Beenakker}},\ }\bibfield
  {title} {\bibinfo {title} {Quantized {{Conductance}} at the {{Majorana Phase
  Transition}} in a {{Disordered Superconducting Wire}}},\ }\href
  {https://doi.org/10.1103/PhysRevLett.106.057001} {\bibfield  {journal}
  {\bibinfo  {journal} {Phys. Rev. Lett.}\ }\textbf {\bibinfo {volume} {106}},\
  \bibinfo {pages} {057001} (\bibinfo {year} {2011})}\BibitemShut {NoStop}%
\bibitem [{\citenamefont {Liu}\ \emph {et~al.}(2012)\citenamefont {Liu},
  \citenamefont {Potter}, \citenamefont {Law},\ and\ \citenamefont
  {Lee}}]{liu2012zerobias}%
  \BibitemOpen
  \bibfield  {author} {\bibinfo {author} {\bibfnamefont {J.}~\bibnamefont
  {Liu}}, \bibinfo {author} {\bibfnamefont {A.~C.}\ \bibnamefont {Potter}},
  \bibinfo {author} {\bibfnamefont {K.~T.}\ \bibnamefont {Law}},\ and\ \bibinfo
  {author} {\bibfnamefont {P.~A.}\ \bibnamefont {Lee}},\ }\bibfield  {title}
  {\bibinfo {title} {Zero-{{Bias Peaks}} in the {{Tunneling Conductance}} of
  {{Spin-Orbit-Coupled Superconducting Wires}} with and without {{Majorana
  End-States}}},\ }\href {https://doi.org/10.1103/PhysRevLett.109.267002}
  {\bibfield  {journal} {\bibinfo  {journal} {Phys. Rev. Lett.}\ }\textbf
  {\bibinfo {volume} {109}},\ \bibinfo {pages} {267002} (\bibinfo {year}
  {2012})}\BibitemShut {NoStop}%
\bibitem [{\citenamefont {Hui}\ \emph {et~al.}(2015)\citenamefont {Hui},
  \citenamefont {Sau},\ and\ \citenamefont {Das~Sarma}}]{hui2015bulk}%
  \BibitemOpen
  \bibfield  {author} {\bibinfo {author} {\bibfnamefont {H.-Y.}\ \bibnamefont
  {Hui}}, \bibinfo {author} {\bibfnamefont {J.~D.}\ \bibnamefont {Sau}},\ and\
  \bibinfo {author} {\bibfnamefont {S.}~\bibnamefont {Das~Sarma}},\ }\bibfield
  {title} {\bibinfo {title} {Bulk disorder in the superconductor affects
  proximity-induced topological superconductivity},\ }\href
  {https://doi.org/10.1103/PhysRevB.92.174512} {\bibfield  {journal} {\bibinfo
  {journal} {Phys. Rev. B}\ }\textbf {\bibinfo {volume} {92}},\ \bibinfo
  {pages} {174512} (\bibinfo {year} {2015})}\BibitemShut {NoStop}%
\bibitem [{\citenamefont {Liu}\ \emph {et~al.}(2017)\citenamefont {Liu},
  \citenamefont {Sau}, \citenamefont {Stanescu},\ and\ \citenamefont
  {Das~Sarma}}]{liu2017andreev}%
  \BibitemOpen
  \bibfield  {author} {\bibinfo {author} {\bibfnamefont {C.-X.}\ \bibnamefont
  {Liu}}, \bibinfo {author} {\bibfnamefont {J.~D.}\ \bibnamefont {Sau}},
  \bibinfo {author} {\bibfnamefont {T.~D.}\ \bibnamefont {Stanescu}},\ and\
  \bibinfo {author} {\bibfnamefont {S.}~\bibnamefont {Das~Sarma}},\ }\bibfield
  {title} {\bibinfo {title} {Andreev bound states versus {{Majorana}} bound
  states in quantum dot-nanowire-superconductor hybrid structures: {{Trivial}}
  versus topological zero-bias conductance peaks},\ }\href
  {https://doi.org/10.1103/PhysRevB.96.075161} {\bibfield  {journal} {\bibinfo
  {journal} {Phys. Rev. B}\ }\textbf {\bibinfo {volume} {96}},\ \bibinfo
  {pages} {075161} (\bibinfo {year} {2017})}\BibitemShut {NoStop}%
\bibitem [{\citenamefont {Haim}\ and\ \citenamefont
  {Stern}(2019)}]{haim2019benefits}%
  \BibitemOpen
  \bibfield  {author} {\bibinfo {author} {\bibfnamefont {A.}~\bibnamefont
  {Haim}}\ and\ \bibinfo {author} {\bibfnamefont {A.}~\bibnamefont {Stern}},\
  }\bibfield  {title} {\bibinfo {title} {Benefits of {{Weak Disorder}} in
  {{One-Dimensional Topological Superconductors}}},\ }\href
  {https://doi.org/10.1103/PhysRevLett.122.126801} {\bibfield  {journal}
  {\bibinfo  {journal} {Phys. Rev. Lett.}\ }\textbf {\bibinfo {volume} {122}},\
  \bibinfo {pages} {126801} (\bibinfo {year} {2019})}\BibitemShut {NoStop}%
\bibitem [{\citenamefont {Das~Sarma}\ \emph {et~al.}(2012)\citenamefont
  {Das~Sarma}, \citenamefont {Sau},\ and\ \citenamefont
  {Stanescu}}]{dassarma2012splitting}%
  \BibitemOpen
  \bibfield  {author} {\bibinfo {author} {\bibfnamefont {S.}~\bibnamefont
  {Das~Sarma}}, \bibinfo {author} {\bibfnamefont {J.~D.}\ \bibnamefont {Sau}},\
  and\ \bibinfo {author} {\bibfnamefont {T.~D.}\ \bibnamefont {Stanescu}},\
  }\bibfield  {title} {\bibinfo {title} {Splitting of the zero-bias conductance
  peak as smoking gun evidence for the existence of the {{Majorana}} mode in a
  superconductor-semiconductor nanowire},\ }\href
  {https://doi.org/10.1103/PhysRevB.86.220506} {\bibfield  {journal} {\bibinfo
  {journal} {Phys. Rev. B}\ }\textbf {\bibinfo {volume} {86}},\ \bibinfo
  {pages} {220506} (\bibinfo {year} {2012})}\BibitemShut {NoStop}%
\bibitem [{\citenamefont {Albrecht}\ \emph {et~al.}(2016)\citenamefont
  {Albrecht}, \citenamefont {Higginbotham}, \citenamefont {Madsen},
  \citenamefont {Kuemmeth}, \citenamefont {Jespersen}, \citenamefont
  {Nyg{\aa}rd}, \citenamefont {Krogstrup},\ and\ \citenamefont
  {Marcus}}]{albrecht2016exponential}%
  \BibitemOpen
  \bibfield  {author} {\bibinfo {author} {\bibfnamefont {S.~M.}\ \bibnamefont
  {Albrecht}}, \bibinfo {author} {\bibfnamefont {A.}~\bibnamefont
  {Higginbotham}}, \bibinfo {author} {\bibfnamefont {M.}~\bibnamefont
  {Madsen}}, \bibinfo {author} {\bibfnamefont {F.}~\bibnamefont {Kuemmeth}},
  \bibinfo {author} {\bibfnamefont {T.~S.}\ \bibnamefont {Jespersen}}, \bibinfo
  {author} {\bibfnamefont {J.}~\bibnamefont {Nyg{\aa}rd}}, \bibinfo {author}
  {\bibfnamefont {P.}~\bibnamefont {Krogstrup}},\ and\ \bibinfo {author}
  {\bibfnamefont {C.}~\bibnamefont {Marcus}},\ }\bibfield  {title} {\bibinfo
  {title} {Exponential protection of zero modes in {{Majorana}} islands},\
  }\href {https://www.nature.com/articles/nature17162} {\bibfield  {journal}
  {\bibinfo  {journal} {Nature (London)}\ }\textbf {\bibinfo {volume} {531}},\
  \bibinfo {pages} {206} (\bibinfo {year} {2016})}\BibitemShut {NoStop}%
\bibitem [{\citenamefont {Chen}\ \emph {et~al.}(2017)\citenamefont {Chen},
  \citenamefont {Yu}, \citenamefont {Stenger}, \citenamefont {Hocevar},
  \citenamefont {Car}, \citenamefont {Plissard}, \citenamefont {Bakkers},
  \citenamefont {Stanescu},\ and\ \citenamefont
  {Frolov}}]{chen2017experimental}%
  \BibitemOpen
  \bibfield  {author} {\bibinfo {author} {\bibfnamefont {J.}~\bibnamefont
  {Chen}}, \bibinfo {author} {\bibfnamefont {P.}~\bibnamefont {Yu}}, \bibinfo
  {author} {\bibfnamefont {J.}~\bibnamefont {Stenger}}, \bibinfo {author}
  {\bibfnamefont {M.}~\bibnamefont {Hocevar}}, \bibinfo {author} {\bibfnamefont
  {D.}~\bibnamefont {Car}}, \bibinfo {author} {\bibfnamefont {S.~R.}\
  \bibnamefont {Plissard}}, \bibinfo {author} {\bibfnamefont {E.~P. A.~M.}\
  \bibnamefont {Bakkers}}, \bibinfo {author} {\bibfnamefont {T.~D.}\
  \bibnamefont {Stanescu}},\ and\ \bibinfo {author} {\bibfnamefont {S.~M.}\
  \bibnamefont {Frolov}},\ }\bibfield  {title} {\bibinfo {title} {Experimental
  phase diagram of zero-bias conductance peaks in superconductor/semiconductor
  nanowire devices},\ }\href {https://doi.org/10.1126/sciadv.1701476}
  {\bibfield  {journal} {\bibinfo  {journal} {Science Advances}\ }\textbf
  {\bibinfo {volume} {3}},\ \bibinfo {pages} {e1701476} (\bibinfo {year}
  {2017})}\BibitemShut {NoStop}%
\bibitem [{\citenamefont {Lee}\ \emph {et~al.}(2019)\citenamefont {Lee},
  \citenamefont {Choi}, \citenamefont {Pendharkar}, \citenamefont {Pennachio},
  \citenamefont {Markman}, \citenamefont {Seas}, \citenamefont {Koelling},
  \citenamefont {Verheijen}, \citenamefont {Casparis}, \citenamefont
  {Petersson}, \citenamefont {Petkovic}, \citenamefont {Schaller},
  \citenamefont {Rodwell}, \citenamefont {Marcus}, \citenamefont {Krogstrup},
  \citenamefont {Kouwenhoven}, \citenamefont {Bakkers},\ and\ \citenamefont
  {Palmstr{\o}m}}]{lee2019selectivearea}%
  \BibitemOpen
  \bibfield  {author} {\bibinfo {author} {\bibfnamefont {J.~S.}\ \bibnamefont
  {Lee}}, \bibinfo {author} {\bibfnamefont {S.}~\bibnamefont {Choi}}, \bibinfo
  {author} {\bibfnamefont {M.}~\bibnamefont {Pendharkar}}, \bibinfo {author}
  {\bibfnamefont {D.~J.}\ \bibnamefont {Pennachio}}, \bibinfo {author}
  {\bibfnamefont {B.}~\bibnamefont {Markman}}, \bibinfo {author} {\bibfnamefont
  {M.}~\bibnamefont {Seas}}, \bibinfo {author} {\bibfnamefont {S.}~\bibnamefont
  {Koelling}}, \bibinfo {author} {\bibfnamefont {M.~A.}\ \bibnamefont
  {Verheijen}}, \bibinfo {author} {\bibfnamefont {L.}~\bibnamefont {Casparis}},
  \bibinfo {author} {\bibfnamefont {K.~D.}\ \bibnamefont {Petersson}}, \bibinfo
  {author} {\bibfnamefont {I.}~\bibnamefont {Petkovic}}, \bibinfo {author}
  {\bibfnamefont {V.}~\bibnamefont {Schaller}}, \bibinfo {author}
  {\bibfnamefont {M.~J.~W.}\ \bibnamefont {Rodwell}}, \bibinfo {author}
  {\bibfnamefont {C.~M.}\ \bibnamefont {Marcus}}, \bibinfo {author}
  {\bibfnamefont {P.}~\bibnamefont {Krogstrup}}, \bibinfo {author}
  {\bibfnamefont {L.~P.}\ \bibnamefont {Kouwenhoven}}, \bibinfo {author}
  {\bibfnamefont {E.~P. A.~M.}\ \bibnamefont {Bakkers}},\ and\ \bibinfo
  {author} {\bibfnamefont {C.~J.}\ \bibnamefont {Palmstr{\o}m}},\ }\bibfield
  {title} {\bibinfo {title} {Selective-area chemical beam epitaxy of in-plane
  {{InAs}} one-dimensional channels grown on {{InP}}(001), {{InP}}(111){{B}},
  and {{InP}}(011) surfaces},\ }\href
  {https://doi.org/10.1103/PhysRevMaterials.3.084606} {\bibfield  {journal}
  {\bibinfo  {journal} {Phys. Rev. Materials}\ }\textbf {\bibinfo {volume}
  {3}},\ \bibinfo {pages} {084606} (\bibinfo {year} {2019})}\BibitemShut
  {NoStop}%
\bibitem [{\citenamefont {Shen}\ \emph {et~al.}(2021)\citenamefont {Shen},
  \citenamefont {Winkler}, \citenamefont {Borsoi}, \citenamefont {Heedt},
  \citenamefont {Levajac}, \citenamefont {Wang}, \citenamefont {{van Driel}},
  \citenamefont {Bouman}, \citenamefont {Gazibegovic}, \citenamefont
  {Op~Het~Veld}, \citenamefont {Car}, \citenamefont {Logan}, \citenamefont
  {Pendharkar}, \citenamefont {Palmstr{\o}m}, \citenamefont {Bakkers},
  \citenamefont {Kouwenhoven},\ and\ \citenamefont {{van
  Heck}}}]{shen2021full}%
  \BibitemOpen
  \bibfield  {author} {\bibinfo {author} {\bibfnamefont {J.}~\bibnamefont
  {Shen}}, \bibinfo {author} {\bibfnamefont {G.~W.}\ \bibnamefont {Winkler}},
  \bibinfo {author} {\bibfnamefont {F.}~\bibnamefont {Borsoi}}, \bibinfo
  {author} {\bibfnamefont {S.}~\bibnamefont {Heedt}}, \bibinfo {author}
  {\bibfnamefont {V.}~\bibnamefont {Levajac}}, \bibinfo {author} {\bibfnamefont
  {J.-Y.}\ \bibnamefont {Wang}}, \bibinfo {author} {\bibfnamefont
  {D.}~\bibnamefont {{van Driel}}}, \bibinfo {author} {\bibfnamefont
  {D.}~\bibnamefont {Bouman}}, \bibinfo {author} {\bibfnamefont
  {S.}~\bibnamefont {Gazibegovic}}, \bibinfo {author} {\bibfnamefont
  {R.~L.~M.}\ \bibnamefont {Op~Het~Veld}}, \bibinfo {author} {\bibfnamefont
  {D.}~\bibnamefont {Car}}, \bibinfo {author} {\bibfnamefont {J.~A.}\
  \bibnamefont {Logan}}, \bibinfo {author} {\bibfnamefont {M.}~\bibnamefont
  {Pendharkar}}, \bibinfo {author} {\bibfnamefont {C.~J.}\ \bibnamefont
  {Palmstr{\o}m}}, \bibinfo {author} {\bibfnamefont {E.~P. A.~M.}\ \bibnamefont
  {Bakkers}}, \bibinfo {author} {\bibfnamefont {L.~P.}\ \bibnamefont
  {Kouwenhoven}},\ and\ \bibinfo {author} {\bibfnamefont {B.}~\bibnamefont
  {{van Heck}}},\ }\bibfield  {title} {\bibinfo {title} {Full parity phase
  diagram of a proximitized nanowire island},\ }\href
  {https://doi.org/10.1103/PhysRevB.104.045422} {\bibfield  {journal} {\bibinfo
   {journal} {Phys. Rev. B}\ }\textbf {\bibinfo {volume} {104}},\ \bibinfo
  {pages} {045422} (\bibinfo {year} {2021})}\BibitemShut {NoStop}%
\bibitem [{\citenamefont {Yu}\ \emph {et~al.}(2021)\citenamefont {Yu},
  \citenamefont {Chen}, \citenamefont {Gomanko}, \citenamefont {Badawy},
  \citenamefont {Bakkers}, \citenamefont {Zuo}, \citenamefont {Mourik},\ and\
  \citenamefont {Frolov}}]{yu2021nonmajorana}%
  \BibitemOpen
  \bibfield  {author} {\bibinfo {author} {\bibfnamefont {P.}~\bibnamefont
  {Yu}}, \bibinfo {author} {\bibfnamefont {J.}~\bibnamefont {Chen}}, \bibinfo
  {author} {\bibfnamefont {M.}~\bibnamefont {Gomanko}}, \bibinfo {author}
  {\bibfnamefont {G.}~\bibnamefont {Badawy}}, \bibinfo {author} {\bibfnamefont
  {E.~P. a.~M.}\ \bibnamefont {Bakkers}}, \bibinfo {author} {\bibfnamefont
  {K.}~\bibnamefont {Zuo}}, \bibinfo {author} {\bibfnamefont {V.}~\bibnamefont
  {Mourik}},\ and\ \bibinfo {author} {\bibfnamefont {S.~M.}\ \bibnamefont
  {Frolov}},\ }\bibfield  {title} {\bibinfo {title} {Non-{{Majorana}} states
  yield nearly quantized conductance in proximatized nanowires},\ }\href
  {https://doi.org/10.1038/s41567-020-01107-w} {\bibfield  {journal} {\bibinfo
  {journal} {Nature Physics}\ }\textbf {\bibinfo {volume} {17}},\ \bibinfo
  {pages} {482} (\bibinfo {year} {2021})}\BibitemShut {NoStop}%
\bibitem [{\citenamefont {Olsson}\ \emph {et~al.}(1996)\citenamefont {Olsson},
  \citenamefont {Andersson}, \citenamefont {H{\aa}kansson}, \citenamefont
  {Kanski}, \citenamefont {Ilver},\ and\ \citenamefont
  {Karlsson}}]{olsson1996charge}%
  \BibitemOpen
  \bibfield  {author} {\bibinfo {author} {\bibfnamefont {L.~{\"O}.}\
  \bibnamefont {Olsson}}, \bibinfo {author} {\bibfnamefont {C.~B.~M.}\
  \bibnamefont {Andersson}}, \bibinfo {author} {\bibfnamefont {M.~C.}\
  \bibnamefont {H{\aa}kansson}}, \bibinfo {author} {\bibfnamefont
  {J.}~\bibnamefont {Kanski}}, \bibinfo {author} {\bibfnamefont
  {L.}~\bibnamefont {Ilver}},\ and\ \bibinfo {author} {\bibfnamefont {U.~O.}\
  \bibnamefont {Karlsson}},\ }\bibfield  {title} {\bibinfo {title} {Charge
  {{Accumulation}} at {{InAs Surfaces}}},\ }\href
  {https://doi.org/10.1103/PhysRevLett.76.3626} {\bibfield  {journal} {\bibinfo
   {journal} {Phys. Rev. Lett.}\ }\textbf {\bibinfo {volume} {76}},\ \bibinfo
  {pages} {3626} (\bibinfo {year} {1996})}\BibitemShut {NoStop}%
\bibitem [{\citenamefont {Weber}\ \emph {et~al.}(2010)\citenamefont {Weber},
  \citenamefont {Janotti},\ and\ \citenamefont {{Van de
  Walle}}}]{weber2010intrinsic}%
  \BibitemOpen
  \bibfield  {author} {\bibinfo {author} {\bibfnamefont {J.~R.}\ \bibnamefont
  {Weber}}, \bibinfo {author} {\bibfnamefont {A.}~\bibnamefont {Janotti}},\
  and\ \bibinfo {author} {\bibfnamefont {C.~G.}\ \bibnamefont {{Van de
  Walle}}},\ }\bibfield  {title} {\bibinfo {title} {Intrinsic and extrinsic
  causes of electron accumulation layers on {{InAs}} surfaces},\ }\href
  {https://doi.org/10.1063/1.3518061} {\bibfield  {journal} {\bibinfo
  {journal} {Appl. Phys. Lett.}\ }\textbf {\bibinfo {volume} {97}},\ \bibinfo
  {pages} {192106} (\bibinfo {year} {2010})}\BibitemShut {NoStop}%
\bibitem [{\citenamefont {Castleton}\ \emph {et~al.}(2013)\citenamefont
  {Castleton}, \citenamefont {H{\"o}glund}, \citenamefont {G{\"o}thelid},
  \citenamefont {Qian},\ and\ \citenamefont {Mirbt}}]{castleton2013hydrogen}%
  \BibitemOpen
  \bibfield  {author} {\bibinfo {author} {\bibfnamefont {C.~W.~M.}\
  \bibnamefont {Castleton}}, \bibinfo {author} {\bibfnamefont {A.}~\bibnamefont
  {H{\"o}glund}}, \bibinfo {author} {\bibfnamefont {M.}~\bibnamefont
  {G{\"o}thelid}}, \bibinfo {author} {\bibfnamefont {M.~C.}\ \bibnamefont
  {Qian}},\ and\ \bibinfo {author} {\bibfnamefont {S.}~\bibnamefont {Mirbt}},\
  }\bibfield  {title} {\bibinfo {title} {Hydrogen on {{III-V}} (110) surfaces:
  {{Charge}} accumulation and {{STM}} signatures},\ }\href
  {https://doi.org/10.1103/PhysRevB.88.045319} {\bibfield  {journal} {\bibinfo
  {journal} {Phys. Rev. B}\ }\textbf {\bibinfo {volume} {88}},\ \bibinfo
  {pages} {045319} (\bibinfo {year} {2013})}\BibitemShut {NoStop}%
\bibitem [{\citenamefont {Winkler}\ \emph {et~al.}(2019)\citenamefont
  {Winkler}, \citenamefont {Antipov}, \citenamefont {{van Heck}}, \citenamefont
  {Soluyanov}, \citenamefont {Glazman}, \citenamefont {Wimmer},\ and\
  \citenamefont {Lutchyn}}]{winkler2019unified}%
  \BibitemOpen
  \bibfield  {author} {\bibinfo {author} {\bibfnamefont {G.~W.}\ \bibnamefont
  {Winkler}}, \bibinfo {author} {\bibfnamefont {A.~E.}\ \bibnamefont
  {Antipov}}, \bibinfo {author} {\bibfnamefont {B.}~\bibnamefont {{van Heck}}},
  \bibinfo {author} {\bibfnamefont {A.~A.}\ \bibnamefont {Soluyanov}}, \bibinfo
  {author} {\bibfnamefont {L.~I.}\ \bibnamefont {Glazman}}, \bibinfo {author}
  {\bibfnamefont {M.}~\bibnamefont {Wimmer}},\ and\ \bibinfo {author}
  {\bibfnamefont {R.~M.}\ \bibnamefont {Lutchyn}},\ }\bibfield  {title}
  {\bibinfo {title} {Unified numerical approach to topological
  semiconductor-superconductor heterostructures},\ }\href
  {https://doi.org/10.1103/PhysRevB.99.245408} {\bibfield  {journal} {\bibinfo
  {journal} {Phys. Rev. B}\ }\textbf {\bibinfo {volume} {99}},\ \bibinfo
  {pages} {245408} (\bibinfo {year} {2019})}\BibitemShut {NoStop}%
\bibitem [{\citenamefont {Escribano}\ \emph {et~al.}(2019)\citenamefont
  {Escribano}, \citenamefont {Levy~Yeyati}, \citenamefont {Oreg},\ and\
  \citenamefont {Prada}}]{escribano2019effects}%
  \BibitemOpen
  \bibfield  {author} {\bibinfo {author} {\bibfnamefont {S.~D.}\ \bibnamefont
  {Escribano}}, \bibinfo {author} {\bibfnamefont {A.}~\bibnamefont
  {Levy~Yeyati}}, \bibinfo {author} {\bibfnamefont {Y.}~\bibnamefont {Oreg}},\
  and\ \bibinfo {author} {\bibfnamefont {E.}~\bibnamefont {Prada}},\ }\bibfield
   {title} {\bibinfo {title} {Effects of the electrostatic environment on
  superlattice {{Majorana}} nanowires},\ }\href
  {https://doi.org/10.1103/PhysRevB.100.045301} {\bibfield  {journal} {\bibinfo
   {journal} {Phys. Rev. B}\ }\textbf {\bibinfo {volume} {100}},\ \bibinfo
  {pages} {045301} (\bibinfo {year} {2019})}\BibitemShut {NoStop}%
\bibitem [{\citenamefont {Woods}\ \emph {et~al.}(2020)\citenamefont {Woods},
  \citenamefont {Das~Sarma},\ and\ \citenamefont
  {Stanescu}}]{woods2020subband}%
  \BibitemOpen
  \bibfield  {author} {\bibinfo {author} {\bibfnamefont {B.~D.}\ \bibnamefont
  {Woods}}, \bibinfo {author} {\bibfnamefont {S.}~\bibnamefont {Das~Sarma}},\
  and\ \bibinfo {author} {\bibfnamefont {T.~D.}\ \bibnamefont {Stanescu}},\
  }\bibfield  {title} {\bibinfo {title} {Subband occupation in
  semiconductor-superconductor nanowires},\ }\href
  {https://doi.org/10.1103/PhysRevB.101.045405} {\bibfield  {journal} {\bibinfo
   {journal} {Phys. Rev. B}\ }\textbf {\bibinfo {volume} {101}},\ \bibinfo
  {pages} {045405} (\bibinfo {year} {2020})}\BibitemShut {NoStop}%
\bibitem [{\citenamefont {Liu}\ \emph {et~al.}(2021)\citenamefont {Liu},
  \citenamefont {Schuwalow}, \citenamefont {Liu}, \citenamefont {Vilkelis},
  \citenamefont {Manesco}, \citenamefont {Krogstrup},\ and\ \citenamefont
  {Wimmer}}]{liu2021electronic}%
  \BibitemOpen
  \bibfield  {author} {\bibinfo {author} {\bibfnamefont {C.-X.}\ \bibnamefont
  {Liu}}, \bibinfo {author} {\bibfnamefont {S.}~\bibnamefont {Schuwalow}},
  \bibinfo {author} {\bibfnamefont {Y.}~\bibnamefont {Liu}}, \bibinfo {author}
  {\bibfnamefont {K.}~\bibnamefont {Vilkelis}}, \bibinfo {author}
  {\bibfnamefont {A.~L.~R.}\ \bibnamefont {Manesco}}, \bibinfo {author}
  {\bibfnamefont {P.}~\bibnamefont {Krogstrup}},\ and\ \bibinfo {author}
  {\bibfnamefont {M.}~\bibnamefont {Wimmer}},\ }\bibfield  {title} {\bibinfo
  {title} {Electronic properties of {{InAs}}/{{EuS}}/{{Al}} hybrid nanowires},\
  }\href {https://doi.org/10.1103/PhysRevB.104.014516} {\bibfield  {journal}
  {\bibinfo  {journal} {Phys. Rev. B}\ }\textbf {\bibinfo {volume} {104}},\
  \bibinfo {pages} {014516} (\bibinfo {year} {2021})}\BibitemShut {NoStop}%
\bibitem [{\citenamefont {Stanescu}\ \emph {et~al.}(2011)\citenamefont
  {Stanescu}, \citenamefont {Lutchyn},\ and\ \citenamefont
  {Das~Sarma}}]{stanescu2011majorana}%
  \BibitemOpen
  \bibfield  {author} {\bibinfo {author} {\bibfnamefont {T.~D.}\ \bibnamefont
  {Stanescu}}, \bibinfo {author} {\bibfnamefont {R.~M.}\ \bibnamefont
  {Lutchyn}},\ and\ \bibinfo {author} {\bibfnamefont {S.}~\bibnamefont
  {Das~Sarma}},\ }\bibfield  {title} {\bibinfo {title} {Majorana fermions in
  semiconductor nanowires},\ }\href
  {https://doi.org/10.1103/PhysRevB.84.144522} {\bibfield  {journal} {\bibinfo
  {journal} {Phys. Rev. B}\ }\textbf {\bibinfo {volume} {84}},\ \bibinfo
  {pages} {144522} (\bibinfo {year} {2011})}\BibitemShut {NoStop}%
\bibitem [{\citenamefont {Lutchyn}\ \emph {et~al.}(2012)\citenamefont
  {Lutchyn}, \citenamefont {Stanescu},\ and\ \citenamefont
  {Das~Sarma}}]{lutchyn2012momentum}%
  \BibitemOpen
  \bibfield  {author} {\bibinfo {author} {\bibfnamefont {R.~M.}\ \bibnamefont
  {Lutchyn}}, \bibinfo {author} {\bibfnamefont {T.~D.}\ \bibnamefont
  {Stanescu}},\ and\ \bibinfo {author} {\bibfnamefont {S.}~\bibnamefont
  {Das~Sarma}},\ }\bibfield  {title} {\bibinfo {title} {Momentum relaxation in
  a semiconductor proximity-coupled to a disordered {$s$}-wave superconductor:
  {{Effect}} of scattering on topological superconductivity},\ }\href
  {https://doi.org/10.1103/PhysRevB.85.140513} {\bibfield  {journal} {\bibinfo
  {journal} {Phys. Rev. B}\ }\textbf {\bibinfo {volume} {85}},\ \bibinfo
  {pages} {140513} (\bibinfo {year} {2012})}\BibitemShut {NoStop}%
\bibitem [{\citenamefont {Rainis}\ \emph {et~al.}(2013)\citenamefont {Rainis},
  \citenamefont {Trifunovic}, \citenamefont {Klinovaja},\ and\ \citenamefont
  {Loss}}]{rainis2013realistic}%
  \BibitemOpen
  \bibfield  {author} {\bibinfo {author} {\bibfnamefont {D.}~\bibnamefont
  {Rainis}}, \bibinfo {author} {\bibfnamefont {L.}~\bibnamefont {Trifunovic}},
  \bibinfo {author} {\bibfnamefont {J.}~\bibnamefont {Klinovaja}},\ and\
  \bibinfo {author} {\bibfnamefont {D.}~\bibnamefont {Loss}},\ }\bibfield
  {title} {\bibinfo {title} {Towards a realistic transport modeling in a
  superconducting nanowire with {{Majorana}} fermions},\ }\href
  {https://doi.org/10.1103/PhysRevB.87.024515} {\bibfield  {journal} {\bibinfo
  {journal} {Phys. Rev. B}\ }\textbf {\bibinfo {volume} {87}},\ \bibinfo
  {pages} {024515} (\bibinfo {year} {2013})}\BibitemShut {NoStop}%
\bibitem [{\citenamefont {DeGottardi}\ \emph {et~al.}(2013)\citenamefont
  {DeGottardi}, \citenamefont {Sen},\ and\ \citenamefont
  {Vishveshwara}}]{degottardi2013majorana}%
  \BibitemOpen
  \bibfield  {author} {\bibinfo {author} {\bibfnamefont {W.}~\bibnamefont
  {DeGottardi}}, \bibinfo {author} {\bibfnamefont {D.}~\bibnamefont {Sen}},\
  and\ \bibinfo {author} {\bibfnamefont {S.}~\bibnamefont {Vishveshwara}},\
  }\bibfield  {title} {\bibinfo {title} {Majorana {{Fermions}} in
  {{Superconducting 1D Systems Having Periodic}}, {{Quasiperiodic}}, and
  {{Disordered Potentials}}},\ }\href
  {https://doi.org/10.1103/PhysRevLett.110.146404} {\bibfield  {journal}
  {\bibinfo  {journal} {Phys. Rev. Lett.}\ }\textbf {\bibinfo {volume} {110}},\
  \bibinfo {pages} {146404} (\bibinfo {year} {2013})}\BibitemShut {NoStop}%
\bibitem [{\citenamefont {Adagideli}\ \emph {et~al.}(2014)\citenamefont
  {Adagideli}, \citenamefont {Wimmer},\ and\ \citenamefont
  {Teker}}]{adagideli2014effects}%
  \BibitemOpen
  \bibfield  {author} {\bibinfo {author} {\bibfnamefont {{\.I}.}~\bibnamefont
  {Adagideli}}, \bibinfo {author} {\bibfnamefont {M.}~\bibnamefont {Wimmer}},\
  and\ \bibinfo {author} {\bibfnamefont {A.}~\bibnamefont {Teker}},\ }\bibfield
   {title} {\bibinfo {title} {Effects of electron scattering on the topological
  properties of nanowires: {{Majorana}} fermions from disorder and
  superlattices},\ }\href {https://doi.org/10.1103/PhysRevB.89.144506}
  {\bibfield  {journal} {\bibinfo  {journal} {Phys. Rev. B}\ }\textbf {\bibinfo
  {volume} {89}},\ \bibinfo {pages} {144506} (\bibinfo {year}
  {2014})}\BibitemShut {NoStop}%
\bibitem [{\citenamefont {Cole}\ \emph {et~al.}(2016)\citenamefont {Cole},
  \citenamefont {Sau},\ and\ \citenamefont {Das~Sarma}}]{cole2016proximity}%
  \BibitemOpen
  \bibfield  {author} {\bibinfo {author} {\bibfnamefont {W.~S.}\ \bibnamefont
  {Cole}}, \bibinfo {author} {\bibfnamefont {J.~D.}\ \bibnamefont {Sau}},\ and\
  \bibinfo {author} {\bibfnamefont {S.}~\bibnamefont {Das~Sarma}},\ }\bibfield
  {title} {\bibinfo {title} {Proximity effect and {{Majorana}} bound states in
  clean semiconductor nanowires coupled to disordered superconductors},\ }\href
  {https://doi.org/10.1103/PhysRevB.94.140505} {\bibfield  {journal} {\bibinfo
  {journal} {Phys. Rev. B}\ }\textbf {\bibinfo {volume} {94}},\ \bibinfo
  {pages} {140505} (\bibinfo {year} {2016})}\BibitemShut {NoStop}%
\bibitem [{\citenamefont {Woods}\ \emph {et~al.}(2019)\citenamefont {Woods},
  \citenamefont {Chen}, \citenamefont {Frolov},\ and\ \citenamefont
  {Stanescu}}]{woods2019zeroenergy}%
  \BibitemOpen
  \bibfield  {author} {\bibinfo {author} {\bibfnamefont {B.~D.}\ \bibnamefont
  {Woods}}, \bibinfo {author} {\bibfnamefont {J.}~\bibnamefont {Chen}},
  \bibinfo {author} {\bibfnamefont {S.~M.}\ \bibnamefont {Frolov}},\ and\
  \bibinfo {author} {\bibfnamefont {T.~D.}\ \bibnamefont {Stanescu}},\
  }\bibfield  {title} {\bibinfo {title} {Zero-energy pinning of topologically
  trivial bound states in multiband semiconductor-superconductor nanowires},\
  }\href {https://doi.org/10.1103/PhysRevB.100.125407} {\bibfield  {journal}
  {\bibinfo  {journal} {Phys. Rev. B}\ }\textbf {\bibinfo {volume} {100}},\
  \bibinfo {pages} {125407} (\bibinfo {year} {2019})}\BibitemShut {NoStop}%
\bibitem [{\citenamefont {Pan}\ and\ \citenamefont
  {Das~Sarma}(2021{\natexlab{b}})}]{pan2021crossover}%
  \BibitemOpen
  \bibfield  {author} {\bibinfo {author} {\bibfnamefont {H.}~\bibnamefont
  {Pan}}\ and\ \bibinfo {author} {\bibfnamefont {S.}~\bibnamefont
  {Das~Sarma}},\ }\bibfield  {title} {\bibinfo {title} {Crossover between
  trivial zero modes in {{Majorana}} nanowires},\ }\href
  {https://doi.org/10.1103/PhysRevB.104.054510} {\bibfield  {journal} {\bibinfo
   {journal} {Phys. Rev. B}\ }\textbf {\bibinfo {volume} {104}},\ \bibinfo
  {pages} {054510} (\bibinfo {year} {2021}{\natexlab{b}})}\BibitemShut
  {NoStop}%
\bibitem [{\citenamefont {Vuik}\ \emph {et~al.}(2016)\citenamefont {Vuik},
  \citenamefont {Eeltink}, \citenamefont {Akhmerov},\ and\ \citenamefont
  {Wimmer}}]{vuik2016effects}%
  \BibitemOpen
  \bibfield  {author} {\bibinfo {author} {\bibfnamefont {A.}~\bibnamefont
  {Vuik}}, \bibinfo {author} {\bibfnamefont {D.}~\bibnamefont {Eeltink}},
  \bibinfo {author} {\bibfnamefont {A.~R.}\ \bibnamefont {Akhmerov}},\ and\
  \bibinfo {author} {\bibfnamefont {M.}~\bibnamefont {Wimmer}},\ }\bibfield
  {title} {\bibinfo {title} {Effects of the electrostatic environment on the
  {{Majorana}} nanowire devices},\ }\href
  {https://doi.org/10.1088/1367-2630/18/3/033013} {\bibfield  {journal}
  {\bibinfo  {journal} {New J. Phys.}\ }\textbf {\bibinfo {volume} {18}},\
  \bibinfo {pages} {033013} (\bibinfo {year} {2016})}\BibitemShut {NoStop}%
\bibitem [{\citenamefont {Antipov}\ \emph {et~al.}(2018)\citenamefont
  {Antipov}, \citenamefont {Bargerbos}, \citenamefont {Winkler}, \citenamefont
  {Bauer}, \citenamefont {Rossi},\ and\ \citenamefont
  {Lutchyn}}]{antipov2018effects}%
  \BibitemOpen
  \bibfield  {author} {\bibinfo {author} {\bibfnamefont {A.~E.}\ \bibnamefont
  {Antipov}}, \bibinfo {author} {\bibfnamefont {A.}~\bibnamefont {Bargerbos}},
  \bibinfo {author} {\bibfnamefont {G.~W.}\ \bibnamefont {Winkler}}, \bibinfo
  {author} {\bibfnamefont {B.}~\bibnamefont {Bauer}}, \bibinfo {author}
  {\bibfnamefont {E.}~\bibnamefont {Rossi}},\ and\ \bibinfo {author}
  {\bibfnamefont {R.~M.}\ \bibnamefont {Lutchyn}},\ }\bibfield  {title}
  {\bibinfo {title} {Effects of {{Gate-Induced Electric Fields}} on
  {{Semiconductor Majorana Nanowires}}},\ }\href
  {https://doi.org/10.1103/PhysRevX.8.031041} {\bibfield  {journal} {\bibinfo
  {journal} {Phys. Rev. X}\ }\textbf {\bibinfo {volume} {8}},\ \bibinfo {pages}
  {031041} (\bibinfo {year} {2018})}\BibitemShut {NoStop}%
\bibitem [{\citenamefont {Mikkelsen}\ \emph {et~al.}(2018)\citenamefont
  {Mikkelsen}, \citenamefont {Kotetes}, \citenamefont {Krogstrup},\ and\
  \citenamefont {Flensberg}}]{mikkelsen2018hybridization}%
  \BibitemOpen
  \bibfield  {author} {\bibinfo {author} {\bibfnamefont {A.~E.~G.}\
  \bibnamefont {Mikkelsen}}, \bibinfo {author} {\bibfnamefont {P.}~\bibnamefont
  {Kotetes}}, \bibinfo {author} {\bibfnamefont {P.}~\bibnamefont {Krogstrup}},\
  and\ \bibinfo {author} {\bibfnamefont {K.}~\bibnamefont {Flensberg}},\
  }\bibfield  {title} {\bibinfo {title} {Hybridization at
  {{Superconductor-Semiconductor Interfaces}}},\ }\href
  {https://doi.org/10.1103/PhysRevX.8.031040} {\bibfield  {journal} {\bibinfo
  {journal} {Phys. Rev. X}\ }\textbf {\bibinfo {volume} {8}},\ \bibinfo {pages}
  {031040} (\bibinfo {year} {2018})}\BibitemShut {NoStop}%
\bibitem [{\citenamefont {Schuwalow}\ \emph {et~al.}(2019)\citenamefont
  {Schuwalow}, \citenamefont {Schroeter}, \citenamefont {Gukelberger},
  \citenamefont {Thomas}, \citenamefont {Strocov}, \citenamefont {Gamble},
  \citenamefont {Chikina}, \citenamefont {Caputo}, \citenamefont {Krieger},
  \citenamefont {Gardner}, \citenamefont {Troyer}, \citenamefont {Aeppli},
  \citenamefont {Manfra},\ and\ \citenamefont {Krogstrup}}]{schuwalow2019band}%
  \BibitemOpen
  \bibfield  {author} {\bibinfo {author} {\bibfnamefont {S.}~\bibnamefont
  {Schuwalow}}, \bibinfo {author} {\bibfnamefont {N.~B.~M.}\ \bibnamefont
  {Schroeter}}, \bibinfo {author} {\bibfnamefont {J.}~\bibnamefont
  {Gukelberger}}, \bibinfo {author} {\bibfnamefont {C.}~\bibnamefont {Thomas}},
  \bibinfo {author} {\bibfnamefont {V.}~\bibnamefont {Strocov}}, \bibinfo
  {author} {\bibfnamefont {J.}~\bibnamefont {Gamble}}, \bibinfo {author}
  {\bibfnamefont {A.}~\bibnamefont {Chikina}}, \bibinfo {author} {\bibfnamefont
  {M.}~\bibnamefont {Caputo}}, \bibinfo {author} {\bibfnamefont
  {J.}~\bibnamefont {Krieger}}, \bibinfo {author} {\bibfnamefont {G.~C.}\
  \bibnamefont {Gardner}}, \bibinfo {author} {\bibfnamefont {M.}~\bibnamefont
  {Troyer}}, \bibinfo {author} {\bibfnamefont {G.}~\bibnamefont {Aeppli}},
  \bibinfo {author} {\bibfnamefont {M.~J.}\ \bibnamefont {Manfra}},\ and\
  \bibinfo {author} {\bibfnamefont {P.}~\bibnamefont {Krogstrup}},\ }\bibfield
  {title} {\bibinfo {title} {Band bending profile and band offset extraction at
  semiconductor-metal interfaces},\ }\href {http://arxiv.org/abs/1910.02735}
  {\bibfield  {journal} {\bibinfo  {journal} {arXiv:1910.02735 [cond-mat,
  physics:quant-ph]}\ } (\bibinfo {year} {2019})}\BibitemShut {NoStop}%
\bibitem [{\citenamefont {Winkler}(2003)}]{winkler2003spinorbit}%
  \BibitemOpen
  \bibfield  {author} {\bibinfo {author} {\bibfnamefont {R.}~\bibnamefont
  {Winkler}},\ }\href@noop {} {\emph {\bibinfo {title} {Spin-Orbit Coupling
  Effects in Two-Dimensional Electron and Hole Systems}}},\ \bibinfo {series}
  {Springer Tracts in Modern Physics}\ No.\ \bibinfo {number} {v. 191}\
  (\bibinfo  {publisher} {{Springer}},\ \bibinfo {address} {{Berlin ; New
  York}},\ \bibinfo {year} {2003})\BibitemShut {NoStop}%
\bibitem [{\citenamefont {Huang}\ \emph {et~al.}(2018)\citenamefont {Huang},
  \citenamefont {Pan}, \citenamefont {Liu}, \citenamefont {Sau}, \citenamefont
  {Stanescu},\ and\ \citenamefont {Das~Sarma}}]{huang2018metamorphosis}%
  \BibitemOpen
  \bibfield  {author} {\bibinfo {author} {\bibfnamefont {Y.}~\bibnamefont
  {Huang}}, \bibinfo {author} {\bibfnamefont {H.}~\bibnamefont {Pan}}, \bibinfo
  {author} {\bibfnamefont {C.-X.}\ \bibnamefont {Liu}}, \bibinfo {author}
  {\bibfnamefont {J.~D.}\ \bibnamefont {Sau}}, \bibinfo {author} {\bibfnamefont
  {T.~D.}\ \bibnamefont {Stanescu}},\ and\ \bibinfo {author} {\bibfnamefont
  {S.}~\bibnamefont {Das~Sarma}},\ }\bibfield  {title} {\bibinfo {title}
  {Metamorphosis of {{Andreev}} bound states into {{Majorana}} bound states in
  pristine nanowires},\ }\href {https://doi.org/10.1103/PhysRevB.98.144511}
  {\bibfield  {journal} {\bibinfo  {journal} {Phys. Rev. B}\ }\textbf {\bibinfo
  {volume} {98}},\ \bibinfo {pages} {144511} (\bibinfo {year}
  {2018})}\BibitemShut {NoStop}%
\bibitem [{\citenamefont {Badawy}\ \emph {et~al.}(2019)\citenamefont {Badawy},
  \citenamefont {Gazibegovic}, \citenamefont {Borsoi}, \citenamefont {Heedt},
  \citenamefont {Wang}, \citenamefont {Koelling}, \citenamefont {Verheijen},
  \citenamefont {Kouwenhoven},\ and\ \citenamefont {Bakkers}}]{badawy2019high}%
  \BibitemOpen
  \bibfield  {author} {\bibinfo {author} {\bibfnamefont {G.}~\bibnamefont
  {Badawy}}, \bibinfo {author} {\bibfnamefont {S.}~\bibnamefont {Gazibegovic}},
  \bibinfo {author} {\bibfnamefont {F.}~\bibnamefont {Borsoi}}, \bibinfo
  {author} {\bibfnamefont {S.}~\bibnamefont {Heedt}}, \bibinfo {author}
  {\bibfnamefont {C.-A.}\ \bibnamefont {Wang}}, \bibinfo {author}
  {\bibfnamefont {S.}~\bibnamefont {Koelling}}, \bibinfo {author}
  {\bibfnamefont {M.~A.}\ \bibnamefont {Verheijen}}, \bibinfo {author}
  {\bibfnamefont {L.~P.}\ \bibnamefont {Kouwenhoven}},\ and\ \bibinfo {author}
  {\bibfnamefont {E.~P. A.~M.}\ \bibnamefont {Bakkers}},\ }\bibfield  {title}
  {\bibinfo {title} {High {{Mobility Stemless InSb Nanowires}}},\ }\href
  {https://doi.org/10.1021/acs.nanolett.9b00545} {\bibfield  {journal}
  {\bibinfo  {journal} {Nano Lett.}\ }\textbf {\bibinfo {volume} {19}},\
  \bibinfo {pages} {3575} (\bibinfo {year} {2019})}\BibitemShut {NoStop}%
\bibitem [{\citenamefont {Chung}\ \emph {et~al.}(2021)\citenamefont {Chung},
  \citenamefont {Villegas~Rosales}, \citenamefont {Baldwin}, \citenamefont
  {Madathil}, \citenamefont {West}, \citenamefont {Shayegan},\ and\
  \citenamefont {Pfeiffer}}]{chung2021ultrahighquality}%
  \BibitemOpen
  \bibfield  {author} {\bibinfo {author} {\bibfnamefont {Y.~J.}\ \bibnamefont
  {Chung}}, \bibinfo {author} {\bibfnamefont {K.~A.}\ \bibnamefont
  {Villegas~Rosales}}, \bibinfo {author} {\bibfnamefont {K.~W.}\ \bibnamefont
  {Baldwin}}, \bibinfo {author} {\bibfnamefont {P.~T.}\ \bibnamefont
  {Madathil}}, \bibinfo {author} {\bibfnamefont {K.~W.}\ \bibnamefont {West}},
  \bibinfo {author} {\bibfnamefont {M.}~\bibnamefont {Shayegan}},\ and\
  \bibinfo {author} {\bibfnamefont {L.~N.}\ \bibnamefont {Pfeiffer}},\
  }\bibfield  {title} {\bibinfo {title} {Ultra-high-quality two-dimensional
  electron systems},\ }\href {https://doi.org/10.1038/s41563-021-00942-3}
  {\bibfield  {journal} {\bibinfo  {journal} {Nat. Mater.}\ }\textbf {\bibinfo
  {volume} {20}},\ \bibinfo {pages} {632} (\bibinfo {year} {2021})}\BibitemShut
  {NoStop}%
\bibitem [{\citenamefont {Das~Sarma}\ \emph {et~al.}(2015)\citenamefont
  {Das~Sarma}, \citenamefont {Hwang}, \citenamefont {Kodiyalam}, \citenamefont
  {Pfeiffer},\ and\ \citenamefont {West}}]{dassarma2015transport}%
  \BibitemOpen
  \bibfield  {author} {\bibinfo {author} {\bibfnamefont {S.}~\bibnamefont
  {Das~Sarma}}, \bibinfo {author} {\bibfnamefont {E.~H.}\ \bibnamefont
  {Hwang}}, \bibinfo {author} {\bibfnamefont {S.}~\bibnamefont {Kodiyalam}},
  \bibinfo {author} {\bibfnamefont {L.~N.}\ \bibnamefont {Pfeiffer}},\ and\
  \bibinfo {author} {\bibfnamefont {K.~W.}\ \bibnamefont {West}},\ }\bibfield
  {title} {\bibinfo {title} {Transport in two-dimensional modulation-doped
  semiconductor structures},\ }\href
  {https://doi.org/10.1103/PhysRevB.91.205304} {\bibfield  {journal} {\bibinfo
  {journal} {Phys. Rev. B}\ }\textbf {\bibinfo {volume} {91}},\ \bibinfo
  {pages} {205304} (\bibinfo {year} {2015})}\BibitemShut {NoStop}%
\bibitem [{\citenamefont {Hwang}\ and\ \citenamefont
  {Das~Sarma}(2013)}]{hwang2013electronic}%
  \BibitemOpen
  \bibfield  {author} {\bibinfo {author} {\bibfnamefont {E.~H.}\ \bibnamefont
  {Hwang}}\ and\ \bibinfo {author} {\bibfnamefont {S.}~\bibnamefont
  {Das~Sarma}},\ }\bibfield  {title} {\bibinfo {title} {Electronic transport in
  two-dimensional {{Si}}:{{P}} {$\ensuremath{\delta}$}-doped layers},\ }\href
  {https://doi.org/10.1103/PhysRevB.87.125411} {\bibfield  {journal} {\bibinfo
  {journal} {Phys. Rev. B}\ }\textbf {\bibinfo {volume} {87}},\ \bibinfo
  {pages} {125411} (\bibinfo {year} {2013})}\BibitemShut {NoStop}%
\bibitem [{\citenamefont {Lutchyn}\ \emph {et~al.}(2018)\citenamefont
  {Lutchyn}, \citenamefont {Bakkers}, \citenamefont {Kouwenhoven},
  \citenamefont {Krogstrup}, \citenamefont {Marcus},\ and\ \citenamefont
  {Oreg}}]{lutchyn2018majorana}%
  \BibitemOpen
  \bibfield  {author} {\bibinfo {author} {\bibfnamefont {R.~M.}\ \bibnamefont
  {Lutchyn}}, \bibinfo {author} {\bibfnamefont {E.~P. A.~M.}\ \bibnamefont
  {Bakkers}}, \bibinfo {author} {\bibfnamefont {L.~P.}\ \bibnamefont
  {Kouwenhoven}}, \bibinfo {author} {\bibfnamefont {P.}~\bibnamefont
  {Krogstrup}}, \bibinfo {author} {\bibfnamefont {C.~M.}\ \bibnamefont
  {Marcus}},\ and\ \bibinfo {author} {\bibfnamefont {Y.}~\bibnamefont {Oreg}},\
  }\bibfield  {title} {\bibinfo {title} {Majorana zero modes in
  superconductor\textendash semiconductor heterostructures},\ }\href
  {https://doi.org/10.1038/s41578-018-0003-1} {\bibfield  {journal} {\bibinfo
  {journal} {Nature Reviews Materials}\ }\textbf {\bibinfo {volume} {3}},\
  \bibinfo {pages} {52} (\bibinfo {year} {2018})}\BibitemShut {NoStop}%
\bibitem [{\citenamefont {Blonder}\ \emph {et~al.}(1982)\citenamefont
  {Blonder}, \citenamefont {Tinkham},\ and\ \citenamefont
  {Klapwijk}}]{blonder1982transition}%
  \BibitemOpen
  \bibfield  {author} {\bibinfo {author} {\bibfnamefont {G.~E.}\ \bibnamefont
  {Blonder}}, \bibinfo {author} {\bibfnamefont {M.}~\bibnamefont {Tinkham}},\
  and\ \bibinfo {author} {\bibfnamefont {T.~M.}\ \bibnamefont {Klapwijk}},\
  }\bibfield  {title} {\bibinfo {title} {Transition from metallic to tunneling
  regimes in superconducting microconstrictions: {{Excess}} current, charge
  imbalance, and supercurrent conversion},\ }\href
  {https://doi.org/10.1103/PhysRevB.25.4515} {\bibfield  {journal} {\bibinfo
  {journal} {Phys. Rev. B}\ }\textbf {\bibinfo {volume} {25}},\ \bibinfo
  {pages} {4515} (\bibinfo {year} {1982})}\BibitemShut {NoStop}%
\bibitem [{\citenamefont {Datta}(1995)}]{datta1995electronic}%
  \BibitemOpen
  \bibfield  {author} {\bibinfo {author} {\bibfnamefont {S.}~\bibnamefont
  {Datta}},\ }\href {https://doi.org/10.1017/CBO9780511805776} {\emph {\bibinfo
  {title} {Electronic {{Transport}} in {{Mesoscopic Systems}}}}},\ Cambridge
  {{Studies}} in {{Semiconductor Physics}} and {{Microelectronic Engineering}}\
  (\bibinfo  {publisher} {{Cambridge University Press}},\ \bibinfo {address}
  {{Cambridge}},\ \bibinfo {year} {1995})\BibitemShut {NoStop}%
\bibitem [{\citenamefont {Anantram}\ and\ \citenamefont
  {Datta}(1996)}]{anantram1996current}%
  \BibitemOpen
  \bibfield  {author} {\bibinfo {author} {\bibfnamefont {M.~P.}\ \bibnamefont
  {Anantram}}\ and\ \bibinfo {author} {\bibfnamefont {S.}~\bibnamefont
  {Datta}},\ }\bibfield  {title} {\bibinfo {title} {Current fluctuations in
  mesoscopic systems with {{Andreev}} scattering},\ }\href
  {https://doi.org/10.1103/PhysRevB.53.16390} {\bibfield  {journal} {\bibinfo
  {journal} {Phys. Rev. B}\ }\textbf {\bibinfo {volume} {53}},\ \bibinfo
  {pages} {16390} (\bibinfo {year} {1996})}\BibitemShut {NoStop}%
\bibitem [{\citenamefont {Groth}\ \emph {et~al.}(2014)\citenamefont {Groth},
  \citenamefont {Wimmer}, \citenamefont {Akhmerov},\ and\ \citenamefont
  {Waintal}}]{groth2014kwant}%
  \BibitemOpen
  \bibfield  {author} {\bibinfo {author} {\bibfnamefont {C.~W.}\ \bibnamefont
  {Groth}}, \bibinfo {author} {\bibfnamefont {M.}~\bibnamefont {Wimmer}},
  \bibinfo {author} {\bibfnamefont {A.~R.}\ \bibnamefont {Akhmerov}},\ and\
  \bibinfo {author} {\bibfnamefont {X.}~\bibnamefont {Waintal}},\ }\bibfield
  {title} {\bibinfo {title} {Kwant: A software package for quantum transport},\
  }\href {http://iopscience.iop.org/article/10.1088/1367-2630/16/6/063065/meta}
  {\bibfield  {journal} {\bibinfo  {journal} {New Journal of Physics}\ }\textbf
  {\bibinfo {volume} {16}},\ \bibinfo {pages} {063065} (\bibinfo {year}
  {2014})}\BibitemShut {NoStop}%
\bibitem [{\citenamefont {Rosdahl}\ \emph {et~al.}(2018)\citenamefont
  {Rosdahl}, \citenamefont {Vuik}, \citenamefont {Kjaergaard},\ and\
  \citenamefont {Akhmerov}}]{rosdahl2018andreev}%
  \BibitemOpen
  \bibfield  {author} {\bibinfo {author} {\bibfnamefont {T.~{\"O}.}\
  \bibnamefont {Rosdahl}}, \bibinfo {author} {\bibfnamefont {A.}~\bibnamefont
  {Vuik}}, \bibinfo {author} {\bibfnamefont {M.}~\bibnamefont {Kjaergaard}},\
  and\ \bibinfo {author} {\bibfnamefont {A.~R.}\ \bibnamefont {Akhmerov}},\
  }\bibfield  {title} {\bibinfo {title} {Andreev rectifier: {{A}} nonlocal
  conductance signature of topological phase transitions},\ }\href
  {https://doi.org/10.1103/PhysRevB.97.045421} {\bibfield  {journal} {\bibinfo
  {journal} {Phys. Rev. B}\ }\textbf {\bibinfo {volume} {97}},\ \bibinfo
  {pages} {045421} (\bibinfo {year} {2018})}\BibitemShut {NoStop}%
\end{thebibliography}%
\appendix
\onecolumngrid
\section{Transport theory}\label{app:A}
In this section, we introduce the Boltzmann formalism used to obtain the transport results in the main text.
The central quantity for the calculation of the transport mobility is the transport relaxation time $\tau_\mathrm{t}$, which, at zero temperature, is directly related to the mobility through the well-known Drude formula $\mu = e \tau_\mathrm{t} / m$. Within the leading order Born approximation, the relaxation time for 2D charge carriers at $T=0$ is given by  
\begin{equation} \label{eq:transport_relaxation_time}
\begin{aligned}
    \frac{1}{\tau_\mathrm{t}^{(\alpha)}}&=\frac{2\pi}{\hbar} \int N_i^{(\alpha)}(z) dz \int  \frac{d^2 k'}{(2\pi)^2} \left| V^{(\alpha)}_{\bm k- \bm k'}(z) \right|^2 \\
    &\times (1-\cos{ \theta_{\bm k, \bm k'} })\delta[E(\bm k)-E(\bm k')]
\end{aligned}
\end{equation}
where $\alpha$ indicates the type of disorder limiting the mobility, $N_i^{(\alpha)}$ is the 3D density of impurities, $V_{\bm k - \bm k'}(z)$ is the electron-impurity interaction, $\theta_{\bm k, \bm k'}$ is the scattering angle between the incoming state ($\bm k$) and outgoing state ($\bm k'$), and $E(\bm k)=\hbar^2k^2/2m$ is the energy dispersion of 2D charge carriers. For short-range scatterers (such as dislocation, point defect, atomic vacancy), the impurity potential $V_{\bm q}$ is extremely localized in real space, and thus is given by a constant in the momentum space, i.e.,  $V_{\bm q}=V_0$. Note that $V_0$ can be factored out of the integral since it is a constant, and thus $V^2_0 N_i^{(\alpha)}$ can be considered as a single parameter in transport calculations. For long-range Coulomb scatterers, the disorder potential is written as
\begin{equation}
    \left|V_{\bm q}(z)\right|^2 = \left| \frac{2\pi e^2}{\kappa [q + q_\mathrm{TF}F(q) ]}\right|^2e^{-2q |z|} F_i(q)
\end{equation}
where $q= |\bm q|$, $q_\mathrm{TF}=2me^2/\kappa\hbar^2$ is the 2D Thomas-Fermi wavevector, $\kappa$ denotes the background static dielectric constant, and $z$ is the separation of the charged impurities from the 2D layer of carriers. Here $F(q)$ and $F_i(q)$ are the form factor of electron-electron interaction and electron-impurity interaction, respectively, arising from the finite thickness of the quantum well associated with the electron confinement wave function $\phi(z)=\sqrt{a/2}\cos{(\pi z/a)}$ for $|z|<a/2$, and given by \cite{hwang2013electronic}
\begin{equation}
    F(q)=\frac{ 3(qa) + 8\pi^2/(qa)}{(qa)^2 + 4\pi^2} - \frac{32\pi^4(1-e^{-qa})}{(qa)^2[(qa)^2 + 4\pi^2 ]^2}
\end{equation}
and
\begin{equation}
    F_i(q)= \frac{4}{qa} \frac{ 2\pi^2(1-e^{-qa/2}) + (qa)^2 }{ (4\pi)^2 + (qa)^2 }.
\end{equation}
We emphasize that all of our transport results presented in the manuscript are realistic in the sense that the finite width effects are taken into account in both the electron-impurity interaction and the electron screening.
 
\section{Percolation analysis}\label{app:B}
\begin{figure*}[!htb]
    \centering
    \includegraphics[width=6.8in]{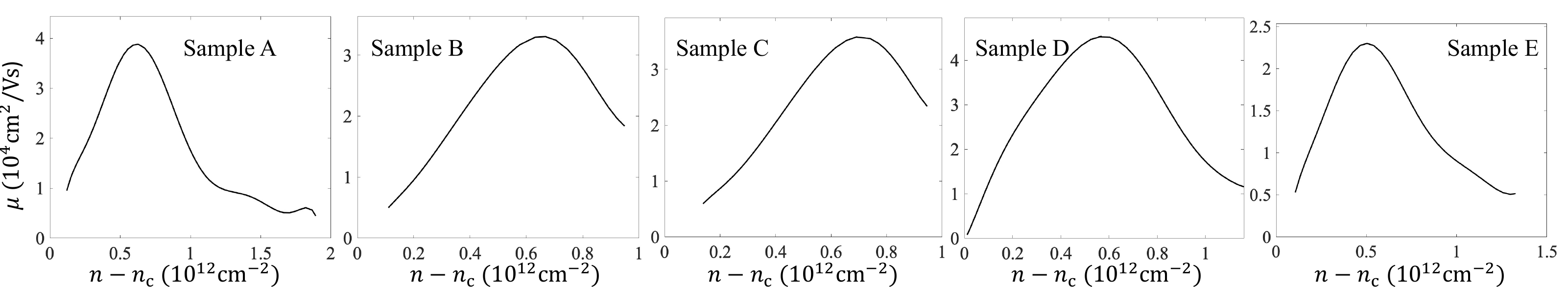}
    \caption{Plot of the experimental mobility as a function of $n-n_\mathrm{c}$.}
    \label{fig:appendix_B_mu}
  \end{figure*}

\begin{figure*}[!htb]
  \centering
  \includegraphics[width=6.8in]{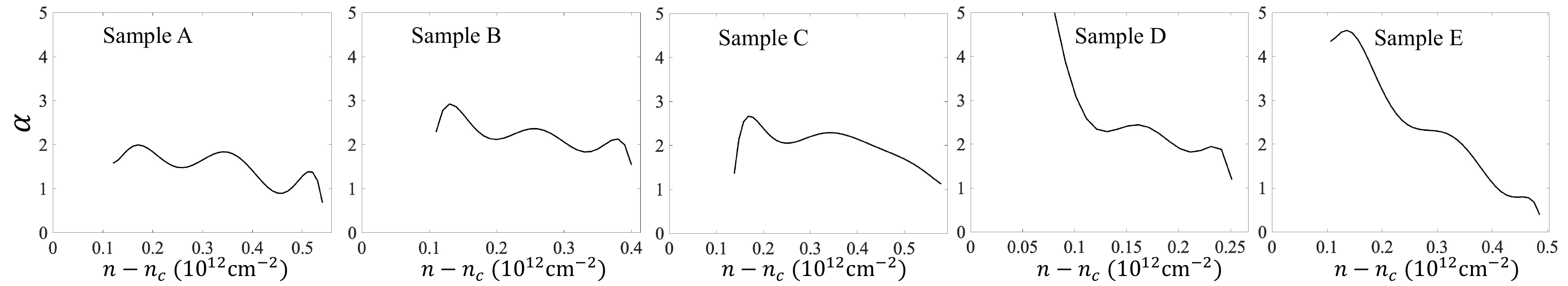}
  \caption{Numerically extracted exponent obtained through $\alpha = \frac{d\ln{\mu}}{d\ln{n}}$ from the InAs transport data from the Copenhagen group of Krogstrup. }
  \label{fig:appendix_B_alpha}
\end{figure*}

\begin{figure*}[!htb]
  \centering
  \includegraphics[width=6.8in]{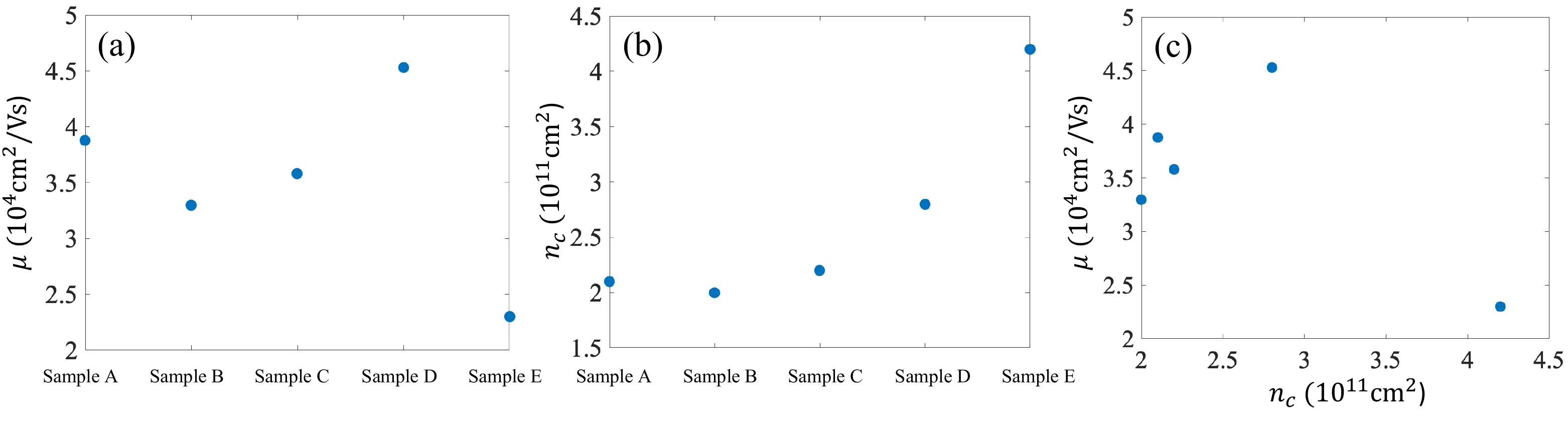}
  \caption{(a) Peak mobility and (b) percolation critical density for different samples. (c) Plot of the peak mobility as a function of the percolation critical density.}
  \label{fig:appendix_B_scatter_plot_critical_density}
\end{figure*}

The mobility near the percolation critical density behaves as $\mu=A(n-n_\mathrm{c})^\delta$, where $n_\mathrm{c}$ is the percolation critical density, $A$ is a constant of proportionality, and $\delta$ is the percolation critical exponent, which is expected to be $4/3$ for 2D by the percolation theory.
Figures~\ref{fig:appendix_B_mu} and \ref{fig:appendix_B_alpha} present the plot of the mobility and its corresponding exponent $\alpha$, respectively, as a function of $n-n_\mathrm{c}$ for various samples. Here $n_\mathrm{c}$ is obtained by extrapolating the measured experimental mobility up to the point where the mobility vanishes, i.e., $\mu(n_\mathrm{c})=0$, and the exponent of the mobility $\alpha$ is numerically extracted from the experimental mobility data through the relation $\alpha = \frac{d\ln{\mu}}{d\ln{n}}$. Our results show that the extracted exponent near the critical density is around $1.5-5.0$, which is somewhat larger than the theoretical 2D percolation exponent $4/3$. This discrepancy is unavoidable due to the experimental inaccuracy, but note that the extracted exponent is close to the theoretical value $4/3$ within a factor of $1$-$4$, which is enough for our purpose of verifying that the percolation transition occurs at around the estimated critical density $n_\mathrm{c}$.
Figure~\ref{fig:appendix_B_scatter_plot_critical_density} shows (a) the peak mobility and (b) the percolation critical density $n_\mathrm{c}$ for each sample. In Fig.~\ref{fig:appendix_B_scatter_plot_critical_density} (c), we plot the peak mobility as a function of $n_\mathrm{c}$ to highlight the correlation between them. The figure shows that the peak mobility decreases with increasing critical percolation density because lower peak mobility implies higher impurity density.

\section{Single-particle level broadening}\label{app:C}
\begin{figure}[!htb]
  \centering
  \includegraphics[width=0.8\linewidth]{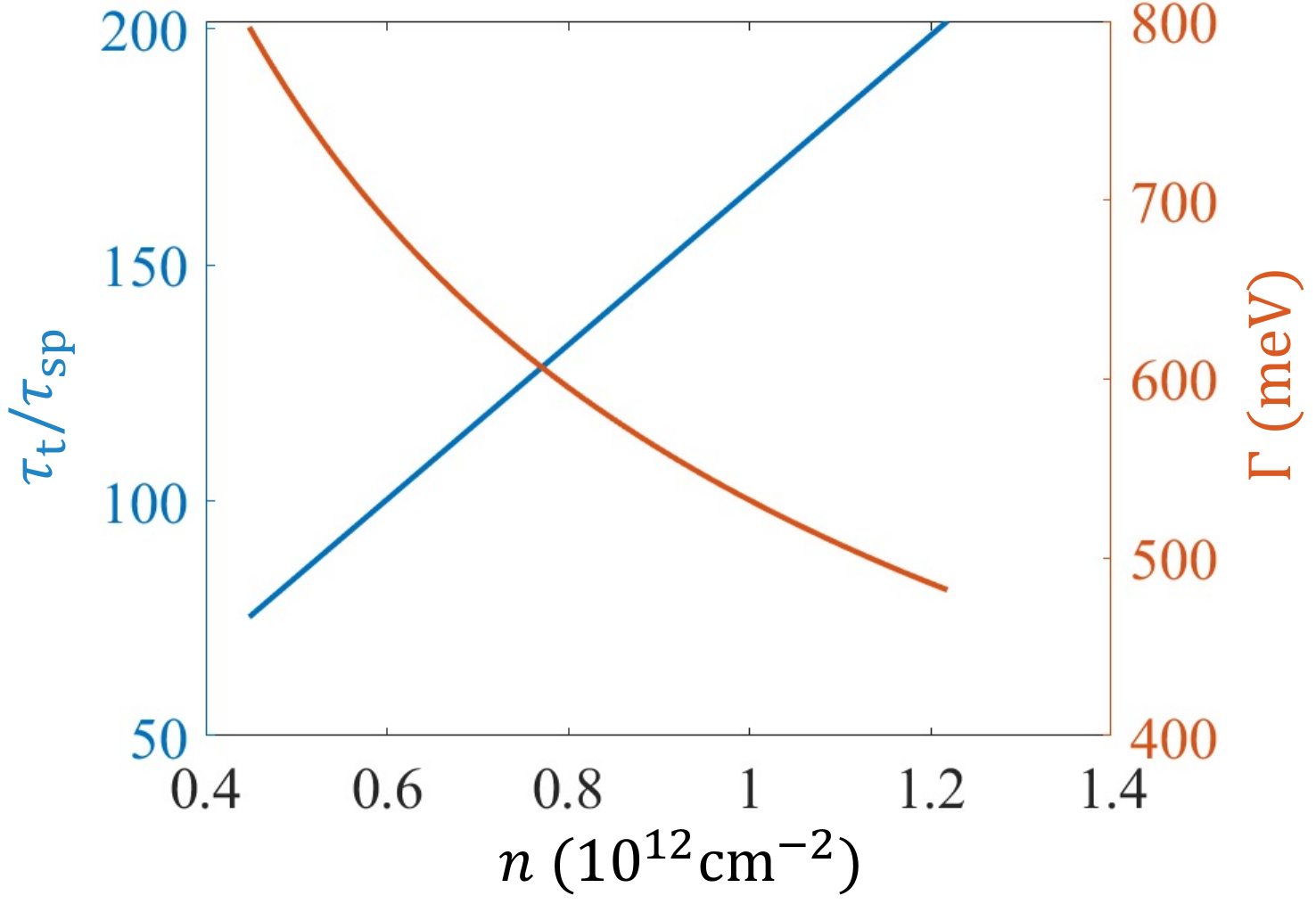}
  \caption{Plot of calculated ratio of the transport time $\tau_\mathrm{t}$ to the single-particle relaxation time $\tau_\mathrm{sp}$ and the single particle level broadening $\Gamma$ as a function of density for the InSb sample from the Bakkers group in Eindhoven. Here we consider only 3D background impurities distributed in the 2D InSb quantum well with the width $a=80\mathrm{nm}$. }
  \label{fig:gamma_InSb}
\end{figure}

\begin{figure*}[!htb]
  \centering
  \includegraphics[width=\linewidth]{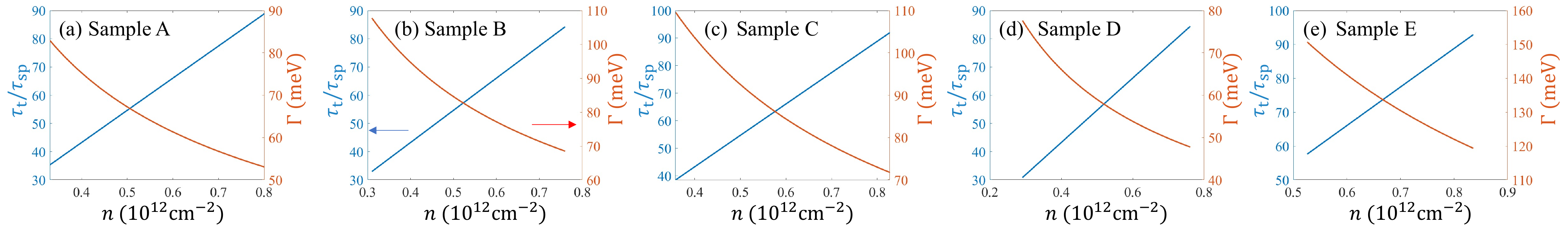}
  \caption{Plot of the calculated ratio of the transport time $\tau_\mathrm{t}$ to the single-particle relaxation time $\tau_\mathrm{sp}$ and the single-particle level broadening $\Gamma$ as a function of density for InAs sample from the Krogstrup group in the Copenhagen. Here we consider only the remote surface impurities at the dielectric interface placed 10 nm away from the surface of the 2D InAs quantum well layer. }
  \label{fig:gamma_InAs1}
\end{figure*}

\begin{figure*}[!htb]
  \centering
  \includegraphics[width=\linewidth]{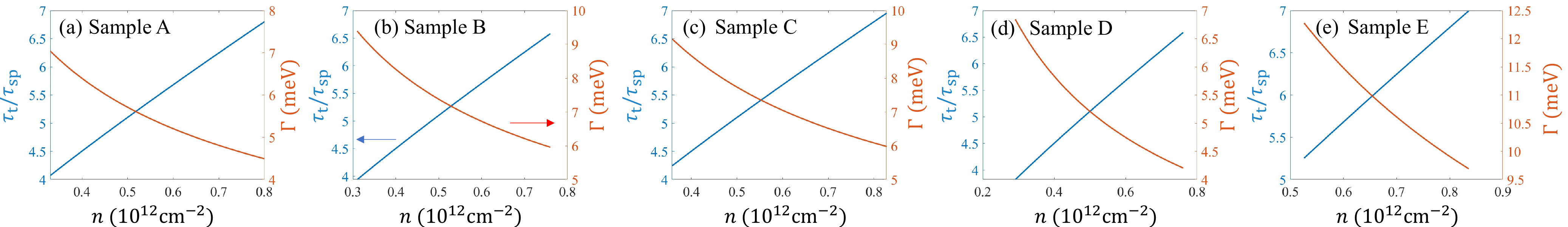}
  \caption{Same as Fig.~\ref{fig:gamma_InAs1}, but here we consider only background impurities uniformly distributed in the 2D InAs quantum well with the width $a=30\mathrm{nm}$. }
  \label{fig:gamma_InAs2}
\end{figure*}

\begin{figure*}[!htb]
  \centering
  \includegraphics[width=\linewidth]{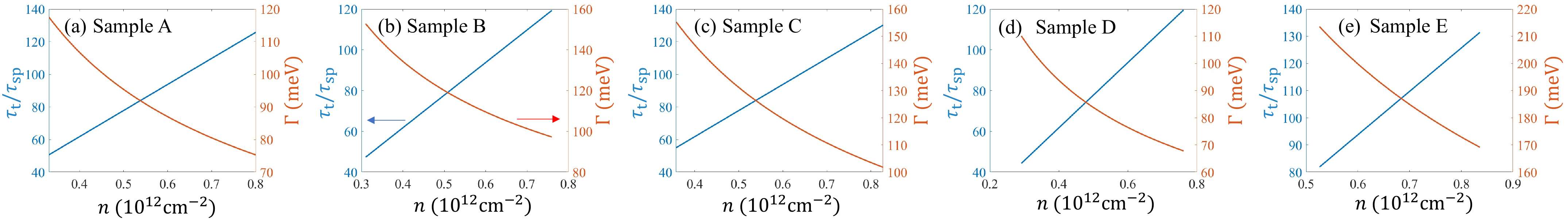}
  \caption{Same as Fig.~\ref{fig:gamma_InAs1}, but here we consider only remote impurities uniformly distributed in the oxide(Al$_2$O$_3$) layer with the width $a_\mathrm{oxide}=8\mathrm{nm}$. }
  \label{fig:gamma_InAs3}
\end{figure*}

The single-particle level broadening can be calculated from the imaginary part of the self-energy in the presence of disorder and given by $\Gamma=\hbar/2\tau_\mathrm{sp}$ where $\tau_\mathrm{sp}$ is the single particle relaxation time. Within the leading-order approximation, $\tau_\mathrm{sp}$ is given by
\begin{equation}
\begin{aligned}
    \frac{1}{\tau_\mathrm{sp}^{(\alpha)}}&=\frac{2\pi}{\hbar} \int N_i^{(\alpha)}(z) dz \int  \frac{d^2 k'}{(2\pi)^2} \left| V^{(\alpha)}_{\bm k- \bm k'}(z) \right|^2 \\
    &\times \delta[E(\bm k)-E(\bm k')],
    \label{eq:single_particle_relaxation_time}
\end{aligned}
\end{equation}
which is the same as the transport relaxation time Eq.~(\ref{eq:transport_relaxation_time}) but without the weighting factor that accounts for the backscattering $1-\cos\theta$ arising from the vertex correction. In Figs.~\ref{fig:gamma_InSb}-\ref{fig:gamma_InAs3}, we present the calculated $\Gamma$ and the ratio of the transport relaxation time $\tau_\mathrm{t}$ to the single particle relaxation time $\tau_\mathrm{sp}$ corresponding to the calculated mobility results present in Figs.~\ref{fig:mobility_fit_InSb_1}(a) and~\ref{fig:mobility_fit_InAs_Purdue_3}-\ref{fig:mobility_fit_InAs_Purdue_5}. It is important to note that $\tau_\mathrm{t}$ is much larger than $\tau_\mathrm{sp}$ by a factor of at least from five to several hundreds, and thus the quantum level broadening determined by $\tau_\mathrm{sp}$ is substantially larger ($\sim$10-200 meV) than that estimated from the measured transport mobility ($\sim$2-4 meV).


\section{Theory of superconductor-semiconductor nanowire}\label{app:D}

In this appendix, we briefly introduce the numerical details of the calculation of the tunnel conductance in the semiconductor-superconductor nanowire. The Bogoliubov-de Gennes Hamiltonian for the minimal model for a finite-length 1D wire is $H=\frac12\int_{0}^L dx \hat{\Psi}^\dagger(x)H_{\text{BdG}}\hat{\Psi}(x)$, where~\cite{lutchyn2010majorana,sau2010nonabelian,sau2010generic,oreg2010helical}
\begin{equation}\label{eq:Ham}
    H_{\text{BdG}}=\left(-\frac{\hbar^2\partial_x^2}{2m^*}-i\alpha\partial_x\sigma_y-\mu\right)\tau_z+V_z\sigma_x-\gamma\frac{\omega+\Delta_0\tau_x}{\sqrt{\Delta_0^2-\omega^2}}+V_{dis}(x)\tau_z
\end{equation}
and $\hat{\Psi}(x)=\left[\hat{\psi}_{\uparrow}(x),\hat{\psi}_{\downarrow}(x),\hat{\psi}_{\downarrow}^\dagger(x),-\hat{\psi}_{\uparrow}^\dagger(x)\right]^\intercal$. In Eq.~\eqref{eq:Ham}, the first term is the Hamiltonian for the pristine semiconductor, the second term represents the Zeeman field, the third term accounts for the proximitized superconductivity, and the last term phenomenologically describes the potential disorder arising from charge impurities and all various gate voltages, which should be zero in the pristine limit. Unless otherwise stated, we follow the parameters in InSb-Al hybrid nanowire~\cite{lutchyn2018majorana}: the effective mass $m^*=0.015m_e$ ($m_e$  is the electron rest mass), Al superconducting gap $\Delta_0=0.2$ meV, chemical potential $\mu$ ranges from -4 to 4 meV, Rashba-type spin-orbit coupling $\alpha=0.5$ eV \AA, superconductor-semiconductor coupling strength $\gamma=0.2$ meV, wire length $L=3~\mu$m or $10~\mu$m, zero temperature, and zero dissipation. 

In this paper, we only focus on the effect of the random disorder in the chemical potential. Thus $V_{dis}(x)$ is a short-range random potential drawn from an uncorrelated Gaussian distribution with zero mean and standard deviation $\sigma_\mu$. Note that all the numerical results in the main text are calculated in the presence of a particular realization of disorder that does not change as we tune the chemical potential or Zeeman field. 

To numerically simulate the tunnel conductance spectra in experiments, we use Blonder-Tinkham-Klapwijk formalism~\cite{blonder1982transition,datta1995electronic,anantram1996current} with the help of a Python package KWANT~\cite{groth2014kwant}. We first attach two semi-infinite normal leads on both sides of the nanowire, where the Hamiltonians take the same form as that of the semiconductor except for the superconducting pairing term (i.e., the third term in Eq.~\eqref{eq:Ham} is absent). Then, we assume a propagating wave in the normal lead by setting its chemical potential to 25 meV. 
To simulate the tunnel gate at the interface, we model an effective barrier height which is located in a few sites at the NS interface. The low (high) value of the barrier height corresponds to the strong (weak) coupling strength between the lead and nanowire, which controls the high (low) transmission transparency. Here, we set the barrier height to 10 or 20 meV depending on the chemical potential in the nanowire. For phase diagrams in Figs.~\ref{fig:21} and~\ref{fig:22}, we particularly keep the relative barrier height with respect to the chemical potential in the nanowire constant as we tune the chemical potential to ensure that the effect of the barrier height on the conductance is qualitative the same~\cite{setiawan2017electron}. Since the system has two normal leads, we can calculate the local conductance from both ends of the wire simultaneously. It is a well-established three-terminal measurement setup, where more details can be found in Ref.~\onlinecite{pan2021threeterminal} and~\onlinecite{rosdahl2018andreev}. 

\section{Details on modeling hybrid structures with surface charge impurities and solving the associated Schr{\"o}dinger-Poisson problem} \label{app:E}
In this appendix, we provide details about the model used in Sec. \ref{sec:SC}, along with some discussion regarding the numerical solution to the Schr{\"o}dinger-Poisson equations. Note that the method used in this work is a slightly altered version of the method we used in Ref.~\onlinecite{woods2021charge}. We, therefore, refer the reader to Ref.~\onlinecite{woods2021charge} for further details regarding the solution method after some preliminary discussion below.

As discussed in the main text in Sec. \ref{sec:SC}, the charge density has contributions from both the free charge density and the immobile surface charge impurities. The impurity charge density is simply the sum of many single charge impurities
\begin{equation}
    \rho_{imp}(\mathbf{r}) = \sum_{j = 1}^{N_{imp}} 
    \rho_j (\mathbf{r}),
\end{equation}
where $N_{imp}$ is the total number of surface charge impurities and $\rho_j$ is the charge density of the j\textsuperscript{th} surface charge impurity. The charge of a single surface charge takes the form,
\begin{align}
    \rho_j (\mathbf{r}) &= \sigma_j(x,y) \lambda_j(z), \\
   \lambda_j(z) &=
    \begin{cases}
    \frac{1}{\ell}, & z_j - \frac{\ell}{2} \leq z \leq z_j + \frac{\ell}{2} \\
    0, & \text{otherwise}
    \end{cases},
\end{align}
where $z_j$ is the $z$-coordinate at the center of the j\textsuperscript{th} impurity, $\ell = 2~\text{nm}$ is the length of the impurity, and $\sigma_j(x,y)$ describes the charge density profile in the transverse direction of the nanowire. The transverse charge density $\sigma_j(x,y)$ is highly localized, completely resides within $2~\text{nm}$ of one of the facets of the InAs nanowire not covered by Al, and is chosen such that $ \int \rho_j(\mathbf{r})\, d\mathbf{r} = +e$, where $e$ is the elementary charge. 
It is useful to break the impurity charge density into two components,
\begin{equation}
    \rho_{imp}(\mathbf{r}) = \bar{\rho}_{imp}(x,y) + \rho_{imp}^\prime(\mathbf{r}), \label{RhoImpDecomp}
\end{equation}    
where $\bar{\rho}_{imp}$ is the average surface charge density, i.e. 
\begin{equation}
    \bar{\rho}_{imp}(x,y) = \frac{1}{L}\int \rho_{imp}(\mathbf{r}) \, dz,
\end{equation}
with $L$ being the total length of the system, while $\rho_{imp}^\prime$ represents fluctuations around the average. We then break down the \textit{total} charge density into three components,
\begin{equation}
    \rho(\mathbf{r}) = \rho_o(x,y) + \rho_{imp}^\prime(\mathbf{r}) + \rho_{red}(\mathbf{r}),
\end{equation}
where $\rho_o$ is the total charge density in the \textit{absence} of surface charge density fluctuations and $\rho_{red}$ accounts for the redistribution of free charge due to the presence of the surface charge density fluctuations. Note that $\rho_o$ is translation invariant along the length of the wire. We stress that $\bar{\rho}_{imp}$ is included in $\rho_o$ such that the effects of translation invariant component of the surface charge density are included in the initial step of our solution method that ignores $\rho_{imp}^\prime$ and $\rho_{red}$. Also note that the redistribution of free charge partially screens the potential nonuniformities arising from the surface charge density fluctuations. Similarly, we break down the electrostatic potential into three terms,
\begin{equation}
    \phi(\mathbf{r}) = \phi_o(x,y) + \phi_{imp}^\prime(\mathbf{r}) + \phi_{red}(\mathbf{r}).
\end{equation}
Each of these components satisfy a Poisson equation of the form,
\begin{equation}
    \nabla \cdot \left[\epsilon(\mathbf{r}) \nabla\phi_i(\mathbf{r})\right] =
    	  -\rho_i(\mathbf{r}), \label{PoisComp}
\end{equation}
where the electrostatic potential and charge density pair satisfy $(\phi_i,\rho_i)\in \{(\phi_o,\rho_o), ~(\phi_{imp}^\prime,\rho_{imp}^\prime), ~(\phi_{red},\rho_{red}) \}$. The Dirichlet boundary conditions for nonzero values of $V_{BG}$, $V_L$, $V_R$ and $V_{SC}$ are imposed on $\phi_o$, while  $\phi_{imp}^\prime$ and $\phi_{red}$ are subject to trivial boundary conditions.

With these aspects of the model discussed, we refer readers to the section entitled ``Self-consistent Schr{\"o}dinger-Poisson scheme'' within our work in Ref.~\onlinecite{woods2021charge} for remaining details regarding the self-consistent solution to the Schr{\"o}dinger-Poisson equations. One can follow the method in Ref.~\onlinecite{woods2021charge} exactly, except $\rho_{imp} \rightarrow \rho_{imp}^\prime$, $\phi_{imp} \rightarrow \phi_{imp}^\prime$, and $\rho_o$ is now the total charge density in the absence of impurity charge density \textit{fluctuations} instead of the free charge density in the absence of all impurities. The final product from this calculation is an effective disorder potential arising from the self-consistent \textit{screened} impurity potential, which is given by
\begin{equation}
    V_{eff}^{\alpha \beta}(z) = 
    \int \varphi_{\alpha}^*(x,y)
    \left(
    \phi_{imp}^\prime(\mathbf{r}) 
    + \phi_{red}(\mathbf{r}) 
    \right)
    \varphi_{\beta}(x,y) \, dxdy, \label{EffPot}
\end{equation}
where $\varphi_{\alpha}$ is the normalized transverse orbital of the $\alpha$ subband satisfying,
\begin{equation}
    \left[
    -\frac{\hbar^2}{2 m^*} 
	\nabla^2_\perp 
	- e \phi_o\left(\mathbf{r}\right)
	\right] \varphi_{\alpha}(x,y) = \varepsilon_{\alpha,o} \varphi_{\alpha}(x,y),
\end{equation}
with $\nabla_\perp^2 = \partial_x^2 + \partial_y^2$. In other words, the clean system without any surface charge density fluctuations allows us to define subbands, where the $\alpha$ subband has a band-edge energy and transverse orbital given by $\varepsilon_{\alpha,o}$ and $\varphi_{\alpha}$, respectively. Note that the effective potential in Eq. (\ref{EffPot}) has both intra-subband coupling ($\alpha = \beta$) and inter-subband coupling ($\alpha \neq \beta$). In the main text, we only include intra-subband coupling in our 1D finite wire calculations, which acts essentially as a disorder potential for the subband near the Fermi level. Including inter-subband couplings only makes the system more disordered \cite{woods2019zeroenergy,pan2020generic}. We are therefore being optimistic with regard to the fate of Majorana physics by neglecting these couplings.

\end{document}